\documentclass[article]{JHEP3}

\usepackage{amsmath,amssymb}
\usepackage[dvips]{graphics}
\usepackage{epsfig}
\usepackage{psfrag}

\newcommand{\re}{{\mathbb{R}}{\mathrm{e}}}
\newcommand{\im}{{\mathbb{I}}{\mathrm{m}}}

\newcommand{\be}{\begin{equation}}
\newcommand{\ee}{\end{equation}}
\newcommand{\bea}{\begin{eqnarray}}
\newcommand{\eea}{\end{eqnarray}}
\newcommand{\bean}{\begin{eqnarray*}}
\newcommand{\eean}{\end{eqnarray*}}

\def\beq{\begin{equation}}

\def\eeq{\end{equation}}

\relax

\def\MM{\boldsymbol{\mu}}
\def\QQ{\boldsymbol{\theta}}
\def\LL{\boldsymbol{\lambda}}

\preprint{{\small \texttt{hep-th/0411267}}}

\title{On the Classification of Asymptotic Quasinormal Frequencies for $d$--Dimensional Black Holes and Quantum Gravity}

\author{Jos\'e Nat\'ario$^{a}$ and Ricardo Schiappa$^{a,b}$
\\
$^{a}$CAMGSD, Departamento de Matem\'atica, Instituto Superior T\'ecnico,\\ 
Av. Rovisco Pais 1, 1049--001 Lisboa, Portugal\\
\\
$^{b}$Faculdade de Engenharia, Universidade Cat\'olica 
Portuguesa,\\
Estrada de Tala\'\i de, 2635--631 Rio de Mouro, Lisboa, Portugal\\
\\
\email{jnatar@math.ist.utl.pt}, \quad 
\email{schiappa@math.ist.utl.pt}
}

\abstract{
We provide a complete classification of asymptotic quasinormal frequencies for static, spherically symmetric black hole spacetimes in $d$ dimensions. This includes all possible types of gravitational perturbations (tensor, vector and scalar type) as described by the Ishibashi--Kodama master equations. The frequencies for Schwarzschild are dimension independent, while for Reissner--Nordstr\"om are dimension dependent (the extremal Reissner--Nordstr\"om case must be considered separately from the non--extremal case). For Schwarzschild de Sitter, there is a dimension independent formula for the frequencies, except in dimension $d=5$ where the formula is different. For Reissner--Nordstr\"om de Sitter there is a dimension dependent formula for the frequencies, except in dimension $d=5$ where the formula is different. Schwarzschild and Reissner--Nordstr\"om Anti--de Sitter black hole spacetimes are simpler: the formulae for the frequencies will depend upon a parameter related to the tortoise coordinate at spatial infinity, and scalar type perturbations in dimension $d=5$ lead to a continuous spectrum for the quasinormal frequencies. We also address non--black hole spacetimes, such as pure de Sitter spacetime---where there are quasinormal modes only in odd dimensions---and pure Anti--de Sitter spacetime---where again scalar type perturbations in dimension $d=5$ lead to a continuous spectrum for the normal frequencies. Our results match previous numerical calculations with great accuracy. Asymptotic quasinormal frequencies have also been applied in the framework of quantum gravity for black holes. Our results show that it is only in the simple Schwarzschild case which is possible to obtain sensible results concerning area quantization or loop quantum gravity. In an effort to keep this paper self--contained we also review earlier results in the literature.
}


\begin{document}



\vfill

\eject


\section{Introduction}


Black holes are an undeniable landmark in the road that connects classical to quantum gravity. Having been first discovered as static solutions of classical general relativity, they were later shown to actually radiate and evaporate once quantum effects were properly taken into account \cite{bch, hawking}. Thus, it seems that in a true quantum theory of gravity, non--extremal black holes will actually be unstable. In this setting, an important question that immediately comes to mind is whether the black hole solution under consideration was really a stable solution of the \textit{classical} theory, to start off with.

Researchers first focused on analyzing the linear stability of four dimensional black hole solutions of general relativity in \cite{regge-wheeler, zerilli-1, zerilli-2}. The linear perturbation theory for the Schwarzschild black hole was set up in \cite{regge-wheeler}, where the classical stability of the solution was also proven. The equation derived in that paper to describe the linear perturbations, and their frequencies, is a Schr\"odinger--like equation and is now known as the Regge--Wheeler (RW) equation. The procedure to derive the RW equation is based on a study of the linearized Einstein equations in the given background and proceeds with a decomposition of the perturbation in tensor spherical harmonics (for spherically symmetric backgrounds) in order to obtain a radial equation describing the propagation of linear perturbations. This procedure, as applied to the analysis of perturbations to the Schwarzschild metric, was brought to firmer grounds in \cite{zerilli-1} with an explicit construction of four dimensional tensor spherical harmonics. This work was further completed in \cite{zerilli-2} with an extension to the Reissner--Nordstr\"om (RN) black hole solution. There, it was shown that the complete Einstein--Maxwell system for each type of multipole (electric or magnetic) could be reduced to two second order Schr\"odinger--like equations, generalizing the RW equation of the Schwarzschild case. 

Part of the physical picture that emerged from this study of linear perturbations to black holes is the following. After the onset of a perturbation, the return to equilibrium of a black hole spacetime is dominated by damped, single frequency oscillations, which are known as the quasinormal modes (we refer the reader to \cite{nollert, kokkotas-schmidt} for recent reviews on quasinormal modes, and a more complete list of references). These modes are quite special: they depend \textit{only} on the parameters of the given black hole spacetime, being \textit{independent} of the details concerning the initial perturbation we started off with. Moreover, modes which damp infinitely fast do \textit{not} radiate at all, and can thus be interpreted as some sort of fundamental oscillations for the black hole spacetime. We shall return to this point in a moment.

It was not until recently that the black hole stability problem was addressed within a $d$--dimensional setting \cite{kodama-ishibashi-1, kodama-ishibashi-2, kodama-ishibashi-3}. These papers tried to be as exhaustive as possible, studying in detail the perturbation theory of static, spherically symmetric black holes in any spacetime dimension $d>3$ and allowing for the possibilities of both electromagnetic charge and a background cosmological constant. The set of equations describing linear perturbations in $d$--dimensions was derived in \cite{kodama-ishibashi-1, kodama-ishibashi-3}. These equations generalize the RW equation and will be denoted as the Ishibashi--Kodama (IK) master equations. The perturbations come in three types: tensor type perturbations, vector type perturbations and scalar type perturbations. It should be noted that this nomenclature refers to the tensorial behavior on the sphere, ${\mathbb{S}}^{d-2}$, of each Einstein--Maxwell gauge invariant type of perturbation, and is \textit{not} related to perturbations associated to external particles. For instance, one should not confuse vector type perturbations with perturbations associated to the propagation of a spin--1 vector particle in the background spacetime, or scalar type perturbations with perturbations associated to the propagation of a spin--0 scalar particle. The IK master equations were used in \cite{kodama-ishibashi-2, kodama-ishibashi-3} to study the stability of $d$--dimensional black holes, and although many known solutions were shown to be stable, stability of some other solutions is still an open problem. In this work we shall make use of the IK master equations to analytically compute asymptotic quasinormal frequencies, and thus focus on the $d$--dimensional Einstein--Maxwell classical theory of gravity.

Having thus acquired a list of stable black hole solutions, the next question to address within this general problem are the quasinormal modes---the damped oscillations which describe the return to the initial configuration. As we have said, modes which damp infinitely fast do not radiate, and they are known as asymptotic quasinormal modes. Besides their natural role in the perturbation theory of general relativity, asymptotic quasinormal modes have recently been focus of much attention following suggestions that they could have a role to play in the quest for a theory of quantum gravity \cite{hod, dreyer}. It was suggested in \cite{hod} that an application of Bohr's correspondence principle to the asymptotic quasinormal frequencies could yield new information about quantum gravity, in particular on the quantization of area at a black hole event horizon. It was further suggested in \cite{dreyer} that asymptotic quasinormal frequencies could help fix certain parameters in loop quantum gravity. Both these suggestions lie deeply on the fact that the real part of the asymptotic quasinormal frequencies is given by the logarithm of an integer number, a fact that was analytically shown to be true, for Schwarzschild black holes in $d$--dimensional spacetime, in \cite{motl, motl-neitzke}. A question of particular relevance that immediately follows is whether the suggestions in \cite{hod, dreyer} are universal or are only applicable to the Schwarzschild solution. Given the mentioned analysis of \cite{kodama-ishibashi-1, kodama-ishibashi-3}, one has at hand all the required information to address this problem and compute asymptotic quasinormal frequencies for $d$--dimensional black holes\footnote{The least damped quasinormal modes associated to the IK master equations were addressed numerically in \cite{konoplya-2}.}. A preliminary clue is already present in \cite{motl-neitzke}, where the analysis of the four dimensional RN solution yielded a negative answer: the asymptotic quasinormal frequencies obeyed a complicated relation which did not seem to have the required form. Another clue was presented in \cite{cns}, where the analysis of the four dimensional Schwarzschild de Sitter (dS) and Schwarzschild Anti--de Sitter (AdS) black holes also yielded a negative answer; again the asymptotic quasinormal frequencies did not seem to have the required form. It is the goal of this paper to carry out an extension of the techniques in \cite{motl-neitzke, cns} to static, spherically symmetric black hole spacetimes in any dimension $d>3$, including both electromagnetic charge and a background cosmological constant. Besides the intrinsic general relativistic interest of classifying these asymptotic quasinormal frequencies, it is also hoped that our results can yield conclusive implications for the proposals of \cite{hod, dreyer}, dealing with the application of quasinormal modes to quantum gravity. We shall later see that it is only in the simple Schwarzschild case which is possible to obtain sensible results concerning area quantization or loop quantum gravity.

It is important to stress that even if the ideas in \cite{hod, dreyer} turn out not to be universal, it is still the case that quasinormal frequencies will most likely have some role to play in the quest for a theory of quantum gravity. Indeed, quasinormal frequencies can also be regarded as the poles in the black hole greybody factors which play a pivotal role in the study of Hawking radiation. Furthermore, the monodromy technique introduced in \cite{motl-neitzke} to analytically compute asymptotic quasinormal frequencies was later extended, in \cite{neitzke}, so that it can also be used in the computation of asymptotic greybody factors. It was first suggested in \cite{neitzke} that the results obtained for these asymptotic greybody factors could be of help in identifying the dual conformal field theory (CFT) which microscopically describes the black hole, and these ideas have been taken one step forward with the work of \cite{krasnov-solodukhin}. It remains to be seen how much asymptotic quasinormal modes and greybody factors can help in understanding quantum gravity.

The organization of this paper is as follows. In section 2 we provide a summary of our results for the reader who wishes to skip the technical details on a first reading. Section 3 represents the main body of the paper, where we use both the analytic continuation of the IK master equations to the complex plane and a method of monodromy matching at the several singularities in the plane, in order to analytically compute asymptotic quasinormal frequencies for static, spherically symmetric black hole spacetimes in dimension $d>3$. This includes a brief review of quasinormal modes, as well as the Schwarzschild case of \cite{motl-neitzke}, for both completeness and pedagogical purposes. In section 4 we address some exact solutions for quasinormal frequencies, dealing with non--black hole spacetimes. We address Rindler, dS and AdS spacetimes, providing full solutions in all cases\footnote{Other exact calculations of quasinormal frequencies can be found in, \textit{e.g.}, \cite{cardoso-lemos-extra, birmingham-extra}.}. It should be noted that there has been some confusion in the literature concerning the computation of quasinormal modes using the monodromy technique, as well as on the computation of quasinormal modes in pure dS spacetime. We hope that this paper will serve to lay these confusions to rest. Section 5 reviews \cite{hod,dreyer} and the applications of asymptotic quasinormal frequencies to area quantization and loop quantum gravity. We show that these applications only seem to work in the simple Schwarzschild setting. In section 6 we conclude, listing some future directions of research. We also include three appendices. In appendix A we present a thorough list of conventions for black holes with mass $M$, charge $Q$ and background cosmological constant $\Lambda$, in $d$--dimensional spacetime. Appendix B includes all required formulae for the IK master equations, and appendix C makes a complete study of the tortoise coordinate in the spacetimes in consideration, alongside with analysis of the IK master equation potentials at several singularities in the complex plane. Throughout we will show that our results match earlier numerical calculations with great accuracy and will review some of the earlier literature on each case considered, as an effort to make this paper self--contained.


\section{Summary of Results}


For the reader who wishes to skip the main calculation on a first reading, we present in the following a summary of our results for the asymptotic quasinormal frequencies, in any spacetime dimension $d>3$. In the appendices we review the black hole spacetime solutions we wish to consider, as well as the perturbation theory for these spacetimes which leads to the quasinormal mode analysis. The set of equations describing quasinormal modes in $d$--dimensions was derived in \cite{kodama-ishibashi-1, kodama-ishibashi-3}, and the perturbations come in three types: tensor type perturbations, vector type perturbations and scalar type perturbations. For each spacetime in consideration we compute asymptotic quasinormal frequencies, given each type of $d$--dimensional perturbation. In the case of black hole spacetimes the computation involves a detailed monodromy analysis, alongside with some simple differential equations. In fact, the IK master equation describing quasinormal modes is of Schr\"odinger type, where the potential is associated to the perturbation under study. When using the monodromy technique one needs to solve the Schr\"odinger--like master equation at points where the potential is usually either zero (yielding simple plane wave solutions) or of Bessel type

\begin{equation}\label{bessel}
- \frac{ d^{2} \Phi}{dx^{2}} (x) + \frac{j^{2}-1}{4 x^{2}} \Phi (x) = \omega^{2} \Phi (x),
\end{equation}

\noindent
where $j$ is determined for each different case (for each type of perturbation and for each background considered). This equation can be solved in terms of Bessel functions, $J_{\nu}(x)$, with the result

\begin{equation}\label{bessel-solution}
\Phi (x) = A_{+}\, \sqrt{2\pi \omega x}\, J_{\frac{j}{2}} \left( \omega x \right) + A_{-}\, \sqrt{2\pi \omega x}\, J_{-\frac{j}{2}} \left( \omega x \right),
\end{equation}

\noindent
where $A_{+}$ and $A_{-}$ are constants. Let us list the values of $j$ which one has to deal with, when considering the black hole singularity region, at the origin of the coordinate frame. For all \textit{uncharged} black hole solutions one finds $j_{\mathsf{T}} = 0$ for tensor type perturbations, $j_{\mathsf{V}} = 2$ for vector type perturbations and $j_{\mathsf{S}} = 0$ for scalar type perturbations. For all \textit{charged} black hole solutions one finds $j_{\mathsf{T}} = \frac{d-3}{2d-5}$ for tensor type perturbations, $j_{\mathsf{V}^{+}} = j_{\mathsf{V}^{-}} = \frac{3d-7}{2d-5}$ for vector type perturbations and $j_{\mathsf{S}^{+}} = j_{\mathsf{S}^{-}}= \frac{d-3}{2d-5}$ for scalar type perturbations. It is furthermore simple to observe that for \textit{all} black hole spacetimes $j_{\mathsf{V}^{+}} = j_{\mathsf{V}^{-}} \equiv j_{\mathsf{V}}$, $j_{\mathsf{S}^{+}} = j_{\mathsf{S}^{-}} \equiv j_{\mathsf{S}}$, $j_{\mathsf{T}} = j_{\mathsf{S}}$, and $j_{\mathsf{T}} + j_{\mathsf{V}} = 2$. This will ultimately imply that the asymptotic quasinormal frequencies will depend \textit{only} on the background, being the same for the three types of perturbations. Let us also recall that one can prove (for spacetimes which are not asymptotically AdS), without explicit computation of the quasinormal frequencies, that $j=0$ and $j=2$ perturbations must have identical quasinormal spectra \cite{chandra}.

For spacetimes without a black hole (but with cosmological constant) the calculation of quasinormal frequencies follows in an analytic way, without any asymptotic restrictions. Here one need not use the monodromy method (even though it can be simply generalized to those cases as well) and we have proceeded by solving the wave equation directly. This is accomplished by first finding an appropriate change of variables that brings the quasinormal master equation to a hypergeometric form. Let us finally list our results on quasinormal frequencies:

\medskip

\noindent
\underline{\textsf{The Schwarzschild Solution:}} This case was first studied in \cite{motl, motl-neitzke} and we address it in this paper for the sake of completeness. For all types of perturbations, tensor, vector and scalar type perturbations, the algebraic equation for the asymptotic quasinormal frequencies is the same and is

$$
e^{\frac{\omega}{T_{H}}} + 3 = 0,
$$

\noindent
where $T_{H}$ is the Hawking temperature in the Schwarzschild spacetime. As is well known, this case is particularly simple and one can moreover solve for the asymptotic quasinormal frequency as

$$
\omega = T_{H} \log 3 + 2 \pi i T_{H} \left( n + \frac{1}{2} \right) \qquad \left( n \in {\mathbb{N}}, \quad n \gg 1 \right).
$$

\noindent
The result is independent of spacetime dimension\footnote{By ``independent'' we mean that there is no explicit dependence on the spacetime dimension, $d$, in the above formula. Of course if one wishes to compute the Hawking temperature in this Schwarzschild background, then one finds that it relates to the mass $M$ via an expression which also involves the dimension $d$.}.

\medskip

\noindent
\underline{\textsf{The RN Solution:}} This case was studied, in the particular $d=4$ case, in \cite{motl-neitzke}. Here we extend those results to arbitrary dimension. For all types of perturbations, tensor, vector and scalar perturbations, the algebraic equation for the asymptotic quasinormal frequencies is the same and is

$$
e^{\frac{\omega}{T_{H}^{+}}} + \Big( 1 + 2 \cos \left( \pi j \right) \Big) + \Big( 2 + 2 \cos \left( \pi j \right) \Big) e^{-\frac{\omega}{T_{H}^{-}}} = 0,
$$

\noindent
where $T_{H}^{\pm}$ are the Hawking temperatures at outer and inner horizons (notice that $T_{H}^{-} < 0$), and where

$$
j = \frac{d-3}{2d-5}.
$$

\noindent
There is no known algebraic solution in $\omega$ for the above equation.

\medskip

\noindent
\underline{\textsf{The Extremal RN Solution:}} It is important to realize that, in general, quasinormal frequencies of extremal solutions \textit{cannot} be obtained from the corresponding expression for the non--extremal solution. In fact, the monodromy technique deployed in this paper is very sensitive to both the location of the complex horizons and the structure of the tortoise at the origin. Thus, as one changes the background solution there will be a change of topology in the complex plane and the solution to the quasinormal mode problem will be different. We present the extremal RN solution as an example, but one should keep this in mind if also interested in extremal solutions with a cosmological constant (which we do not address in this paper, but list the possibilities in appendix A). For all types of perturbations, tensor, vector and scalar perturbations, the algebraic equation for the asymptotic quasinormal frequencies is the same and is

$$
\sin \left( \frac{\pi j}{2} \right) e^{\frac{\omega}{T}} - \sin \left( \frac{5\pi j}{2} \right) = 0,
$$

\noindent
where $j$ is as in the previous RN case and $T$ is \textit{not} a temperature---in fact in the extremal case there is no Hawking emission. Rather, it is given by

$$
T = \frac{d-3}{d-2}\ \left( \frac{d-3}{4\pi\MM^{\frac{1}{d-3}}} \right),
$$

\noindent
where $\MM$ is related to the black hole mass (see appendix A). It is simple to solve for the asymptotic quasinormal frequency as

$$
\omega = T \log \left( \frac{\sin \left( \frac{5\pi j}{2} \right)}{\sin \left( \frac{\pi j}{2} \right)} \right) + 2\pi i n T \qquad \left( n \in {\mathbb{N}}, \quad n \gg 1 \right).
$$

\noindent
Observe that in dimension $d=5$ there is no solution for $\omega$. This is in fact the only dimension where there is no solution for the asymptotic quasinormal frequencies of the extremal RN geometry.

\medskip

\noindent
\underline{\textsf{The Schwarzschild dS Solution:}} This case was studied, in the particular $d=4$ case, in \cite{cns}. Here we extend those results to arbitrary dimension. For all types of perturbations, tensor, vector and scalar perturbations, the algebraic equation for the asymptotic quasinormal frequencies is the same and is

$$
\cosh \left( \frac{\omega}{2T_{H}} - \frac{\omega}{2T_{C}} \right) + 3 \cosh \left( \frac{\omega}{2T_{H}} + \frac{\omega}{2T_{C}} \right) = 0,
$$

\noindent
where $T_{H}$ is the Hawking temperature at the black hole event horizon and $T_{C}$ is the [negative] Hawking temperature at the cosmological horizon. There is no known algebraic solution in $\omega$ for the above equation. The result is independent of spacetime dimension, except in dimension $d=5$ where the formula above must be replaced by:

$$
\sinh \left( \frac{\omega}{2T_{H}} - \frac{\omega}{2T_{C}} \right) - 3 \sinh \left( \frac{\omega}{2T_{H}} + \frac{\omega}{2T_{C}} \right) = 0.
$$

\noindent
Observe that for this solution one can actually take the Schwarzschild limit $\LL \to 0$ without provoking any topology change in the complex plane where the monodromy analysis is performed. Thus, the result obtained for Schwarzschild dS includes the pure Schwarzschild solution once one sets the cosmological constant to vanish (both expressions have the same, correct, Schwarzschild limit).

\medskip

\noindent
\underline{\textsf{The RN dS Solution:}} For all types of perturbations, tensor, vector and scalar perturbations, the algebraic equation for the asymptotic quasinormal frequencies is the same and is

$$
\cosh \left( \frac{\omega}{2T_{H}^{+}} - \frac{\omega}{2T_{C}} \right) + \left( 1 + 2 \cos \left( \pi j \right) \right) \cosh \left( \frac{\omega}{2T_{H}^{+}} + \frac{\omega}{2T_{C}} \right) + \left( 2 + 2 \cos \left( \pi j \right) \right) \cosh \left( \frac{\omega}{T_{H}^{-}} + \frac{\omega}{2T_{H}^{+}} + \frac{\omega}{2T_{C}} \right) = 0,
$$

\noindent
where $j$ is as in the previous RN case, $T_{H}^{\pm}$ are the Hawking temperatures at outer and inner black hole event horizons and $T_{C}$ is the Hawking temperature at the cosmological horizon. There is no known algebraic solution in $\omega$ for the above equation. While this result explicitly depends on the spacetime dimension (because $j$ depends on $d$), the formula is not valid in dimension $d=5$, where it must be replaced by:

$$
\sinh \left( \frac{\omega}{2T_{C}} - \frac{\omega}{2T_{H}^{+}} \right) + \frac{1+\sqrt{5}}{2} \sinh \left( \frac{\omega}{2T_{H}^{+}} + \frac{\omega}{2T_{C}} \right) + \frac{3+\sqrt{5}}{2} \sinh \left( \frac{\omega}{T_{H}^{-}} + \frac{\omega}{2T_{H}^{+}} + \frac{\omega}{2T_{C}} \right) = 0.
$$

\noindent
Similarly to the Schwarzschild dS case, one can take the pure RN limit $\LL \to 0$ without provoking any topology change in the complex plane where the monodromy analysis is performed. Thus, the result obtained for RN dS includes the pure RN solution once one sets the cosmological constant to vanish (both expressions have the same, correct, RN limit).

\medskip

\noindent
\underline{\textsf{The Schwarzschild AdS Solution:}} This case was studied, in the particular case of $d=4$ and large black hole, in \cite{cns}. It was further studied, in the particular case of $d=5$ and large black hole, in \cite{starinets, nunez-starinets, fhks, musiri-siopsis} using a variety of different analytical methods. Here we extend those results for arbitrary dimension and away from the large black hole approximation. For all types of perturbations, tensor, vector and scalar perturbations, the algebraic equation for the asymptotic quasinormal frequencies is the same and is

$$
\tan \left( \frac{\pi}{4} \left( d + 1 \right) - \omega x_{0} \right) = \frac{i}{3}, 
$$

\noindent
where $x_{0}$ is a parameter related to the tortoise coordinate at spatial infinity and is given by

$$
x_{0} = \sum_{n=1}^{d-1} \frac{1}{2k_{n}} \log \left( - \frac{1}{R_{n}} \right),
$$

\noindent
where the $R_{n}$ are the $(d-1)$ complex horizons and the $k_{n}$ the surface gravities at each of these complex horizons. There is no general analytic solution for $x_{0}$. However, in some cases, it can be computed exactly. For instance, for \textit{large} black holes we compute it to be

$$
\frac{1}{x_{0}} = 4 T_{H} \sin \left( \frac{\pi}{d-1} \right) \exp \left( \frac{i\pi}{d-1} \right)
$$

\noindent
where $T_{H}$ is the Hawking temperature in the Schwarzschild AdS spacetime. In spite of not having a general analytic solution for $x_{0}$ one can still solve for the asymptotic quasinormal frequency as

$$
\omega x_{0} = \frac{\pi}{4} \left( d + 1 \right) - \arctan \left( \frac{i}{3} \right) + n \pi \qquad \left( n \in {\mathbb{N}}, \quad n \gg 1 \right).
$$

\noindent
If one concentrates on large Schwarzschild AdS black holes, of particular relevance  to describe thermal gauge theories within the AdS/CFT framework, then the formulae above lead to the following analytical result for the leading term, as $n \to + \infty$, in the asymptotic quasinormal frequencies:

$$
\omega = 4 \pi n T_{H} \sin \left( \frac{\pi}{d-1} \right) \exp \left( \frac{i\pi}{d-1} \right) + \cdots .
$$

\noindent
For the most popular AdS/CFT dimensions, $d=4$, $d=5$ and $d=7$, one obtains

$$
\omega_{d=4} \sim 2 \sqrt{3} \pi n T_{H} e^{\frac{i\pi}{3}}, \qquad \omega_{d=5} \sim 2 \sqrt{2} \pi n T_{H} e^{\frac{i\pi}{4}} \qquad {\mathrm{and}} \qquad \omega_{d=7} \sim 2 \pi n T_{H} e^{\frac{i\pi}{6}}.
$$

\noindent
Notice that the above formulae are \textit{not} valid for \textit{scalar} type perturbations in dimensions four and five. Instead, one finds for these perturbations in dimension $d=4$

$$
\omega x_{0} = \frac{3\pi}{4} - \arctan \left( \frac{i}{3} \right) + n \pi \qquad \left( n \in {\mathbb{N}}, \quad n \gg 1 \right),
$$

\noindent
and more surprisingly, a continuous spectrum in five dimensions,

$$
\omega \in {\mathbb{C}}.
$$

\medskip

\noindent
\underline{\textsf{The RN AdS Solution:}} For all types of perturbations, tensor, vector and scalar perturbations, the algebraic equation for the asymptotic quasinormal frequencies is the same and is

$$
\sin \left( \pi j \right) e^{i \left( \frac{\pi}{4} \left( d-1 \right) - \omega x_{0} \right)} + \sin \left( \frac{\pi j}{2} \right) e^{- i \left( \frac{\pi}{4} \left( d-1 \right) - \omega x_{0} \right)} = 0,
$$

\noindent
where $x_{0}$ is as before (only one should recall that this time around there are $(2d-4)$ complex horizons as we are in a charged situation) and $j$ is as in the previous RN cases. Again, there is no general analytic solution for $x_{0}$. In spite of this, one can still solve for the asymptotic quasinormal frequency as

$$
\omega x_{0} = \frac{\pi}{4} \left( d + 1 \right) - \frac{i}{2} \log \left( 2 \cos \left( \frac{\pi j}{2} \right) \right) + n \pi \qquad \left( n \in {\mathbb{N}}, \quad n \gg 1 \right).
$$

\noindent
Notice that the above formula is \textit{not} valid for \textit{scalar} type perturbations in dimensions four and five. Instead, one finds for these perturbations in dimension $d=4$

$$
\omega x_{0} = \frac{3\pi}{4} - \frac{i}{2} \log \left( 2 \cos \left( \frac{\pi j}{2} \right) \right) + n \pi \qquad \left( n \in {\mathbb{N}}, \quad n \gg 1 \right),
$$

\noindent
and more surprisingly, a continuous spectrum in five dimensions,

$$
\omega \in {\mathbb{C}}.
$$

\medskip

Besides the previous results, concerning black hole spacetimes, we have also addressed the case of spacetimes without a black hole but with a cosmological constant, \textit{i.e.}, the cases of AdS (where there are only \textit{normal} modes) and of dS (where one finds quasinormal modes only in odd spacetime dimensions). Let us list those results as well:

\medskip

\noindent
\underline{\textsf{The AdS Solution:}} This case was studied before in \cite{burgess-lutken} for a massless scalar field, and in \cite{ckl} for $d=4$ tensor and vector type perturbations. Here we extend those results to arbitrary dimension and perturbation. The first thing to notice is that AdS spacetime acts as an infinite potential well and the Schr\"odinger--like equation yields real frequencies only. In other words, there are no quasinormal modes in pure AdS, only \textit{normal} modes. The normal frequencies one finds are (no asymptotic restrictions here, this is an exact result)

$$
\omega_{n} = \sqrt{|\LL|} \left( 2n + d + \ell - j \right), \qquad n \in {\mathbb{N}},
$$

\noindent
where $\ell$ is the angular momentum quantum number (eigenvalue of the spherical laplacian) and $j$ is $j=1$ for tensor type perturbations, $j=2$ for vector type perturbations and $j=3$ for scalar type perturbations. Thus, for fixed $\ell$, tensor and scalar type perturbations yield the same normal frequencies, although different from the vector type frequencies. Notice that the above formula is \textit{not} valid for \textit{scalar} type perturbations in dimensions four and five. Instead, one finds for these perturbations in dimension $d=4$

$$
\omega_{n} = \sqrt{|\LL|} \left( 2n + \ell + 2 \right), \qquad n \in {\mathbb{N}},
$$

\noindent
and more surprisingly, a continuous spectrum in five dimensions,

$$
\omega \in {\mathbb{R}}.
$$

\medskip

\noindent
\underline{\textsf{The dS Solution:}} This case was studied in several different works, discussed in the main text, all with contradictory results. Here, we solve this apparent confusion in the literature with the following results. There are \textit{no} dS quasinormal modes when the spacetime dimension is \textit{even}. However, when the spacetime dimension is \textit{odd}, there are quasinormal modes, with the quasinormal frequencies (no asymptotic restrictions here, this is an exact result)

$$
\omega_{n} = i \sqrt{\LL} \left( 2n + \ell + \frac{j-1}{2} \right), \qquad n \in {\mathbb{N}},
$$

\noindent
where $\ell$ is the angular momentum quantum number (eigenvalue of the spherical laplacian) and $j$ is $j=1$ for tensor type perturbations, $j=3$ for vector type perturbations and $j=5$ for scalar type perturbations. Thus, for fixed $\ell$, tensor and scalar type perturbations yield the same quasinormal frequencies, although different from the vector type frequencies. The above formula is independent of spacetime (odd) dimension.

\medskip

Finally, we have studied the Rindler solution where we found---without great surprise---that there are no quasinormal modes. A detailed discussion concerning applications of these results to quantum gravity is included in the text, and we shall make no attempt of summarizing it in here.


\section{Asymptotic Quasinormal Frequencies}


In this section we first wish to review and set notation on both the perturbation theory for spherically symmetric, static $d$--dimensional black holes\footnote{Throughout this work we consider only dimension $d>3$.}, with mass $M$, charge $Q$ and background cosmological constant $\Lambda$, and the computation of quasinormal modes and quasinormal frequencies. We refer the reader to appendix A for a full list of conventions on the black hole spacetimes we shall consider. In the general $(M,Q,\Lambda)$ case, there are two different types of fields which can be excited: these are the electromagnetic vector field $A_{\mu}$ and the gravitational metric tensor field $g_{\mu\nu}$, and we shall study perturbations to both these fields. Because there is no scalar field present in the Einstein--Maxwell system, there are no scalar field perturbations to consider. Nevertheless, we shall start by studying the scalar wave equation in our black hole backgrounds, in order to set notation on quasinormal modes.

Consider a massless, uncharged, scalar field, $\phi$, in a background spacetime described by a metric $g_{\mu\nu}$. Its wave equation is well known

\begin{equation} \label{wave}
\frac{1}{\sqrt{-g}} \partial_{\mu} \left( \sqrt{-g} g^{\mu\nu} \partial_{\nu} \phi \right) = 0,
\end{equation}

\noindent
where $g$ is the determinant of $g_{\mu\nu}$. The question we wish to address is what is the form of the wave equation for a background spacetime metric of the type

$$
g = - f(r)\ dt \otimes dt + {f(r)}^{-1}\ dr \otimes dr + r^{2} d \Omega_{d-2}^{2},
$$

\noindent
\textit{i.e.}, in a $d$--dimensional spherically symmetric background. This issue was addressed in \cite{cdl}, with the following result. First perform a harmonic decomposition of the scalar field as $\phi = \sum_{\ell,m} r^{\frac{2-d}{2}}\ \psi_{\ell} (r,t)\ Y_{\ell m} \left( \theta_{i} \right)$, where the $\theta_{i}$ are the $(d-2)$ angles and the $Y_{\ell m} ( \theta_{i} )$ are the $d$--dimensional spherical harmonics. Then, if 

$$
\psi_{\ell} (r,t) = \Phi_{\omega} (r) e^{i \omega t}
$$ 

\noindent
is the [time] Fourier decomposition of the scalar field, the wave equation can be recast in a Schr\"odinger--like form as

\begin{equation} \label{schrodinger}
- \frac{ d^{2} \Phi_{\omega}}{dx^{2}} (x) + V (x) \Phi_{\omega} (x) = \omega^{2} \Phi_{\omega} (x),
\end{equation}

\noindent
where $x$ is the so--called tortoise coordinate and $V(x)$ is the potential, both determined from the function $f(r)$ in the background metric. The tortoise coordinate is defined so that \cite{nollert, kokkotas-schmidt}

$$
f(r) \frac{d}{dr} \left( f(r) \frac{d}{dr} \right) = \frac{d^{2}}{dx^{2}} 
$$

\noindent
and is thus given by

\begin{equation} \label{tortoise}
dx = \frac{dr}{f(r)}.
\end{equation}

\noindent
This new coordinate $x$ keeps infinity (or the cosmological horizon, $R_{C}$, in the dS case) at $x = + \infty$ and sends the black hole event horizon, $R_{H}$, to $x = - \infty$ (in the charged cases this refers to the outer horizon). The region of positivity of $f(r)$ thus becomes the real line when in tortoise coordinates: $f (x) > 0$ for $x \in {\mathbb{R}}$. The potential in the Schr\"odinger--like equation  can be determined as one moves from the general form of the wave equation, (\ref{wave}), to its Schr\"odinger--like form, (\ref{schrodinger}), and is given by \cite{cdl}

$$
V(r) = f(r) \left( \frac{\ell \left( \ell + d - 3 \right)}{r^{2}} + \frac{\left( d-2 \right) \left( d-4 \right) f(r)}{4r^{2}} + \frac{\left( d-2 \right) f'(r)}{2r} \right).
$$

\noindent
Here, $\ell \left( \ell + d - 3 \right)$ (with $\ell \in {\mathbb{N}}$) are the eigenvalues of the Laplacian  on the ${\mathbb{S}}^{d-2}$ sphere, and the potential still needs to be re--written in terms of the tortoise coordinate in order to be used in the Schr\"odinger--like equation. Once all this is done, the question still remains on what are the allowed values for $\omega$, \textit{i.e.}, what is the spectrum of the Schr\"odinger--like operator above. It turns out that the spectrum will contain both a continuous and a discrete part, this last one being found when imposing ``out going'' boundary conditions: nothing arrives neither from infinity nor from within the black hole horizon. The [spherical] waves are out going at both extrema in $x$:

\begin{eqnarray*}
\Phi_{\omega} (x) &\sim& e^{i\omega x}\;\, {\mathrm{as}}\;\, x \to - \infty, \\
\Phi_{\omega} (x) &\sim& e^{-i\omega x}\;\, {\mathrm{as}}\;\, x \to + \infty.
\end{eqnarray*}

\noindent
Solutions to the Schr\"odinger--like equation with the previous ``out going'' boundary conditions lead to a discrete set of allowed frequencies, the quasinormal frequencies, with corresponding solutions, the quasinormal modes \cite{nollert, kokkotas-schmidt}. These quasinormal frequencies, $\omega$, are complex numbers, the real part representing the actual frequency of oscillation, the imaginary part representing the damping. It can be further shown for many cases that frequencies of quasinormal modes with negative imaginary part do not exist, meaning that these solutions are actually stable. Our interest here is to study the asymptotic behavior of quasinormal frequencies, \textit{i.e.}, the case where the imaginary part of $\omega$ grows to infinity. It turns out that in some cases the real part of the frequency approaches a finite limit, while the imaginary parts grow linearly without bound. These generics on quasinormal modes will obviously hold true for the vector field perturbations and metric tensor field perturbations, the main difference being the change in the potential of the Schr\"odinger--like equation. A complete description of the required potentials to describe the most general situation, as derived in \cite{kodama-ishibashi-1, kodama-ishibashi-3}, is presented in appendix B, to which we refer the reader for further details.

For a four dimensional Schwarzschild black hole, one has the asymptotic quasinormal frequencies

$$
\lim_{n \to + \infty} \omega_{n} \sim [ {\mathrm{offset}} ] + i n [ {\mathrm{gap}} ] + {\mathcal{O}} \left( \frac{1}{\sqrt{n}} \right),
$$

\noindent
where the real part of the offset is the frequency of the emitted radiation, and the gap are the quantized increments in the inverse relaxation time. Here, the gap is given by the surface gravity. One can try to extend this analysis to more general situations and also include spacetimes with two horizons, but then generic results become much harder to obtain \cite{mmv-1, padmanabhan, mmv-2, choudhury-padmanabhan} (curiously, for spacetimes with multiple horizons, there is a unique definition of temperature only when the ratio of surface gravities is in ${\mathbb{Q}}$ \cite{choudhury-padmanabhan, choudhury-padmanabhan-2}). Another property of quasinormal modes is a reflection symmetry $\omega \leftrightarrow - \bar{\omega}$ which changes the sign of ${\mathbb{R}}{\mathrm{e}} (\omega)$. Indeed, $\bar{\Phi}_{\omega}$ is a quasinormal mode corresponding to $-\bar{\omega}$.

We have chosen the time dependence for the perturbation to be $e^{i\omega t}$, so that ${\mathbb{I}}{\mathrm{m}} (\omega) > 0$ for stable solutions (implying that the perturbation vanishes exponentially in time). This implies that $\re(\pm i \omega x) \to + \infty$ as $x \to \mp \infty$ and thus, while solutions can oscillate, they must exponentially increase with $|x|$. In other words, $\Phi_{\omega}(x)$ is not a normalizable wave function and this is, ultimately, the reason why we speak of quasinormal modes rather than normal modes (which would form a complete set of stationary eigenfunctions for the IK master differential operator).

Before embarking in the actual calculation, let us address the general strategy of the method introduced in \cite{motl-neitzke}, which begins with the question of how to properly define and impose quasinormal boundary conditions. This is known to be somewhat complicated, at least at an operational level, because as long as we restrict $x \in {\mathbb{R}}$ the quasinormal boundary conditions above amount to distinguishing between an exponentially vanishing term and an exponentially growing term. The idea of \cite{motl-neitzke} is to do an analytic continuation to the complex plane, taking both $r \in {\mathbb{C}}$ and $x \in {\mathbb{C}}$. We will see that in the complex plane one can impose quasinormal boundary conditions in a completely new way. Indeed, if one picks the complex contour $\im (\omega x) = 0$ in ${\mathbb{C}}$, then the asymptotic behavior of $e^{\pm i \omega x}$ is always oscillatory on this line and there will never be any problems with exponentially growing versus exponentially vanishing terms. One should thus restrict to studying the boundary conditions on the so--called Stokes line, $\im ( \omega x ) = 0$.

For Schwarzschild, RN, Schwarzschild dS and RN dS black hole spacetimes, numerical tests generically indicate that the asymptotic quasinormal frequencies are such that $\im (\omega) \gg \re (\omega)$. In other words, one should have $\omega = \omega_{R} + i n \omega_{I}$, with $\omega_{R},\omega_{I} \in {\mathbb{R}}$, $\omega_{I} > 0$ and $n \to + \infty$, in the asymptotic case. Thus, in the $n \sim + \infty$ limit, the contour $\im ( \omega x) = 0$ can be approximated by the curve $\re ( x ) = 0$ which is immediate to plot in ${\mathbb{C}}$. This replacement of contours is selecting asymptotic conditions for the quasinormal modes. For Schwarzschild AdS and RN AdS things are different, as numerical tests generically indicate that asymptotic quasinormal frequencies behave as $\im ( \omega ) \sim \re ( \omega )$, and thus one should have $\omega = n ( \omega_{R} + i \omega_{I} ) + \omega_{0}$, with $\omega_{R},\omega_{I} \in {\mathbb{R}}$, $\omega_{I} > 0$ and $n \to + \infty$, in the asymptotic case. We shall later see with greater detail how to plot the Stokes lines for each black hole spacetime under consideration.

To fully understand the advantage of the analytic continuation in the exact computation of asymptotic quasinormal frequencies, let us first address regions of the complex $r$--plane around an event horizon (be it a black hole horizon or a cosmological horizon). The horizons themselves are defined by the zeroes of $f(r)$, \textit{i.e.}, $f(R_H)=0$, and besides the physical real horizons $R_{H} \in {\mathbb{R}}$ there can be other, non--physical complex horizons $R_{H} \in {\mathbb{C}}$. We shall denote horizons which are not real as fictitious horizons. Power series expansion near any of these horizons simply yields $f(r) \simeq \left( r-R_H \right) f' (R_H) + \cdots$, and it follows for the tortoise coordinate

\begin{equation}
x = \int \frac{dr}{f(r)} \sim \int \frac{dr}{(r-R_{H}) f'(R_{H})} = \frac{1}{f'(R_{H})} \log (r-R_{H}) = \frac{1}{2k_{H}} \log (r-R_{H}) = \frac{1}{4 \pi T_{H}} \log (r-R_{H}), \label{tortoise_horizon}
\end{equation}

\noindent
locally near the chosen horizon (provided that the horizon is nondegenerate, \textit{i.e.}, that $R_H$ is a simple zero of $f(r)$). Here $k_{H}$ is the surface gravity and $T_{H}$ is the Hawking temperature.

One learns that around any nondegenerate horizon the tortoise coordinate will thus be multivalued and it makes sense to ask for the monodromy of tortoise plane waves in clockwise contours around \textit{any} given horizon $f(R_{H})=0$. Let us consider an horizon $R_{H} \in {\mathbb{C}}$, and a clockwise contour $\gamma \subset {\mathbb{C}}$ centered at $R_{H}$ and not including any other horizon. As we take the tortoise coordinate around $\gamma$ it increases by $- \frac{i\pi}{k_{H}}$. This immediately implies the monodromy of the plane waves along the selected contour:

$$
{\mathfrak{M}}_{\gamma,R_{H}} \left[ e^{\pm i \omega x} \right] = e^{\pm \frac{\pi \omega}{k_{H}}}.
$$

\noindent
This result is quite interesting as it now allows one to recast quasinormal boundary conditions as monodromy conditions \cite{motl-neitzke}; if one wants, as quasinormal mode monodromy conditions. At the black hole event horizon the quasinormal boundary condition is 

$$
\Phi_{\omega} (x) \sim e^{i\omega x}\;\, {\mathrm{as}}\;\, x \to - \infty,
$$

\noindent
where $x \to - \infty$ as $r \to R_{H}$, with $R_{H}$ the black hole horizon. One immediately re--writes this boundary condition as a monodromy condition for the solution of the master equation at the black hole event horizon:

$$
{\mathfrak{M}}_{\gamma,R_{H}} \left[ \Phi_{\omega} (x) \right] = e^{\frac{\pi\omega}{k_{H}}},
$$

\noindent
with $\gamma$ a clockwise contour. The other quasinormal boundary condition,

$$
\Phi_{\omega} (x) \sim e^{-i\omega x}\;\, {\mathrm{as}}\;\, x \to + \infty,
$$

\noindent
lives at $r \sim + \infty$, as $x \to + \infty$ when $r \to + \infty$ (for all spacetimes except asymptotically dS spacetimes). Thus, this boundary condition \textit{cannot} be recast as a quasinormal monodromy condition. Instead, if one considers asymptotically dS spacetimes, the above boundary condition is located at the cosmological event horizon, as $x \to + \infty$ when $r \to R_{C}$. Then, it is immediate to recast this boundary condition as a monodromy condition for the solution of the master equation at the cosmological event horizon:

$$
{\mathfrak{M}}_{\gamma,R_{C}} \left[ \Phi_{\omega} (x) \right] = e^{-\frac{\pi\omega}{k_{C}}},
$$

\noindent
with $\gamma$ a clockwise contour. We shall see in the following how these simple ideas about boundary conditions will allow for analytic calculations of asymptotic quasinormal frequencies in all static, spherically symmetric black hole spacetimes.


\subsection{Asymptotically Flat Spacetimes}


\cite{kodama-ishibashi-3} discusses the stability of black holes in asymptotically flat spacetimes to tensor, vector and scalar perturbations. For black holes without charge, all types of perturbations are stable in any dimension. For charged black holes, tensor and vector perturbations are stable in any dimension. Scalar perturbations are stable in four and five dimensions but there is no proof of stability in dimension $d \ge 6$. As we work in generic dimension $d$ we are thus not guaranteed to always have a stable solution. Our results will apply if and only if the spacetime in consideration is stable.


\subsubsection{The Schwarzschild Solution}


For completeness, we first present a computation of the asymptotic quasinormal frequencies for the Schwarzschild $d$--dimensional black hole, using the monodromy method. This calculation was first done in \cite{motl-neitzke}. The interesting result shown in \cite{motl-neitzke} is that, if one is only interested in the \textit{asymptotic} quasinormal modes, there is no need to solve the IK master  equation exactly. Rather, there is a method which explores the analytic continuation of the master equation to the complex plane and demands only for approximate solutions near infinity, near the origin, and near the black hole event horizon. Knowledge of the solutions in these regions, together with monodromy matching along a specially chosen contour, then yields the quasinormal frequencies. Let us carefully explain this method in the simplest Schwarzschild example, as we shall employ it several times in the following.

We consider solutions of the Schr\"odinger--like master equation (\ref{schrodinger})

$$
- \frac{ d^{2} \Phi}{dx^{2}} (x) + V \big[ r(x) \big] \Phi (x) = \omega^{2} \Phi (x)
$$

\noindent
in the complex $r$--plane. Let us begin at infinity. Since the potential $V(r)$ vanishes for $r \sim +\infty$, we will have

$$
\Phi(x) \sim A_+ e^{i \omega x} + A_- e^{-i \omega x}
$$

\noindent
in this region. The boundary condition for quasinormal modes at infinity is then

\begin{equation}\label{Schwarz_inf}
A_+=0.
\end{equation}

\noindent
Next we study the behavior of $\Phi(x)$ near the singularity $r=0$. In this region, the tortoise coordinate is

$$
x \sim - \frac{r^{d-2}}{2(d-2)\MM},
$$

\noindent
and the potential for tensor and scalar type perturbations is

$$
V \big[ r(x) \big] \sim -\frac1{4x^2} = \frac{j^2 - 1}{4x^2},
$$

\noindent
with $j=0$ (see appendix \ref{appendixC}). The Schr\"odinger--like master equation approximates to

$$
- \frac{ d^{2} \Phi}{dx^{2}} (x) + \frac{j^2 - 1}{4x^2} \Phi (x) = \omega^{2} \Phi (x)
$$

\noindent
whose solution is (this is the solution for $j \neq 0$---more on this in a moment)

$$
\Phi (x) \sim B_+ \sqrt{2\pi\omega x}\ J_{\frac{j}{2}} \left( \omega x \right) + B_- \sqrt{2\pi\omega x}\ J_{-\frac{j}{2}} \left( \omega x \right),
$$

\noindent
where $J_\nu(x)$ represents a Bessel function of the first kind and $B_\pm$ are (complex) integration constants. One would next like to link this solution at the origin with the solution at infinity.

For the Schwarzschild asymptotic quasinormal modes one has $\im (\omega) \gg \re (\omega)$, with $\im (\omega) \to + \infty$, and hence $\omega$ is very large and approximately purely imaginary. Consequently, one has $\omega x \in \mathbb{R}$ for $x \in i\mathbb{R}$; in a neighborhood of the origin, the above relation between $x$ and $r$ tells us that this happens for

$$
r=\rho\ e^{\frac{i\pi}{2(d-2)}+\frac{in\pi}{d-2}},
$$

\noindent
with $\rho>0$ and $n=0,1,\ldots, 2d-5$. These are half--lines starting at the origin, equally spaced by an angle of $\frac{\pi}{d-2}$. Notice that the sign of $\omega x$ on these half--lines is $(-1)^n$; in other words, starting with the half--line corresponding to $n=0$, the sign of $\omega x$ is alternately positive and negative as one goes anti--clockwise around the origin.

Precisely because we are interested in these asymptotic modes, we may consider the following asymptotic expansion of the Bessel functions

\begin{equation}
J_\nu(z) \sim \sqrt{\frac{2}{\pi z}} \cos\left(z-\frac{\nu \pi}{2}-\frac{\pi}4 \right), \qquad z \gg 1,  \label{Bessel+}
\end{equation}

\noindent
from where we learn that

\begin{eqnarray}
\Phi (x) & \sim & 2 B_+ \cos \left( \omega x - \alpha_+ \right) + 2 B_- \cos \left( \omega x - \alpha_- \right) \nonumber \\
& = & \left( B_+ e^{-i\alpha_+} + B_- e^{-i\alpha_-}\right) e^{i \omega x} + \left( B_+ e^{i\alpha_+} + B_- e^{i\alpha_-}\right) e^{-i \omega x} \label{Schwarz_0}
\end{eqnarray}

\noindent
in any one of the lines corresponding to positive $\omega x$, and where we have defined

$$
\alpha_\pm = \frac{\pi}4 (1 \pm j).
$$

\noindent
This asymptotic expression for $\Phi$ near the origin is ideal to make the matching with its asymptotic expression at infinity. This matching must however be done along the so--called Stokes line, defined by $\omega x \in \mathbb{R}$ (or $\im ( \omega x ) = 0$), so that neither of the exponentials $e^{\pm i\omega x}$ dominates the other. In this Schwarzschild case the Stokes line definition corresponds to $x \in i\mathbb{R}$ or $\re(x)=0$.

To trace out the Stokes line $\re(x)=0$ let us first observe that we already know its behavior near the origin. Furthermore, this is the only singular point of this curve: indeed, since $x$ is a holomorphic function of $r$, the critical points of the function $\re(x)$ are the zeros of

$$
\frac{dx}{dr} = \frac{1}{f(r)} = \frac{r^{d-3}}{r^{d-3}-2\MM}
$$

\noindent
(\textit{i.e.}, $r=0$ only). We have an additional problem that $x$ is a multivalued function: each of the ``horizons''

$$
R_{n} = \left| \left( 2\MM \right)^{\frac{1}{d-3}} \right| \exp \left( \frac{2\pi i}{d-3}\ n \right), \qquad n=0,1,\ldots,d-4,
$$

\noindent
is a branch point. In fact, near such points one has (see (\ref{tortoise_horizon}))

$$
x \sim \frac{1}{f'(R_{n})} \log (r-R_{n}).
$$

\noindent
Thus we see that although the function $\re(x)$ is well defined around $R_n$ with $n=0$ and $n=\frac{d-3}{2}$ (if $d$ is odd), as $f'(R_n)$ is real in these cases, it will be multivalued around all the other fictitious horizons. 

For $r \sim \infty$ one has $x \sim r$ (see appendix \ref{appendixC}). Consequently, $x$ is holomorphic at infinity and we can choose the branch cuts to cancel out among themselves. Therefore $\re(x)$ is well defined in a neighborhood of infinity, and moreover $\re(x)=0$ will be approximately parallel to $\re(r)=0$ in this neighborhood. Two of the $2(d-2)$ branches of the Stokes line starting out at the origin must therefore be unbounded. The remaining can either connect to another branch or end up in a branch cut. On the other hand, it is easy to see that the Stokes line must intersect the positive real axis exactly in one point, greater than $R_0$. Using this information plus elementary considerations of symmetry and the sign of $\re(x)$, one can deduce that the Stokes line must be of the form indicated in Figure \ref{StokesS}. These results are moreover verified by the numerical computation of the same Stokes lines, as indicated in Figure \ref{NumericalStokesS}.

\FIGURE[ht]{\label{StokesS}
	\centering
	\psfrag{A}{$A$}
	\psfrag{B}{$B$}
	\psfrag{r0=1}{$R_0$}
	\psfrag{r1}{$R_1$}
	\psfrag{r2}{$R_2$}
	\psfrag{Re}{$\re$}
	\psfrag{Im}{$\im$}
	\psfrag{contour}{contour}
	\psfrag{Stokes line}{Stokes line}
	\psfrag{ramification line}{branch cut}
	\epsfxsize=.6\textwidth
	\leavevmode
	\epsfbox{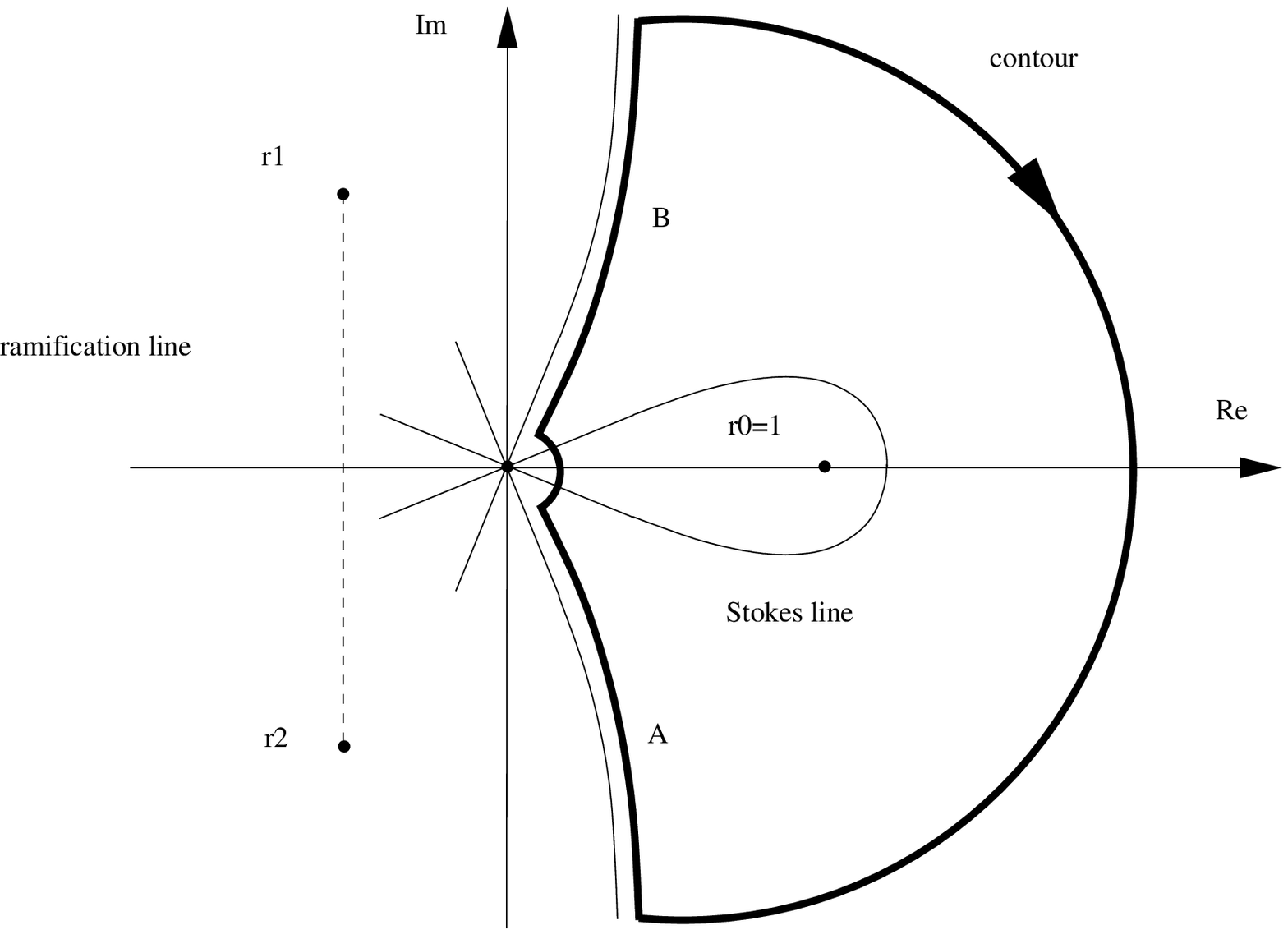}
\caption{Stokes line for the Schwarzschild black hole, along with the chosen contour for monodromy matching, in the case of dimension $d=6$.}
}

\FIGURE[ht]{\label{NumericalStokesS}
	\centering
	\epsfig{file=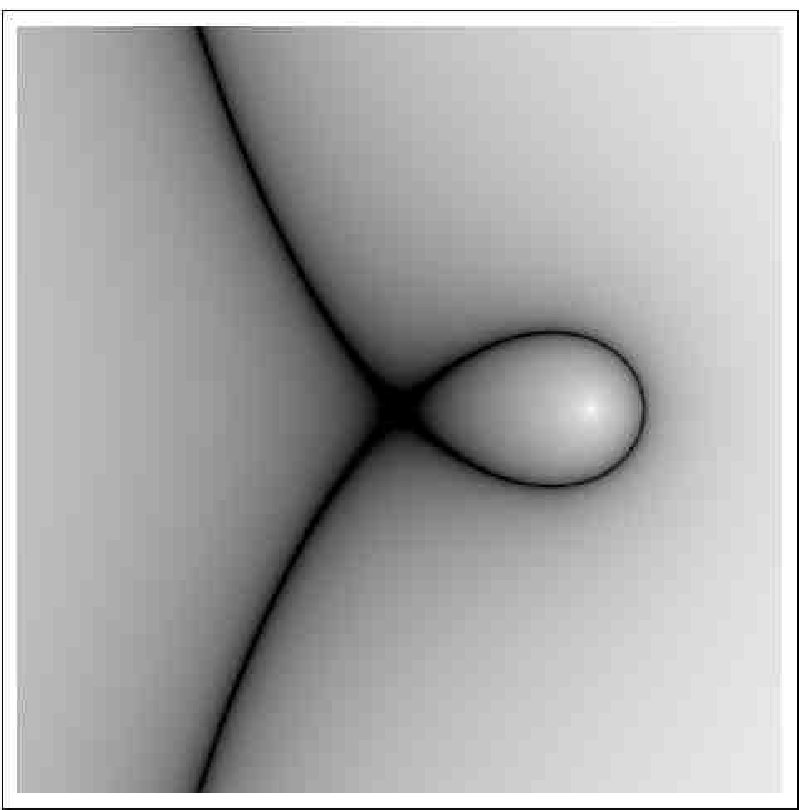, width=1.5in, height=1.5in}
    $\quad$
    \epsfig{file=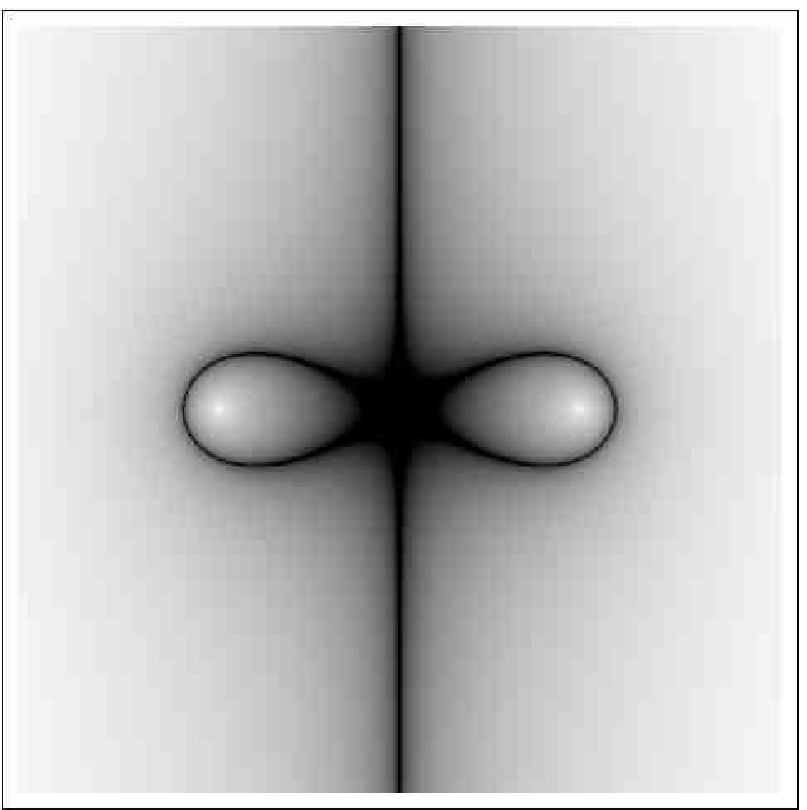, width=1.5in, height=1.5in}
    $\quad$
    \epsfig{file=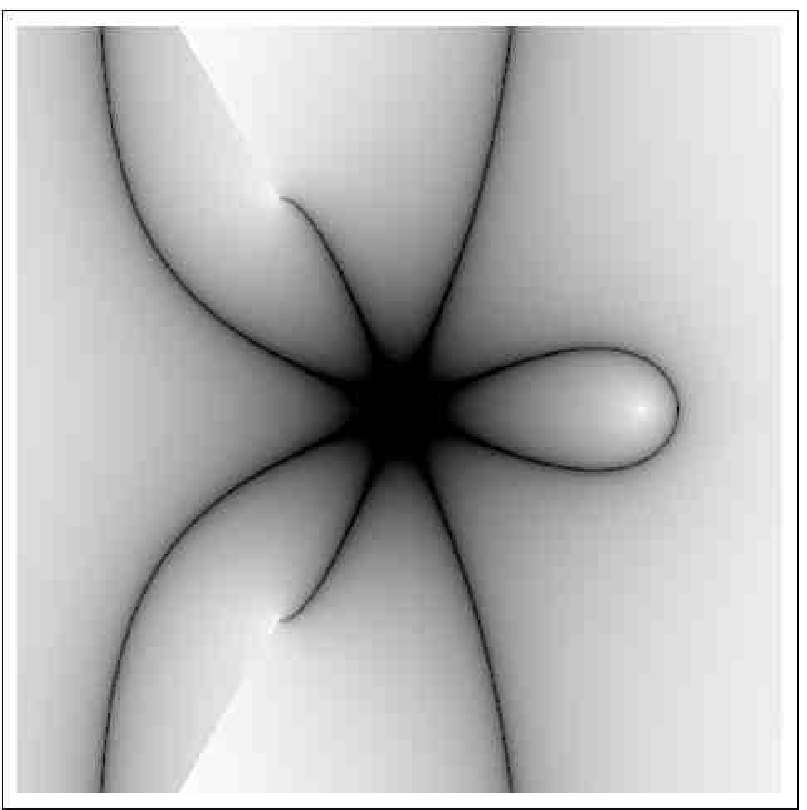, width=1.5in, height=1.5in}
    $\quad$
    \epsfig{file=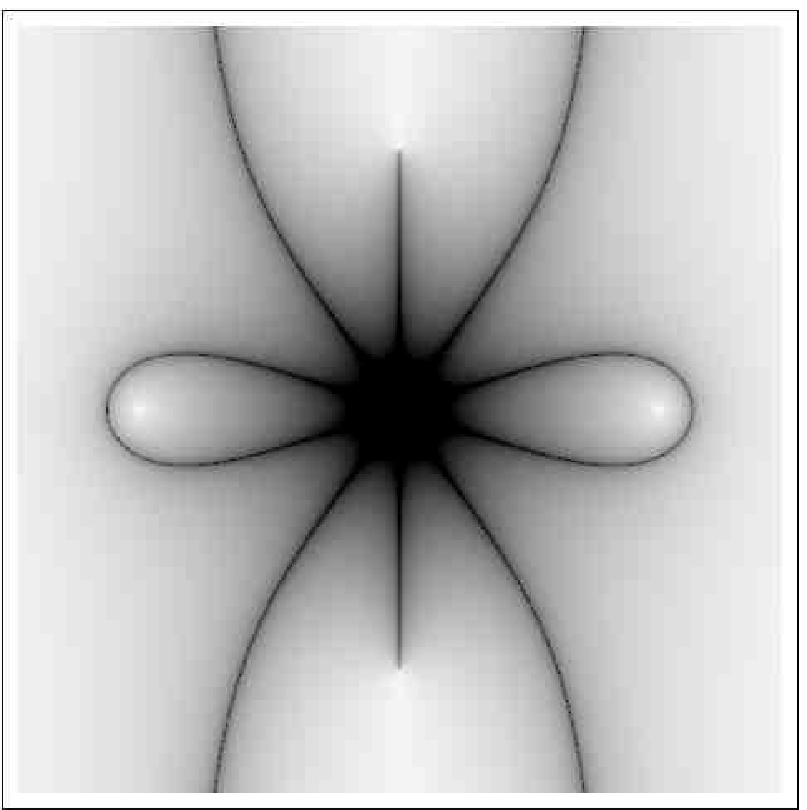, width=1.5in, height=1.5in}
\caption{Numerical calculation of the Stokes lines for the Schwarzschild black hole in dimensions $d=4$, $d=5$, $d=6$ and $d=7$. The different shadings also illustrate the various horizon singularities and branch cuts (note that these branch cuts are not necessarily equal to the ones used for the calculation in the main text).}
}

Let us now consider the contour obtained by closing the unlimited portions of the Stokes line near $r\sim \infty$, as shown in Figure \ref{StokesS}. At point $A$ we have $\omega x \gg 0$, and therefore the expansion (\ref{Schwarz_0}) holds at this point. Imposing condition (\ref{Schwarz_inf}) one obtains

\begin{equation}\label{S_1}
B_+ e^{-i\alpha_+} + B_- e^{-i\alpha_-} = 0.
\end{equation}

\noindent
For $z \sim 0$ one has the expansion

\begin{equation}
J_\nu(z)=z^\nu w(z),  \label{Bessel0}
\end{equation}

\noindent
where $w$ is an even holomorphic function. Consequently, as one rotates from the branch containing point $A$ to the branch containing point $B$, through an angle of $\frac{3\pi}{d-2}$, $x\sim - \frac{r^{d-2}}{2(d-2)\MM}$ rotates through an angle of $3\pi$, and since

$$
\sqrt{2\pi e^{3\pi i} \omega x}\ J_{\pm\frac{j}{2}} \left( e^{3\pi i} \omega x \right) = e^{\frac{3\pi i}2 (1 \pm j)}\sqrt{2\pi \omega x}\ J_{\pm\frac{j}{2}} \left( \omega x \right) \sim 2 e^{6i\alpha_\pm} \cos(\omega x - \alpha_\pm),
$$

\noindent
one has (notice that $e^{3\pi i} \omega x=-\omega x$)

\begin{eqnarray*}
\Phi (x) & \sim & 2 B_+ e^{6i\alpha_+} \cos \left( -\omega x - \alpha_+ \right) + 2 B_- e^{6i\alpha_-} \cos \left( -\omega x - \alpha_- \right) \\
& = & \left( B_+ e^{7i\alpha_+} + B_- e^{7i\alpha_-}\right) e^{i \omega x} + \left( B_+ e^{5i\alpha_+} + B_- e^{5i\alpha_-}\right) e^{-i \omega x}
\end{eqnarray*}

\noindent
at point $B$. As one closes the contour near $r \sim \infty$, one has $x \sim r$ and hence $\re(x)>0$; since $\im(\omega) \gg 0$, it follows that $e^{i\omega x}$ is exponentially small in this part of the contour, and therefore only the coefficient of $e^{-i\omega x}$ should be trusted. As one completes the contour, this coefficient gets multiplied by

$$
\frac{B_+ e^{5i\alpha_+} + B_- e^{5i\alpha_-}}{B_+ e^{i\alpha_+} + B_- e^{i\alpha_-}}.
$$

\noindent
On the other hand, the monodromy of $e^{-i\omega x}$ going clockwise around this contour is $e^{-\frac{\pi\omega}{k}}$, where 

$$
k = \frac12 f'(R_0)
$$

\noindent
is the surface gravity at the horizon. Since

$$
x \sim \frac1{f'(R_0)}\log(r-R_0)
$$

\noindent
for $r\sim R_0$, $x$ will increase by $- \frac{2\pi i}{f'(R_0)} = - \frac{\pi i}{k}$ as one goes clockwise around $R_0$, and hence $e^{-i\omega x}$ will get multiplied by

$$
e^{- i\omega \left(-\frac{2\pi i}{f'(R_0)}\right)} = e^{-\frac{\pi\omega}{k}}.
$$

\noindent
Thus the clockwise monodromy of $\Phi$ around the contour depicted in Figure \ref{StokesS} is

$$
\frac{B_+ e^{5i\alpha_+} + B_- e^{5i\alpha_-}}{B_+ e^{i\alpha_+} + B_- e^{i\alpha_-}} e^{-\frac{\pi\omega}{k}}.
$$

\noindent
The important point to realize now is that one can deform this chosen contour---without crossing any singularities---so that it becomes a small clockwise circle around the black hole event horizon $r = R_0$. Near the horizon $r = R_0$, we have again $V(r)\sim 0$, and hence again

$$
\Phi(x) \sim C_+ e^{i \omega x} + C_- e^{-i \omega x}.
$$

\noindent
The condition for quasinormal modes at the horizon is therefore $C_- = 0$. Again using the fact that

$$
x \sim \frac1{f'(R_0)}\log(r-R_0)
$$

\noindent
for $r\sim R_0$, we can restate this boundary condition as the condition that the monodromy of $\Phi$ going clockwise around the contour should be

$$
e^{i\omega \left(-\frac{2\pi i}{f'(R_0)}\right)} = e^{\frac{\pi\omega}{k}}.
$$

\noindent
Because the monodromy is invariant under this deformation of the contour, the condition for quasinormal modes at the horizon follows as

\begin{equation}\label{S_2}
\frac{B_+ e^{5i\alpha_+} + B_- e^{5i\alpha_-}}{B_+ e^{i\alpha_+} + B_- e^{i\alpha_-}} e^{-\frac{\pi\omega}{k}} = e^{\frac{\pi\omega}{k}}.
\end{equation}

The final condition for equations (\ref{S_1}) and (\ref{S_2}), which reflect quasinormal mode boundary conditions both at infinity and at the black hole horizon, to have nontrivial solutions $(B_+,B_-)$ is simply

\begin{equation} \label{Sdeterminant}
\left| 
\begin{array}{ccc}
e^{-i\alpha_+} & \quad & e^{-i\alpha_-} \\ & & \\
e^{5i\alpha_+} e^{-\frac{\pi\omega}{k}} - e^{i\alpha_+} e^{\frac{\pi\omega}{k}}  &  & e^{5i\alpha_-} e^{-\frac{\pi\omega}{k}}- e^{i\alpha_-} e^{\frac{\pi\omega}{k}}
\end{array}
\right| = 0 \quad
\Leftrightarrow \quad
\left| 
\begin{array}{ccc}
e^{-4i\alpha_+} & & e^{-4i\alpha_-} \\ & & \\
\sin\left(2\alpha_+ + \frac{i\pi\omega}{k} \right) & \quad & \sin\left( 2\alpha_- + \frac{i\pi\omega}{k} \right)
\end{array}
\right| = 0.
\end{equation}

\noindent
This equation is automatically satisfied for $j=0$. This is to be expected, as for $j=0$ the Bessel functions $J_{\pm\frac{j}2}$ coincide and do not form a basis for the space of solutions of the Schr\"odinger--like master equation near the origin. Following \cite{motl-neitzke}, we consider this equation for $j$ nonzero and take the limit as $j\to 0$. This amounts to writing the equation as a power series in $j$ and equating to zero the first non--vanishing coefficient, which in this case is the coefficient of the linear part. Thus we just have to require that the derivative of the determinant above with respect to $j$ be zero for $j=0$:

\begin{eqnarray*}
&& 
\left|
\begin{array}{ccc}
-i\pi e^{-i\pi} & \quad & i\pi e^{-i\pi} \\ & & \\
\sin\left( \frac{\pi}2 + \frac{i\pi\omega}{k}\right) & \quad & \sin\left( \frac{\pi}2 + \frac{i\pi\omega}{k}\right)
\end{array}
\right| + \left|
\begin{array}{ccc}
e^{-i\pi} & & e^{-i\pi} \\ & \quad & \\
\frac{\pi}2 \cos\left( \frac{\pi}2 + \frac{i\pi\omega}{k}\right) & \quad & -\frac{\pi}2 \cos\left( \frac{\pi}2 + \frac{i\pi\omega}{k}\right)
\end{array}
\right| = 0 \\
&&
\Leftrightarrow \quad 2\cosh\left( \frac{\pi\omega}{k}\right) - \sinh\left( \frac{\pi\omega}{k}\right)=0 \quad \Leftrightarrow \quad e^{\frac{2\pi\omega}{k}} = -3 \quad \Leftrightarrow \quad \frac{2\pi\omega}{k} = \log 3 + i(2n + 1)\pi, \quad n \in \mathbb{N}.
\end{eqnarray*}

For vector type perturbations, the potential for the Schr\"odinger--like master equation near the origin is of the form

$$
V \big[ r(x) \big] \sim \frac{j^2 - 1}{4x^2},
$$

\noindent
with $j=2$ (see appendix \ref{appendixC}). Repeating the same argument, one ends  with the same equation as (\ref{Sdeterminant}), except that now $2\alpha_\pm=\frac{\pi}2 \pm \pi$, $4\alpha_\pm=\pi \pm 2\pi$. This equation is exactly the same as in the $j=0$ case, for which $2\alpha_\pm=\frac{\pi}2$, $4\alpha_\pm=\pi$, and consequently we end up with the precise same quasinormal frequencies.

In the calculation above we have obtained the asymptotic quasinormal frequencies $\omega$ with $\re(\omega)>0$. However, one knows that if $\omega$ is a quasinormal frequency then so is $-\overline{\omega}$. Consequently $\frac{2\pi\omega}{k} = -\log 3 + i(2n + 1)\pi$ (for $n \in \mathbb{N}$) must also be a solution for the asymptotic quasinormal frequencies. To understand how these comes about, and why we did not obtain them above, we recall that we imposed condition (\ref{Schwarz_inf}) at point $A$, where $ix \to +\infty$. Notice that we could instead have imposed it at point $B$, where $ix \to -\infty$. It is in choosing one of these points that we automatically choose the sign of $\re(\omega)$. Indeed, if $\re(\omega)>0$ then $e^{i\omega x} \gg e^{-i\omega x}$ at point $A$ and $e^{i\omega x} \ll e^{-i\omega x}$ at point $B$. Consequently, only at point $A$ is it meaningful to impose condition (\ref{Schwarz_inf}). On the other hand, if $\re(\omega)<0$ then $e^{i\omega x} \ll e^{-i\omega x}$ at point $A$ and $e^{i\omega x} \gg e^{-i\omega x}$ at point $B$, and only at point $B$ is it meaningful to impose condition (\ref{Schwarz_inf}). Had we done this, we would have obtained the second set of asymptotic quasinormal frequencies.

The results above have been thoroughly checked in the literature. Actually, an  analytic calculation of the asymptotic quasinormal frequencies was first done in \cite{motl}, in four dimensions, and then extended to $d$--dimensions in \cite{motl-neitzke}, at least for tensor type perturbations (and already using the monodromy method). The result was shown to be dimension independent and equal to $T_{H} \log 3$. It was already expected that these $d$--dimensional frequencies would scale linearly with the Hawking temperature, from the earlier results in \cite{kunstatter}. Later, in \cite{birmingham}, it was shown that the same result of $T_{H} \log 3$ holds for both vector and scalar perturbations. All these results were later subject to a detailed numerical check in \cite{cly, konoplya-1}, with fully positive results.


\subsubsection{The Reissner--Nordstr\"om Solution}


Here we compute the asymptotic quasinormal modes of the RN $d$--dimensional black hole using the monodromy method. This calculation was done for $d=4$ in \cite{motl-neitzke}. We consider solutions of the Schr\"odinger--like equation (\ref{schrodinger}) in the complex $r$--plane. Since the potential $V$ vanishes for $r \sim +\infty$, we will again have

$$
\Phi(x) \sim A_+ e^{i \omega x} + A_- e^{-i \omega x}
$$

\noindent
in this region, the boundary condition for quasinormal modes at infinity being

\begin{equation}\label{RN_inf}
A_+=0.
\end{equation}

\noindent
Next we study the behavior of $\Phi(x)$ near the singularity $r=0$. In this region, the tortoise coordinate is

$$
x \sim \frac{r^{2d-5}}{\left( 2d-5 \right) \QQ^{2}},
$$

\noindent
and the potential for tensor type and scalar type perturbations is

$$
V \big[ r(x) \big] \sim \frac{j^2 - 1}{4x^2},
$$

\noindent
with $j=\frac{d-3}{2d-5}$ (see appendix \ref{appendixC}). The solution of the Schr\"odinger--like equation in this region is therefore well approximated by

$$
\Phi (x) \sim B_+ \sqrt{2\pi\omega x}\ J_{\frac{j}{2}} \left( \omega x \right) + B_- \sqrt{2\pi\omega x}\ J_{-\frac{j}{2}} \left( \omega x \right),
$$

\noindent
where $J_\nu(x)$ represents a Bessel function of the first kind and $B_\pm$ are (complex) integration constants.

For the asymptotic quasinormal modes one has $\im (\omega) \gg \re (\omega)$, and hence $\omega$ is approximately purely imaginary. Consequently, one has $\omega x \in \mathbb{R}$ for $x \in i\mathbb{R}$; in a neighborhood of the origin, this happens for

$$
r=\rho\ e^{\frac{i\pi}{2(2d-5)}+\frac{in\pi}{2d-5}},
$$

\noindent
with $\rho>0$ and $n=0,1,\ldots, 4d-11$. These are half--lines starting at the origin, equally spaced by an angle of $\frac{\pi}{2d-5}$. Notice that the sign of $\omega x$ on these lines is $(-1)^{n+1}$; in other words, starting with the line corresponding to $n=0$, the sign of $\omega x$ is alternately negative and positive as one goes anti--clockwise around the origin.

From the asymptotic expansion (\ref{Bessel+}) we see that

\begin{eqnarray} \label{RN_0}
\Phi (x) & \sim & 2 B_+ \cos \left( \omega x - \alpha_+ \right) + 2 B_- \cos \left( \omega x - \alpha_- \right) \nonumber \\
& = & \left( B_+ e^{-i\alpha_+} + B_- e^{-i\alpha_-}\right) e^{i \omega x} + \left( B_+ e^{i\alpha_+} + B_- e^{i\alpha_-}\right) e^{-i \omega x} 
\end{eqnarray}

\noindent
in any one of the lines corresponding to positive $\omega x$, where again we define

$$
\alpha_\pm = \frac{\pi}4 (1 \pm j).
$$

\noindent
This asymptotic expression for $\Phi$ near the origin is to be matched with its asymptotic expression at infinity. This matching must again be done along the Stokes line $\omega x \in \mathbb{R}$ (that is, $x \in i\mathbb{R}$), so that neither of the exponentials $e^{\pm i\omega x}$ dominates the other.

To trace out the Stokes line $\re(x)=0$ we observe that we already know its behavior near the origin. Furthermore, this is the only singular point of this curve: indeed, since $x$ is a holomorphic function of $r$, the critical points of the function $\re(x)$ are the zeros of

$$
\frac{dx}{dr} = \frac{1}{f(r)} = \frac{r^{2d-6}}{r^{2d-6}-2\MM r^{d-3}+ \QQ^2}
$$

\noindent
(\textit{i.e.}, $r=0$ only). We have the additional problem that $x$ is a multivalued function: each of the ``horizons''

$$
r=R^\pm_n, \qquad n=0, 1, \ldots, d-4
$$

\noindent
(see appendix \ref{appendixC}) is a branch point. From (\ref{tortoise_horizon}), we see that although the function $\re(x)$ is still well defined around $R^\pm_n$ with $n=0$ and $n=\frac{d-3}{2}$ (if $d$ is odd), as $f'(R^\pm_n)$ is real in these cases, it will be multivalued around all the other fictitious horizons.

For $r \sim \infty$ one has $f (r) \sim 1$ and hence $x \sim r$. Consequently $x$ is holomorphic at infinity, and we can choose the branch cuts to cancel out among themselves. Therefore $\re(x)$ is well defined in a neighborhood of infinity, and moreover $\re(x)=0$ will be approximately parallel to $\re(r)=0$ in this neighborhood. Two of the $2(2d-5)$ branches of the Stokes line starting out at the origin must therefore be unbounded. The remaining can either connect to another branch or end up in a branch cut. On the other hand, it is easy to see that the Stokes line must intersect the positive real axis exactly in two points, one in each of the intervals $(R_0^-,R_0^+)$ and $(R_0^+, +\infty)$. Using this information plus elementary considerations of symmetry and the sign of $\re(x)$, one can deduce that the Stokes line must be of the form indicated in Figure \ref{StokesRN}. These results are moreover verified by the numerical computation of the same Stokes lines, as indicated in Figure \ref{NumericalStokesRN}.

\FIGURE[ht]{\label{StokesRN}
	\centering
	\psfrag{A}{$A$}
	\psfrag{B}{$B$}
	\psfrag{r0}{$R_0^-$}
	\psfrag{r1}{$R_1^-$}
	\psfrag{r2}{$R_2^-$}
	\psfrag{R0}{$R_0^+$}
	\psfrag{R1}{$R_1^+$}
	\psfrag{R2}{$R_2^+$}
	\psfrag{Re}{$\re$}
	\psfrag{Im}{$\im$}
	\psfrag{contour}{contour}
	\psfrag{Stokes line}{Stokes line}
	\epsfxsize=.6\textwidth
	\leavevmode
	\epsfbox{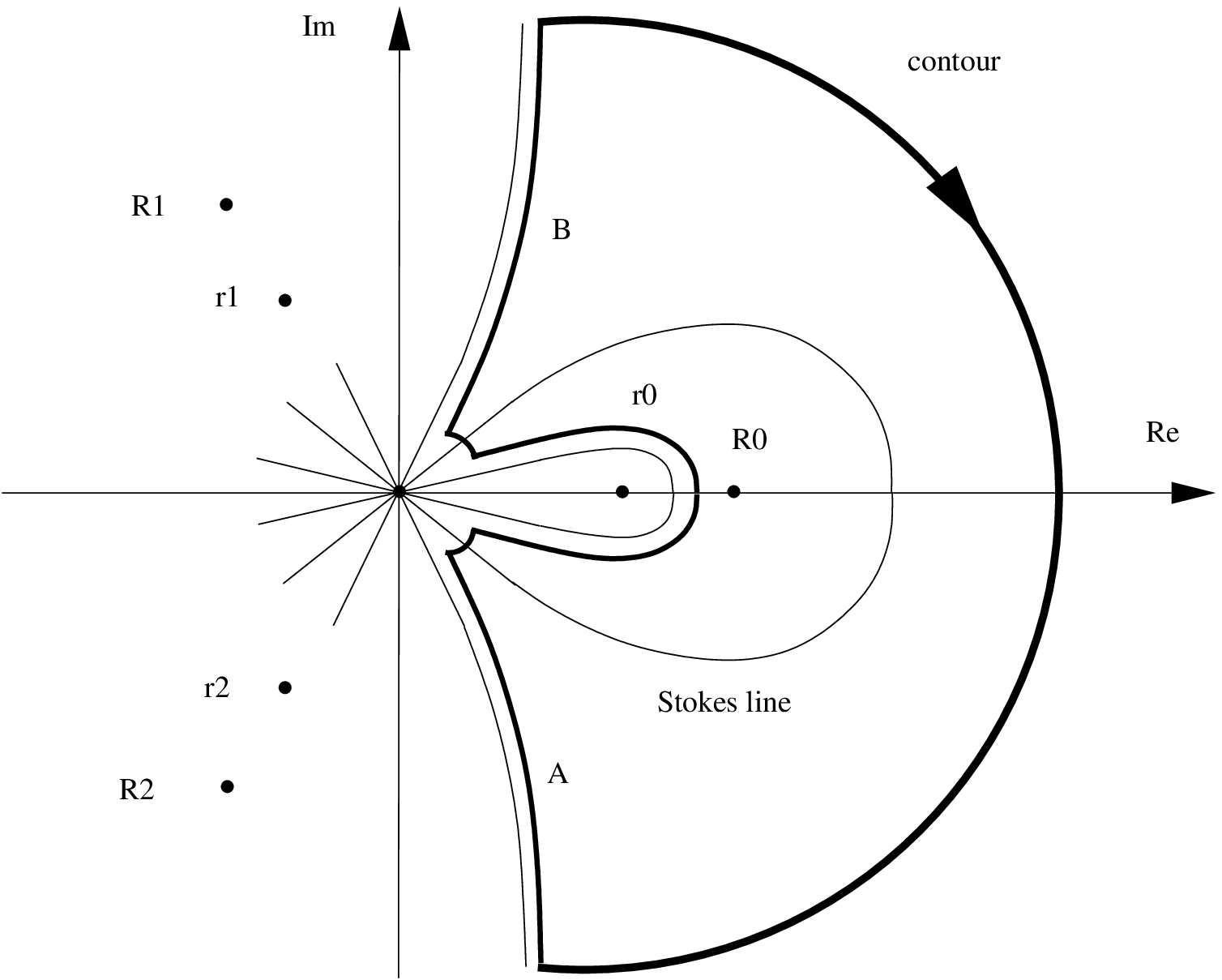}
\caption{Stokes line for the Reissner--Nordstr\"om black hole, along with the chosen contour for monodromy matching, in the case $d=6$.}
}

\FIGURE[ht]{\label{NumericalStokesRN}
	\centering
	\epsfig{file=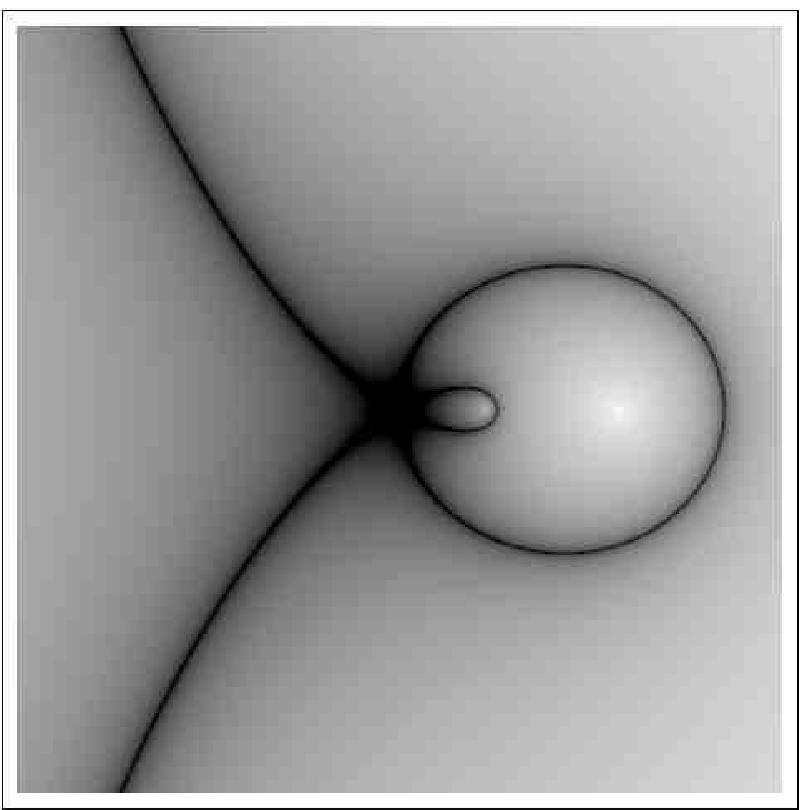, width=1.5in, height=1.5in}
    $\quad$
    \epsfig{file=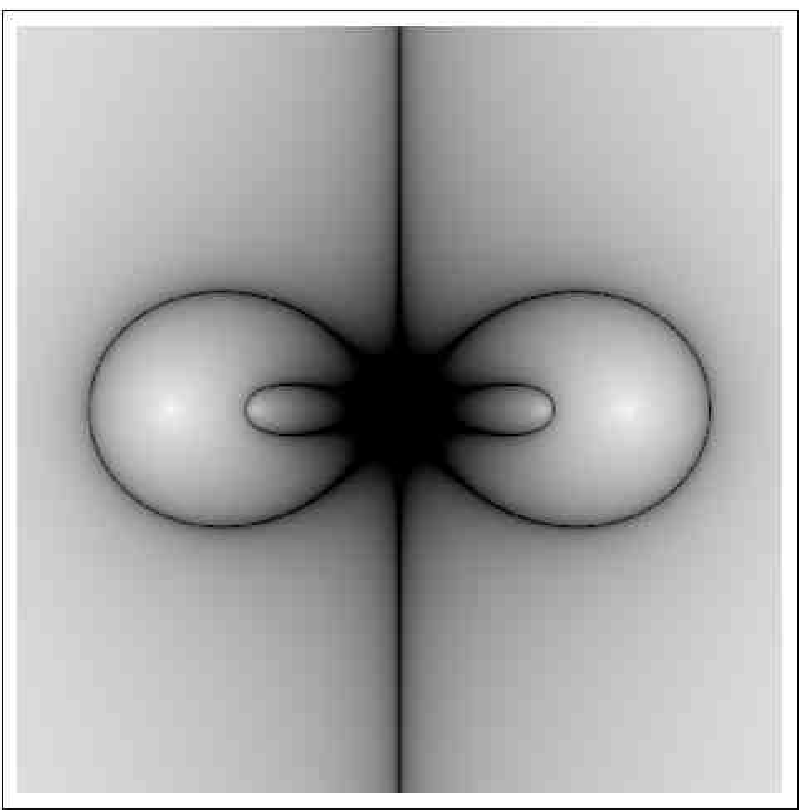, width=1.5in, height=1.5in}
    $\quad$
    \epsfig{file=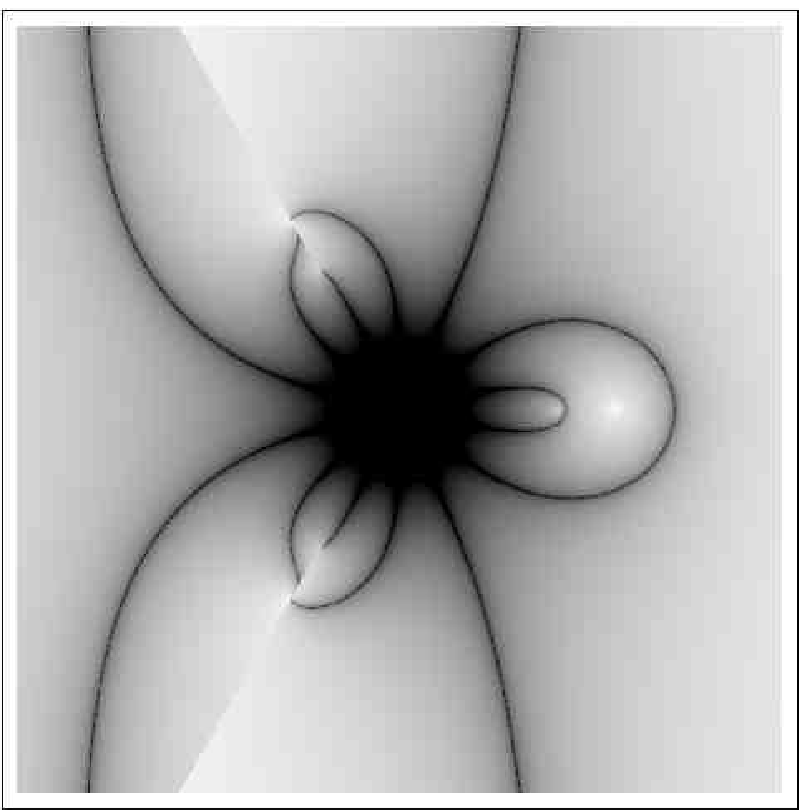, width=1.5in, height=1.5in}
    $\quad$
    \epsfig{file=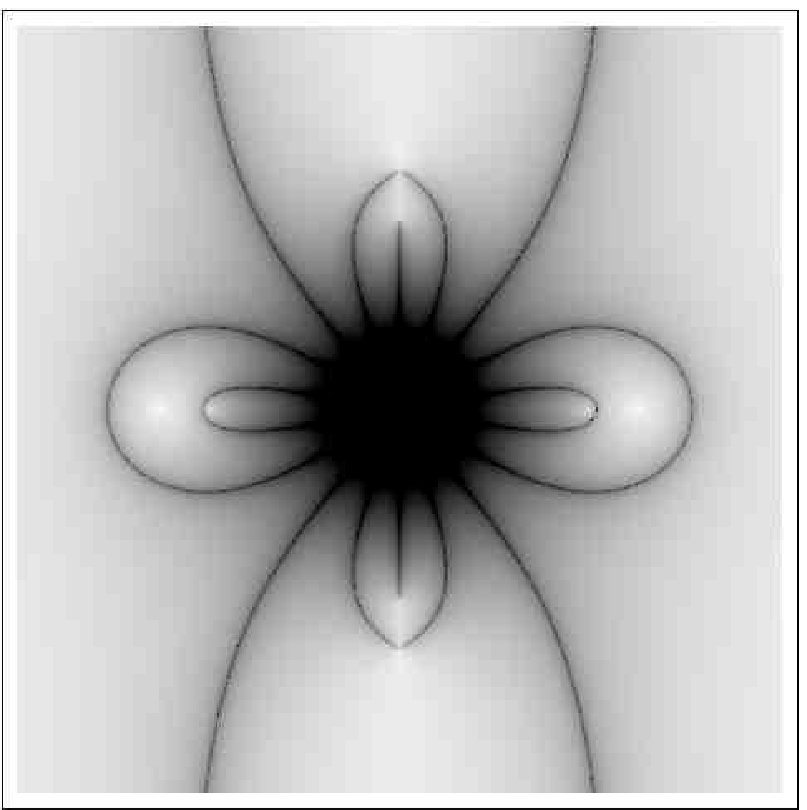, width=1.5in, height=1.5in}
\caption{Numerical calculation of the Stokes lines for the Reissner--Nordstr\"om black hole in dimensions $d=4$, $d=5$, $d=6$ and $d=7$. The different shadings also illustrate the various horizon singularities and branch cuts.}
}

We consider the contour obtained by closing the unlimited portions of the Stokes line near $r \sim \infty$ and avoiding the enclosure of the inner horizon, as shown in Figure \ref{StokesRN}. At point $A$ we have $\omega x>0$, and therefore the expansion (\ref{RN_0}) holds at this point. Imposing condition (\ref{RN_inf}) one obtains

\begin{equation} \label{RN_1}
B_+ e^{-i\alpha_+} + B_- e^{-i\alpha_-} = 0.
\end{equation}

\noindent
For $z \sim 0$ one has the expansion (\ref{Bessel0}). Consequently, as one rotates from the branch containing point $A$ to the next branch in the contour (through an angle of  $\frac{2\pi}{2d-5}$), $x\sim \frac{r^{2d-5}}{\left( 2d-5 \right) \QQ^{2}}$ rotates through an angle of $2\pi$, and since

$$
\sqrt{2\pi e^{2\pi i} \omega x}\ J_{\pm\frac{j}{2}} \left( e^{2\pi i} \omega x \right) = e^{\frac{2\pi i}2 (1 \pm j)}\sqrt{2\pi \omega x}\ J_{\pm\frac{j}{2}} \left( \omega x \right) \sim 2 e^{4i\alpha_\pm} \cos(\omega x - \alpha_\pm)
$$

\noindent
one has

\begin{eqnarray}\label{RN_00}
\Phi (x) & \sim & 2 B_+ e^{4i\alpha_+} \cos \left( \omega x - \alpha_+ \right) + 2 B_- e^{4i\alpha_-} \cos \left( \omega x - \alpha_- \right) \nonumber \\
& = & \left( B_+ e^{3i\alpha_+} + B_- e^{3i\alpha_-}\right) e^{i \omega x} + \left( B_+ e^{5i\alpha_+} + B_- e^{5i\alpha_-}\right) e^{-i \omega x} 
\end{eqnarray}

\noindent
on that branch. As one follows the contour around the inner horizon $r=R_0^-$, it is easily seen from (\ref{tortoise_horizon}) that $x$ approaches the point

$$
\delta = \frac{2\pi i}{f'(R_0^-)} = \frac{\pi i}{k^-},
$$

\noindent
where

$$
k^- = \frac12 f'(R_0^-)
$$

\noindent
is the surface gravity at the inner horizon (notice that $k^- < 0$). Consequently, after going around the inner horizon, the solution will be of the form

$$
\Phi (x) \sim C_+ \sqrt{2\pi\omega (x-\delta)}\ J_{\frac{j}{2}} \left( \omega (x-\delta) \right) + C_- \sqrt{2\pi\omega (x-\delta)}\ J_{-\frac{j}{2}} \left( \omega (x-\delta) \right),
$$

\noindent
as one approaches the origin. Since $\omega(x-\delta)$ is negative on this branch, from the asymptotic expansion

\begin{equation}
J_\nu(z)=\sqrt{\frac{2}{\pi z}} \cos\left(z+\frac{\nu \pi}{2}+\frac{\pi}4 \right) , \qquad z \ll -1,  \label{Bessel-}
\end{equation}

\noindent
we have

\begin{eqnarray*}
\Phi (x) & \sim & 2 C_+ \cos \left( \omega (x-\delta) + \alpha_+ \right) + 2 C_- \cos \left( \omega (x - \delta) + \alpha_- \right)\\
& = & \left( C_+ e^{i\alpha_+} e^{-i\omega\delta} + C_- e^{i\alpha_-} e^{-i\omega\delta}\right) e^{i \omega x} + \left( C_+ e^{-i\alpha_+} e^{i\omega\delta} + C_- e^{-i\alpha_-} e^{i\omega\delta}\right) e^{-i \omega x} 
\end{eqnarray*}

\noindent
on this branch. This must be matched to the expression (\ref{RN_00}) for $\Phi(x)$, and hence we must have

\begin{eqnarray}
&& B_+ e^{3i\alpha_+} + B_- e^{3i\alpha_-} = C_+ e^{i\alpha_+} e^{-i\omega\delta} + C_- e^{i\alpha_-} e^{-i\omega\delta} \label{RN_2}\\
&& B_+ e^{5i\alpha_+} + B_- e^{5i\alpha_-} = C_+ e^{-i\alpha_+} e^{i\omega\delta} + C_- e^{-i\alpha_-} e^{i\omega\delta} \label{RN_3}
\end{eqnarray}

\noindent
Finally, we must rotate to the branch containing point $B$. Again $x-\delta$ rotates through an angle of $2\pi$, and since

$$
\sqrt{2\pi e^{2\pi i} \omega (x-\delta)}\ J_{\pm\frac{j}{2}} \left( e^{2\pi i} \omega (x-\delta) \right) = e^{\frac{2\pi i}2 (1 \pm j)}\sqrt{2\pi \omega (x-\delta)}\ J_{\pm\frac{j}{2}} \left( \omega (x-\delta) \right) \sim 2 e^{4i\alpha_\pm} \cos(\omega (x-\delta) + \alpha_\pm)
$$

\noindent
(as $\omega(x-\delta) < 0$ on the branch containing point $B$), one has

\begin{eqnarray*}
\Phi (x) & \sim & 2 C_+ e^{4i\alpha_+} \cos \left( \omega (x-\delta) + \alpha_+ \right) + 2 C_- e^{4i\alpha_-} \cos \left( \omega (x-\delta) + \alpha_- \right) \\
& = & \left( C_+ e^{5i\alpha_+} e^{-i\omega\delta} + C_- e^{5i\alpha_-} e^{-i\omega\delta}\right) e^{i \omega x} + \left( C_+ e^{3i\alpha_+} e^{i\omega\delta} + C_- e^{3i\alpha_-}e^{i\omega\delta}\right) e^{-i \omega x}
\end{eqnarray*}

\noindent
on that branch. As one closes the contour near $r \sim \infty$, one has $x \sim r$ and hence $\re(x)>0$; since $\im(\omega) \gg 0$, it follows that $e^{i\omega x}$ is exponentially small in this part of the contour, and therefore only the coefficient of $e^{-i\omega x}$ should be trusted. As one follows the contour, this coefficient is multiplied by

$$
\frac{C_+ e^{3i\alpha_+} e^{i\omega\delta} + C_- e^{3i\alpha_-} e^{i\omega\delta}}{B_+ e^{i\alpha_+} + B_- e^{i\alpha_-}}.
$$

\noindent
On the other hand, the monodromy of $e^{-i\omega x}$ going clockwise around this contour is $e^{-\frac{\pi\omega}{k^+}}$. 

We can now deform the contour to a small clockwise circle around the black hole event horizon. Near the outer horizon $r=R_0^+$, we have again $V\sim 0$, and hence again

$$
\Phi(x) \sim D_+ e^{i \omega x} + D_- e^{-i \omega x}.
$$

\noindent
The condition for quasinormal modes at the horizon is therefore $D_- = 0$. Using (\ref{tortoise_horizon}), we restate this condition as the condition that the monodromy of $\Phi$ going clockwise around the contour should be

$$
e^{i\omega \left(-\frac{2\pi i}{f'(R_0^+)}\right)} = e^{\frac{\pi\omega}{k^+}},
$$

\noindent
where

$$
k^+ = \frac12 f'(R_0^+)
$$

\noindent
is the surface gravity at the outer horizon. Therefore, one finally concludes that  the condition for quasinormal modes at the horizon is

\begin{equation}\label{RN_4}
\frac{C_+ e^{3i\alpha_+} e^{i\omega\delta} + C_- e^{3i\alpha_-} e^{i\omega\delta}}{B_+ e^{i\alpha_+} + B_- e^{i\alpha_-}} e^{-\frac{\pi\omega}{k^+}} = e^{\frac{\pi\omega}{k^+}}.
\end{equation}

The final condition for equations (\ref{RN_1}), (\ref{RN_2}), (\ref{RN_3}) and (\ref{RN_4}), which reflect quasinormal mode boundary conditions both at infinity and at the black hole event horizon, to have nontrivial solutions $(B_+,B_-,C_+,C_-)$ is

\begin{eqnarray} \label{resultsRN}
&&
\left| 
\begin{array}{ccccccc}
e^{-i\alpha_+} & \ & e^{-i\alpha_-} & \ & 0 & \ & 0 \\
e^{3i\alpha_+} & \ & e^{3i\alpha_-} & \ & - e^{i\alpha_+} e^{-i\omega\delta} & \ &  - e^{i\alpha_-} e^{-i\omega\delta} \\
e^{5i\alpha_+} & \ & e^{5i\alpha_-} & \ & - e^{-i\alpha_+} e^{i\omega\delta} & \ &  - e^{-i\alpha_-} e^{i\omega\delta} \\
e^{i\alpha_+} e^{\frac{2\pi\omega}{k^+}} & \ &  e^{i\alpha_-} e^{\frac{2\pi\omega}{k^+}} & \ & -e^{3i\alpha_+} e^{i\omega\delta} & \ & -e^{3i\alpha_-} e^{i\omega\delta}
\end{array}
\right| = 0 \quad \Leftrightarrow \quad \left| 
\begin{array}{ccccccc}
e^{-2i\alpha_+} & \ & e^{-2i\alpha_-} & \ & 0 & \ & 0 \\
e^{2i\alpha_+} & \ & e^{2i\alpha_-} & \ & e^{\frac{2\pi\omega}{k^-}} & \ & e^{\frac{2\pi\omega}{k^-}} \\
e^{4i\alpha_+} & \ & e^{4i\alpha_-} & \ & e^{-2i\alpha_+} & \ & e^{-2i\alpha_-} \\
e^{\frac{2\pi\omega}{k^+}} & \ & e^{\frac{2\pi\omega}{k^+}} & \ & e^{2i\alpha_+} & \ & e^{2i\alpha_-} 
\end{array}
\right| = 0 \nonumber \\
&& \Leftrightarrow \quad
(1-\cos(\pi j)) e^{\frac{2\pi\omega}{k^+}} + (\cos(\pi j)-\cos(2\pi j)) + (1-\cos(2\pi j)) e^{-\frac{2\pi\omega}{k^-}} = 0 \nonumber \\
&& \Leftrightarrow \quad
e^{\frac{2\pi\omega}{k^+}} = - (1 + 2\cos(\pi j)) - ( 2 + 2\cos(\pi j)) e^{-\frac{2\pi\omega}{k^-}}.
\end{eqnarray}

\noindent
As in the Schwarzschild case, the fact that we imposed condition (\ref{RN_inf}) at point point $A$ means that equation (\ref{resultsRN}) yields the asymptotic quasinormal modes $\omega$ with $\re(\omega)>0$. The remaining modes are of the form $-\overline{\omega}$, and could be obtained by imposing condition (\ref{RN_inf}) at point point $B$.

For vector type perturbations, the potential near the origin is of the form

$$
V \big[ r(x) \big]  \sim  \frac{j^2 - 1}{4x^2},
$$

\noindent
with $j=\frac{3d-7}{2d-5}$. Repeating the same argument, one ends up with the same equation (\ref{resultsRN}) for this new value of $j$. Since 

$$
\frac{3d-7}{2d-5} = 2 - \frac{d-3}{2d-5},
$$

\noindent
we see that this equation is exactly the same as in the $j=\frac{d-3}{2d-5}$ case, and consequently we end up with the same quasinormal modes.

One may wonder if it is possible to recover the Schwarzschild quasinormal frequencies from our result above. This would arise from the $\QQ \to 0$ limit of (\ref{resultsRN}). In this limit we have $R_{0}^{-} \to 0$ and hence

$$
k^{-} \sim - \frac{(d-3)\MM}{\left( R_0^- \right)^{d-2}} \to - \infty,
$$

\noindent
while

$$
k^{+} \sim \frac{(d-3)\MM}{\left( R_0^+ \right)^{d-2}}
$$

\noindent
remains fixed. In this case it is simple to check that the RN quasinormal equation reduces to 

$$
e^{\frac{2\pi\omega}{k^+}} = - \left( 3 + 4 \cos \left( \pi j \right) \right),
$$

\noindent
which is \textit{not} the quasinormal equation for the Schwarzschild spacetime. For example, in $d=4$, this formula yields $e^{\frac{2\pi\omega}{k^+}} = - 5$. The reason why this is so traces back to the Stokes line. Comparing the Stokes lines of the Schwarzschild and RN spacetimes one immediately realizes that there is a change in the topology of this curve at the origin, thus invalidating the limit above. Another limit one could try to obtain concerns the quasinormal frequencies of the extremal RN black hole, when $\QQ \to \MM$. This time around the computation is slightly trickier and in the end one will obtain the equation in the extremal limit as

$$
e^{\frac{2\pi\omega}{k}} = - \left( 2 + 2 \cos \left( \pi j \right) \right),
$$

\noindent
where we have defined

$$
k = \frac{(d-3)^2}{2(d-2)R_0}
$$

\noindent
with $R_0 = \MM^{\frac{1}{d-3}}$. As we shall see in the following this equation yields the wrong quasinormal frequencies for the extremal RN spacetime (although it actually predicts the correct gap). In this case there is no change in the topology of the Stokes line, but it turns out that one \textit{cannot} deform the monodromy contour of the RN black hole into the monodromy contour of the extremal RN black hole without crossing horizon singularities in the complex plane, thus invalidating the limit above. Interestingly enough, the same parameter $k$ as above will also appear when performing the extremal RN calculation from scratch. The lesson one should retain is that in many cases one must be careful when taking limits of the final equation for the asymptotic quasinormal frequencies.

Let us briefly review the literature concerning asymptotic quasinormal frequencies in the RN spacetime, as we would like to compare our results to what has been previously accomplished on this subject. Quasinormal frequencies for near extremal RN solutions were first studied numerically in \cite{andersson-onozawa}, focusing on the four dimensional case. The highly damped modes were found to have a peculiar behavior, as black holes with different charge apparently shared a specific mode frequency, and some of the modes seemed to be multivalued. The first analytic solution, in four dimensions and for non--extremal RN solutions, was given in \cite{motl-neitzke, neitzke}. These authors found a formula (which we have generalized above) for the asymptotic quasinormal frequencies, but they have not managed to find its general solution. Also, simple numeric analysis seemed to point towards the peculiarities previously found. This is in full agreement with what we have found, and just seems to be pointing in the direction that there is no simple solution for the frequencies in this case, unlike the previous Schwarzschild solution. Numerical checks to the formula in \cite{motl-neitzke} were performed in \cite{berti-kokkotas-2, andersson-howls}. \cite{berti-kokkotas-2} again found that very highly damped quasinormal modes of RN black holes have an oscillatory behavior as a function of the charge. \cite{andersson-howls} found that RN asymptotic quasinormal frequencies are typically \textit{not} periodic in $\im (\omega)$ (in contrast with the Schwarzschild black hole case). They also understood why the $\QQ \to 0$ limit of RN quasinormal frequencies does not yield Schwarzschild quasinormal frequencies: the limit is a singular one as it involves topology change at the level of the contours in the complex plane (the same thing happening in the attempt to go extremal, $\QQ^{2} \to \MM^{2}$, as we shall see in detail in the following). In summary, our results completely match what was previously known in the literature. However, it would be very interesting to produce further numerical data concerning higher dimensional RN quasinormal modes to match against all our analytical results.


\subsubsection{The Extremal Reissner--Nordstr\"om Solution}


We now compute the asymptotic quasinormal modes of the $d$--dimensional extremal RN  black hole using the monodromy method. This example is illustrative of the fact that the monodromy calculation is very sensitive to the location of the singularities in the complex plane, as well as to the topology of the Stokes line at the origin. Thus, when taking limits of the parameters, one should always have in mind whether one is crossing singularities or changing the topology of the contour. In an affirmative case, the limit on the parameters will not be valid and one will have to address the calculation from scratch.

Again we consider solutions of the Schr\"odinger--like equation (\ref{schrodinger}) in the complex $r$--plane. Since the potential $V$ vanishes for $r \sim +\infty$, we will have

$$
\Phi(x) \sim A_+ e^{i \omega x} + A_- e^{-i \omega x}
$$

\noindent
in this region. The boundary condition for quasinormal modes at infinity is then

\begin{equation}\label{RNE_inf}
A_+=0.
\end{equation}

\noindent
Next we study the behavior of $\Phi(x)$ near the singularity $r=0$. In this region, the tortoise coordinate is still

$$
x \sim \frac{r^{2d-5}}{(2d-5)\QQ^2},
$$

\noindent
and the potential for scalar type and tensor type perturbations is still

$$
V \big[ r(x) \big] \sim \frac{j^2 - 1}{4x^2},
$$

\noindent
with $j=\frac{d-3}{2d-5}$ (see appendix \ref{appendixC}). Therefore the solution of the Schr\"odinger--like equation in this region approximates

$$
\Phi (x) \sim B_+ \sqrt{2\pi\omega x}\ J_{\frac{j}{2}} \left( \omega x \right) + B_- \sqrt{2\pi\omega x}\ J_{-\frac{j}{2}} \left( \omega x \right),
$$

\noindent
where $J_\nu(x)$ represents a Bessel function of the first kind and $B_\pm$ are (complex) integration constants.

For the asymptotic quasinormal modes one has $\im (\omega) \gg \re (\omega)$, and hence $\omega$ is approximately purely imaginary. Consequently, one has $\omega x \in \mathbb{R}$ for $x \in i\mathbb{R}$; in a neighborhood of the origin, this happens for

$$
r=\rho\ e^{\frac{i\pi}{2(2d-5)}+\frac{in\pi}{2d-5}},
$$

\noindent
with $\rho>0$ and $n=0,1,\ldots, 4d-11$. These are half--lines starting at the origin, equally spaced by an angle of $\frac{\pi}{2d-5}$. Notice that the sign of $\omega x$ on these lines is $(-1)^{n+1}$; in other words, starting with the line corresponding to $n=0$, the sign of $\omega x$ is alternately negative and positive as one goes anti--clockwise around the origin.

From the asymptotic expansion (\ref{Bessel+}) we see that

\begin{eqnarray}\label{RNE_0}
\Phi (x) & \sim & 2 B_+ \cos \left( \omega x - \alpha_+ \right) + 2 B_- \cos \left( \omega x - \alpha_- \right) \nonumber \\
& = & \left( B_+ e^{-i\alpha_+} + B_- e^{-i\alpha_-}\right) e^{i \omega x} + \left( B_+ e^{i\alpha_+} + B_- e^{i\alpha_-}\right) e^{-i \omega x},
\end{eqnarray}

\noindent
in any one of the lines corresponding to positive $\omega x$, where again

$$
\alpha_\pm = \frac{\pi}4 (1 \pm j).
$$

\noindent
This asymptotic expression for $\Phi$ near the origin is to be matched with its asymptotic expression at infinity. This matching must again be done along the Stokes line $\omega x \in \mathbb{R}$ (that is, $x \in i\mathbb{R}$), so that neither of the exponentials $e^{\pm i\omega x}$ dominates the other.

To trace out the Stokes line $\re(x)=0$ we observe that we already know its behavior near the origin. Furthermore, this is the only singular point of this curve: indeed, since $x$ is a holomorphic function of $r$, the critical points of the function $\re(x)$ are the zeros of

$$
\frac{dx}{dr} = \frac{1}{f(r)} = \frac{r^{2d-6}}{\left(r^{d-3}-\QQ\right)^2}
$$

\noindent
(\textit{i.e.}, $r=0$ only). We have the additional problem that $x$ is a multivalued function: each of the ``horizons''

$$
r=R_n, \qquad n=0, 1, \ldots, d-4
$$

\noindent
(see appendix \ref{appendixC}) is a branch point. From (\ref{tortoise_horizon}) we see that although the function $\re(x)$ is still well defined around $R_k$ with $n=0$ and $n=\frac{d-3}{2}$ (if $d$ is odd), it will be multivalued around all the other fictitious horizons. In analogy with the non--extremal case, we are led to define

$$
k  = \frac{(d-3)^{2}}{2(d-2)R_0}
$$

For $r \sim \infty$ one has $f (r) \sim 1$ and hence $x \sim r$. Consequently $x$ is holomorphic at infinity, and we can choose the branch cuts to cancel out among themselves. Therefore $\re(x)$ is well defined in a neighborhood of infinity, and moreover $\re(x)=0$ will be approximately parallel to $\re(r)=0$ in this neighborhood. Two of the $2(2d-5)$ branches of the Stokes line starting out at the origin must therefore be unbounded. The remaining can either connect to another branch or end up in a branch cut. On the other hand, it is easy to see that the Stokes line must intersect the positive real axis exactly in two points, the horizon $r=R_0$ and a point on the interval $(R_0, +\infty)$. Using this information plus elementary considerations on symmetry and the sign of $\re(x)$, one can deduce that the Stokes line must be of the form indicated in Figure \ref{StokesERN}. These results are moreover verified by the numerical computation of the same Stokes lines, as indicates in Figure \ref{NumericalStokesERN}.

\FIGURE[ht]{\label{StokesERN}
	\centering
	\psfrag{A}{$A$}
	\psfrag{B}{$B$}
	\psfrag{r0}{$R_0$}
	\psfrag{r1}{$R_1$}
	\psfrag{r2}{$R_2$}
	\psfrag{Re}{$\re$}
	\psfrag{Im}{$\im$}
	\psfrag{contour}{contour}
	\psfrag{Stokes line}{Stokes line}
	\epsfxsize=.6\textwidth
	\leavevmode
	\epsfbox{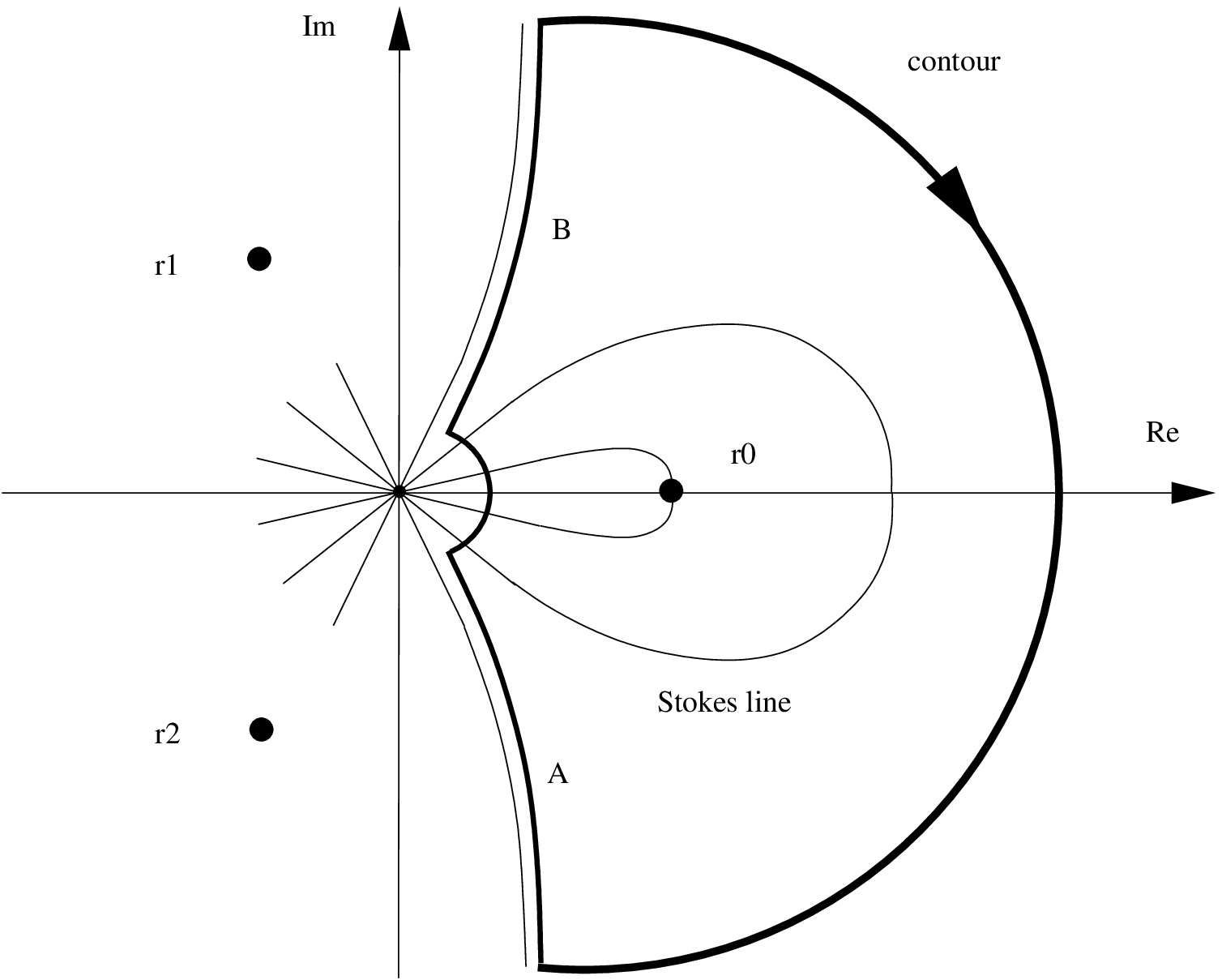}
\caption{Stokes line for the extremal Reissner--Nordstr\"om black hole, along with the chosen contour for monodromy matching, in the case $d=6$.}
}

\FIGURE[ht]{\label{NumericalStokesERN}
	\centering
	\epsfig{file=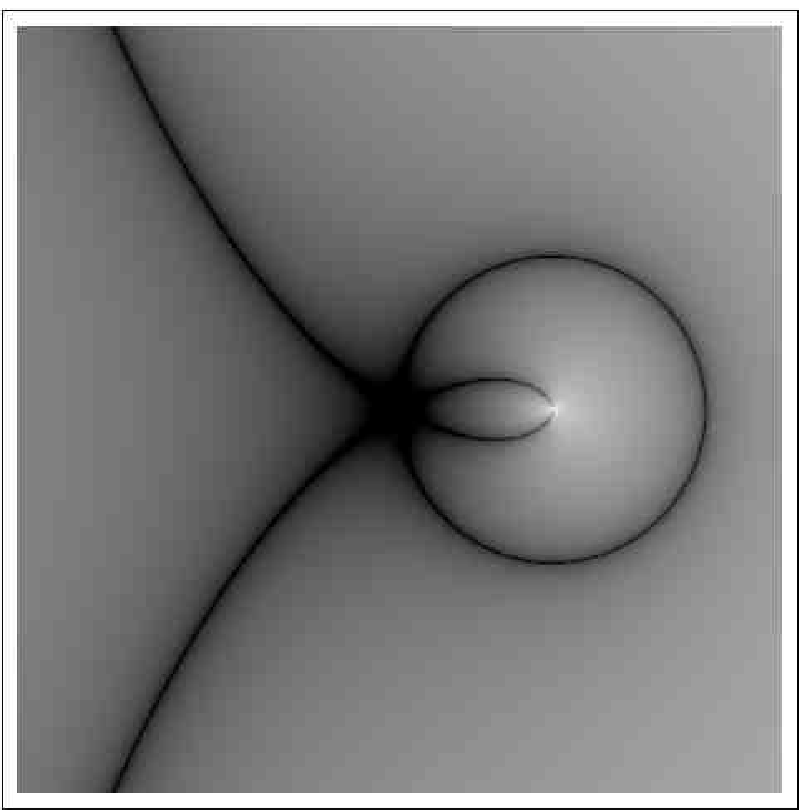, width=1.5in, height=1.5in}
    $\quad$
    \epsfig{file=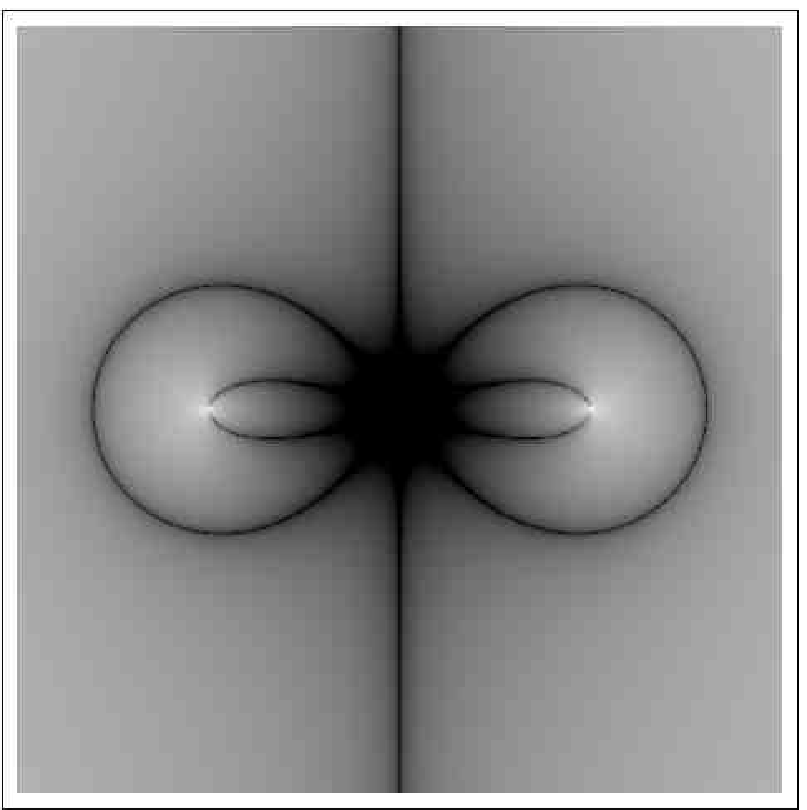, width=1.5in, height=1.5in}
    $\quad$
    \epsfig{file=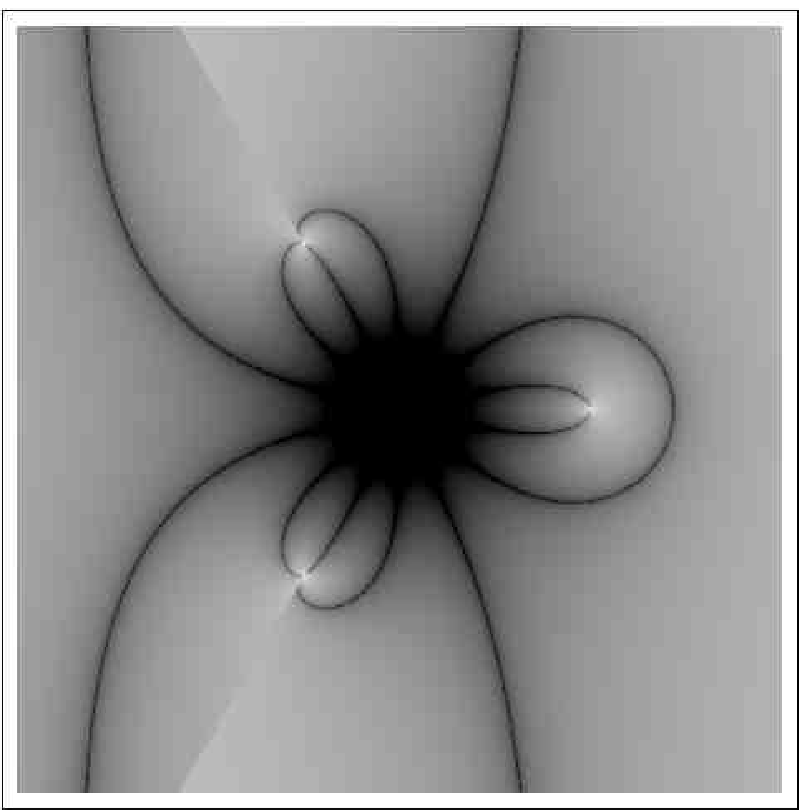, width=1.5in, height=1.5in}
    $\quad$
    \epsfig{file=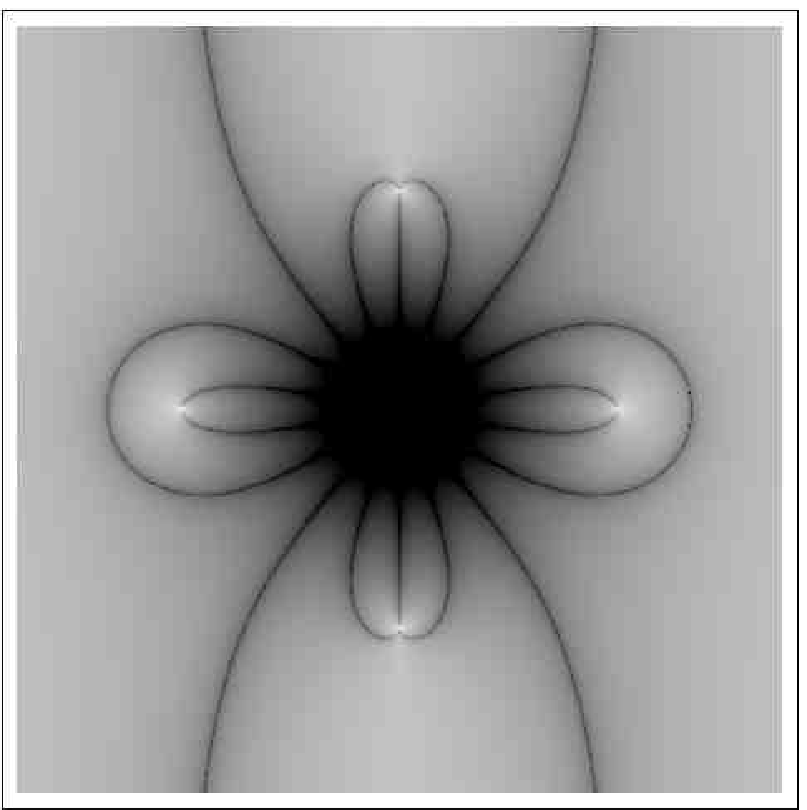, width=1.5in, height=1.5in}
	\caption{Numerical calculation of the Stokes lines for the extremal  Reissner--Nordstr\"om black hole in dimensions $d=4$, $d=5$, $d=6$ and $d=7$. Different shadings also illustrate the various horizon singularities and branch cuts.}
}

We consider the contour obtained by closing the unlimited portions of the Stokes line near $r\sim \infty$, as shown in Figure \ref{StokesERN}. At point $A$ we have $\omega x>0$, and therefore the expansion (\ref{RNE_0}) holds at this point. Imposing condition (\ref{RNE_inf}) one obtains

\begin{equation}\label{RNE_1}
B_+ e^{-i\alpha_+} + B_- e^{-i\alpha_-} = 0.
\end{equation}

\noindent
For $z \sim 0$ one has the expansion (\ref{Bessel0}). Consequently, as on rotates from the branch containing point $A$ to the branch containing point $B$ (through an angle of $\frac{5\pi}{2d-5}$), $x\sim \frac{r^{2d-5}}{2d-5}$ rotates through an angle of $5\pi$, and since

$$
\sqrt{2\pi e^{5\pi i} \omega x}\ J_{\pm\frac{j}{2}} \left( e^{5\pi i} \omega x \right) = e^{\frac{5\pi i}2 (1 \pm j)}\sqrt{2\pi \omega x}\ J_{\pm\frac{j}{2}} \left( \omega x \right) \sim 2 e^{10i\alpha_\pm} \cos(\omega x - \alpha_\pm),
$$

\noindent
one has (notice that $e^{5\pi i} \omega x=-\omega x$)

\begin{eqnarray*}
\Phi (x) & \sim & 2 B_+ e^{10i\alpha_+} \cos \left( -\omega x - \alpha_+ \right) + 2 B_- e^{10i\alpha_-} \cos \left( -\omega x - \alpha_- \right) \nonumber \\
& = & \left( B_+ e^{11i\alpha_+} + B_- e^{11i\alpha_-}\right) e^{i \omega x} + \left( B_+ e^{9i\alpha_+} + B_- e^{9i\alpha_-}\right) e^{-i \omega x}
\end{eqnarray*}

\noindent
at point $B$. As one closes the contour near $r \sim \infty$, one has $x \sim r$ and hence $\re(x)>0$; since $\im(\omega) \gg 0$, it follows that $e^{i\omega x}$ is exponentially small in this part of the contour, and therefore only the coefficient of $e^{-i\omega x}$ should be trusted. As one follows the contour, this coefficient is multiplied by

$$
\frac{B_+ e^{9i\alpha_+} + B_- e^{9i\alpha_-}}{B_+ e^{i\alpha_+} + B_- e^{i\alpha_-}}.
$$

\noindent
On the other hand, the monodromy of $e^{-i\omega x}$ going clockwise around this contour is $e^{-\frac{\pi\omega}{k}}$. We can now deform the contour to a small clockwise circle around the black hole event horizon. Near the horizon $r = R_0$, we have again $V \sim 0$, and hence again

$$
\Phi(x) \sim C_+ e^{i \omega x} + C_- e^{-i \omega x}.
$$

\noindent
The condition for quasinormal modes at the horizon is therefore $C_- = 0$. We restate this condition as the condition that the monodromy of $\Phi$ going clockwise around the contour should be $e^{\frac{\pi\omega}{k}}$. Therefore the condition for quasinormal modes at the horizon is

\begin{equation}\label{RNE_2}
\frac{B_+ e^{9i\alpha_+} + B_- e^{9i\alpha_-}}{B_+ e^{i\alpha_+} + B_- e^{i\alpha_-}} e^{-\frac{\pi\omega}{k}} = e^{\frac{\pi\omega}{k}}.
\end{equation}

\noindent
The final condition for equations (\ref{RNE_1}), (\ref{RNE_2}), reflecting quasinormal boundary conditions, to have nontrivial solutions $(B_+,B_-)$ is

\begin{equation} \label{RNEdeterminant}
\left| 
\begin{array}{ccc}
e^{-i\alpha_+} & \ & e^{-i\alpha_-} \\ & & \\
e^{9i\alpha_+} - e^{i\alpha_+} e^{\frac{2\pi\omega}{k}}  & \ & e^{9i\alpha_-} - e^{i\alpha_-} e^{\frac{2\pi\omega}{k}}
\end{array}
\right| = 0 \quad \Leftrightarrow \quad 
e^{\frac{2\pi\omega}{k}} = \frac{\sin\left(\frac{5\pi j}2\right)}{\sin\left(\frac{\pi j}2\right)}.
\end{equation}

\noindent
As before, the fact that we imposed condition (\ref{RNE_inf}) at point point $A$ means that equation (\ref{RNEdeterminant}) yields the asymptotic quasinormal modes $\omega$ with $\re(\omega)>0$. The remaining modes are of the form $-\bar{\omega}$, and could have been obtained by imposing condition (\ref{RNE_inf}) at point $B$. This result also illustrates the fact, which we have advertised earlier, that the extremal limit of the RN quasinormal equation does \textit{not} yield the quasinormal equation for the extremal RN spacetime.

For vector type perturbations, the potential for the Schr\"odinger--like equation near the origin is of the form

$$
V \big[ r(x) \big] \sim \frac{j^2 - 1}{4x^2},
$$

\noindent
with $j=\frac{3d-7}{2d-5}$. Repeating the same argument, one ends up with the same equation (\ref{RNEdeterminant}) for this new value of $j$. Since 

$$
\frac{3d-7}{2d-5} = 2 - \frac{d-3}{2d-5},
$$

\noindent
we see that this equation is exactly the same as in the $j=\frac{d-3}{2d-5}$ case, and consequently we end up with the same the quasinormal modes.

As one reviews the literature searching for earlier numerical checks on our results, one observes that there has not been much work on this matter. Quasinormal frequencies for extremal RN solutions were studied numerically in \cite{omoi}, focusing on the four dimensional case. However, the interest of that paper lied in the least damped modes (opposite to what we want). The authors found a coincidence in quasinormal frequencies for gravitational perturbations with multipole index $\ell$ and those for electromagnetic perturbations with index $\ell-1$ (at extremality). This fact was later explained in \cite{oomi} as arising from the [hidden] supersymmetry of the extremal solution. So, in this extremal limit, the black hole responds to the perturbation of each field in the same manner. This is certainly in agreement with our results, but as we have seen for tensor, vector and scalar type perturbations this is actually a very generic feature. An attempt at computing asymptotic quasinormal frequencies for the extremal RN black hole, using the monodromy method of \cite{motl-neitzke}, was recently performed in \cite{das-shankaranarayanan}. Unfortunately, these authors have misidentified the topology of the tortoise coordinate at the origin and have thus wrongly computed the monodromies for their black holes, obtaining an incorrect result for the quasinormal frequencies. An attempt at computing numerically quasinormal frequencies of the $d=4$ extremal RN black hole was recently done in \cite{berti}, but unfortunately the numerical method was not stable enough to find the asymptotic values for the frequencies. It would thus be very interesting to produce further numerical data concerning extremal RN quasinormal modes in order to match against our analytical results.


\subsection{Asymptotically de Sitter Spacetimes}


\cite{kodama-ishibashi-3} discusses the stability of black holes in asymptotically dS spacetimes to tensor, vector and scalar perturbations. For black holes without charge, tensor and vector perturbations are stable in any dimension. Scalar perturbations are stable up to dimension six but there is no proof of stability in dimension $d \ge 7$. For charged black holes, tensor and vector perturbations are stable in any dimension. Scalar perturbations are stable in four and five dimensions but there is no proof of stability in dimension $d \ge 6$. As we work in generic dimension $d$ we are thus not guaranteed to always have a stable solution. Our results will apply if and only if the spacetime in consideration is stable.

Quantization of a scalar field in dS space was first addressed in \cite{gibbons-hawking}. While these authors found that the cosmological event horizon is stable, they also found that there is an isotropic background of thermal radiation. The emitted particles are, however, observer dependent, as is the ``cosmological sphere'' of dS (we refer the reader to the recent review \cite{ssv}, for more details and a list of current open problems). Analysis of the wave equation in dS spacetime also led to the natural boundary conditions on quasinormal modes: incoming waves at the black hole horizon and outgoing waves at the cosmological horizon.


\subsubsection{The Schwarzschild de Sitter Solution}


We now compute the quasinormal modes of the Schwarzschild dS $d$--dimensional black hole. This calculation was first done in \cite{cns} for the particular case of $d=4$. Again we consider solutions of the Schr\"odinger--like equation (\ref{schrodinger}) in the complex $r$--plane. We start by studying the behavior of $\Phi(x)$ near the singularity $r=0$. In this region, the tortoise coordinate is

$$
x \sim - \frac{r^{d-2}}{2(d-2)\MM},
$$

\noindent
and the potential for tensor and scalar type perturbations is

$$
V \big[ r(x) \big] \sim \frac{j^2 - 1}{4x^2},
$$

\noindent
with $j=0$ (see appendix \ref{appendixC}). The solution of the Schr\"odinger--like equation in this region (with $j \neq 0$) is therefore

$$
\Phi (x) \sim B_+ \sqrt{2\pi\omega x}\ J_{\frac{j}{2}} \left( \omega x \right) + B_- \sqrt{2\pi\omega x}\ J_{-\frac{j}{2}} \left( \omega x \right),
$$

\noindent
where $J_\nu(x)$ represents a Bessel function of the first kind and $B_\pm$ are (complex) integration constants. 

For the asymptotic quasinormal modes one has $\im (\omega) \gg \re (\omega)$, with $\im (\omega) \to + \infty$, and hence $\omega$ is approximately purely imaginary. Consequently, $\omega x \in \mathbb{R}$ for $x \in i\mathbb{R}$; in a neighborhood of the origin, the above relation between $x$ and $r$ tells us that this happens for

$$
r=\rho\ e^{\frac{i\pi}{2(d-2)}+\frac{in\pi}{d-2}},
$$

\noindent
with $\rho>0$ and $n=0,1,\ldots, 2d-5$. These are half--lines starting at the origin, equally spaced by an angle of $\frac{\pi}{d-2}$. Notice that the sign of $\omega x$ on these half--lines is $(-1)^n$; in other words, starting with the half--line corresponding to $n=0$, the sign of $\omega x$ is alternately positive and negative as one goes anti--clockwise around the origin.

From the asymptotic expansion (\ref{Bessel+}) we see that 

\begin{eqnarray}
\Phi (x) & \sim & 2 B_+ \cos \left( \omega x - \alpha_+ \right) + 2 B_- \cos \left( \omega x - \alpha_- \right) \nonumber \\
& = & \left( B_+ e^{-i\alpha_+} + B_- e^{-i\alpha_-}\right) e^{i \omega x} + \left( B_+ e^{i\alpha_+} + B_- e^{i\alpha_-}\right) e^{-i \omega x} \label{SdS_0}
\end{eqnarray}

\noindent
in any one of the lines corresponding to positive $\omega x$, where again

$$
\alpha_\pm = \frac{\pi}4 (1 \pm j).
$$

\noindent
We shall use this asymptotic expression to make the monodromy matching. This matching must be done along the Stokes line $\omega x \in \mathbb{R}$ (or $x \in i\mathbb{R}$), so that neither of the exponentials $e^{\pm i\omega x}$ dominates the other.

To trace out the Stokes line $\re(x)=0$ let us first observe that we already know its behavior near the origin. Furthermore, this is the only singular point of this curve: indeed, since $x$ is a holomorphic function of $r$, the critical points of the function $\re(x)$ are the zeros of

$$
\frac{dx}{dr} = \frac{1}{f(r)} = \frac{r^{d-3}}{-\LL r^{d-1}+r^{d-3}-2\MM}
$$

\noindent
(\textit{i.e.}, $r=0$ only). We have an additional problem that $x$ is a multivalued function: each of the ``horizons'' $R_{n}$ ($n=1,\ldots,d-1$) is a branch point. From (\ref{tortoise_horizon}) we see that although the function $\re(x)$ is still well defined around the real horizons, it will be multivalued around all the other fictitious horizons.

For $r \sim \infty$ one has $x \sim x_0 + \frac{1}{\LL r}$ (see appendix \ref{appendixC}). Thus, $x$ is holomorphic at infinity, and we can choose the branch cuts to cancel out among themselves. Therefore $\re(x)$ is well defined in a neighborhood of infinity, and moreover $\re(x)=0$ cannot extend out to infinity as, generically, $x_0$ is not real. Thus the Stokes line branches starting at the origin must either connect to another branch or end up in a branch cut (and one would expect them to connect in the region where $\re(x)$ is well defined). On the other hand, it is easy to see that the Stokes line must intersect the positive real axis exactly in two points: one between the black hole horizon $R_H$ and the cosmological horizon $R_C$, the other between the cosmological horizon and infinity. Using this information plus elementary considerations on symmetry and the sign of $\re(x)$, one can deduce that the Stokes line must be of the form indicated in Figure \ref{StokesSdS}. These results are moreover verified by the numerical computation of the same Stokes lines, as indicated in Figure \ref{NumericalStokesSdS}.

\FIGURE[ht]{\label{StokesSdS}
	\centering
	\psfrag{A}{$A$}
	\psfrag{B}{$B$}
	\psfrag{RH}{$R_H$}
	\psfrag{RC}{$R_C$}
	\psfrag{R}{$\widetilde{R}$}
	\psfrag{g}{$\gamma$}
	\psfrag{og}{$\overline{\gamma}$}
	\psfrag{Re}{$\re$}
	\psfrag{Im}{$\im$}
	\psfrag{contour}{contour}
	\psfrag{Stokes line}{Stokes line}
	\epsfxsize=.6\textwidth
	\leavevmode
	\epsfbox{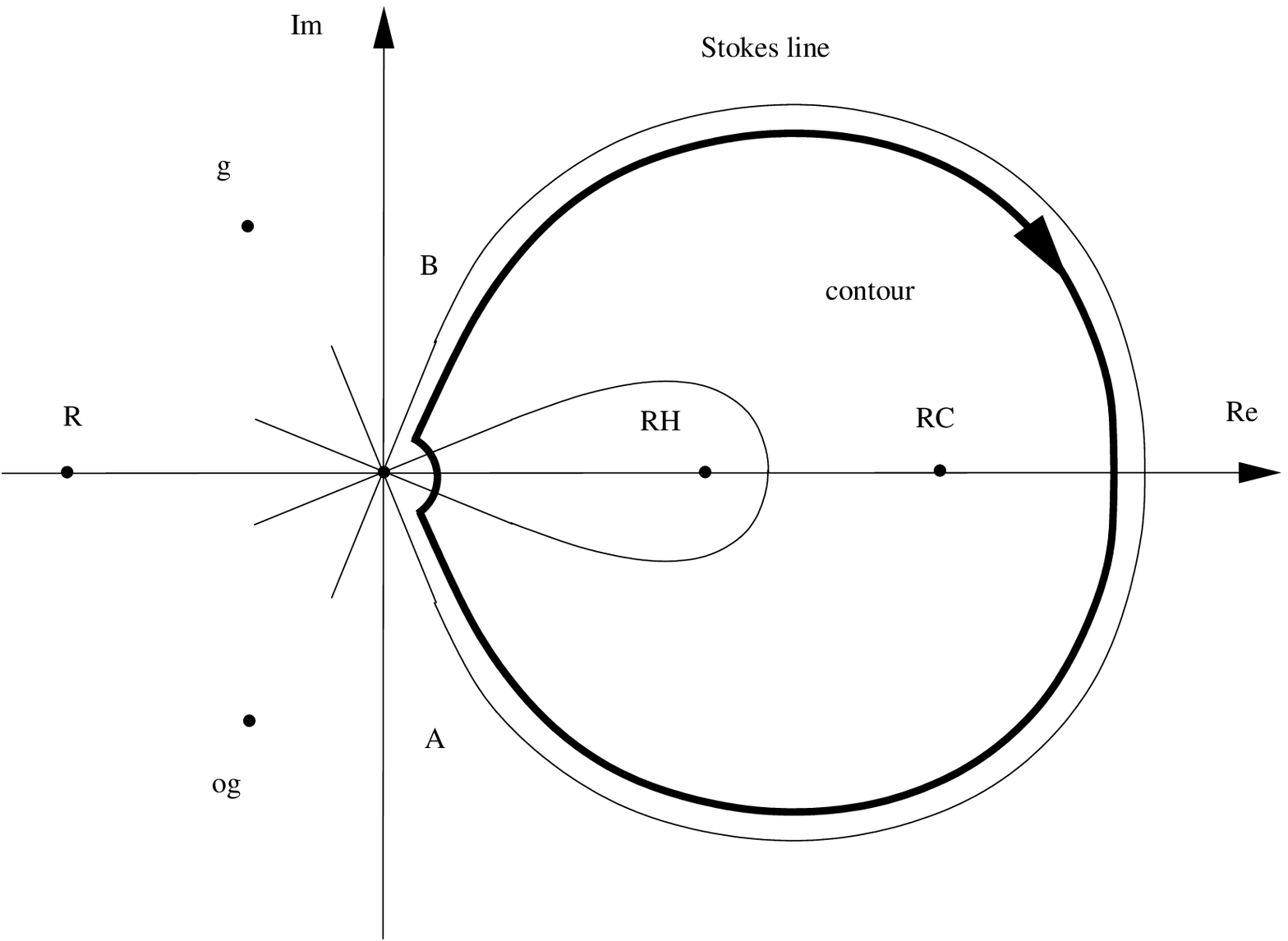}
\caption{Stokes line for the Schwarzschild de Sitter black hole, along with the chosen contour for monodromy matching, in the case of dimension $d=6$.}
}

\FIGURE[ht]{\label{NumericalStokesSdS}
	\centering
	\epsfig{file=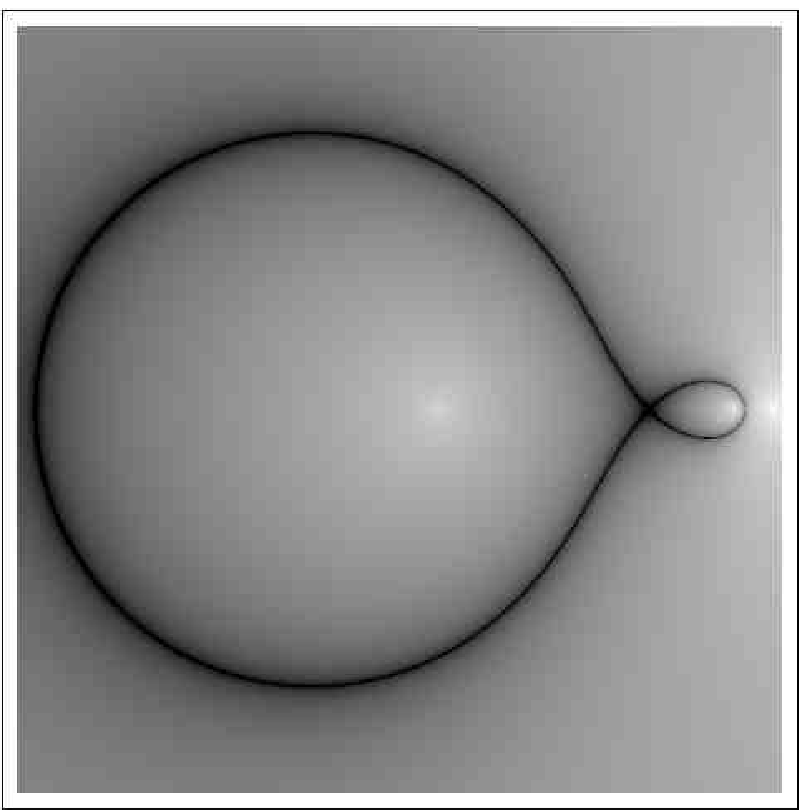, width=1.5in, height=1.5in}
    $\quad$
    \epsfig{file=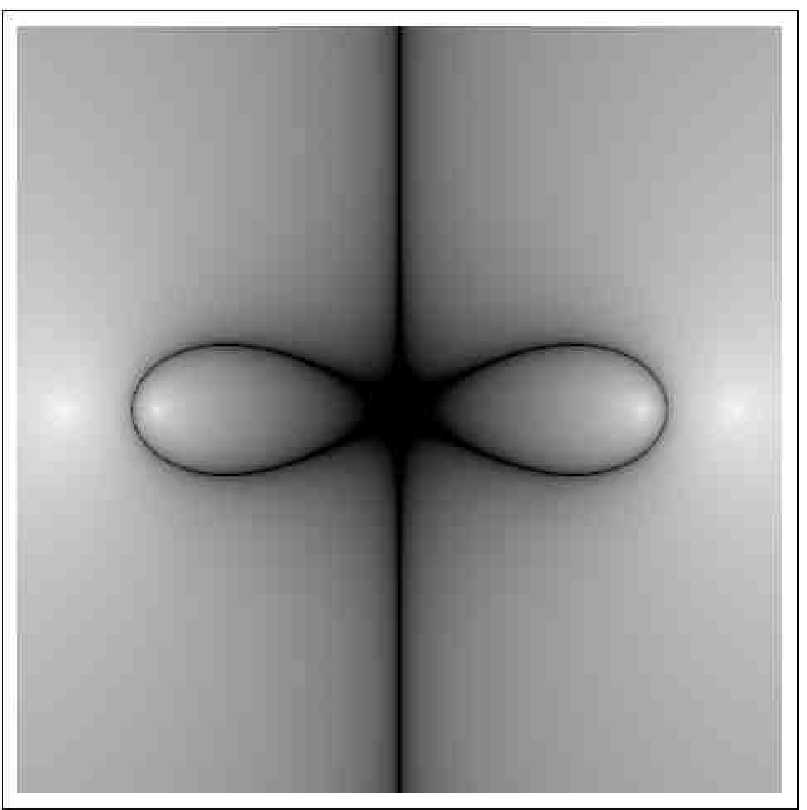, width=1.5in, height=1.5in}
    $\quad$
    \epsfig{file=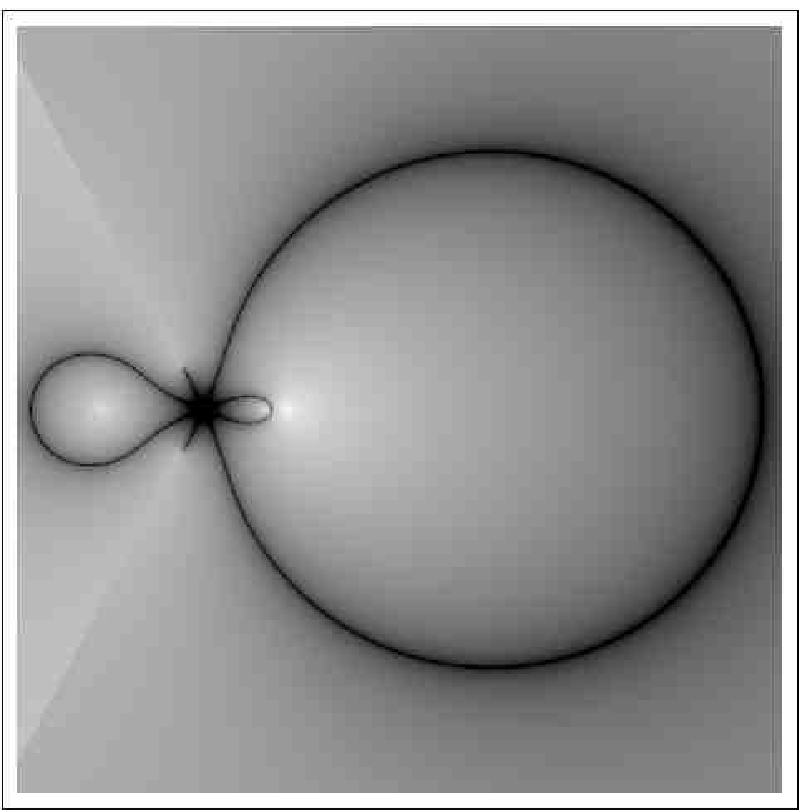, width=1.5in, height=1.5in}
    $\quad$
    \epsfig{file=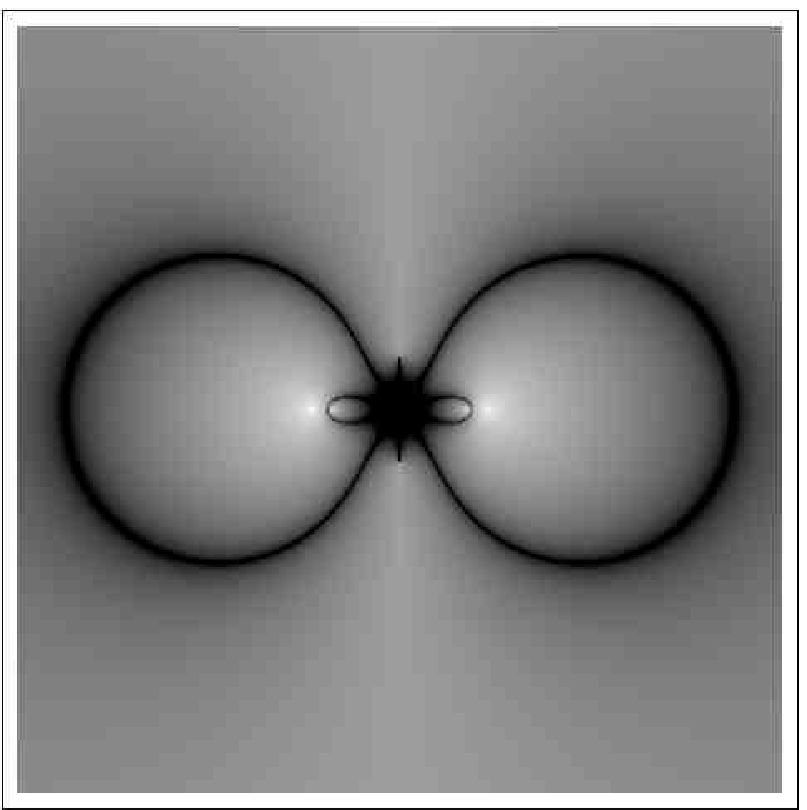, width=1.5in, height=1.5in}
\caption{Numerical calculation of the Stokes lines for the Schwarzschild de Sitter black hole in dimensions $d=4$, $d=5$, $d=6$ and $d=7$. The different shadings also illustrate the various horizon singularities and branch cuts (note that these branch cuts are not necessarily equal to the ones used for the calculation in the main text).}
}

Let us now consider the contour defined by the closed part of the Stokes line (contained in the region  where $\re(x)$ is well defined), as shown in Figure \ref{StokesSdS}. At point $A$ we have $\omega x \gg 1$, and therefore the expansion (\ref{SdS_0}) holds at this point. For $z \sim 0$ one has the expansion (\ref{Bessel0}). Consequently, as one rotates from the branch containing point $A$ to the branch containing point $B$, through an angle of $\frac{3\pi}{d-2}$, $x\sim - \frac{r^{d-2}}{2(d-2)\MM}$ rotates through an angle of $3\pi$, and since

$$
\sqrt{2\pi e^{3\pi i} \omega x}\ J_{\pm\frac{j}{2}} \left( e^{3\pi i} \omega x \right) = e^{\frac{3\pi i}2 (1 \pm j)}\sqrt{2\pi \omega x}\ J_{\pm\frac{j}{2}} \left( \omega x \right) \sim 2 e^{6i\alpha_\pm} \cos(\omega x - \alpha_\pm),
$$

\noindent
one has at point $B$ (notice that $e^{3\pi i} \omega x=-\omega x$)

\begin{eqnarray}\label{dsphi}
\Phi (x) & \sim & 2 B_+ e^{6i\alpha_+} \cos \left( -\omega x - \alpha_+ \right) + 2 B_- e^{6i\alpha_-} \cos \left( -\omega x - \alpha_- \right) \nonumber \\
& = & \left( B_+ e^{7i\alpha_+} + B_- e^{7i\alpha_-}\right) e^{i \omega x} + \left( B_+ e^{5i\alpha_+} + B_- e^{5i\alpha_-}\right) e^{-i \omega x}.
\end{eqnarray}

If $\Phi$ corresponds to a quasinormal mode, its clockwise monodromy around $r=R_H$ must be the same as the clockwise monodromy of $e^{i\omega x}$, that is, $e^{\frac{\pi \omega}{k_H}}$, where

$$
k_H = \frac12 f'(R_H)
$$

\noindent
is the surface gravity at the black hole horizon. Similarly, its clockwise monodromy around $r=R_C$ must be the same as the clockwise monodromy of $e^{-i\omega x}$, that is, $e^{-\frac{\pi \omega}{k_C}}$, where

$$
k_C = \frac12 f'(R_C)
$$

\noindent
is the surface gravity at the cosmological horizon (notice that we are taking $k_C < 0$). Since these are the only singularities of $\Phi$ inside the contour, the monodromy of $\Phi$ around the contour must be

$$
e^{\frac{\pi \omega}{k_H}-\frac{\pi \omega}{k_C}}.
$$

\noindent
On the other hand, the clockwise monodromy of $e^{i \omega x}$ around the contour is $e^{\frac{\pi \omega}{k_H}+\frac{\pi \omega}{k_C}}$. Moreover, we have just learned from (\ref{SdS_0}) and (\ref{dsphi}) that as one goes around the contour the coefficient of $e^{i \omega x}$ in the asymptotic expansion of $\Phi$ gets multiplied by

$$
\frac{B_+ e^{7i\alpha_+} + B_- e^{7i\alpha_-}}{B_+ e^{-i\alpha_+} + B_- e^{-i\alpha_-}}.
$$

\noindent
For this term to have the required monodromy we must impose

$$
\frac{B_+ e^{7i\alpha_+} + B_- e^{7i\alpha_-}}{B_+ e^{-i\alpha_+} + B_- e^{-i\alpha_-}} e^{\frac{\pi \omega}{k_H}+\frac{\pi \omega}{k_C}} = e^{\frac{\pi \omega}{k_H}-\frac{\pi \omega}{k_C}} \quad \Leftrightarrow  \quad \frac{B_+ e^{7i\alpha_+} + B_- e^{7i\alpha_-}}{B_+ e^{-i\alpha_+} + B_- e^{-i\alpha_-}} = e^{-\frac{2\pi \omega}{k_C}}.
$$

\noindent
Similarly, for the term in $e^{-i \omega x}$ we get the condition

$$
\frac{B_+ e^{5i\alpha_+} + B_- e^{5i\alpha_-}}{B_+ e^{i\alpha_+} + B_- e^{i\alpha_-}} e^{-\frac{\pi \omega}{k_H}-\frac{\pi \omega}{k_C}} = e^{\frac{\pi \omega}{k_H}-\frac{\pi \omega}{k_C}} \quad 
\Leftrightarrow \quad \frac{B_+ e^{5i\alpha_+} + B_- e^{5i\alpha_-}}{B_+ e^{i\alpha_+} + B_- e^{i\alpha_-}} = e^{\frac{2\pi \omega}{k_H}}.
$$

\noindent
The condition for these equations to have nontrivial solutions $(B_+,B_-)$ is then

\begin{align*}
\left| 
\begin{array}{ccc}
e^{7i\alpha_+} - e^{-\frac{2\pi \omega}{k_C}} e^{-i\alpha_+} & \,\,\,\, & e^{7i\alpha_-} - e^{-\frac{2\pi \omega}{k_C}} e^{-i\alpha_-} \\ & & \\
e^{5i\alpha_+} - e^{\frac{2\pi \omega}{k_H}} e^{i\alpha_+}   &  & e^{5i\alpha_-} - e^{\frac{2\pi \omega}{k_H}} e^{i\alpha_-}
\end{array}
\right| = 0 
\quad \Leftrightarrow \quad
\left| 
\begin{array}{ccc}
\sin\left( 4\alpha_+ - \frac{i\pi\omega}{k_C}\right) & & \sin\left( 4\alpha_- - \frac{i\pi\omega}{k_C}\right) \\ & & \\
\sin\left( 2\alpha_+ + \frac{i\pi\omega}{k_H}\right) & & \sin\left( 2\alpha_- + \frac{i\pi\omega}{k_H}\right)
\end{array}
\right| = 0.
\end{align*}

\noindent
As in the Schwarzschild case, this equation is automatically satisfied for $j=0$. This is to be expected, as for $j=0$ the Bessel functions $J_{\pm\frac{j}2}$ coincide and do not form a basis for the space of solutions of the Schr\"odinger--like  equation near the origin. As in \cite{motl-neitzke}, we consider this equation for $j$ nonzero and take the limit as $j\to 0$. This amounts to writing the equation as a power series in $j$ and equating to zero the first nonvanishing coefficient, which in this case is the coefficient of the linear part. Thus, we just have to require that the derivative of the determinant above with respect to $j$ be zero for $j=0$. This amounts to

\begin{align*}
& \left|
\begin{array}{ccc}
\pi \cos\left( \pi - \frac{i\pi\omega}{k_C}\right) & & -\pi \cos\left( \pi - \frac{i\pi\omega}{k_C}\right) \\ & & \\
\sin\left( \frac{\pi}2 + \frac{i\pi\omega}{k_H}\right) & & \sin\left( \frac{\pi}2 + \frac{i\pi\omega}{k_H}\right)
\end{array}
\right|+
\left|
\begin{array}{ccc}
\sin\left( \pi - \frac{i\pi\omega}{k_C}\right) & & \sin\left( \pi - \frac{i\pi\omega}{k_C}\right) \\ & & \\
\frac{\pi}2 \cos\left( \frac{\pi}2 + \frac{i\pi\omega}{k_H}\right) & & -\frac{\pi}2 \cos\left( \frac{\pi}2 + \frac{i\pi\omega}{k_H}\right)
\end{array}
\right| = 0,
\end{align*}

\noindent
from where we obtain our final result as

\begin{equation}\label{resultsdS}
\cosh\left( \frac{\pi\omega}{k_H}-\frac{\pi\omega}{k_C}\right) + 3 \cosh\left( \frac{\pi\omega}{k_H}+\frac{\pi\omega}{k_C}\right)=0.
\end{equation}

\noindent
Notice that if $\omega$ is a solution of this equation then so is $-\bar{\omega}$, as must be the case with quasinormal modes.

From the expression above it is possible to recover the Schwarzschild quasinormal frequencies. We first write the equation as

$$
e^{\frac{\pi\omega}{k_H}-\frac{\pi\omega}{k_C}} + e^{-\frac{\pi\omega}{k_H}+\frac{\pi\omega}{k_C}} + 3 e^{ \frac{\pi\omega}{k_H}+\frac{\pi\omega}{k_C}} + 3 e^{ -\frac{\pi\omega}{k_H}-\frac{\pi\omega}{k_C}} = 0,
$$

\noindent
and take the limit as $\LL \to 0^+$. In this limit, we have $R_C \sim \LL^{-\frac12}$, and hence $R_C \to + \infty$; therefore

$$
k_C = \frac12 f'(R_C) \sim - \LL R_C = - \frac1{R_C} \to 0^-.
$$

\noindent
If we assume that ${\mathbb{R}}{\mathrm{e}} (\omega) > 0$, we see that the two middle terms are exponentially small, and hence the quasinormal equation reduces to

$$
e^{\frac{\pi \omega}{k_H}}+3e^{-\frac{\pi \omega}{k_H}}=0 \quad \Leftrightarrow \quad e^{\frac{2\pi \omega}{k_H}}=-3,
$$

\noindent
which is exactly the equation obtained in \cite{motl, motl-neitzke}. Therefore, the Schwarzschild black hole is not a singular limit of the Schwarzschild dS black hole as far as quasinormal modes are concerned, unlike what happens with the RN black hole solution. The reason for this is clear from the monodromy calculation: whereas the structure of the tortoise near the singularity $r=0$ in the RN solution depends crucially on whether the charge is zero or not, in the Schwarzschild dS case it does not depend on $\LL$. Thus, as $R\to+\infty$, the cosmological horizon approaches the point at infinity and the contour approaches the contour first used in \cite{motl-neitzke}. One could also try to derive the $\MM \to 0$ limit, trying to obtain quasinormal frequencies for pure dS spacetime. In this case $R_H \to 0$ and

$$
k_H \sim \frac{d-3}{2} R_H \to 0^+,
$$

\noindent
while $k_C \to - \sqrt{\LL}$. The quasinormal equation for the Schwarzschild dS spacetime becomes

$$
e^{-\frac{2\pi\omega}{k_C}} = -3
$$

\noindent
from where one obtains

$$
\omega = i \sqrt{\LL} \left( n + \frac{1}{2} + \frac{\log 3}{2\pi i} \right).
$$

\noindent
Unfortunately, as we shall later see in section 4, this is the incorrect value for the dS quasinormal frequencies. While there was no expectation that the offset would come out right, one could have hoped that the gap could have been correct. We shall later learn that the correct gap in pure dS spacetime is $2\sqrt{\LL}$ and not $\sqrt{\LL}$ as obtained above. Thus, while the Schwarzschild spacetime is a good limit of the Schwarzschild dS black hole, pure dS spacetime is not, at least as far as asymptotic quasinormal frequencies are concerned.

For vector type perturbations, the potential for the Schr\"odinger--like master equation near the origin is of the form

$$
V \big[ r(x) \big] \sim \frac{j^2 - 1}{4x^2},
$$

\noindent
with $j=2$ (see appendix \ref{appendixC}). Repeating the same argument, one ends up with the same equation, except that now $2\alpha_\pm=\frac{\pi}2 \pm \pi$, $4\alpha_\pm=\pi \pm 2\pi$. This equation is therefore exactly the same as in the $j=0$ case, for which $2\alpha_\pm=\frac{\pi}2$, $4\alpha_\pm=\pi$, and consequently we end up with the precise same quasinormal frequencies.

Some remarks are due for $d=4$ and $d=5$, as in these cases the Stokes lines are not quite of the general form we assumed in the previous calculation. We start by studying the behavior of solutions to the Schr\"odinger--like equation (\ref{schrodinger}) near infinity. For $r \sim \infty$ one has $x \sim x_0$ for a (complex) constant $x_0$, and the potential is

$$
V \sim \frac{{j_{\infty}}^2-1}{4(x-x_0)^2}
$$

\noindent
with ${j_{\infty}}=d-1,d-3,d-5$ for tensor type, vector type and scalar type perturbations (see appendix \ref{appendixC}). Hence,

$$
\Phi (x) \sim A_+ \sqrt{2\pi\omega (x-x_0)}\ J_{\frac{{j_{\infty}}}{2}} \left( \omega (x-x_0) \right) + A_- \sqrt{2\pi\omega (x-x_0)}\ J_{-\frac{{j_{\infty}}}{2}} \left( \omega (x-x_0) \right)
$$

\noindent
in this region, where $J_\nu(x)$ represents a Bessel function of the first kind and $A_\pm$ are (complex) integration constants. Given expansion (\ref{Bessel0}), we see that $\Phi$ has monodromy $(-1)^d$ around infinity. The $d=4$ case was actually done in \cite{cns}. In this case, one can take advantage of the fact that $\Phi$ has zero monodromy at infinity to conclude that as one follows the Stokes line the (anti--clockwise) monodromy of $\Phi$ around the fictitious horizon $r=\widetilde{R}$ equals the (clockwise) monodromy around the real horizons $r=R_H$ and $r=R_C$; since the same is true for the plane wave functions $e^{\pm i\omega x}$, the whole calculation reduces to what was done above, and we get the same formula. Another way to see this is that since $x$, $\Phi$ are regular at infinity one can deform the $d=4$ contour into, say, the $d=6$ contour in the Riemann sphere $\mathbb{C} \cup \{ \infty \}$.

For $d=5$ the situation is more subtle. As shown in Figure \ref{NumericalStokesSdS}, the Stokes line approaches infinity, and hence one must close the contour near infinity. In the branch containing point B, the solution is then of the form

\begin{align*}
\Phi (x) &\sim A_+ \sqrt{2\pi\omega (x-x_0)}\ J_{\frac{{j_{\infty}}}{2}} \left( \omega (x-x_0) \right) + A_- \sqrt{2\pi\omega (x-x_0)}\ J_{-\frac{{j_{\infty}}}{2}} \left( \omega (x-x_0) \right) \\
&\sim \left( A_+ e^{-i\beta_+} + A_- e^{-i\beta_-}\right) e^{i \omega (x-x_0)} + \left( A_+ e^{i\beta_+} + A_- e^{i\beta_-}\right) e^{-i \omega (x-x_0)},
\end{align*}

\noindent
where $\beta_\pm = \frac{\pi}4 (1 \pm {j_{\infty}})$, with ${j_{\infty}}=4,2,0$ for tensor type, vector type and scalar type perturbations. As $r$ rotates clockwise towards the branch containing point $A$, by $\frac{\pi}2$, $x-x_0 \sim \frac{1}{\LL r}$ rotates anti--clockwise by the same amount, and hence

$$
\Phi (x) \sim \left( A_+ e^{3i\beta_+} + A_- e^{3i\beta_-}\right) e^{i \omega (x-x_0)} + \left( A_+ e^{i\beta_+} + A_- e^{i\beta_-}\right) e^{-i \omega (x-x_0)}
$$

\noindent
at point $B$. Whereas the coefficient of $e^{-i \omega x}$ is unchanged in this process, it is easy to see that the coefficient of $e^{i \omega x}$ reverses sign for all three values of ${j_{\infty}}$. Therefore, there appears an extra minus sign in the corresponding monodromy matching equation, and the condition for nontrivial solutions $(B_+,B_-)$ changes slightly. Carrying out the computation, the condition for quasinormal modes in $d=5$ turns out to be

$$
\sinh\left( \frac{\pi\omega}{k_H}-\frac{\pi\omega}{k_C}\right) - 3 \sinh\left( \frac{\pi\omega}{k_H}+\frac{\pi\omega}{k_C}\right)=0.
$$

\noindent
It is easily seen that this equation still has the correct Schwarzschild limit.

Let us now review the literature concerning asymptotic quasinormal frequencies in the Schwarzschild dS spacetime, so that we can compare our results to what has been previously accomplished on this subject. For Schwarzschild dS, early results on quasinormal modes were studied in \cite{bclp}, without great emphasis on the asymptotic case. The first analytical results in $d=4$ were derived in the near--extremal situation, where event and cosmological horizons are nearly coincident \cite{cardoso-lemos}, but the approximation used therein is not expected to hold in the asymptotic limit, at least on what concerns the real part of the asymptotic frequencies. These results were later extended to $d$--dimensions (and to RN dS black holes) in \cite{molina}, with similar results. Further approximations were studied in \cite{suneeta}, in a limit where the black hole mass is much smaller than the spacetime radius of curvature, but focusing explicitly on the time \textit{dependent} transient situation. An attempt at an analytic solution for the asymptotic quasinormal frequencies, using the monodromy technique of \cite{motl-neitzke}, was done in \cite{castellobranco-abdalla}. However, an erroneous identification of the relevant contours led these authors to a wrong result---in fact, the authors assumed the monodromy contour to be the same as in the Schwarzschild case, while as we have seen the contour changes dramatically in non--asymptotically flat spacetimes (see also \cite{medved-martin}, where other arguments were given trying to explain the failure of \cite{castellobranco-abdalla} to reproduce available numerical data). Focusing on $d=4$, there are  some interesting analytical results in \cite{choudhury-padmanabhan}. These authors find that because there are two different surface gravities, there are also two sets of solutions for ${{\mathbb{I}}{\mathrm{m}}} \left( \omega \right)$ when the horizons are \textit{widely} spaced, namely ${{\mathbb{I}}{\mathrm{m}}} \left( \omega \right)$ equally spaced with spacing equal to $k_{H}$ or ${{\mathbb{I}}{\mathrm{m}}} \left( \omega \right)$ equally spaced with spacing equal to $k_{C}$. It was further claimed in \cite{choudhury-padmanabhan} that this lack of consensus on quasinormal frequencies was due to the fact that there is no global definition of temperature in this spacetime. Our results appear to confirm this expectation: in the limit where the cosmological radius goes to infinity, and one recovers the Schwarzschild modes, we found the spacing to be equal to $k_{H}$. In the limit where the black hole radius is very small, our final formula yields modes with spacing equal to $k_{C}$.

The four dimensional case was first solved analytically in \cite{cns}. As discussed in that paper, the analytical results are in complete agreement, and to a large degree of accuracy, with the numerical results provided (for this case of $d=4$ Schwarzschild dS) in \cite{yoshida-futamase, konoplya-zhidenko}. Alongside with the fact that our expressions yield the correct Schwarzschild limit in any dimension, we believe that there is strong evidence supporting our general results. However, it would be very interesting to produce further numerical data concerning higher dimensional Schwarzschild dS quasinormal modes to match against all our analytical results.


\subsubsection{The Reissner--Nordstr\"om de Sitter Solution}


We now compute the quasinormal modes of the RN dS $d$--dimensional black hole. Again we consider solutions of the Schr\"odinger--like equation (\ref{schrodinger}) in the complex $r$--plane. We start by studying the behavior of $\Phi(x)$ near the singularity $r=0$. In this region, the tortoise coordinate is

$$
x \sim \frac{r^{2d-5}}{\left( 2d-5 \right) \QQ^{2}},
$$

\noindent
and the potential for tensor and scalar type perturbations is

$$
V \big[ r(x) \big] \sim \frac{j^2 - 1}{4x^2},
$$

\noindent
with $j=\frac{d-3}{2d-5}$ (see appendix \ref{appendixC}). The solution of the Schr\"odinger--like equation in this region (with $j \neq 0$) is therefore

$$
\Phi (x) \sim B_+ \sqrt{2\pi\omega x}\ J_{\frac{j}{2}} \left( \omega x \right) + B_- \sqrt{2\pi\omega x}\ J_{-\frac{j}{2}} \left( \omega x \right),
$$

\noindent
where $J_\nu(x)$ represents a Bessel function of the first kind and $B_\pm$ are (complex) integration constants. 

For the asymptotic quasinormal modes one has $\im (\omega) \gg \re (\omega)$, with $\im (\omega) \to + \infty$, and hence $\omega$ is approximately purely imaginary. Consequently, $\omega x \in \mathbb{R}$ for $x \in i\mathbb{R}$; in a neighborhood of the origin, the above relation between $x$ and $r$ tells us that this happens for

$$
r=\rho\ e^{\frac{i\pi}{2(2d-5)}+\frac{in\pi}{2d-5}},
$$

\noindent
with $\rho>0$ and $n=0,1,\ldots, 4d-11$. These are half--lines starting at the origin, equally spaced by an angle of $\frac{\pi}{2d-5}$. Notice that the sign of $\omega x$ on these lines is $(-1)^{n+1}$; in other words, starting with the line corresponding to $n=0$, the sign of $\omega x$ is alternately negative and positive as one goes anti--clockwise around the origin.

From the asymptotic expansion (\ref{Bessel+}) we see that 

\begin{eqnarray}
\Phi (x) & \sim & 2 B_+ \cos \left( \omega x - \alpha_+ \right) + 2 B_- \cos \left( \omega x - \alpha_- \right) \nonumber \\
& = & \left( B_+ e^{-i\alpha_+} + B_- e^{-i\alpha_-}\right) e^{i \omega x} + \left( B_+ e^{i\alpha_+} + B_- e^{i\alpha_-}\right) e^{-i \omega x} \label{RNdS_0}
\end{eqnarray}

\noindent
in any one of the lines corresponding to positive $\omega x$, where again

$$
\alpha_\pm = \frac{\pi}4 (1 \pm j).
$$

\noindent
We shall use this asymptotic expression to make the monodromy matching. This matching must be done along the Stokes line $\omega x \in \mathbb{R}$ (or $x \in i\mathbb{R}$), so that neither of the exponentials $e^{\pm i\omega x}$ dominates the other.

To trace out the Stokes line $\re(x)=0$ let us first observe that we already know its behavior near the origin. Furthermore, this is the only singular point of this curve: indeed, since $x$ is a holomorphic function of $r$, the critical points of the function $\re(x)$ are the zeros of

$$
\frac{dx}{dr} = \frac{1}{f(r)} = \frac{r^{2d-6}}{-\LL r^{2d-4}+r^{2d-6}-2\MM r^{d-3} + \QQ^2}
$$

\noindent
(\textit{i.e.}, $r=0$ only). We have an additional problem that $x$ is a multivalued function: each of the ``horizons'' $R_{n}$ ($n=1,\ldots,2d-4$) is a branch point. From (\ref{tortoise_horizon}) we see that although the function $\re(x)$ is still well defined around the real horizons, it will be multivalued around all the other fictitious horizons.

For $r \sim \infty$ one has $x \sim x_0 + \frac{1}{\LL r}$ (see appendix \ref{appendixC}). Consequently, $x$ is holomorphic at infinity, and we can choose the branch cuts to cancel out among themselves. Therefore $\re(x)$ is well defined in a neighborhood of infinity, and moreover $\re(x)=0$ cannot extend out to infinity, as generically $x_0$ is not real. Thus the Stokes line branches starting out at the origin must either connect to another branch or end up in a branch cut (and one would expect them to connect in the region where $\re(x)$ is well defined). On the other hand, it is easy to see that the Stokes line must intersect the positive real axis exactly in three points, one in each of the intervals $(R_{H}^{-}, R_{H}^{+})$, $(R_{H}^{+}, R_{C})$ and $(R_{C}, + \infty)$. Using this information plus elementary considerations on symmetry and the sign of $\re(x)$, one can deduce that the Stokes line must be of the form indicated in Figure \ref{StokesRNdS}. These results are moreover verified by the numerical computation of the same Stokes lines, as indicated in Figures \ref{NumericalStokesRNdS} and \ref{NumericalStokesRNdS_CloseUp}.

\FIGURE[ht]{\label{StokesRNdS}
	\centering
	\psfrag{A}{$A$}
	\psfrag{B}{$B$}
	\psfrag{r0}{$R_H^-$}
	\psfrag{R0}{$R_H^+$}
	\psfrag{RC}{$R_C$}
	\psfrag{R}{$\widetilde{R}$}
	\psfrag{r1}{$\gamma_1$}
	\psfrag{r2}{$\overline{\gamma}_1$}
	\psfrag{R1}{$\gamma_2$}
	\psfrag{R2}{$\overline{\gamma}_2$}
	\psfrag{Re}{$\re$}
	\psfrag{Im}{$\im$}
	\psfrag{contour}{contour}
	\psfrag{Stokes line}{Stokes line}
	\epsfxsize=.6\textwidth
	\leavevmode
	\epsfbox{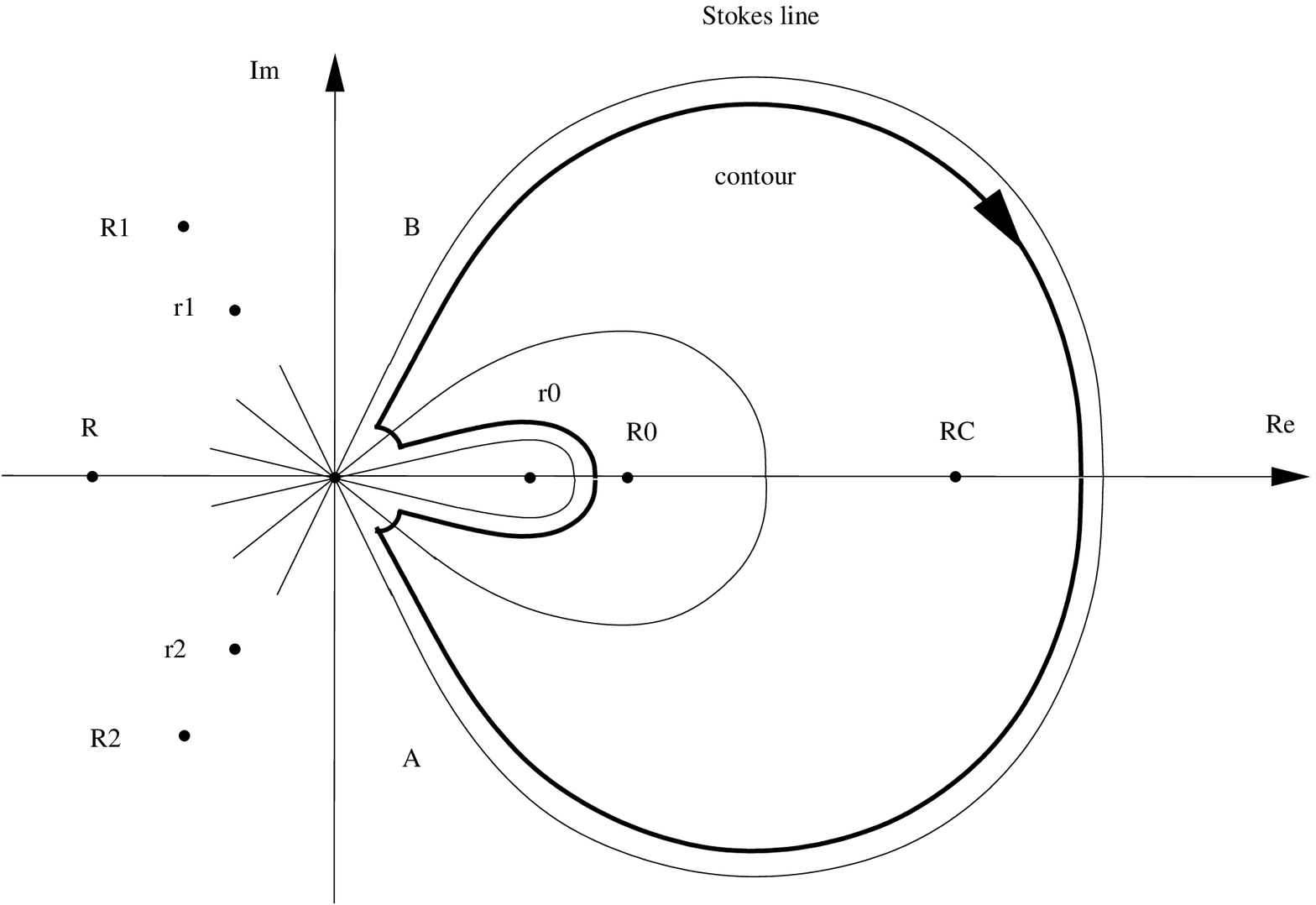}
\caption{Stokes line for the Reissner--Nordstr\"om de Sitter black hole, along with the chosen contour for monodromy matching, in the case of dimension $d=6$.}
}

\FIGURE[ht]{\label{NumericalStokesRNdS}
	\centering
	\epsfig{file=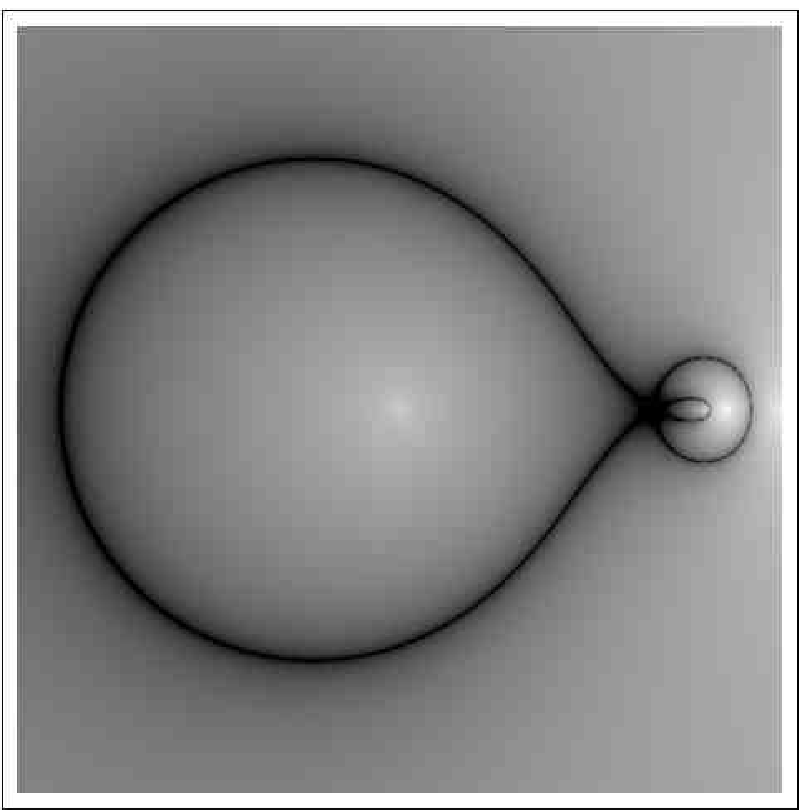, width=1.5in, height=1.5in}
    $\quad$
    \epsfig{file=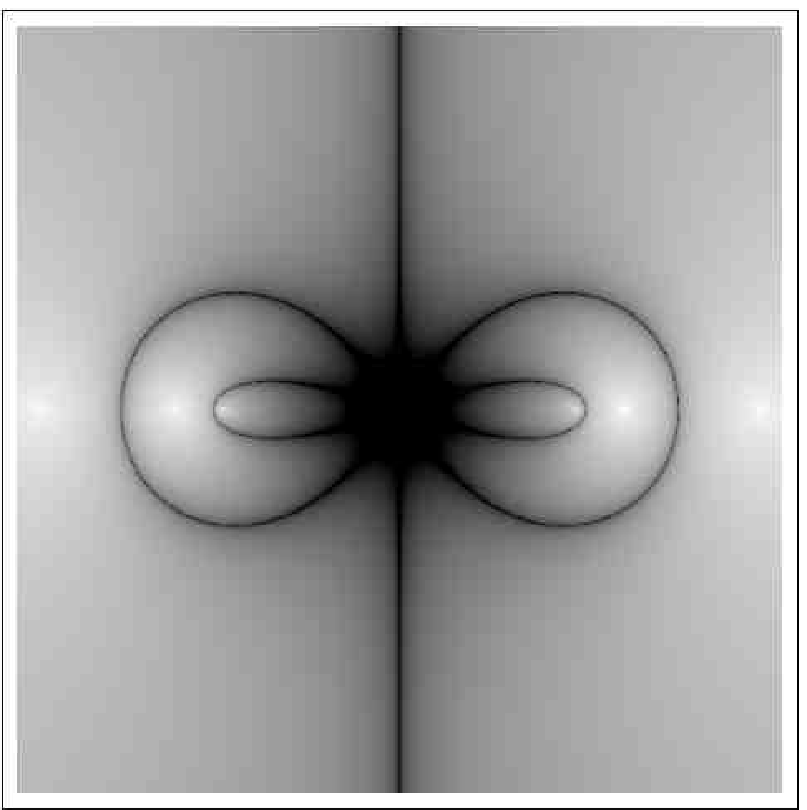, width=1.5in, height=1.5in}
    $\quad$
    \epsfig{file=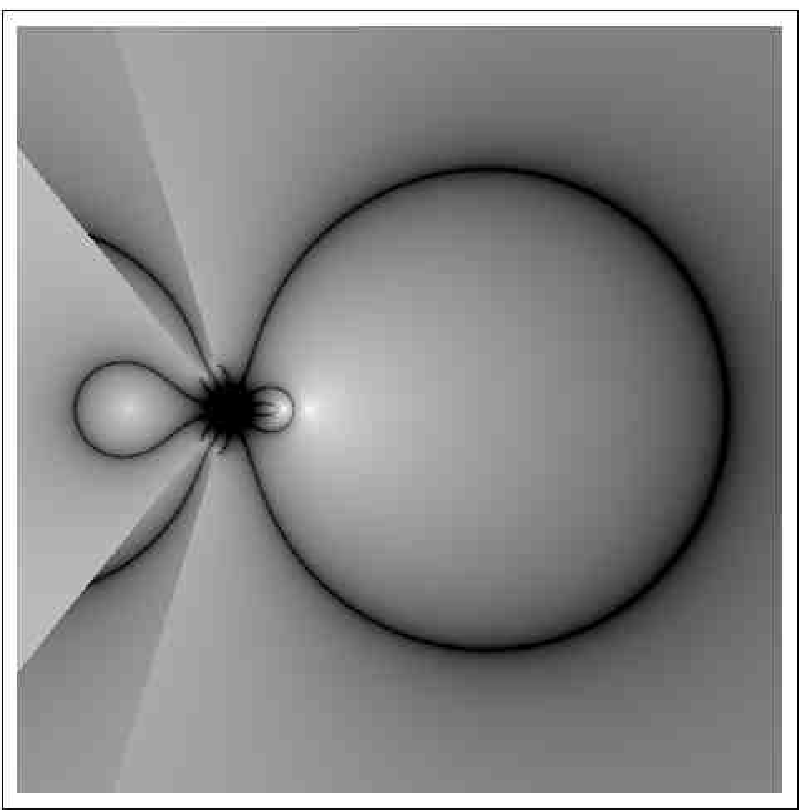, width=1.5in, height=1.5in}
    $\quad$
    \epsfig{file=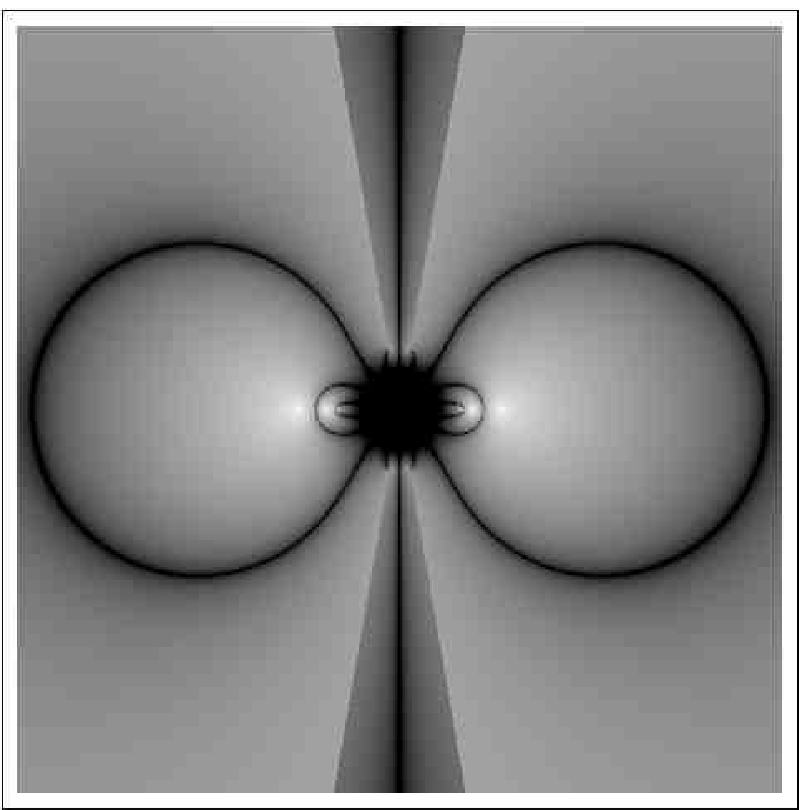, width=1.5in, height=1.5in}
\caption{Numerical calculation of the Stokes lines for the Reissner--Nordstr\"om de Sitter black hole in dimensions $d=4$, $d=5$, $d=6$ and $d=7$. Different shadings also illustrate the various horizon singularities and branch cuts (note that these branch cuts are not necessarily equal to the ones used for the calculation in the main text).}
}

\FIGURE[ht]{\label{NumericalStokesRNdS_CloseUp}
        \centering
        \epsfig{file=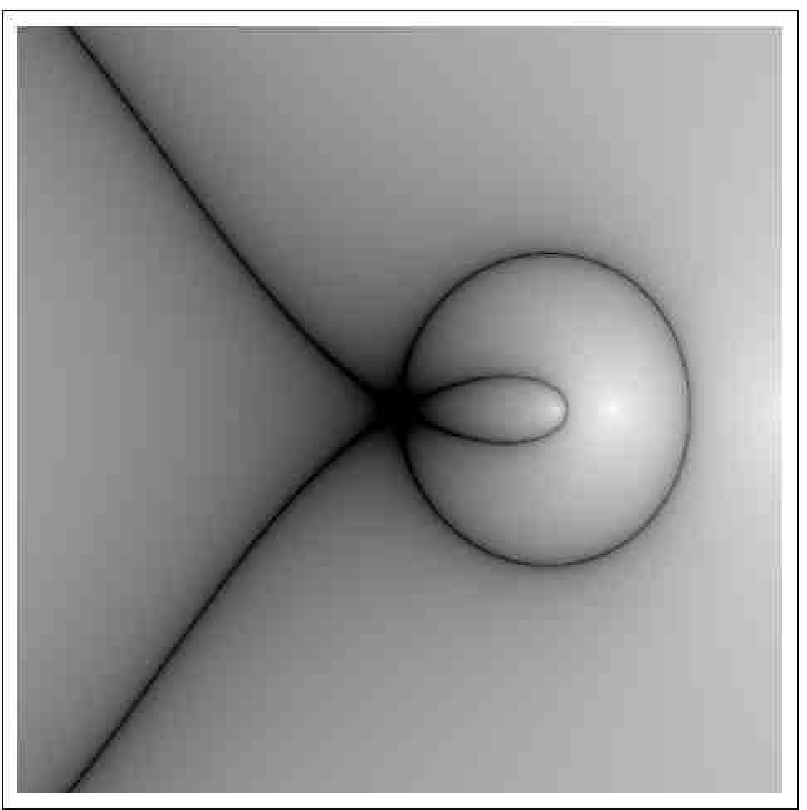, width=1.5in, height=1.5in}
    $\qquad$
    \epsfig{file=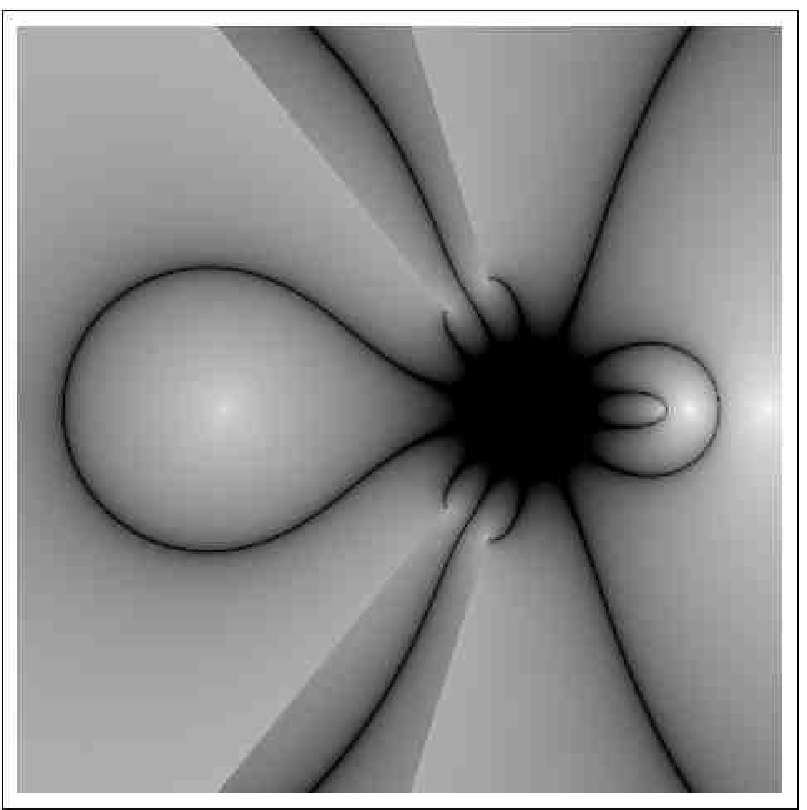, width=1.5in, height=1.5in}
    $\qquad$
    \epsfig{file=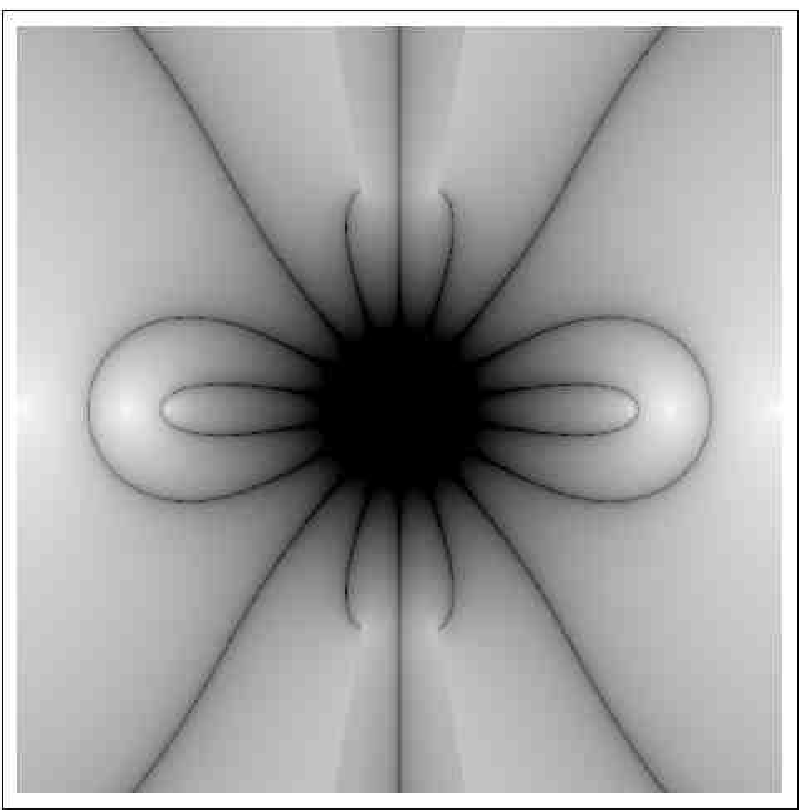, width=1.5in, height=1.5in}
        \caption{Close--up on the numerical calculation of the Stokes lines for the Reissner--Nordstr\"om de Sitter black hole in $d=4$, $d=6$ and $d=7$. Locally near the origin these lines resemble the corresponding ones for the Reissner--Nordstr\"om black hole in an asymptotically flat spacetime (see Figure \ref{NumericalStokesRN}).}
}

Let us now consider the contour defined by the closed portions of the Stokes line in such a way that one avoids enclosing the inner horizon, as shown in Figure \ref{StokesRNdS} (this contour is contained in the region where $\re(x)$ is well defined). At point $A$ we have $\omega x>0$, and therefore the expansion (\ref{RNdS_0}) holds at this point. For $z \sim 0$ one has the expansion (\ref{Bessel0}). Consequently, as one rotates from the branch containing point $A$ to the next branch in the contour (through an angle of  $\frac{2\pi}{2d-5}$), $x\sim \frac{r^{2d-5}}{\left( 2d-5 \right) \QQ^{2}}$ rotates through an angle of $2\pi$, and since

$$
\sqrt{2\pi e^{2\pi i} \omega x}\ J_{\pm\frac{j}{2}} \left( e^{2\pi i} \omega x \right) = e^{\frac{2\pi i}2 (1 \pm j)}\sqrt{2\pi \omega x}\ J_{\pm\frac{j}{2}} \left( \omega x \right) \sim 2 e^{4i\alpha_\pm} \cos(\omega x - \alpha_\pm)
$$

\noindent
one has

\begin{eqnarray}\label{RNdS_00}
\Phi (x) & \sim & 2 B_+ e^{4i\alpha_+} \cos \left( \omega x - \alpha_+ \right) + 2 B_- e^{4i\alpha_-} \cos \left( \omega x - \alpha_- \right) \nonumber \\
& = & \left( B_+ e^{3i\alpha_+} + B_- e^{3i\alpha_-}\right) e^{i \omega x} + \left( B_+ e^{5i\alpha_+} + B_- e^{5i\alpha_-}\right) e^{-i \omega x} 
\end{eqnarray}

\noindent
on that branch. As one follows the contour around the inner horizon $r=R_H^-$, it is easily seen from (\ref{tortoise_horizon}) that $x$ approaches the point

$$
\delta = \frac{2\pi i}{f'(R_H^-)} = \frac{\pi i}{k^-},
$$

\noindent
where

$$
k^- = \frac12 f'(R_H^-) 
$$

\noindent
is the surface gravity at the inner horizon (notice that $k^- < 0$). Consequently, after going around the inner horizon, the solution will be of the form

$$
\Phi (x) \sim C_+ \sqrt{2\pi\omega (x-\delta)}\ J_{\frac{j}{2}} \left( \omega (x-\delta) \right) + C_- \sqrt{2\pi\omega (x-\delta)}\ J_{-\frac{j}{2}} \left( \omega (x-\delta) \right),
$$

\noindent
as one approaches the origin. Since $\omega(x-\delta)$ is negative on this branch, from the asymptotic expansion (\ref{Bessel-}) we have

\begin{eqnarray*}
\Phi (x) & \sim & 2 C_+ \cos \left( \omega (x-\delta) + \alpha_+ \right) + 2 C_- \cos \left( \omega (x - \delta) + \alpha_- \right)\\
& = & \left( C_+ e^{i\alpha_+} e^{-i\omega\delta} + C_- e^{i\alpha_-} e^{-i\omega\delta}\right) e^{i \omega x} + \left( C_+ e^{-i\alpha_+} e^{i\omega\delta} + C_- e^{-i\alpha_-} e^{i\omega\delta}\right) e^{-i \omega x} 
\end{eqnarray*}

\noindent
on this branch. This must be matched to the expression (\ref{RNdS_00}) for $\Phi(x)$, and hence we must have

\begin{eqnarray}
&& B_+ e^{3i\alpha_+} + B_- e^{3i\alpha_-} = C_+ e^{i\alpha_+} e^{-i\omega\delta} + C_- e^{i\alpha_-} e^{-i\omega\delta} \label{RNdS_2}\\
&& B_+ e^{5i\alpha_+} + B_- e^{5i\alpha_-} = C_+ e^{-i\alpha_+} e^{i\omega\delta} + C_- e^{-i\alpha_-} e^{i\omega\delta} \label{RNdS_3}
\end{eqnarray}

\noindent
Finally, we must rotate to the branch containing point $B$. Again $x-\delta$ rotates through an angle of $2\pi$, and since

$$
\sqrt{2\pi e^{2\pi i} \omega (x-\delta)}\ J_{\pm\frac{j}{2}} \left( e^{2\pi i} \omega (x-\delta) \right) = e^{\frac{2\pi i}2 (1 \pm j)}\sqrt{2\pi \omega (x-\delta)}\ J_{\pm\frac{j}{2}} \left( \omega (x-\delta) \right) \sim 2 e^{4i\alpha_\pm} \cos(\omega (x-\delta) + \alpha_\pm)
$$

\noindent
(as $\omega(x-\delta) < 0$ on the branch containing point $B$), one has

\begin{eqnarray*}
\Phi (x) & \sim & 2 C_+ e^{4i\alpha_+} \cos \left( \omega (x-\delta) + \alpha_+ \right) + 2 C_- e^{4i\alpha_-} \cos \left( \omega (x-\delta) + \alpha_- \right) \\
& = & \left( C_+ e^{5i\alpha_+} e^{-i\omega\delta} + C_- e^{5i\alpha_-} e^{-i\omega\delta}\right) e^{i \omega x} + \left( C_+ e^{3i\alpha_+} e^{i\omega\delta} + C_- e^{3i\alpha_-}e^{i\omega\delta}\right) e^{-i \omega x}
\end{eqnarray*}

\noindent
on that branch.

If $\Phi$ corresponds to a quasinormal mode, its clockwise monodromy around $r=R_H^+$ must be the same as the clockwise monodromy of $e^{i\omega x}$, that is, $e^{\frac{\pi \omega}{k^+}}$, where

$$
k^+ = \frac12 f'(R_H^+)
$$

\noindent
is the surface gravity at the black hole horizon. Similarly, its clockwise monodromy around $r=R_C$ must be the same as the clockwise monodromy of $e^{-i\omega x}$, that is, $e^{-\frac{\pi \omega}{k_C}}$, where

$$
k_C = \frac12 f'(R_C)
$$

\noindent
is the surface gravity at the cosmological horizon (notice that we are taking $k_C < 0$). Since these are the only singularities of $\Phi$ inside the contour, the monodromy of $\Phi$ around the contour must be

$$
e^{\frac{\pi \omega}{k^+}-\frac{\pi \omega}{k_C}}.
$$

\noindent
On the other hand, the clockwise monodromy of $e^{i \omega x}$ around the contour is $e^{\frac{\pi \omega}{k^+}+\frac{\pi \omega}{k_C}}$. Moreover, we have just learned that as one goes around the contour the coefficient of $e^{i \omega x}$ in the asymptotic expansion of $\Phi$ gets multiplied by

$$
\frac{C_+ e^{5i\alpha_+}e^{-i\omega\delta} + C_- e^{5i\alpha_-}e^{-i\omega\delta}}{B_+ e^{-i\alpha_+} + B_- e^{-i\alpha_-}}.
$$

\noindent
For this term to have the required monodromy we must impose

\begin{equation} \label{RNdS_4}
\frac{C_+ e^{5i\alpha_+}e^{-i\omega\delta} + C_- e^{5i\alpha_-}e^{-i\omega\delta}}{B_+ e^{-i\alpha_+} + B_- e^{-i\alpha_-}} e^{\frac{\pi \omega}{k^+}+\frac{\pi \omega}{k_C}} = e^{\frac{\pi \omega}{k^+}-\frac{\pi \omega}{k_C}} \quad \Leftrightarrow  \quad \frac{C_+ e^{5i\alpha_+}e^{-i\omega\delta} + C_- e^{5i\alpha_-}e^{-i\omega\delta}}{B_+ e^{-i\alpha_+} + B_- e^{-i\alpha_-}} = e^{-\frac{2\pi \omega}{k_C}}.
\end{equation}

\noindent
Similarly, for the term in $e^{-i \omega x}$ we get the condition

\begin{equation} \label{RNdS_5}
\frac{C_+ e^{3i\alpha_+}e^{i\omega\delta} + C_- e^{3i\alpha_-}e^{i\omega\delta}}{B_+ e^{i\alpha_+} + B_- e^{i\alpha_-}} e^{-\frac{\pi \omega}{k^+}-\frac{\pi \omega}{k_C}} = e^{\frac{\pi \omega}{k^+}-\frac{\pi \omega}{k_C}} \quad 
\Leftrightarrow \quad \frac{C_+ e^{3i\alpha_+}e^{i\omega\delta} + C_- e^{3i\alpha_-}e^{i\omega\delta}}{B_+ e^{i\alpha_+} + B_- e^{i\alpha_-}} = e^{\frac{2\pi \omega}{k^+}}.
\end{equation}

\noindent
The condition for equations (\ref{RNdS_2}), (\ref{RNdS_3}), (\ref{RNdS_4}), (\ref{RNdS_5}), reflecting quasinormal boundary conditions both at cosmological and outer black hole horizons, to have nontrivial solutions $(B_+,B_-,C_+,C_-)$ is then

\begin{eqnarray*}
&&
\left| 
\begin{array}{ccccccc}
e^{3i\alpha_+} & & e^{3i\alpha_-} & & -e^{i\alpha_+}e^{-i\omega\delta} & & -e^{i\alpha_-}e^{-i\omega\delta} \\
e^{5i\alpha_+} & & e^{5i\alpha_-} & & -e^{i\alpha_+}e^{i\omega\delta} & & -e^{i\alpha_-}e^{i\omega\delta} \\
e^{-i\alpha_+}e^{-\frac{2\pi \omega}{k_C}} & & e^{-i\alpha_-}e^{-\frac{2\pi \omega}{k_C}} & & -e^{5i\alpha_+}e^{-i\omega\delta} & & -e^{5i\alpha_-}e^{-i\omega\delta} \\
e^{i\alpha_+}e^{\frac{2\pi \omega}{k^+}} & & e^{i\alpha_-}e^{\frac{2\pi \omega}{k^+}} & & -e^{3i\alpha_+}e^{i\omega\delta} & & -e^{3i\alpha_-}e^{i\omega\delta}
\end{array}
\right| = 0 \\
&& \Leftrightarrow \quad
\left| 
\begin{array}{ccccccc}
e^{4i\alpha_+} & & e^{4i\alpha_-} & & e^{2i\alpha_+}e^{\frac{2\pi \omega}{k^-}} & & e^{2i\alpha_-}e^{\frac{2\pi \omega}{k^-}} \\
e^{6i\alpha_+} & & e^{6i\alpha_-} & & 1 & & 1 \\
e^{-\frac{2\pi \omega}{k_C}} & & e^{-\frac{2\pi \omega}{k_C}} & & e^{6i\alpha_+}e^{\frac{2\pi \omega}{k^-}} & & e^{6i\alpha_-}e^{\frac{2\pi \omega}{k^-}} \\
e^{2i\alpha_+}e^{\frac{2\pi \omega}{k^+}} & & e^{2i\alpha_-}e^{\frac{2\pi \omega}{k^+}} & & e^{4i\alpha_+} & & e^{4i\alpha_-}
\end{array}
\right| = 0
\end{eqnarray*}

\noindent
from where we obtain our final result as

\begin{equation}\label{resultsRNdS}
\cosh\left( \frac{\pi\omega}{k^+}-\frac{\pi\omega}{k_C}\right) + (1+2\cos(\pi j)) \cosh\left( \frac{\pi\omega}{k^+}+\frac{\pi\omega}{k_C}\right) + (2+2\cos(\pi j)) \cosh\left( \frac{2\pi\omega}{k^-}+\frac{\pi\omega}{k^+}+\frac{\pi\omega}{k_C}\right)=0.
\end{equation}

\noindent
Notice that if $\omega$ is a solution of this equation then so is $-\bar{\omega}$, as must be the case with quasinormal modes.

To recover the RN quasinormal frequencies we first write out the equation as

$$
1+e^{\frac{2\pi\omega}{k^+}-\frac{2\pi\omega}{k_C}} + (1+2\cos(\pi j))\left( e^{\frac{2\pi\omega}{k^+}}+e^{-\frac{2\pi\omega}{k_C}} \right) + (2+2\cos(\pi j)) \left( e^{\frac{2\pi\omega}{k^-}+\frac{2\pi\omega}{k^+}} + e^{-\frac{2\pi\omega}{k^-}-\frac{2\pi\omega}{k_C}} \right) = 0
$$

\noindent
and next take the limit as $\LL \to 0^+$. In this limit, we have $R_C \sim \LL^{-\frac12}$, and hence $R_C \to + \infty$; therefore

$$
k_C = \frac12 f'(R_C) \sim - \LL R_C = - \frac1{R_C} \to 0^-.
$$

\noindent
If we assume that ${\mathbb{R}}{\mathrm{e}} (\omega) > 0$, we see that the equation reduces to

$$
e^{\frac{2\pi\omega}{k^+}}+(1+2\cos(\pi j))+(2+2\cos(\pi j))e^{-\frac{2\pi\omega}{k^-}}=0,
$$

\noindent
which is exactly equation (\ref{resultsRN}). Therefore, the RN black hole is not a singular limit of the RN dS black hole, as far as the quasinormal modes are concerned. The reason for this is basically the same as for the Schwarzschild dS case: the topology of the tortoise at the origin does not change as we take the cosmological constant to vanish.

For vector type perturbations, the potential for the Schr\"odinger--like master equation near the origin is of the form

$$
V \big[ r(x) \big]  \sim  \frac{j^2 - 1}{4x^2},
$$

\noindent
with $j=\frac{3d-7}{2d-5}$. Repeating the same argument, one ends up with the same equation (\ref{resultsRNdS}) for this new value of $j$. Since 

$$
\frac{3d-7}{2d-5} = 2 - \frac{d-3}{2d-5},
$$

\noindent
we see that this equation is exactly the same as in the $j=\frac{d-3}{2d-5}$ case, and consequently we end up with the same quasinormal modes.

Some remarks are due for $d=4$ and $d=5$, as in these cases the Stokes lines are not quite of the general form we assumed in the previous calculation. We start by studying the behavior of the solutions of the Schr\"odinger--like equation (\ref{schrodinger}) near infinity. For $r \sim \infty$ one has $x \sim x_0$ for a (complex) constant $x_0$, and the potential is

$$
V \sim \frac{{j_{\infty}}^2-1}{4(x-x_0)^2}
$$

\noindent
with ${j_{\infty}}=d-1,d-3,d-5$ for tensor type, vector type and scalar type perturbations (see appendix \ref{appendixC}). Hence,

$$
\Phi (x) \sim A_+ \sqrt{2\pi\omega (x-x_0)}\ J_{\frac{{j_{\infty}}}{2}} \left( \omega (x-x_0) \right) + A_- \sqrt{2\pi\omega (x-x_0)}\ J_{-\frac{{j_{\infty}}}{2}} \left( \omega (x-x_0) \right)
$$

\noindent
in this region, where $J_\nu(x)$ represents a Bessel function of the first kind and $A_\pm$ are (complex) integration constants. Given expansion (\ref{Bessel0}), we see that $\Phi$ has monodromy $(-1)^d$ around infinity. In the $d=4$ case, one can take advantage of the fact that $\Phi$ has zero monodromy at infinity to conclude that as one follows the Stokes line the (anti--clockwise) monodromy of $\Phi$ around the fictitious horizon $r=\widetilde{R}$ equals the (clockwise) monodromy around the real horizons $r = R_{H}^{-}$, $r = R_{H}^{+}$ and $r = R_{C}$; since the same is true for the plane wave functions $e^{\pm i\omega x}$, the whole calculation reduces to what was done above, and we get the same formula. Another way to see this is that since $x$, $\Phi$ are regular at infinity one can deform the $d=4$ contour into, say, the $d=6$ contour in the Riemann sphere $\mathbb{C} \cup \{ \infty \}$.

For $d=5$ the situation is more subtle. As shown in Figure \ref{NumericalStokesRNdS}, the Stokes line approaches infinity, and hence one must close the contour near infinity. In the branch containing point B, the solution is then of the form

\begin{align*}
\Phi (x) &\sim A_+ \sqrt{2\pi\omega (x-x_0)}\ J_{\frac{{j_{\infty}}}{2}} \left( \omega (x-x_0) \right) + A_- \sqrt{2\pi\omega (x-x_0)}\ J_{-\frac{{j_{\infty}}}{2}} \left( \omega (x-x_0) \right) \\
&\sim \left( A_+ e^{-i\beta_+} + A_- e^{-i\beta_-}\right) e^{i \omega (x-x_0)} + \left( A_+ e^{i\beta_+} + A_- e^{i\beta_-}\right) e^{-i \omega (x-x_0)},
\end{align*}

\noindent
where $\beta_\pm = \frac{\pi}4 (1 \pm {j_{\infty}})$, with ${j_{\infty}}=4,2,0$ for tensor type, vector type and scalar type perturbations. As $r$ rotates clockwise towards the branch containing point $A$, by $\frac{\pi}2$, $x-x_0 \sim \frac{1}{\LL r}$ rotates anti--clockwise by the same amount, and hence

$$
\Phi (x) \sim \left( A_+ e^{3i\beta_+} + A_- e^{3i\beta_-}\right) e^{i \omega (x-x_0)} + \left( A_+ e^{i\beta_+} + A_- e^{i\beta_-}\right) e^{-i \omega (x-x_0)}
$$

\noindent
at point $B$. Whereas the coefficient of $e^{-i \omega x}$ is unchanged in this process, it is easy to see that the coefficient of $e^{i \omega x}$ reverses sign for all three values of ${j_{\infty}}$. Therefore, there appears an extra minus sign in the corresponding monodromy matching equation, and the condition for nontrivial solutions $(B_+,B_-,C_+,C_-)$ changes slightly. Carrying out the computation, the condition for quasinormal modes in $d=5$ turns out to be

$$
\sinh \left( \frac{\pi\omega}{k_C} - \frac{\pi\omega}{k^+} \right) + ( 1 + 2 \cos( \frac{2\pi}{5} ) ) \sinh \left( \frac{\pi\omega}{k^+} + \frac{\pi\omega}{k_C} \right) + ( 2 + 2 \cos( \frac{2\pi}{5} ) ) \sinh \left( \frac{2\pi\omega}{k^-} + \frac{\pi\omega}{k^+} + \frac{\pi\omega}{k_C} \right) = 0.
$$

\noindent
It is easily seen that this equation still has the correct RN limit.

As one searches the literature in the quest for earlier numerical results which could serve as checks on our own results, we find that there has not been much work on this matter. In the particular near--extremal situation, where the outer black hole event horizon and the cosmological horizon are nearly coincident, an approximation scheme for the $d$--dimensional RN dS solution was developed in \cite{molina}. However, the approximation used therein is not expected to hold in the asymptotic limit. Quasinormal modes of spinor perturbations in RN dS spacetime were studied in \cite{jing}, but this is a type of perturbation we have not addressed at all. Because our expressions yield the correct RN limit in any dimension, we believe this is strong evidence favoring our general results. However, it would be very interesting to produce numerical data concerning higher dimensional RN dS quasinormal modes to match against our analytical results.


\subsection{Asymptotically Anti--de Sitter Spacetimes}


\cite{kodama-ishibashi-3} discusses the stability of black holes in asymptotically AdS spacetimes to tensor, vector and scalar perturbations. For black holes without charge, tensor and vector perturbations are stable in any dimension. Scalar perturbations are stable in dimension four but there is no proof of stability in dimension $d \ge 5$. For charged black holes, tensor and vector perturbations are stable in any dimension. Scalar perturbations are stable in four dimensions but there is no proof of stability in dimension $d \ge 5$. As we work in generic dimension $d$ we are thus not guaranteed to always have a stable solution. Our results will apply if and only if the spacetime in consideration is stable.
    
Quantization of a scalar field in AdS was first addressed in \cite{ais}, where considerable attention was given to the question of what are the AdS boundary conditions. In fact, in AdS light rays can reach spatial infinity and return to the origin in finite time, as measured by the observer at the origin (crossing AdS within half the natural period). This would apparently indicate ``reflecting'' boundary conditions. However, the ``walls'' of AdS are at [timelike] spatial infinity, and so the concept of ``reflecting'' boundary conditions is actually somewhat obscure. In spite of this, embeddings of AdS spacetime in the Einstein Static Universe may lead to Dirichlet or Neumann boundary conditions (on the counterparts of the AdS fields) which, however, on the AdS fields themselves, will look rather complicated (these are not ``reflecting'' boundary conditions from a pure AdS point of view, see \cite{ais}). As it turns out, the only sensible boundary condition  to impose on quasinormal modes is the usual incoming waves at the black hole event horizon and the new requirement of vanishing of the wave function at infinity.


\subsubsection{The Schwarzschild Anti--de Sitter Solution}


We now compute the quasinormal modes of the Schwarzschild AdS $d$--dimensional black hole. This calculation was first done, for the case of dimension $d=4$ and \textit{large} black holes, in \cite{cns}. Again we consider solutions of the Schr\"odinger--like equation (\ref{schrodinger}) in the complex $r$--plane. We start by studying the behavior of $\Phi(x)$ near the singularity $r=0$. In this region, the tortoise coordinate is

$$
x \sim - \frac{r^{d-2}}{2(d-2)\MM},
$$

\noindent
and the potential for tensor and scalar type perturbations is

$$
V \big[ r(x) \big] \sim \frac{j^2 - 1}{4x^2},
$$

\noindent
with $j=0$ (see appendix \ref{appendixC}). The solution of the Schr\"odinger--like equation in this region (with $j \neq 0$) is therefore

$$
\Phi (x) \sim B_+ \sqrt{2\pi\omega x}\ J_{\frac{j}{2}} \left( \omega x \right) + B_- \sqrt{2\pi\omega x}\ J_{-\frac{j}{2}} \left( \omega x \right),
$$

\noindent
where $J_\nu(x)$ represents a Bessel function of the first kind and $B_\pm$ are (complex) integration constants. 

We now wish to examine the Stokes line, \textit{i.e.}, the curve $\im (\omega x)=0$. For the the Schwarzschild AdS solution, however, it is no longer true that $\im (\omega) \gg \re (\omega)$ for the asymptotic quasinormal modes: instead, one finds  that $\omega x_0$ is asymptotically \textit{real}, where $x \sim x_0$ for $r \sim \infty$ (see appendix \ref{appendixC}; this quantity is well--defined, as $x$ has zero monodromy around infinity, and hence one can choose the branch cuts arising from the branch points at the horizons to cancel among themselves). Interestingly, this is exactly the condition that the Stokes line $\im (\omega x)=0$ should extend out to infinity. From the general expression for the tortoise (see appendix \ref{appendixC}),

$$
x[r] = \sum_{n=1}^{d-1} \frac{1}{2k_{n}}\, \log \left( 1 - \frac{r}{R_{n}} \right),
$$

\noindent
one finds

$$
x_0 = \sum_{n=1}^{d-1} \frac{1}{2k_{n}}\, \log \left( - \frac{1}{R_{n}} \right)
$$

\noindent
(with an appropriate choice of branch cuts). For instance, for large black holes, where one has the approximate expression

$$
R_{n} = \left| \left( \frac{2\MM}{|\LL|} \right)^{\frac{1}{d-1}} \right| \exp \left( \frac{2\pi i}{d-1} (n-1) \right),
$$

\noindent
one finds

$$
x_0 = \frac{1}{4 T_{H} \sin \left( \frac{\pi}{d-1} \right)} \exp\left( -\frac{i\pi}{d-1} \right),
$$

\noindent
where $T_{H}$ is the Hawking temperature for large Schwarzschild AdS black holes. In general, however, there is no analytic solution for $x_0$. Because $\omega x_0$ is asymptotically real we will have $\arg(\omega) = -\arg(x_0) \equiv - \theta_0$ and therefore the function $\im(\omega x)$ will in general be multivalued around {\em all} horizons. To bypass this problem, we choose a particular branch and simply trace out the Stokes curve $\im(\omega x)$ shifting the branch cuts so that it never hits them. Note that the behavior of $e^{\pm i\omega x}$ will still be oscillatory along the curve. In a neighborhood of the origin, the above relation between $x$ and $r$ tells us that $\im(\omega x)=0$ for

$$
r=\rho\ e^{\frac{i\theta_0}{d-2}+\frac{in\pi}{d-2}},
$$

\noindent
with $\rho>0$ and $n=0,1,\ldots, 2d-5$. These are half--lines starting at the origin, equally spaced by an angle of $\frac{\pi}{d-2}$. Notice that the sign of $\omega x$ on these half--lines is $(-1)^{n+1}$; in other words, starting with the half--line corresponding to $n=0$, the sign of $\omega x$ is alternately negative and positive as one goes anti--clockwise around the origin.

From the asymptotic expansion (\ref{Bessel+}) we see that 

\begin{eqnarray}
\Phi (x) & \sim & 2 B_+ \cos \left( \omega x - \alpha_+ \right) + 2 B_- \cos \left( \omega x - \alpha_- \right) \nonumber \\
& = & \left( B_+ e^{-i\alpha_+} + B_- e^{-i\alpha_-}\right) e^{i \omega x} + \left( B_+ e^{i\alpha_+} + B_- e^{i\alpha_-}\right) e^{-i \omega x} \label{SAdS_0}
\end{eqnarray}

\noindent
in any one of the lines corresponding to positive $\omega x$, where again we have defined

$$
\alpha_\pm = \frac{\pi}4 (1 \pm j).
$$

\noindent
We shall use this asymptotic expression for matching. This matching must be done along the Stokes line $\omega x \in \mathbb{R}$, so that neither of the exponentials $e^{\pm i\omega x}$ dominates the other.

To trace out the Stokes line $\im(\omega x)=0$ let us first observe that we already know its behavior near the origin. Furthermore, this is the only singular point of this curve: indeed, since $x$ is a holomorphic function of $r$, the critical points of the function $\re(x)$ are the zeros of

$$
\frac{dx}{dr} = \frac{1}{f(r)} = \frac{r^{d-3}}{-\LL r^{d-1}+r^{d-3}-2\MM}
$$

\noindent
(\textit{i.e.}, $r=0$ only). To understand the behavior of the Stokes line near the singularities, we notice that if one follows our procedure of shifting the branch cuts, the curve

$$
\im \left( \alpha \log (z) \right) = 0
$$

\noindent
is the curve

$$
\alpha \log(z) = \rho \quad \Leftrightarrow \quad z = e^{\xi \rho} e^{i \eta \rho} \quad (\rho \in \mathbb{R})
$$

\noindent
(with $\alpha = \frac1{\xi + i \eta}$). This is a spiral that approaches the singularity $z=0$, except in the case where $\xi = 0$, \textit{i.e.}, when $\alpha$ is purely imaginary. Therefore, generically one expects the Stokes line to hit all $d-1$ horizons in this spiraling fashion. Recall however that at least one branch must extend out to infinity. In our choice of branch cuts, we have arranged things so that this branch corresponds to $n=1$; therefore the preceding branch (corresponding to $n=0$) will hit the black hole horizon $r=R_H$. Notice also that the formula

$$
\omega x \sim \omega x_0 - \frac{\omega}{|\LL|r}
$$

\noindent
for $r \sim \infty$ implies that the argument of $r$ along the branch which extends to infinity must be asymptotically equal to the argument of $\omega$, \textit{i.e.}, $-\theta_0$. Therefore, one would guess that the Stokes line is as depicted in Figure \ref{StokesSAdS}. We have verified this guess with a numerical computation of the same Stokes line, and this is indicated in Figures \ref{NumericalStokesSAdS} and \ref{NumericalStokesSAdS_Large}. It should be noted that due to the branch cuts (which can be readily identified in the figure) the spirals at the singularities are not so clearly depicted in the numerical result.

\FIGURE[ht]{\label{StokesSAdS}
	\centering
	\psfrag{A}{$A$}
	\psfrag{B}{$B$}
	\psfrag{RH}{$R_H$}
	\psfrag{g1}{$\gamma_1$}
	\psfrag{og1}{$\overline{\gamma}_1$}
	\psfrag{g2}{$\gamma_2$}
	\psfrag{og2}{$\overline{\gamma}_2$}
	\psfrag{Re}{$\re$}
	\psfrag{Im}{$\im$}
	\psfrag{contour}{contour}
	\psfrag{Stokes line}{Stokes line}
	\epsfxsize=.6\textwidth
	\leavevmode
	\epsfbox{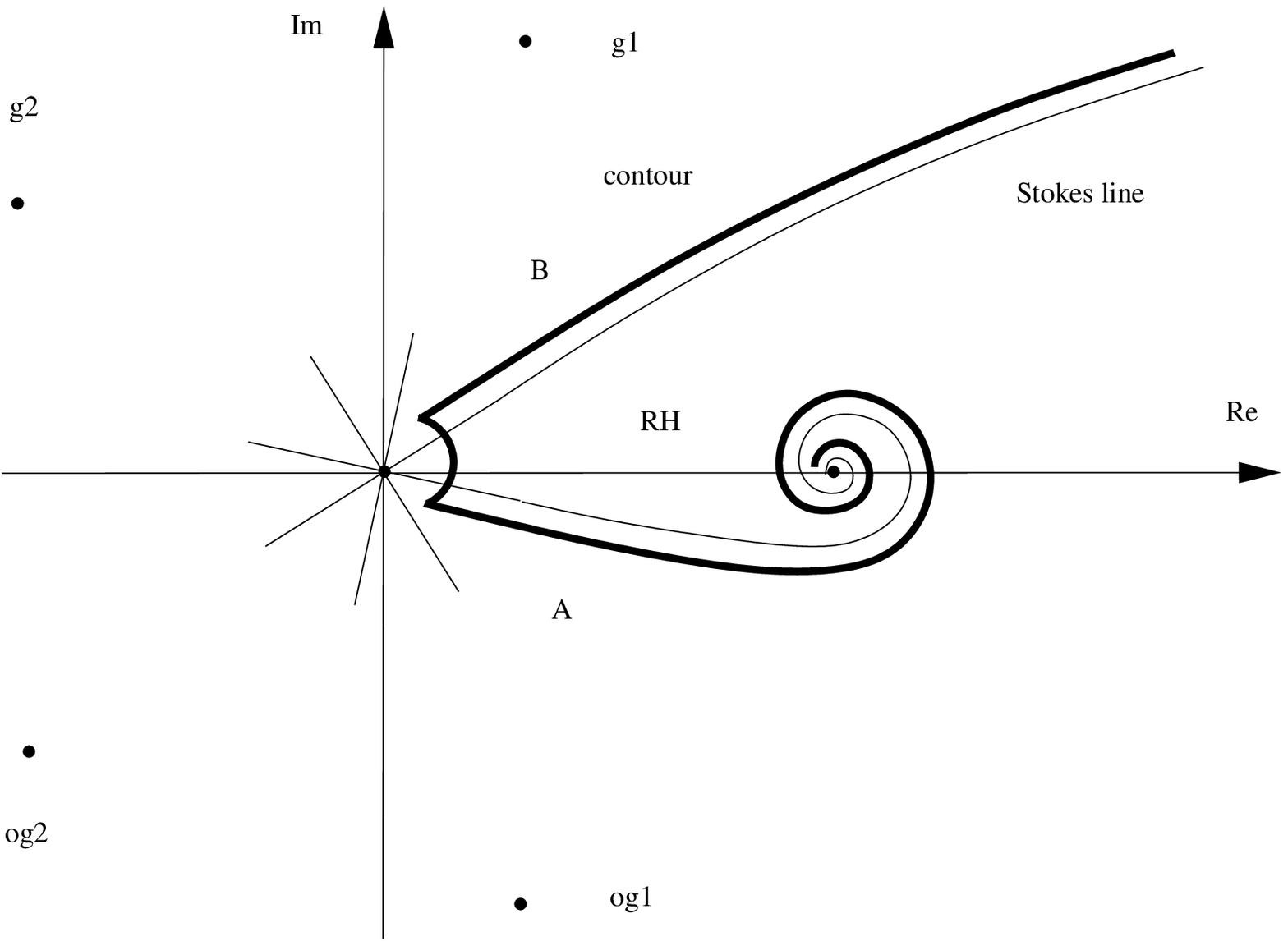}
\caption{Stokes line for the Schwarzschild Anti--de Sitter black hole black hole, along with the chosen contour for monodromy matching, in the case of dimension $d=6$.}
}

\FIGURE[ht]{\label{NumericalStokesSAdS}
	\centering
	\epsfig{file=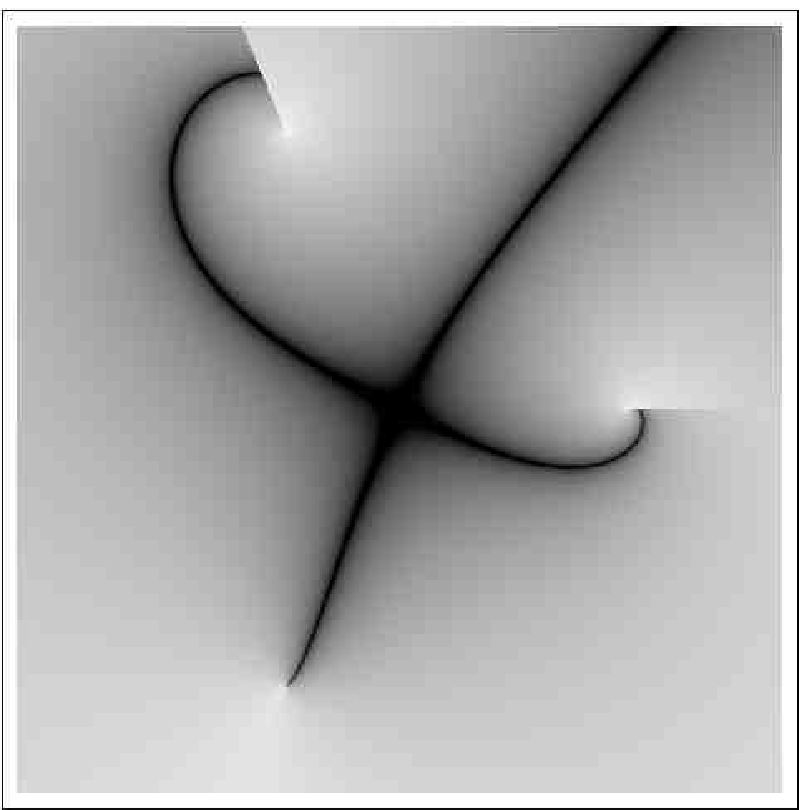, width=1.5in, height=1.5in}
    $\quad$
    \epsfig{file=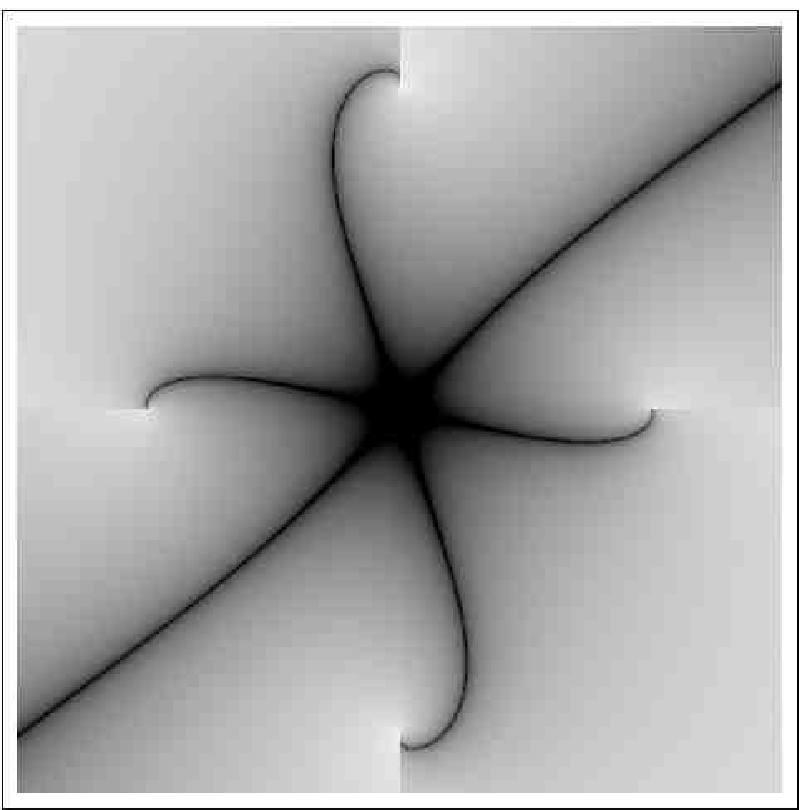, width=1.5in, height=1.5in}
    $\quad$
    \epsfig{file=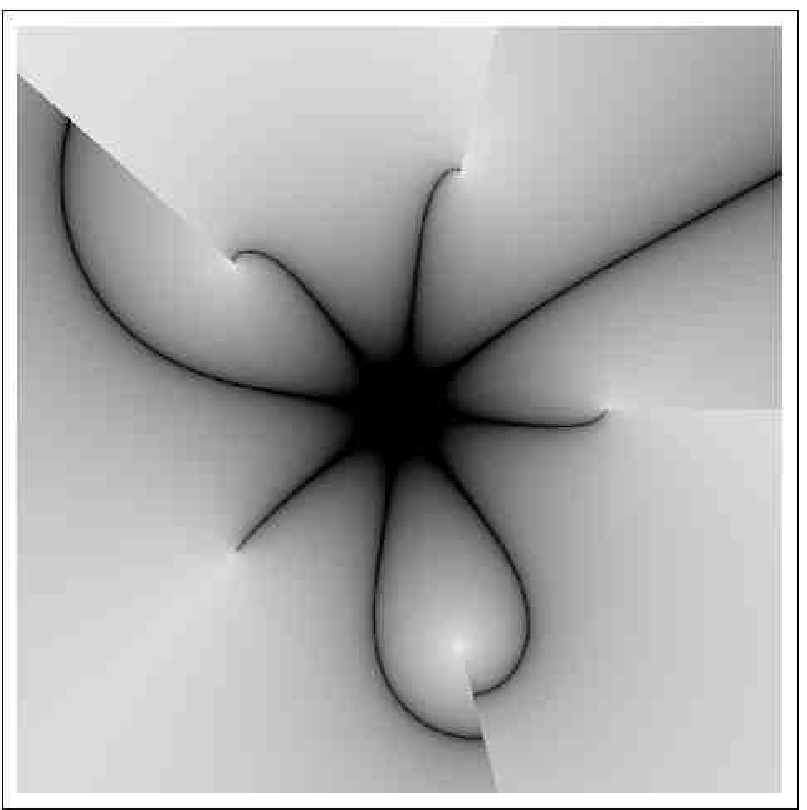, width=1.5in, height=1.5in}
    $\quad$
    \epsfig{file=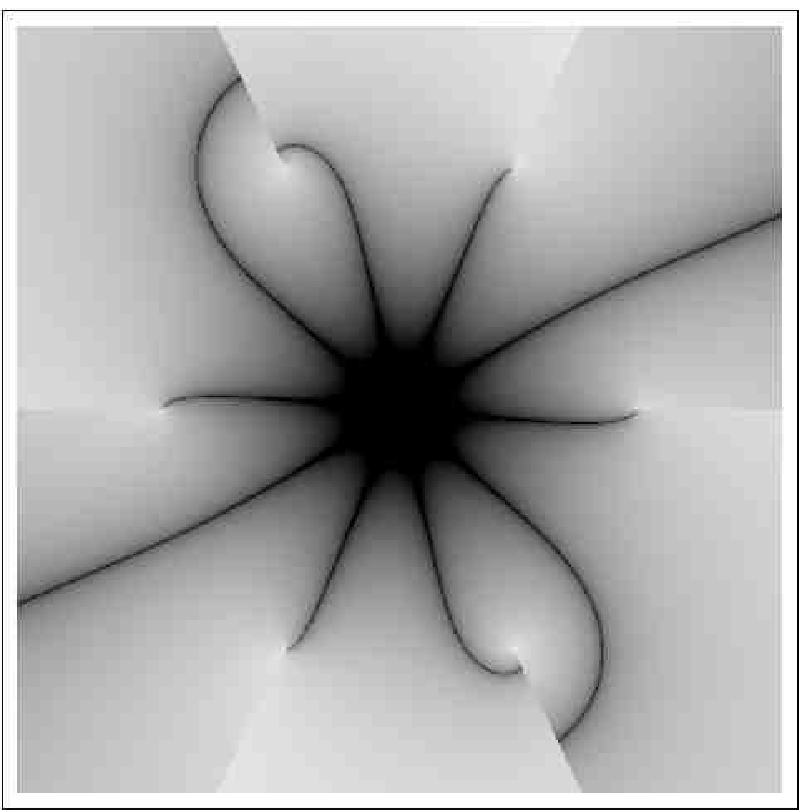, width=1.5in, height=1.5in}
\caption{Numerical calculation of the Stokes lines for the Schwarzschild Anti--de Sitter black hole in dimensions $d=4$, $d=5$, $d=6$ and $d=7$. Different shadings also illustrate the various horizon singularities and branch cuts (note that these branch cuts are not necessarily equal to the ones used for the calculation in the main text).}
}

\FIGURE[ht]{\label{NumericalStokesSAdS_Large}
	\centering
	\epsfig{file=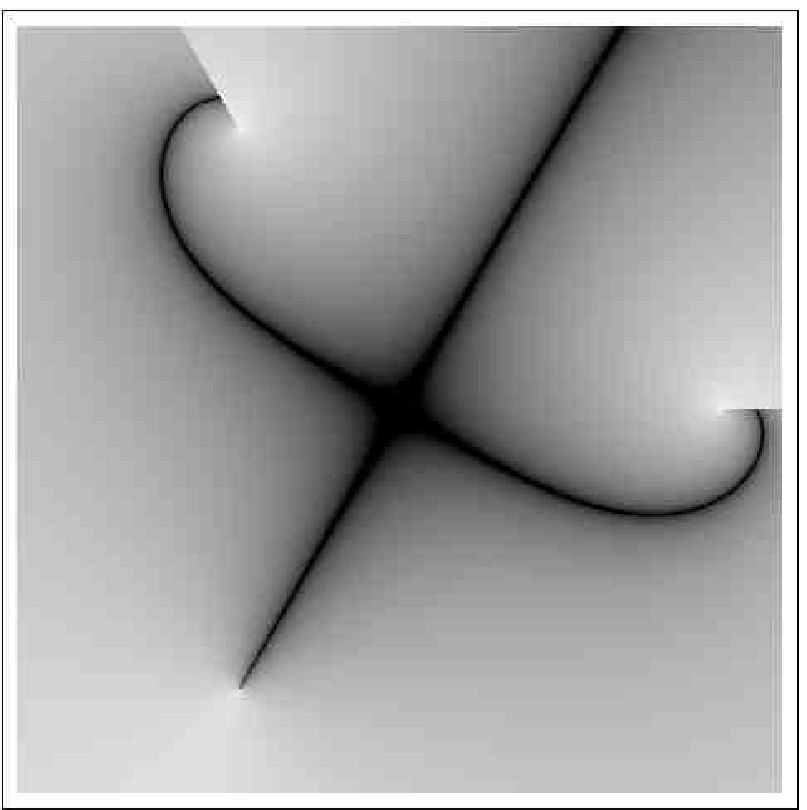, width=1.5in, height=1.5in}
    $\quad$
    \epsfig{file=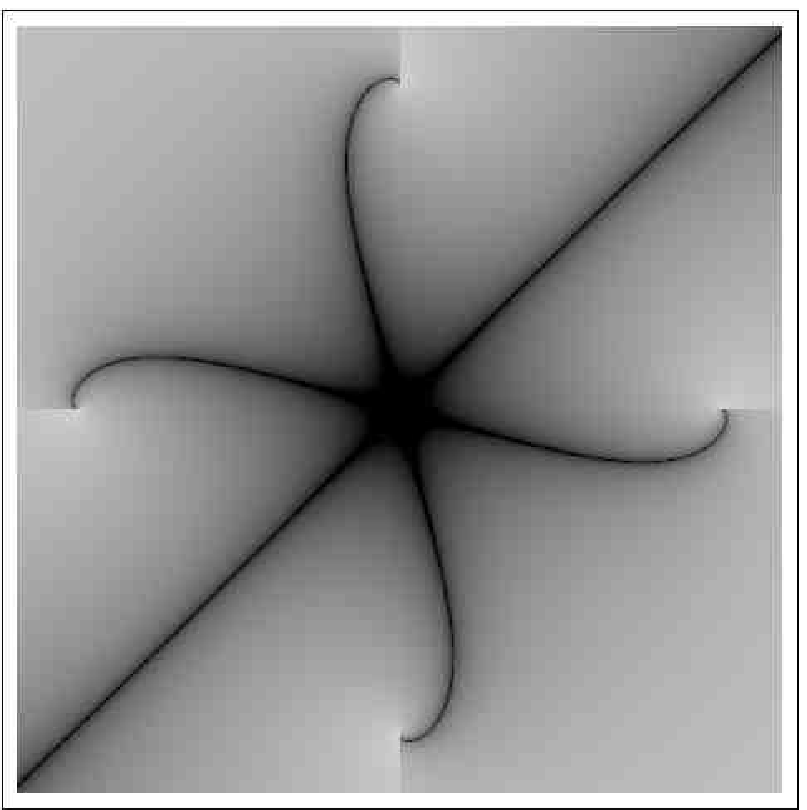, width=1.5in, height=1.5in}
    $\quad$
    \epsfig{file=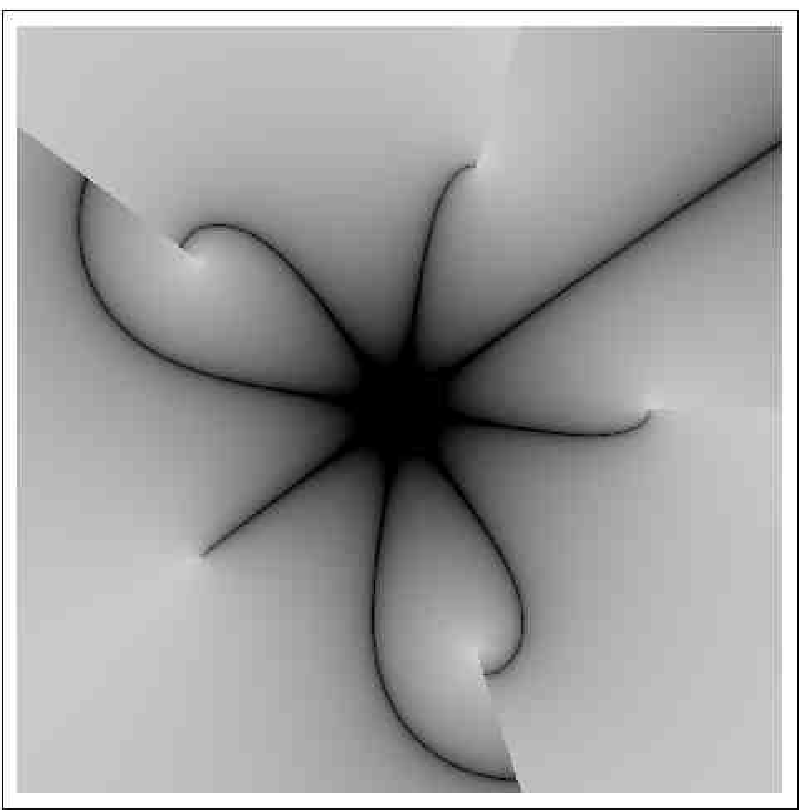, width=1.5in, height=1.5in}
    $\quad$
    \epsfig{file=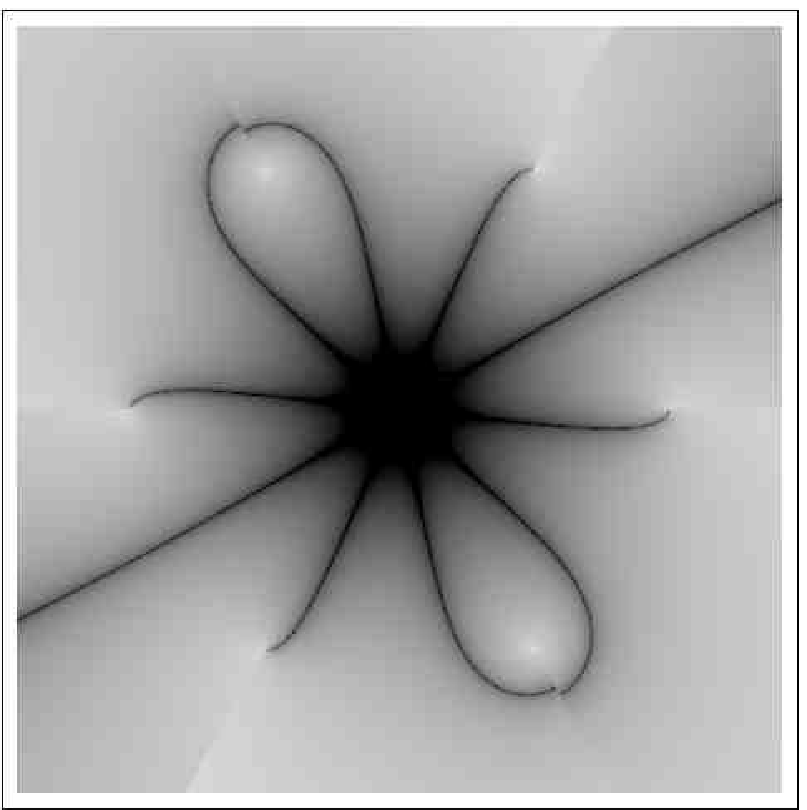, width=1.5in, height=1.5in}
\caption{Numerical calculation of the Stokes lines for \textit{large} Schwarzschild Anti--de Sitter black holes in $d=4$, $d=5$, $d=6$ and $d=7$. The lines are as before, but horizon singularities are now located at $(d-1)$ roots of unity.}
}

Now, for $r \sim \infty$ we have 

$$
V \sim  \frac{{j_{\infty}}^2-1}{4(x-x_0)^2}
$$

\noindent 
where ${j_{\infty}}=d-1,d-3,d-5$ for tensor type, vector type and scalar type perturbations (see appendix \ref{appendixC}). Consequently,

$$
\Phi (x) \sim C_+ \sqrt{2\pi\omega (x-x_0)}\ J_{\frac{{j_{\infty}}}{2}} \left( \omega (x-x_0) \right) + C_- \sqrt{2\pi\omega (x-x_0)}\ J_{-\frac{{j_{\infty}}}{2}} \left( \omega (x-x_0) \right)
$$

\noindent
for $r \sim \infty$. The boundary condition $\Phi = 0$ at $r = \infty$ (discussed at the beginning of this section) requires that $C_-=0$. Hence,

$$
\Phi (x) \sim C_+ \sqrt{2\pi\omega (x-x_0)}\ J_{\frac{{j_{\infty}}}{2}} \left( \omega (x-x_0) \right)
$$

\noindent
for $r \sim \infty$. This is to be matched with expansion (\ref{SAdS_0}) along the branch of the Stokes line extending out to infinity. To do so, we notice that for $r \sim \infty$ one has

$$
\omega(x-x_0) \sim - \frac{\omega}{|\LL|r},
$$

\noindent
and that this quantity is {\em negative} on this branch. Consequently, expansion (\ref{Bessel-}) yields

$$
\Phi (x) \sim  C_+ e^{i\beta_+} e^{i \omega (x-x_0)} + C_+ e^{-i\beta_+} e^{-i \omega (x-x_0)} \\
$$

\noindent
in the same limit, where

$$
\beta_+ = \frac{\pi}4 (1 + {j_{\infty}}).
$$

\noindent
We conclude that $B_+, B_-$ must satisfy

\begin{equation} \label{SAdSinfty}
\left( B_+ e^{-i\alpha_+} + B_- e^{-i\alpha_-}\right) e^{i \omega x_0} e^{-i\beta_+} = \left( B_+ e^{i\alpha_+} + B_- e^{i\alpha_-}\right) e^{-i \omega x_0} e^{i\beta_+}.
\end{equation}

Expansion (\ref{Bessel0}) implies that as one rotates from the branch containing point $B$ to the branch containing point $A$ (see Figure \ref{StokesSAdS}) (through an angle of $-\frac{\pi}{d-2}$ in $r$, corresponding to an angle $-\pi$ in $x$), one has from

$$
\sqrt{2\pi e^{-\pi i} \omega x}\ J_{\pm\frac{j}{2}} \left( e^{-\pi i} \omega x \right) = e^{\frac{-\pi i}2 (1 \pm j)}\sqrt{2\pi \omega x}\ J_{\pm\frac{j}{2}} \left( \omega x \right) \sim 2 e^{-2i\alpha_\pm} \cos(\omega x - \alpha_\pm)
$$

\noindent
that $\Phi$ changes to

\begin{align*}
\Phi (x) & \sim 2 B_+ e^{-2i\alpha_+} \cos \left( -\omega x - \alpha_+ \right) + 2 B_- e^{-2i\alpha_-} \cos \left( -\omega x - \alpha_- \right) = \\
& = \left( B_+ e^{-i\alpha_+} + B_- e^{-i\alpha_-}\right) e^{i \omega x} + \left( B_+ e^{-3i\alpha_+} + B_- e^{-3i\alpha_-}\right) e^{-i \omega x}.
\end{align*}

\noindent
This form of the solution can be propagated along the corresponding branch of the Stokes line which approaches the event horizon, and where we know that $\Phi (x)$ must behave as $e^{i \omega x}$. Consequently we obtain the second condition on $B_+, B_-$, as

$$
B_+ e^{-3i\alpha_+} + B_- e^{-3i\alpha_-} = 0.
$$

\noindent
The two conditions on these coefficients can only have nontrivial solutions if and only if

\begin{align*}
&\left| 
\begin{array}{ccc}
e^{-3i\alpha_+} & \,\,\,\, & e^{-3i\alpha_-} \\ & & \\
e^{-i\alpha_+} e^{-i\beta_+} e^{i \omega x_0} - e^{i\alpha_+} e^{i\beta_+} e^{-i \omega x_0} &  & e^{-i\alpha_-}e^{-i\beta_+}  e^{i \omega x_0} - e^{i\alpha_-} e^{i\beta_+} e^{-i \omega x_0}
\end{array}
\right| = 0 
\\
&\Leftrightarrow \quad
\left| 
\begin{array}{ccc}
e^{-3i\alpha_+} & \,\,\,\, & e^{-3i\alpha_-} \\ & & \\
\sin\left( \alpha_+ + \beta_+ - \omega x_0 \right) & & \sin\left( \alpha_- + \beta_+ - \omega x_0 \right)
\end{array}
\right| = 0.
\end{align*}

\noindent
Again, this equation is automatically satisfied for $j=0$. We must therefore take $j$ nonzero and then compute the limit as $j\to 0$. This amounts to writing the equation as a power series in $j$ and equating to zero the first nonvanishing coefficient, which in this case is the coefficient of the linear part. Thus we just have to require that the derivative of the determinant above with respect to $j$ be zero for $j=0$. This is

\begin{align*}
& \left|
\begin{array}{ccc}
-\frac{3i\pi}4 e^{-\frac{3i\pi}4} & & \frac{3i\pi}4 e^{-\frac{3i\pi}4} \\ & & \\
\sin\left( \frac{\pi}4 + \beta_+ - \omega x_0 \right) & & \sin\left( \frac{\pi}4 +\beta_+ - \omega x_0 \right)
\end{array}
\right|+
\left|
\begin{array}{ccc}
e^{-\frac{3i\pi}4} & & e^{-\frac{3i\pi}4} \\ & & \\
\frac{\pi}4 \cos \left( \frac{\pi}4 + \beta_+ - \omega x_0 \right) & & - \frac{\pi}4 \cos \left( \frac{\pi}4 + \beta_+ - \omega x_0 \right)
\end{array}
\right|=0
\end{align*}

\noindent
from where we obtain our final result as

\begin{equation} \label{resultsSAdS}
\tan \left( \frac{\pi}4+\beta_+-\omega x_0 \right) = \frac{i}3 \quad \Leftrightarrow \quad \omega x_0 = \frac{\pi}4 + \beta_+ - \arctan \left( \frac{i}3 \right) + n \pi \quad (n \in \mathbb{N}).
\end{equation}

\noindent
Notice that in this asymptotically AdS case there was no need to do a monodromy calculation, all we had to do was follow the Stokes line from infinity to the black hole event horizon.

For vector type perturbations, the potential for the Schr\"odinger--like master equation near the origin is of the form

$$
V \big[ r(x) \big] \sim \frac{j^2 - 1}{4x^2},
$$

\noindent
with $j=2$ (see appendix \ref{appendixC}). Repeating the same argument, one ends up with the same equation, except that now $\alpha_\pm=\frac{\pi}4 \pm \frac{\pi}2$. Since $\alpha_+-\alpha_-=\pi$, this equation is still automatically satisfied for $j=2$; taking the limit as $j \to 2$ one obtains

\begin{equation} \label{resultsSAdS2}
\omega x_0 = \frac{3\pi}4 + \beta_+ - \arctan \left( \frac{i}3 \right) + n \pi \quad (n \in \mathbb{N}).
\end{equation}

\noindent
The difference of $\frac{\pi}2$ between this equation and equation (\ref{resultsSAdS}) exactly compensates the difference between the values of $\beta_+$ for the different types of perturbations, so that {\em all} types of perturbation yield exactly the same asymptotic quasinormal frequencies, as given by

$$
\omega x_0 = \frac{\pi}4 \left( d+1 \right) - \arctan \left( \frac{i}3 \right) + n \pi \quad (n \in \mathbb{N}).
$$

Some remarks are due for \textit{scalar} type perturbations in dimensions $d=4$ and $d=5$. Indeed, in the previous calculation we have assumed that $j_{\infty}$ is positive. This is not true for scalar type perturbations in dimensions four and five. In dimension $d=4$ the calculation still goes through as long as one sets $j_{\infty} = + 1$ (rather than $-1$) as all one needs is $j_{\infty}^2=1$. Consequently the formula for asymptotic quasinormal frequencies of scalar type perturbations in dimension $d=4$ gets changed to

$$
\omega x_0 = \frac{3\pi}4 - \arctan \left( \frac{i}3 \right) + n \pi \quad (n \in \mathbb{N}).
$$

\noindent
In dimension $d=5$ the situation is more subtle. Here $j_{\infty} = 0$ and strictly speaking the two Bessel functions in the solution at $r \sim \infty$ are not linearly independent. As usual, we choose $j_{\infty}$ slightly positive and carry through with the calculation, taking the limit $j_{\infty} \to 0$. In this case the solution at $r \sim \infty$ automatically vanishes as $\frac{1 \pm j_{\infty}}{2}$ is positive. Thus, there is no constraint arising from infinity on the solution to the master equation, and all one is left with is the constraint at the black hole horizon. This condition alone is not enough to quantize the asymptotic frequencies. Thus, in five dimensions, the asymptotic spectrum of Schwarzschild AdS scalar type perturbations is \textit{continuous}, \textit{i.e.}, $\omega \in {\mathbb{C}}$. This is a rather unexpected result. We believe that the consequences of this result should deserve further studies.

We now need to review the literature concerning the calculation of asymptotic quasinormal frequencies in the Schwarzschild AdS spacetime, in order to compare our results to earlier work done on this subject. Quasinormal modes of Schwarzschild AdS black holes were first addressed in \cite{chan-mann, horowitz-hubeny, cardoso-lemos-2, giammatteo-jing} although the emphasis lied in the lowest frequency modes (with \cite{giammatteo-jing} focusing on spinor perturbations). It was shown in \cite{horowitz-hubeny} that quasinormal frequencies of Schwarzschild AdS black holes have a direct interpretation in terms of the dual CFT: large AdS black holes correspond to CFT thermal states, so that quasinormal frequencies correspond to poles in the retarded Green's function of the dual CFT. Because in this context the decay of the perturbations is describing the return to thermal equilibrium, the inverse of the gap is naturally associated with the time scale for approaching this  thermal equilibrium and the inverse of the offset is naturally associated with the oscillation time scale. It was moreover shown in \cite{horowitz-hubeny} that these  time scales for the approach of thermal equilibrium are universal, in the sense that all scalar fields with the same angular dependence will decay at this rate. Due to the AdS/CFT correspondence, this work sparkled a series of investigations on AdS asymptotic quasinormal frequencies, which naturally concentrated in the five dimensional case (see, \textit{e.g.}, \cite{starinets, nunez-starinets, musiri-siopsis, siopsis, siopsis-2}).

For the case of dimension $d=4$, the first numerical results for the asymptotic quasinormal frequencies were published in \cite{berti-kokkotas-1}. These authors found that scalar perturbations are isospectral with both odd and even parity gravitational perturbations, and they also found the existence of modes with purely imaginary frequency. Later, an extensive study of asymptotic quasinormal frequencies for Schwarzschild AdS black holes in $d=4$ was done in \cite{ckl}, and numerically produced some numbers which exactly match our analytical prediction. While the authors of \cite{ckl} found that the real part of the frequency mode increases with the overtone number, $n$, in what seems to be a characteristic particular to AdS space, they also found the following numerical data (having set $|\LL|=1$). For \textit{large} black holes, with $R_{H} \gg 1$, the result for scalar field perturbations (corresponding to tensor type perturbations) was found to be (we have translated the results in \cite{ckl} to our own conventions on quasinormal boundary conditions)

$$
\omega_n \sim R_{H} \left[ \left( 1.299 + 2.25 i \right) n + \left( 0.557 + 0.423 i \right) \right],
$$

\noindent
while for gravitational odd perturbations (corresponding to vector type perturbations) they found

$$
\omega_n \sim R_{H} \left[ \left( 1.299 + 2.25 i \right) n + \left( 0.58 + 0.42 i \right) \right],
$$

\noindent
and for gravitational even perturbations (corresponding to scalar type perturbations)

$$
\omega_n \sim R_{H} \left[ \left( 1.299 + 2.25 i \right) n + \left( 0.581 + 0.41 i \right) \right].
$$

\noindent
For intermediate black holes, with $R_{H} \sim 1$, the result for scalar field perturbations (corresponding to tensor type perturbations) was found to be

$$
\omega_n \sim \left( 1.97 + 2.35 i \right) n + \left( 0.79 + 0.35 i \right),
$$

\noindent
while for gravitational odd perturbations (corresponding to vector type perturbations) they found

$$
\omega_n \sim \left( 1.97 + 2.35 i \right) n + \left( 0.93 + 0.32 i \right),
$$

\noindent
and for gravitational even perturbations (corresponding to scalar type perturbations)

$$
\omega_n \sim \left( 1.96 + 2.35 i \right) n + \left( 2.01 + 1.5 i \right).
$$

\noindent
For small black holes $R_{H} \ll 1$ and in this $R_{H} \to 0$ limit the frequencies were shown to approach the pure AdS frequencies (which we address later in the
present paper), a result which had been first shown in \cite{konoplya-3}. The numerical results for scalar field perturbations (corresponding to tensor type perturbations) were found to be

$$
\omega_n \sim \left( 1.69 + 0.57 i \right) n + \left( 2.29 + 0.46 i \right),
$$

\noindent
while for gravitational odd perturbations (corresponding to vector type perturbations) they found

$$
\omega_n \sim \left( 1.69 + 0.59 i \right) n + \left( 2.49 + 0.06 i \right),
$$

\noindent
and for gravitational even perturbations (corresponding to scalar type perturbations)

$$
\omega_n \sim \left( 1.61 + 0.6 i \right) n + \left( 2.7 + 0.37 i \right).
$$

\noindent
Let us also recall that the case of large Schwarzschild AdS black holes in dimension $d=4$ was first solved analytically in \cite{cns}, fully agreeing with the numerical results in \cite{ckl}, as listed above.

For the case of dimension $d=5$, different methods were applied in \cite{starinets, nunez-starinets, musiri-siopsis} to study scalar field perturbations (corresponding to tensor type perturbations). The analysis of \cite{starinets, nunez-starinets} relied on the AdS/CFT correspondence, their goal being to find poles in the retarded gauge theory thermal correlators, which in turn is equivalent to computing quasinormal frequencies in the dual gravitational background. Their results were obtained numerically, via use of the Heun equation. The asymptotic quasinormal frequencies found for \textit{large} AdS black holes were (we have translated the result to our own conventions on quasinormal boundary conditions)

$$
\omega = 2 \pi n T_{H} \left( 1 + i \right) + \omega_{0},
$$

\noindent
where $\omega_{0} = \pi T_{H} \left( 1.2139 + 0.7775 i \right)$ and $T_{H}$ is the Hawking temperature in this spacetime. The analysis of \cite{musiri-siopsis} was similar, in the sense that they also computed asymptotic quasinormal frequencies for \textit{large} AdS black holes in dimensions $d=3$ and $d=5$ via an approximation of the wave equation by the hypergeometric equation, and in the high frequency regime. However, they differed from \cite{starinets, nunez-starinets} as they used a method of monodromy matching rather than a numerical scheme. In dimension $d=5$ they find a result which is in complete agreement with the results of \cite{starinets, nunez-starinets}.

Let us now compare with our own results. As we have seen, our Schwarzschild AdS quasinormal equation depends on a parameter, $x_0$, for which there is no general analytic solution. However, one can find an analytical formula for $x_0$ in the case of \textit{large} black holes, and we shall begin there. For these black holes the asymptotic quasinormal frequencies are given by (notice that one needs to be careful with the choice of logarithmic branch cuts when computing $x_0$)

$$
\omega = 4 \pi n T_{H} \sin \left( \frac{\pi}{d-1} \right) e^{\frac{i\pi}{d-1}} + 4 T_{H} \sin \left( \frac{\pi}{d-1} \right) e^{\frac{i\pi}{d-1}} \left( \frac{\pi}{4} (d+1) - \arctan \left( \frac{i}{3} \right) \right),
$$

\noindent
so that their leading term is located along the direction of a $(d-1)$ root of unity. In dimension $d=4$, and using the relation $T_{H} = \frac{3}{4\pi} |\LL| R_{H}$, it is immediate to write down

\begin{eqnarray*}
\frac{\omega}{|\LL| R_{H}} &=& \frac{3\sqrt{3}}{4} \left( 1 + i \sqrt{3} \right) n + \frac{3\sqrt{3}}{4\pi} \left( 1+i\sqrt{3} \right) \left( \frac{\pi}{4} - \arctan \left( \frac{i}{3} \right) \right) \\
&=& \left( 1.299 + 2.250 i \right) n + \left( 0.573 + 0.419 i \right)
\end{eqnarray*}

\noindent
for tensor (corresponding to scalar field) and vector type perturbations, and

\begin{eqnarray*}
\frac{\omega}{|\LL| R_{H}} &=& \frac{3\sqrt{3}}{4} \left( 1 + i \sqrt{3} \right) n + \frac{3\sqrt{3}}{4\pi} \left( 1+i\sqrt{3} \right) \left( \frac{3\pi}{4} - \arctan \left( \frac{i}{3} \right) \right) \\
&=& \left( 1.299 + 2.250 i \right) n + \left( 1.222 + 1.544 i \right)
\end{eqnarray*}

\noindent
for scalar type perturbations. Similarly, in dimension $d=5$ it is immediate to write down for tensor and vector type perturbations:

\begin{eqnarray*}
\frac{\omega}{T_{H}} &=& 2 \pi n \left( 1 + i \right) + \pi \left( 1 + i \right) \left( 1 - \frac{2}{\pi} \arctan \left( \frac{i}{3} \right) \right) \\
&=& 2 \pi n \left( 1 + i \right) + \pi \left( 1.22064 + 0.77936 i \right).
\end{eqnarray*}

\noindent
In either case, for large Schwarzschild AdS black holes, our analytical results are in full and complete agreement with earlier numerical calculations to a great degree of accuracy, as long as one restricts to tensor (scalar field) or vector type perturbations. For scalar type perturbations in dimension $d=4$ our results yield complete agreement with the gap, but not the offset. We believe further numerical studies should be performed in order to obtain better estimates on the offsets.

Let us next turn to the case of intermediate black holes as discussed in \cite{ckl}, where $R_H \sim 1$. Here, we first proceed numerically in order to determine $x_0$ (paying special attention to the choice of logarithmic branch cuts) and then plug the numerical result for $x_0$ into our general expression for the asymptotic quasinormal frequencies. Using the values of \cite{ckl} for cosmological constant, black hole mass and horizon radius it is immediate to obtain, from the Schwarzschild AdS quasinormal equation,

$$
\omega = \left( 1.969 + 2.350 i \right) n + \left( 0.752 + 0.370 i \right)
$$

\noindent
for tensor (corresponding to scalar field) and vector type perturbations, and

$$
\omega = \left( 1.969 + 2.350 i \right) n + \left( 1.737 + 1.545 i \right)
$$

\noindent
for scalar type perturbations. This result is again in full and complete agreement with the numerical calculations of intermediate black holes in \cite{ckl}, to a great degree of accuracy and for \textit{all} types of perturbations.

Finally, we consider the case of small black holes, where $R_H \to 0$. If we take this limit directly in the Schwarzschild AdS quasinormal equation, it is not too hard to see that only the complex horizons will contribute in the calculation of $x_0$ which becomes

$$
x_0 \sim \frac{\pi}{2\sqrt{|\LL|}}
$$

\noindent
with the asymptotic frequencies immediately following as

$$
\omega = 2 \sqrt{|\LL|} \left( n + \frac{d+1}{4} - \frac{1}{\pi} \arctan \left( \frac{i}{3} \right) \right).
$$

\noindent
As we shall later show, in section 4, this yields the correct value for the gap of pure AdS normal frequencies (while the offset comes out incorrect, but there is no reason why it should be preserved in this limit). Once again this confirms the numerical expectations in \cite{ckl, konoplya-3} where it was observed that the Schwarzschild AdS asymptotic frequencies approached the pure AdS frequencies in the $R_H \to 0$ limit. We can moreover check the numerical prediction of our analytical formula. Using the values of \cite{ckl} for cosmological constant, black hole mass and horizon radius, it is immediate to obtain for small black holes

$$
\omega = \left( 1.696 + 0.571 i \right) n + \left( 2.182 + 0.529 i \right)
$$

\noindent
for tensor (corresponding to scalar field) and vector type perturbations, and

$$
\omega = \left( 1.696 + 0.571 i \right) n + \left( 3.030 + 0.815 i \right)
$$

\noindent
for scalar type perturbations. This result is again in full and complete agreement with the numerical calculations for the gap of small black holes in \cite{ckl}, to a great degree of accuracy and all types of perturbations. The agreement with the numerical calculations for the offset is reasonable, where we believe further numerical studies should be performed in order to obtain better estimates on the offsets.

In summary, our results completely match what was previously known in the literature. However, it would be very interesting to produce further numerical data concerning higher dimensional Schwarzschild AdS quasinormal modes to match against all our analytical results.


\subsubsection{The Reissner--Nordstr\"om Anti--de Sitter Solution}


We now compute the quasinormal modes of the RN AdS $d$--dimensional black hole. Again we consider solutions of the Schr\"odinger--like equation (\ref{schrodinger}) in the complex $r$--plane. We start by studying the behavior of $\Phi(x)$ near the singularity $r=0$. In this region, the tortoise coordinate is

$$
x \sim \frac{r^{2d-5}}{\left( 2d-5 \right) \QQ^{2}},
$$

\noindent
and the potential for tensor and scalar type perturbations is

$$
V \big[ r(x) \big] \sim -\frac1{4x^2} = \frac{j^2 - 1}{4x^2},
$$

\noindent
with $j=\frac{d-3}{2d-5}$ (see appendix \ref{appendixC}). The solution of the Schr\"odinger--like equation in this region (with $j \neq 0$) is therefore

$$
\Phi (x) \sim B_+ \sqrt{2\pi\omega x}\ J_{\frac{j}{2}} \left( \omega x \right) + B_- \sqrt{2\pi\omega x}\ J_{-\frac{j}{2}} \left( \omega x \right),
$$

\noindent
where $J_\nu(x)$ represents a Bessel function of the first kind and $B_\pm$ are (complex) integration constants. 

We now wish to examine the Stokes line, \textit{i.e.}, the curve $\im (\omega x)=0$. For the the RN AdS solution, however, it is no longer true that $\im (\omega) \gg \re (\omega)$ for the asymptotic quasinormal modes: instead, one finds that $\omega x_0$ is asymptotically real, where $x \sim x_0$ for $r \sim \infty$ (see appendix \ref{appendixC}; this quantity is well--defined, as $x$ has zero monodromy around infinity, and hence one can choose the branch cuts arising from the branch points at the horizons to cancel among themselves). Interestingly, this is exactly the condition that the Stokes line $\im (\omega x)=0$ should extend out to infinity. From the general expression for the tortoise (see appendix \ref{appendixC}),

$$
x[r] = \sum_{n=1}^{2d-4} \frac{1}{2k_{n}}\, \log \left( 1 - \frac{r}{R_{n}} \right),
$$

\noindent
one finds

$$
x_0 = \sum_{n=1}^{2d-4} \frac{1}{2k_{n}}\, \log \left( - \frac{1}{R_{n}} \right)
$$

\noindent
(with an appropriate choice of branch cuts). Thus, because $\omega x_0$ is asymptotically real, we will have $\arg(\omega) = -\arg(x_0) \equiv - \theta_0$ and  therefore the function $\im(\omega x)$ will in general be multivalued around {\em all} horizons. To bypass this problem, we choose a particular branch and simply trace out the Stokes curve $\im(\omega x)$ shifting the branch cuts so that it never hits them. Note that the behavior of $e^{\pm i\omega x}$ will still be oscillatory along the curve. In a neighborhood of the origin, the above relation between $x$ and $r$ tells us that $\im(\omega x)=0$ for

$$
r=\rho\ e^{\frac{i\theta_0}{2d-5}+\frac{in\pi}{2d-5}},
$$

\noindent
with $\rho>0$ and $n=0,1,\ldots, 4d-11$. These are half--lines starting at the origin, equally spaced by an angle of $\frac{\pi}{2d-5}$. Notice that the sign of $\omega x$ on these lines is $(-1)^n$; in other words, starting with the line corresponding to $n=0$, the sign of $\omega x$ is alternately positive and negative as one goes anti--clockwise around the origin.

From the asymptotic expansion (\ref{Bessel+}) we see that 

\begin{eqnarray}
\Phi (x) & \sim & 2 B_+ \cos \left( \omega x - \alpha_+ \right) + 2 B_- \cos \left( \omega x - \alpha_- \right) \nonumber \\
& = & \left( B_+ e^{-i\alpha_+} + B_- e^{-i\alpha_-}\right) e^{i \omega x} + \left( B_+ e^{i\alpha_+} + B_- e^{i\alpha_-}\right) e^{-i \omega x} \label{RNAdS_0}
\end{eqnarray}

\noindent
in any one of the lines corresponding to positive $\omega x$, where again

$$
\alpha_\pm = \frac{\pi}4 (1 \pm j).
$$

\noindent
We shall use this asymptotic expression for matching. This matching must be done along the Stokes line $\omega x \in \mathbb{R}$, so that neither of the exponentials $e^{\pm i\omega x}$ dominates the other.

To trace out the Stokes line $\im(\omega x)=0$ let us first observe that we already know its behavior near the origin. Furthermore, this is the only singular point of this curve: indeed, since $x$ is a holomorphic function of $r$, the critical points of the function $\re(x)$ are the zeros of

$$
\frac{dx}{dr} = \frac{1}{f(r)} = \frac{r^{2d-6}}{-\LL r^{2d-4}+r^{2d-6}-2\MM r^{d-3} + \QQ^2}
$$

\noindent
(\textit{i.e.}, $r=0$ only). Generically one expects the Stokes line to hit all $d-1$ horizons in a spiraling fashion. Recall however that at least one branch must extend out to infinity. In our choice of branch cuts, we have arranged things so that this branch corresponds to $n=2$; therefore the preceding branch (corresponding to $n=1$) will hit the black hole horizon $r=R_H^+$. Notice also that the formula

$$
\omega x \sim \omega x_0 - \frac{\omega}{|\LL|r}
$$

\noindent
for $r \sim \infty$ implies that the argument of $r$ along the branch which extends to infinity must be asymptotically equal to the argument of $\omega$, \textit{i.e.}, $-\theta_0$. Therefore, one would guess that the Stokes line is as depicted in Figure \ref{StokesRNAdS}. We have verified this guess with a numerical computation of the same Stokes line, and this is indicated in Figures \ref{NumericalStokesRNAdS} and \ref{NumericalStokesRNAdS_CloseUp}. It should be noted that due to the branch cuts (which can be readily identified in the figure) the spirals at the singularities are not so clearly depicted in the numerical result.

\FIGURE[ht]{\label{StokesRNAdS}
	\centering
	\psfrag{A}{$A$}
	\psfrag{B}{$B$}
	\psfrag{R+}{$R_H^+$}
	\psfrag{R-}{$R_H^-$}
	\psfrag{g1}{$\gamma_1$}
	\psfrag{og1}{$\overline{\gamma}_1$}
	\psfrag{g2}{$\gamma_2$}
	\psfrag{og2}{$\overline{\gamma}_2$}
	\psfrag{g3}{$\gamma_3$}
	\psfrag{og3}{$\overline{\gamma}_3$}
	\psfrag{Re}{$\re$}
	\psfrag{Im}{$\im$}
	\psfrag{contour}{contour}
	\psfrag{Stokes line}{Stokes line}
	\epsfxsize=.6\textwidth
	\leavevmode
	\epsfbox{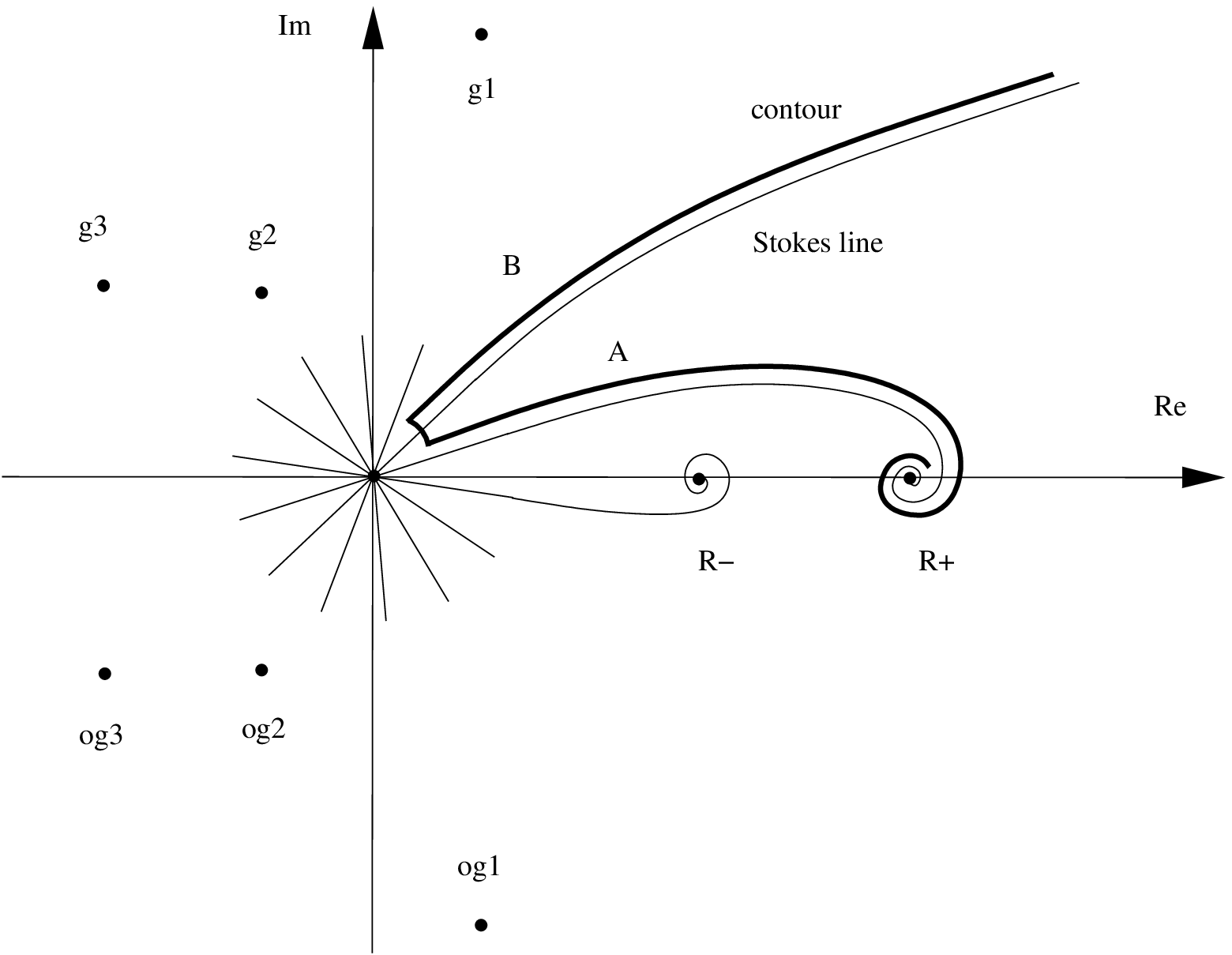}
\caption{Stokes line for the Reissner--Nordstr\"om Anti--de Sitter black hole, along with the chosen contour for monodromy matching, in the case of dimension $d=6$.}
}

\FIGURE[ht]{\label{NumericalStokesRNAdS}
	\centering
	\epsfig{file=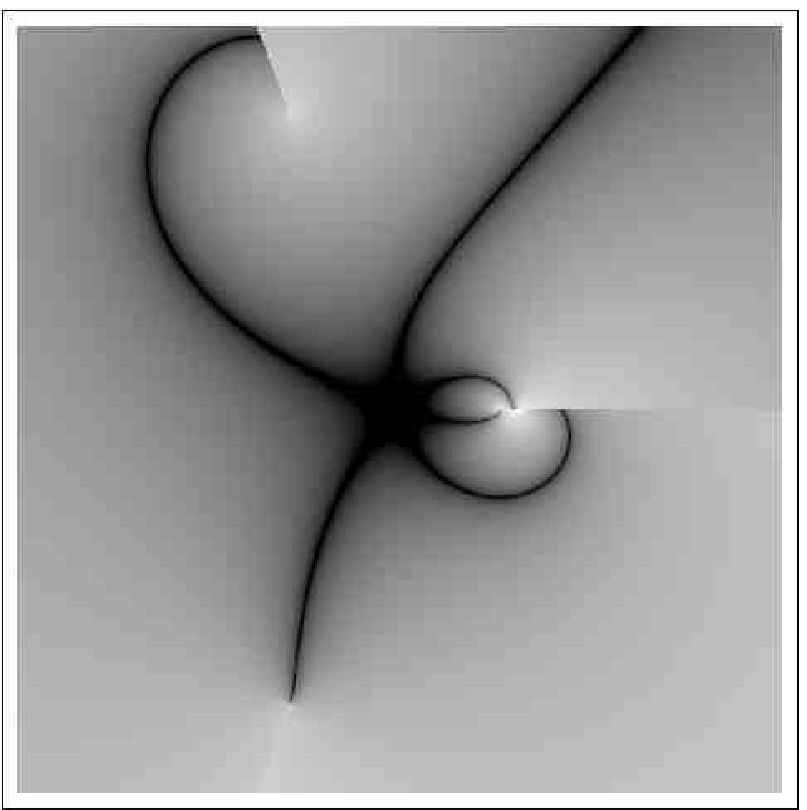, width=1.5in, height=1.5in}
    $\quad$
    \epsfig{file=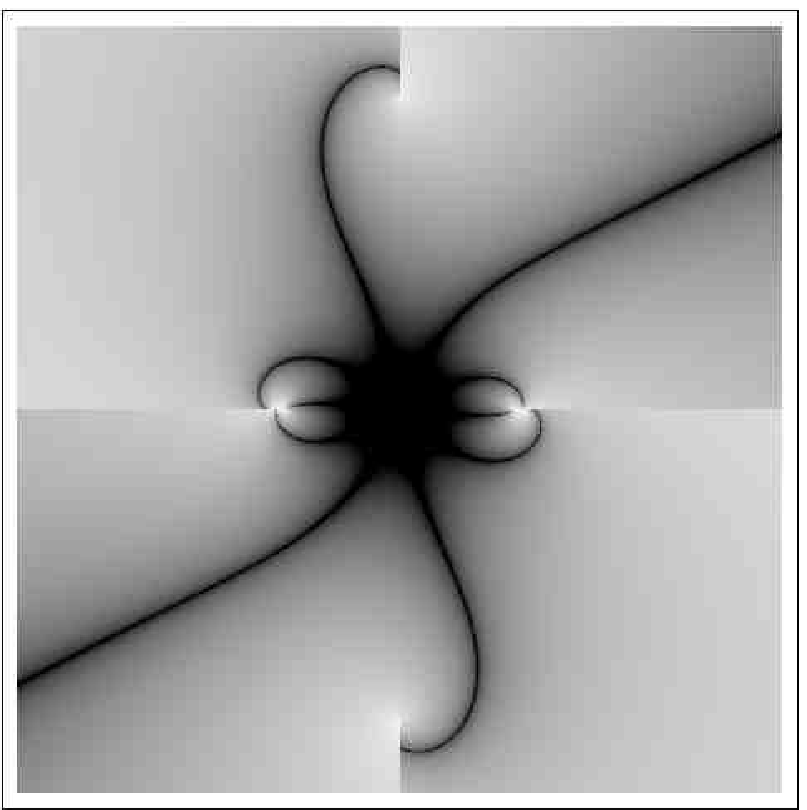, width=1.5in, height=1.5in}
    $\quad$
    \epsfig{file=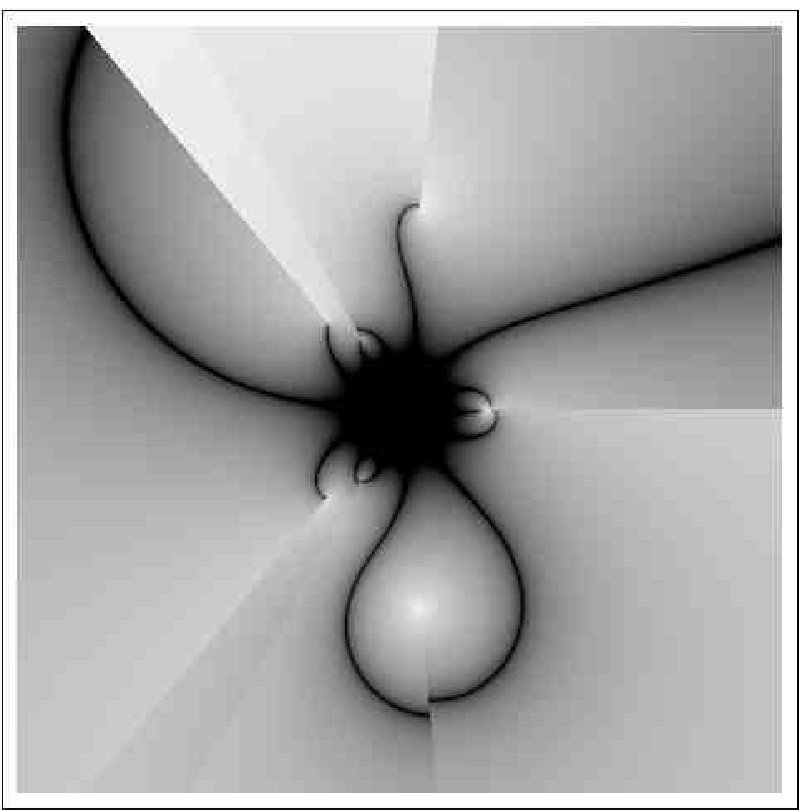, width=1.5in, height=1.5in}
    $\quad$
    \epsfig{file=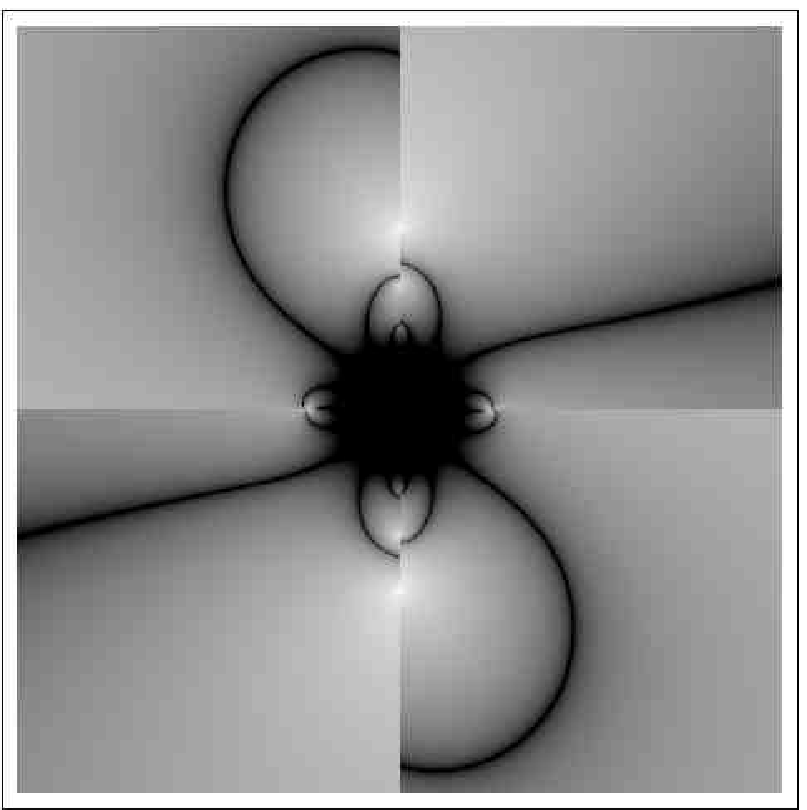, width=1.5in, height=1.5in}
\caption{Numerical calculation of the Stokes lines for the Reissner--Nordstr\"om Anti--de Sitter black hole in dimensions $d=4$, $d=5$, $d=6$ and $d=7$. Different shadings illustrate various horizon singularities and branch cuts (note that these branch cuts are not necessarily equal to the ones used for the calculation in the main text).}
}

\FIGURE[ht]{\label{NumericalStokesRNAdS_CloseUp}
	\centering
	\epsfig{file=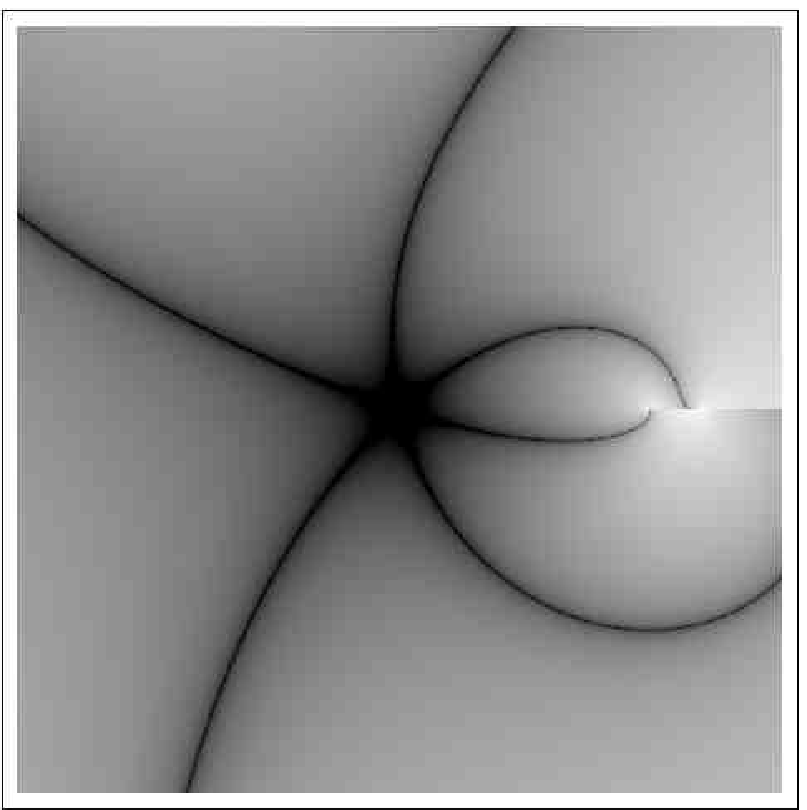, width=1.5in, height=1.5in}
    $\quad$
	\epsfig{file=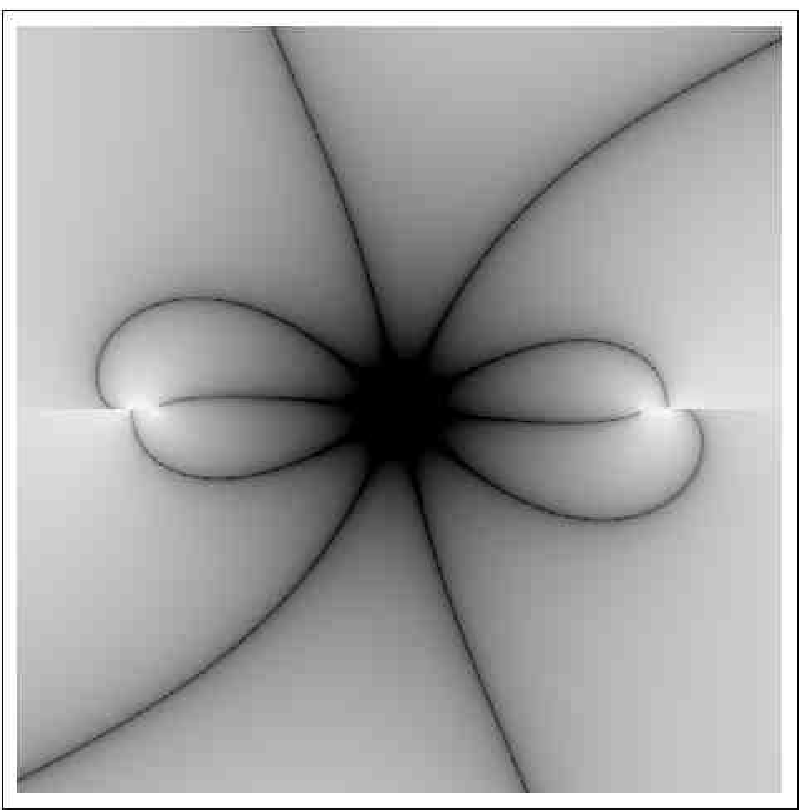, width=1.5in, height=1.5in}
    $\quad$
    \epsfig{file=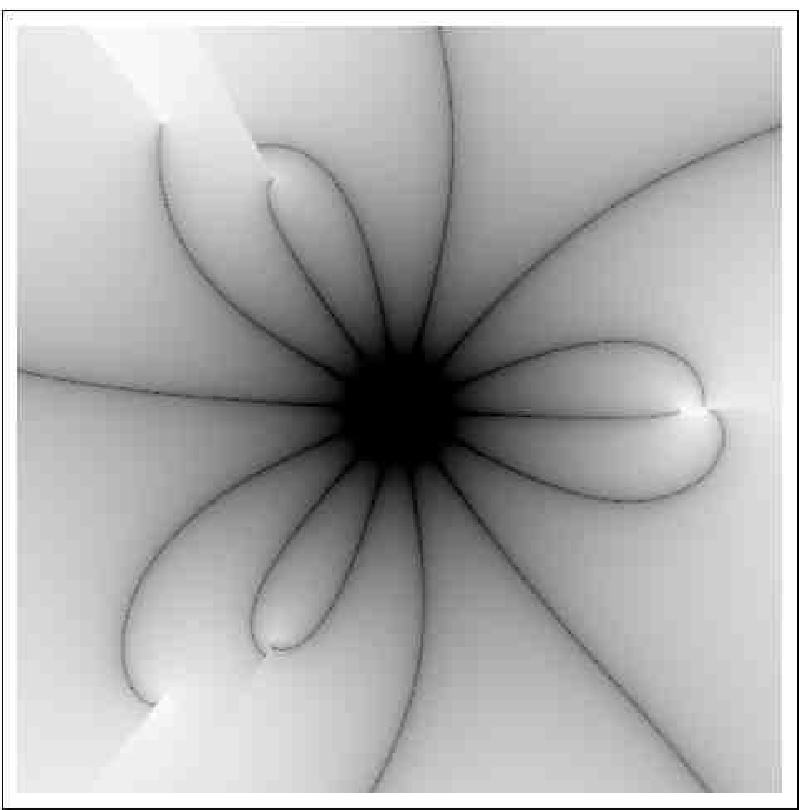, width=1.5in, height=1.5in}
    $\quad$
    \epsfig{file=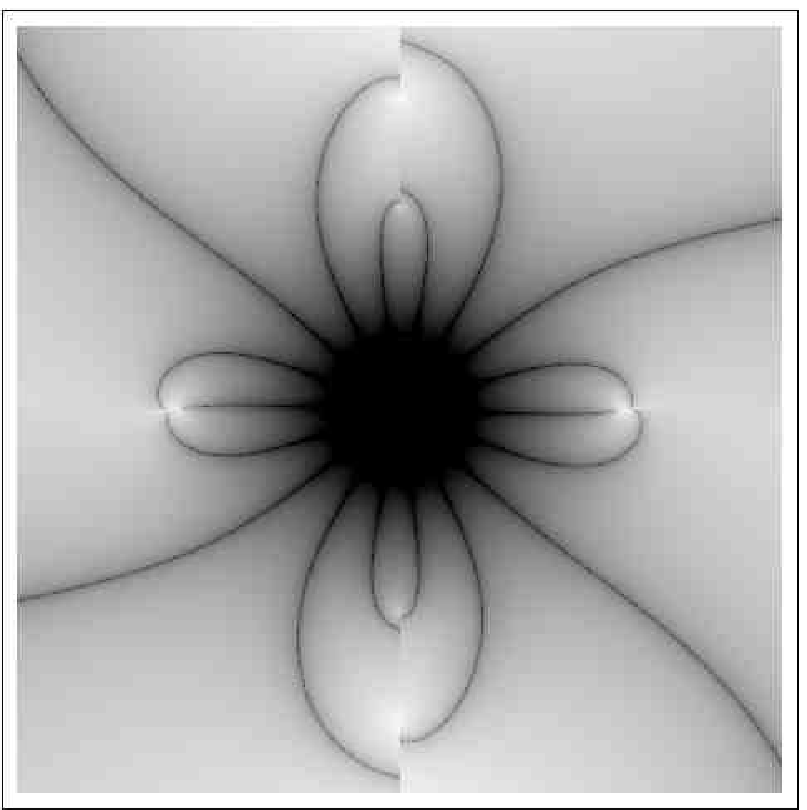, width=1.5in, height=1.5in}
\caption{Close--up, near the origin of the complex plane, on the numerical calculation of the Stokes lines for the Reissner--Nordstr\"om Anti--de Sitter black hole in dimensions $d=4$, $d=5$, $d=6$ and $d=7$.}
}

Now, for $r \sim \infty$ we have 

$$
V \sim  \frac{{j_{\infty}}^2-1}{4(x-x_0)^2}
$$

\noindent 
where ${j_{\infty}}=d-1,d-3,d-5$ for tensor type, vector type and scalar type perturbations (see appendix \ref{appendixC}). Consequently,

$$
\Phi (x) \sim C_+ \sqrt{2\pi\omega (x-x_0)}\ J_{\frac{{j_{\infty}}}{2}} \left( \omega (x-x_0) \right) + C_- \sqrt{2\pi\omega (x-x_0)}\ J_{-\frac{{j_{\infty}}}{2}} \left( \omega (x-x_0) \right)
$$

\noindent
for $r \sim \infty$. The boundary condition $\Phi = 0$ at $r = \infty$ (discussed at the beginning of this section) requires that $C_-=0$. Hence,

$$
\Phi (x) \sim C_+ \sqrt{2\pi\omega (x-x_0)}\ J_{\frac{{j_{\infty}}}{2}} \left( \omega (x-x_0) \right)
$$

\noindent
for $r \sim \infty$. This is to be matched with expansion (\ref{RNAdS_0}) along the branch of the Stokes line extending out to infinity. To do so, we notice that for $r \sim \infty$ one has

$$
\omega(x-x_0) \sim - \frac{\omega}{|\LL|r},
$$

\noindent
and that this quantity is {\em negative} on this branch. Consequently, expansion (\ref{Bessel-}) yields

$$
\Phi (x) \sim  C_+ e^{i\beta_+} e^{i \omega (x-x_0)} + C_+ e^{-i\beta_+} e^{-i \omega (x-x_0)} \\
$$

\noindent
in the same limit, where

$$
\beta_+ = \frac{\pi}4 (1 + {j_{\infty}}).
$$

\noindent
We conclude that $B_+, B_-$ must satisfy

\begin{equation} \label{RNAdSinfty}
\left( B_+ e^{-i\alpha_+} + B_- e^{-i\alpha_-}\right) e^{i \omega x_0} e^{-i\beta_+} = \left( B_+ e^{i\alpha_+} + B_- e^{i\alpha_-}\right) e^{-i \omega x_0} e^{i\beta_+}.
\end{equation}

Expansion (\ref{Bessel0}) implies that as one rotates from the branch containing point $B$ to the branch containing point $A$ (see Figure \ref{StokesRNAdS}) (through an angle of $-\frac{\pi}{2d-5}$ in $r$, corresponding to an angle $-\pi$ in $x$), one has from

$$
\sqrt{2\pi e^{-\pi i} \omega x}\ J_{\pm\frac{j}{2}} \left( e^{-\pi i} \omega x \right) = e^{\frac{-\pi i}2 (1 \pm j)}\sqrt{2\pi \omega x}\ J_{\pm\frac{j}{2}} \left( \omega x \right) \sim 2 e^{-2i\alpha_\pm} \cos(\omega x - \alpha_\pm)
$$

\noindent
that $\Phi$ changes to

\begin{align*}
\Phi (x) & \sim 2 B_+ e^{-2i\alpha_+} \cos \left( -\omega x - \alpha_+ \right) + 2 B_- e^{-2i\alpha_-} \cos \left( -\omega x - \alpha_- \right) = \\
& = \left( B_+ e^{-i\alpha_+} + B_- e^{-i\alpha_-}\right) e^{i \omega x} + \left( B_+ e^{-3i\alpha_+} + B_- e^{-3i\alpha_-}\right) e^{-i \omega x}.
\end{align*}

\noindent
This form of the solution can be propagated along the corresponding branch of the Stokes line which approaches the outer event horizon, and where we know that $\Phi (x)$ must behave as $e^{i \omega x}$. Consequently we obtain the second condition on $B_+, B_-$, as

$$
B_+ e^{-3i\alpha_+} + B_- e^{-3i\alpha_-} = 0.
$$

\noindent
The two conditions on these coefficients can only have nontrivial solutions if and only if

\begin{align*}
&\left| 
\begin{array}{ccc}
e^{-3i\alpha_+} & \,\,\,\, & e^{-3i\alpha_-} \\ & & \\
e^{-i\alpha_+} e^{-i\beta_+} e^{i \omega x_0} - e^{i\alpha_+} e^{i\beta_+} e^{-i \omega x_0} &  & e^{-i\alpha_-}e^{-i\beta_+}  e^{i \omega x_0} - e^{i\alpha_-} e^{i\beta_+} e^{-i \omega x_0}
\end{array}
\right| = 0 
\\
&\Leftrightarrow \quad
e^{\frac{i\pi {j_{\infty}}}4} \sin( \pi j ) e^{-i\omega x_0} + e^{-\frac{i\pi {j_{\infty}}}4} \sin\left( \frac{\pi j}2 \right) e^{i\omega x_0} = 0,
\end{align*}

\noindent
from where we obtain our final result as

\begin{equation} \label{resultsRNAdS}
\omega x_0 = \frac{1}{2i} \log\left( 2\cos\left( \frac{\pi j}2 \right) \right) + \frac{\pi}4 + \beta_+ + n \pi \quad (n \in \mathbb{N}).
\end{equation}

\noindent
Notice that in this asymptotically AdS case there was no need to do a monodromy calculation, all we had to do was follow the Stokes line from infinity to the black hole event horizon.

For vector type perturbations, one has

$$
j = \frac{3d-7}{2d-5} = 2 - \frac{d-3}{2d-5}
$$

\noindent
(see appendix \ref{appendixC}). Therefore it is easily seen that the values of $\omega x_0 - \beta_+$ for the corresponding quasinormal modes are shifted by $\frac{\pi}2$ relative to the quasinormal modes for tensor and scalar type perturbations. This difference exactly compensates the difference between the values of $\beta_+$ for the different types of perturbations, so that {\em all} types of perturbation yield exactly the same asymptotic quasinormal frequencies, as given by

$$
\omega x_0 = \frac{\pi}{4} \left( d+1 \right) + \frac{1}{2i} \log\left( 2\cos\left( \frac{\pi j}2 \right) \right) + n \pi \quad (n \in \mathbb{N}).
$$

Some remarks are due for \textit{scalar} type perturbations in dimensions $d=4$ and $d=5$. Indeed, in the previous calculation we have assumed that $j_{\infty}$ is positive. This is not true for scalar type perturbations in dimensions four and five. In dimension $d=4$ the calculation still goes through as long as one sets $j_{\infty} = + 1$ (rather than $-1$) as all one needs is $j_{\infty}^2=1$. Consequently the formula for asymptotic quasinormal frequencies of scalar type perturbations in dimension $d=4$ gets changed to

$$
\omega x_0 = \frac{3\pi}{4} + \frac{1}{2i} \log\left( 2\cos\left( \frac{\pi j}2 \right) \right) + n \pi \quad (n \in \mathbb{N}).
$$

\noindent
In dimension $d=5$ the situation is more subtle. Here $j_{\infty} = 0$ and strictly speaking the two Bessel functions in the solution at $r \sim \infty$ are not linearly independent. As usual, we choose $j_{\infty}$ slightly positive and carry through with the calculation, taking the limit $j_{\infty} \to 0$. In this case the solution at $r \sim \infty$ automatically vanishes as $\frac{1 \pm j_{\infty}}{2}$ is positive. Thus, there is no constraint arising from infinity on the solution to the master equation, and all one is left with is the constraint at the black hole horizon. This condition alone is not enough to quantize the asymptotic frequencies. Thus, in five dimensions, the asymptotic spectrum of RN AdS scalar type perturbations is \textit{continuous}, \textit{i.e.}, $\omega \in {\mathbb{C}}$. This is a rather unexpected result. We believe that the consequences of this result should deserve further studies.

Let us now review the literature on asymptotic quasinormal frequencies for the RN AdS spacetime, so that we can compare our results with earlier work done on this subject. The stability of RN AdS black holes was addressed in \cite{gubser-mitra-1, gubser-mitra-2}, in a very different way from the analysis in \cite{kodama-ishibashi-3}. These authors focused on the $d=4$ case and found that, while instabilities in the linearized theory are harder to find in this RN AdS case, one can still decide on the stability of the solution if one tries to uncover thermodynamic instabilities. Such instabilities would correspond to the onset of Bose condensation (or other sort of phase transitions) in the dual field theory. Interestingly enough, at the perturbation theory level unstable modes of the RN AdS$_{4}$ black hole are not found in metric fluctuations at linear order, but rather in the gauge fields and scalars of ${\mathcal{N}}=8$ gauged supergravity. \cite{gubser-mitra-1, gubser-mitra-2} argue that this instability is in the same universality class as the one of instabilities where the horizon does fluctuate. As to the metric, it also fluctuates, but only at subleading order. The thermodynamic instabilities can only be studied for large black holes and, in this limit, they do coincide with the dynamical instabilities. For a discussion of this sort of instabilities, in a broader context, we refer the reader to \cite{hubeny-rangamani}. All in all, one should have the above discussion of instabilities in mind when addressing RN AdS black holes.

The first numerical results for the asymptotic quasinormal frequencies of RN AdS black holes were computed in \cite{berti-kokkotas-1}, in the particular case of dimension $d=4$ and focusing on scalar field, electromagnetic field and gravitational field perturbations. The most significant finding of these authors was that purely damped RN AdS modes behave in a very peculiar way: their damping seems to go to infinity in the extremal black hole case, suggesting the possibility that extremally charged RN AdS black holes may be marginally unstable to electromagnetic and gravitational perturbations (unless, of course, the mode amplitude goes to zero in this extremal limit). This question was later solved in \cite{wlm}. Indeed, \cite{wlm} presents an extensive numerical study of quasinormal frequencies for massless scalar fields in $d=4$ RN AdS spacetime, corresponding to tensor type perturbations in our results. They find that as one increases the charge, at first the scalar field decays exponentially and oscillates but for $\QQ > \QQ_{\mathrm{critical}} = 0.3895 \QQ_{\mathrm{max}}$ the decay becomes purely exponential. So, for $\QQ > \QQ_{\mathrm{critical}}$ the asymptotic scalar perturbations are dominated by non--oscillatory modes and $| {\mathbb{I}}{\mathrm{m}}\, \omega |$ keeps decreasing monotonically and smoothly up to $\QQ = \QQ_{\mathrm{max}}$. As the black hole becomes extremal, $| {\mathbb{I}}{\mathrm{m}}\, \omega | \to 0$. However, in the extremal limit the decay of the scalar field perturbation changes from exponential to power law, and the extremal RN AdS black hole is thus stable to scalar field perturbations (not marginally unstable as worried in \cite{berti-kokkotas-1} and mentioned above). For the higher modes both ${\mathbb{R}}{\mathrm{e}}\, \omega$ and ${\mathbb{I}}{\mathrm{m}}\, \omega$ increase with overtone number, $n$, although as one increases charge $\QQ$, ${\mathbb{R}}{\mathrm{e}}\, \omega$ increases slower and ${\mathbb{I}}{\mathrm{m}}\, \omega$ increases faster, with $n$. In the large black hole regime, with $|\LL| = 1$ and $R_{H}^{+} = 100$, asymptotic frequencies become evenly spaced and as $n \to \infty$ (we have translated the results adapting the conventions of \cite{wlm} to our own)

\begin{eqnarray*}
\omega_{n+1} - \omega_{n} &\sim& 129.9 + 225 i, \qquad \QQ=0, \\
\omega_{n+1} - \omega_{n} &\sim& 120.6 + 233 i, \qquad \QQ=0.15\QQ_{\mathrm{max}}, \\
\omega_{n+1} - \omega_{n} &\sim& 111.1 + 256 i, \qquad \QQ=0.3\QQ_{\mathrm{max}}.
\end{eqnarray*}

\noindent
As one increases charge, the real part of the spacing decreases and the imaginary part of the spacing increases. We can now match these predictions to our RN AdS quasinormal equation. This is done in a similar fashion to what we did in the case of Schwarzschild AdS black holes: one computes $x_0$ numerically according to its general expression (as there is no analytical solution for $x_0$, and being careful with the choice of branch cuts for the several logarithms) and then plugs into our analytical formula for the frequencies in order to obtain (using the values of \cite{wlm} for cosmological constant, black hole mass, charge and horizon radius)

$$
\omega = \left( 122.362 + 231.978 i \right) n + \left( 50.871 + 47.297 i \right),
$$

\noindent
when $\QQ=0.15\QQ_{\mathrm{max}}$, and also

$$
\omega = \left( 110.973 + 255.82 i \right) n + \left( 50.108 + 54.253 i \right),
$$

\noindent
when $\QQ=0.3\QQ_{\mathrm{max}}$. Our results are in full and complete agreement with the numerical calculations of \cite{wlm}, to a great degree of accuracy. However, it would be very interesting to produce further numerical data concerning higher dimensional RN AdS quasinormal modes to match against all our analytical results. In particular, for the results above, we have produced both gap and offset, where there is only numerical data for the gap. A point to mention concerns the critical charge $\QQ_{\mathrm{critical}}$ mentioned above. Unfortunately asymptotic quasinormal modes cannot see this critical charge: as described in \cite{wlm}, the critical charge is found by following the frequencies of a \textit{specific} mode, as one varies the charge. For the asymptotic frequencies $n \to + \infty$ and the choice of focusing on a specific mode becomes somewhat obscure. Nevertheless, we believe this matter should be further explored within our general setting.

Let us make one final remark concerning the case above where $\QQ=0$ and where the result exactly matches the Schwarzschild AdS result. One can prove that this must be the case by analyzing the limit of the RN AdS quasinormal frequencies when $\QQ \to 0$. Here, it is not too hard to check that the $(2d-4)$ RN AdS complex horizons have the following limit: $(d-1)$ of these horizons go to the $(d-1)$ complex horizons of Schwarzschild AdS, while the remaining $(d-3)$ horizons go to zero as 

$$
\left( \frac{\QQ^{2}}{2\MM} \right)^{\frac{1}{d-3}}
$$

\noindent
when $\QQ \to 0$. This limit is such that these remaining $(d-3)$ horizons will have zero contribution to $x_0$. Thus, in the $\QQ \to 0$ limit the RN AdS $x_0$ reduces to the Schwarzschild AdS $x_0$. In this case, the gap of the asymptotic quasinormal frequencies of RN AdS spacetime will reduce to the gap of the asymptotic quasinormal frequencies of Schwarzschild AdS spacetime (although the offset comes out incorrect, as expected in this sort of limits). This shows, in particular, why the numerical result of \cite{wlm} must match the Schwarzschild AdS result when $\QQ = 0$.


\section{Exact Solutions for (Quasi)normal Frequencies}


While most black hole solutions do not allow for an exact calculation of
quasinormal frequencies, there are some spacetimes where one can actually
perform such an analytic calculation (and without restricting oneself to
the asymptotic case). We shall address some of these examples in this 
section; in particular, we shall deal with the case where the parameters
$\MM$ and $\QQ$ are both zero but $\LL$ is unconstrained (thus being in a 
situation where there is no black hole), and while still remaining 
in the $d$--dimensional realm. Let us stress that we present these results 
for the sake of completeness, as they do not describe full fledged black
holes. Still, they are valid solutions to Einstein's equations and are 
definitely worth studying within the classification problem we address. 
The asymptotically flat case we shall consider is Rindler space, and it 
is without surprise that we shall find a trivial solution for the quasinormal frequencies (after all, Rindler is but flat space). The 
spacetimes with non--vanishing cosmological constant we will consider are, obviously, AdS and dS. Using the analysis in \cite{kodama-ishibashi-1}, 
we write down the equations for the several types of perturbations and 
show how they can be exactly solved with quasinormal mode boundary conditions. For AdS, we find an unexpected continuous spectrum for scalar perturbations in five dimensions. For dS, we find that there are only quasinormal frequencies 
in odd spacetime dimensions.


\subsection{The Rindler Solution}


Let us begin with the Rindler spacetime solution. It actually happens 
that this solution does not belong to the class of backgrounds analyzed 
in \cite{kodama-ishibashi-1}, and one thus cannot use their perturbation 
theory formulas. Instead of developing a full perturbation theory for this 
spacetime, we shall simply calculate the quasinormal frequencies for a 
massless scalar field (meaning a field obeying the scalar wave equation 
in this spacetime). As we shall see, the solution to this problem is rather 
simple and does not justify by itself an extension of the analysis in 
\cite{kodama-ishibashi-1}. The $d$--dimensional Rindler spacetime is 
described by the metric

$$
g = - \rho^{2} dt \otimes dt + d\rho \otimes d\rho + dz_{1} \otimes dz_{1} 
+ \cdots + dz_{d-2} \otimes dz_{d-2},
$$

\noindent
where the coordinate $\rho$ must take positive values. In this geometry, it 
is immediate to find the massless Klein--Gordon equation for a scalar field $\phi$. We Fourier decompose the field in both time and transverse space (meaning the transverse coordinates $\vec{z} = \left( z_{1}, \dots, z_{d-2} \right)$) as

$$
\phi (\rho, \vec{z}, t) = \Phi (\rho) e^{i \omega t} e^{i \vec{k} \cdot \vec{z}},
$$

\noindent
with $\vec{k}$ the wave vector, and use this decomposition in the wave function. It then simply follows

\begin{equation}\label{rindlereqmotion}
\rho^{2} \frac{d^{2}\Phi}{d\rho^{2}} + \rho \frac{d\Phi}{d\rho} + \left( 
\omega^{2} - \rho^{2} k^{2} \right) \Phi = 0,
\end{equation}

\noindent
with $k^{2} = \vec{k} \cdot \vec{k}$. The general solution to 
(\ref{rindlereqmotion}) is given in terms of the modified Bessel functions 
of the first and second kind, $I_{\nu} (z)$ and $K_{\nu} (z)$, as ($C_{1}, C_{2} \in {\mathbb{C}}$ are constants)

$$
\Phi (\rho) = C_{1}\ I_{i\omega} (k\rho) + C_{2}\ K_{iw} (k\rho).
$$

Next, in order to define quasinormal mode boundary conditions, one first finds the tortoise coordinate. This is a particularly simple case, where $x = \log \rho$. 
Starting with the behavior near $x = - \infty$, one observes that the effective potential $V(x) = k^{2} e^{2x}$ vanishes and the Schr\"odinger--like wave equation is simply solved by plane wave solutions of the type

$$
\Phi (x) \sim e^{\pm i \omega x}.
$$

\noindent
Even though there is no black hole, we shall nevertheless demand the standard 
quasinormal mode boundary conditions, \textit{i.e.}, an in--going wave near the Rindler horizon $x = - \infty$, \textit{i.e.}, $\Phi (x) \sim e^{i \omega x}$. Let us see what this implies in our general solution above. Near the origin, $\rho = 0$, the modified Bessel functions behave as 

\begin{eqnarray*}
I_{i \omega} ( k \rho ) &\sim& \rho^{i \omega}, \\
K_{i \omega} ( k \rho ) &\sim& C_{+} \rho^{i \omega} + C_{-} \rho^{-i \omega},
\end{eqnarray*}

\noindent
where $C_{+}, C_{-} \in {\mathbb{C}}$ are specific constants. Because the boundary condition is $\Phi (\rho) \sim \rho^{i \omega}$ as $\rho \to 0$, one discards the modified Bessel function of the second kind $K_{i \omega} (k\rho)$,  keeping only $I_{i \omega} (k\rho)$ as an allowed solution. Of course one must still analyze whether this is a good solution near the other boundary. At $x = + \infty$ the effective potential blows up and the wave equation is certainly not solved by plane waves. The natural requirement is to demand for regularity, which in this case translates to the vanishing of the wave function. However, the modified Bessel function of the first kind we have kept satisfies

$$
I_{i \omega} ( k \rho ) \sim \frac{1}{\sqrt{\rho}}\ e^{k \rho} \to + \infty \qquad  (\rho \to + \infty).
$$

\noindent
One thus comes to the conclusion that this function is not a good solution, as it does not satisfy the boundary conditions. The one modified Bessel function satisfying the necessary requirements at infinity describes both in--going and  out--going waves at the origin, while the one describing the in--going wave at the origin blows up at infinity. In other words, there are no solutions with quasinormal mode boundary conditions in Rindler space. This is not an entirely unexpected result as Rindler space is but flat space, where we know that there are no quasinormal frequencies (although these are different boundary conditions to the usual ones for Minkowski spacetime).


\subsection{The Anti--de Sitter Solution}


We now turn our attention to the pure AdS spacetime solution. As a Schr\"odinger--type problem, this background effectively behaves as a bounding box, a sort of an infinite well potential with finite extent. This characteristic raises the question of what boundary conditions one should use; these were also earlier discussed in section 3. This question was first addressed in \cite{ais} within the context of quantization of scalar fields in a four dimensional AdS background. Even though an embedding of AdS spacetime in the Einstein Static Universe can lead to both Dirichlet or Neumann boundary conditions (on counterparts of AdS scalar fields), the only sensible conditions found in \cite{ais} were vanishing of the fields at infinity. This issue was further addressed in \cite{mezincescu-townsend, burgess-lutken}, within the study of the $d$--dimensional wave equation for a massive scalar field in AdS. There, it was first realized that in $d$--dimensions the AdS wave equation reduces to a hypergeometric equation and is thus exactly solvable. The authors also realized that only normal modes exist. Due to the finite extent of spacetime in the tortoise coordinate, it is actually not surprising to find that any field evolving in an AdS geometry may be decomposed in a normal mode harmonic analysis (the basic difference between normal and quasinormal modes being that the former are a complete set of stationary eigenfunctions, while the latter are not). Observe that the normal mode problem is different from the quasinormal mode one: for normal modes one requires the vanishing of the wave function at both boundaries, while for quasinormal modes one requires plane wave solutions at these same boundaries. In what follows, we shall describe how to find both normal modes and normal frequencies for $d$--dimensional AdS spacetime. Let us note that this problem has been previously addressed for a massless scalar field in \cite{burgess-lutken}, and for tensor and vector type perturbations of $d=4$ AdS spacetime in \cite{ckl}. Our results naturally extend these, to arbitrary perturbations and arbitrary dimension.

\FIGURE[ht]{\label{5dadst}
	\epsfig{file=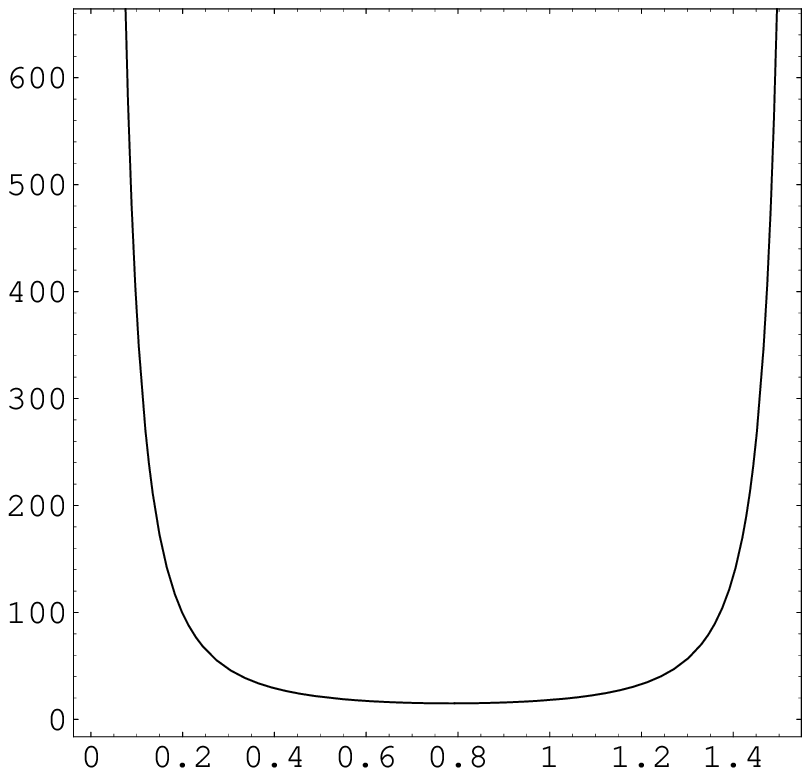, width=1.5in, height=1.5in}
\caption{Potential for AdS tensor type perturbations in $d=5$.}
}

Let us begin with a ``tour'' of the different Schr\"odinger--type effective potentials one can find in AdS, in order to set the stage for the results we shall later find. The AdS solution has parameter $\LL < 0$ (with $\MM = 0 = 
\QQ$), so that $f(r) = 1 - \LL r^{2} = 1 + | \LL | r^{2}$. In any dimension $d$ one can thus simply find the tortoise coordinate as

\begin{eqnarray*}
x &=& \int \frac{d\rho}{f(\rho)} = \frac{1}{\sqrt{|\LL|}} \arctan 
\left( \sqrt{|\LL|}\ r \right), \\
r &=& \frac{1}{\sqrt{|\LL|}} \tan \left( \sqrt{|\LL|}\ x \right).
\end{eqnarray*}

\noindent
The first thing one observes is that [infinite] space becomes a finite interval, when described via the tortoise, as $\lim_{r \to 0} x[r] = 0$ and $\lim_{r \to + \infty} x[r] = \frac{\pi}{2\sqrt{|\LL|}}$. The potential for tensor type perturbations is thus written as (see the appendix)

$$
V_{\mathsf{T}} (x) = \frac{|\LL| \left( 2 \ell \left( \ell+d-3 
\right) + \left( d-2 \right)^{2} + 2 \Big( \ell \left( \ell+d-3 
\right) - \left( d-2 \right) \Big) \cos \left( 2 \sqrt{|\LL|}\ x 
\right) \right)}{4 \sin^{2} \left( \sqrt{|\LL|}\ x \right) \cos^{2} 
\left( \sqrt{|\LL|}\ x \right)}.
$$

\noindent
In its range, $x \in\ ( 0, \frac{\pi}{2\sqrt{|\LL|}} )$, this potential looks like an infinite well potential with rounded edges. In Figure \ref{5dadst} we present an illustrative plot, where we take $|\LL|=1$, $d=5$ and $\ell=1$ (as one should recall that for tensor type perturbations $\ell \geq 1$). The shape of the potential does not change much as we change its parameters. For vector type perturbations, the potential is (see the appendix)

$$
V_{\mathsf{V}} (x) = \frac{|\LL| \left( 2 \ell \left( \ell+d-3 \right) + 
\left( d-2 \right) \left( d-4 \right) + 2 \ell \left( \ell+d-3 \right) 
\cos \left( 2 \sqrt{|\LL|}\ x \right) \right)}{4 \sin^{2} \left( 
\sqrt{|\LL|}\ x \right) \cos^{2} \left( \sqrt{|\LL|}\ x \right)}.
$$

\FIGURE[ht]{\label{45dadsv}
	$\qquad$ 
    \epsfig{file=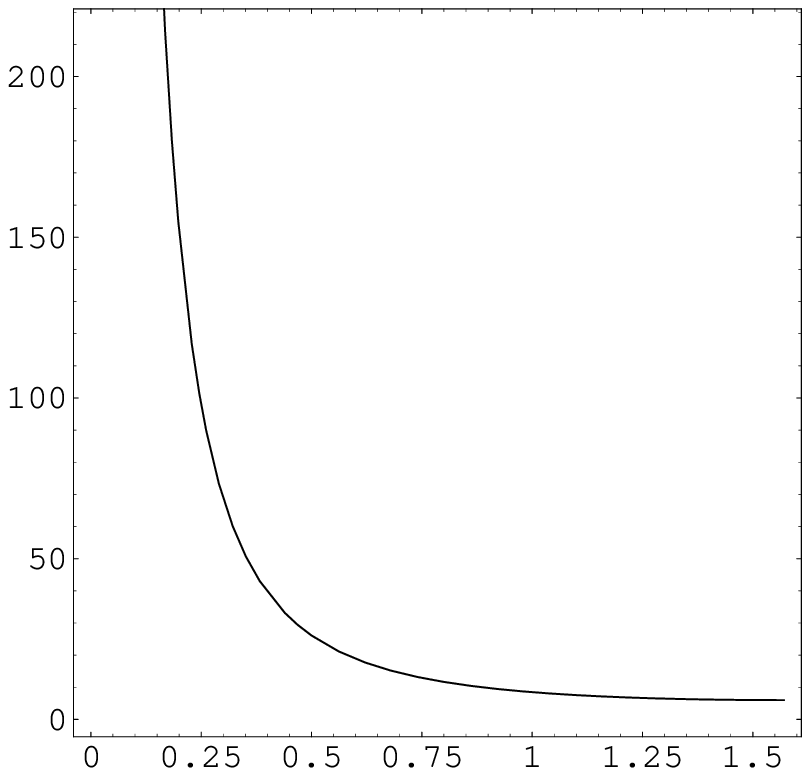, width=1.5in, height=1.5in}
    $\qquad$
    \epsfig{file=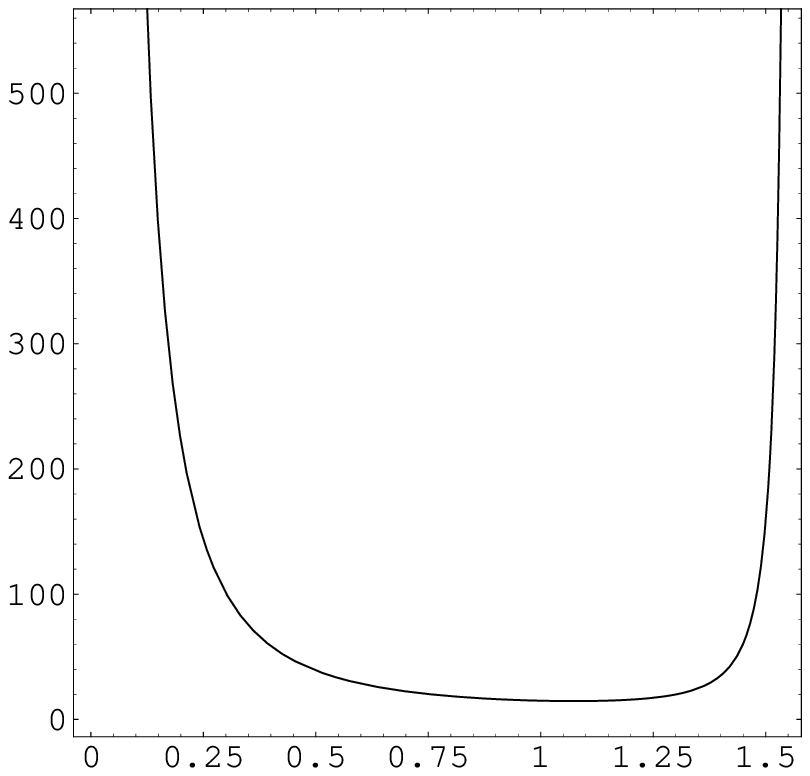, width=1.5in, height=1.5in}
    $\qquad$
\caption{Potentials for AdS vector type perturbations in $d=4$ and $d=5$.}
}

\noindent
This time around, the potential looks like the infinite well potential with rounded edges \textit{only} at dimension $d \geq 5$. In $d=4$ the potential is actually different. In Figure \ref{45dadsv} we present two illustrative plots, where we take $|\LL|=1$ and $\ell=2$ (as one should recall that for vector type perturbations $\ell \geq 2$), and sequentially $d=4$ and $d=5$. Observe that as $x \to \frac{\pi}{2\sqrt{|\LL|}}$, in four dimensions, the potential does \textit{not} go to zero but rather to $|\LL|\ \ell \left( \ell + 1 \right)$. Except for the change at dimension $d=4$, again the shape of the potential will not change much as we change its parameters. We will see that even though the potential in $d=4$ is different, there will be no substantial change in the normal modes. The same will not happen as we turn to scalar type perturbations, where the potential is (see the appendix) 

$$
V_{\mathsf{S}} (x) = \frac{|\LL| \left( 2 \ell \left( \ell+d-3 
\right) + \left( d-4 \right)^{2} + 2 \Big( \ell \left( \ell+d-3 
\right) + \left( d-4 \right) \Big) \cos \left( 2 \sqrt{|\LL|}\ x 
\right) \right)}{4 \sin^{2} \left( \sqrt{|\LL|}\ x \right) \cos^{2} 
\left( \sqrt{|\LL|}\ x \right)}.
$$

\FIGURE[ht]{\label{4567dadss}
	\centering
    \epsfig{file=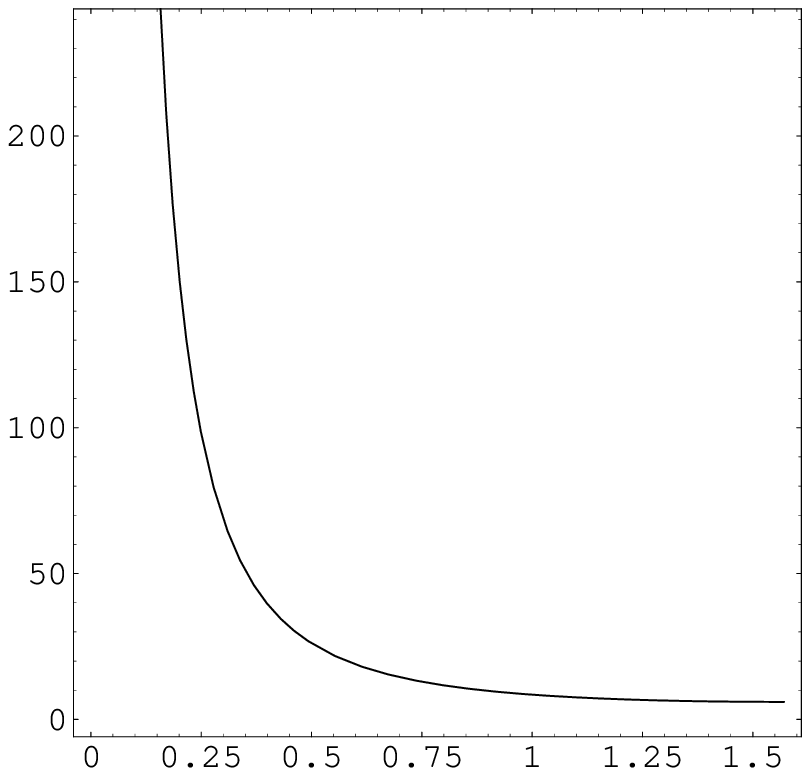, width=1.5in, height=1.5in}
    $\quad$
    \epsfig{file=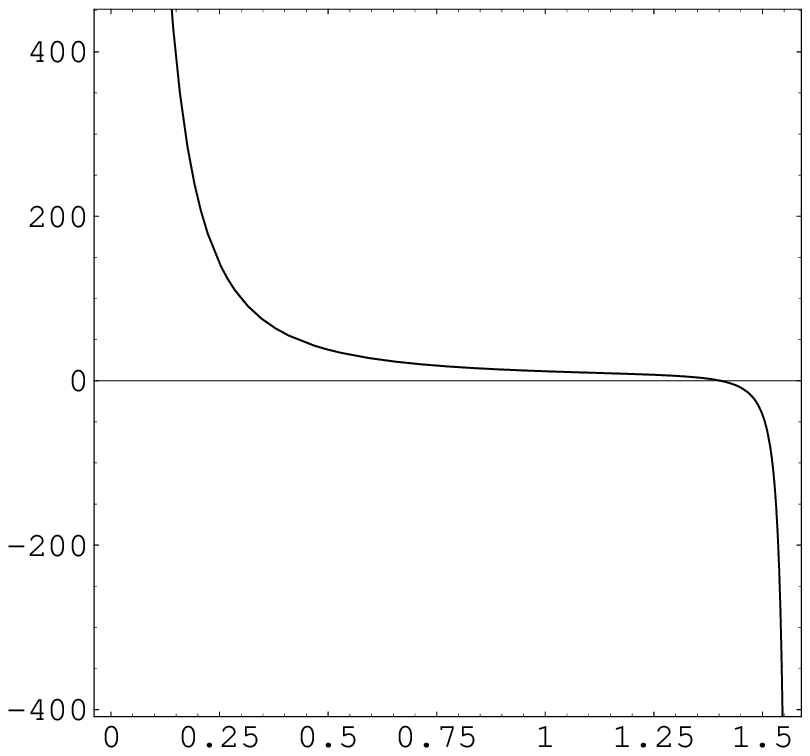, width=1.5in, height=1.5in}
    $\quad$
    \epsfig{file=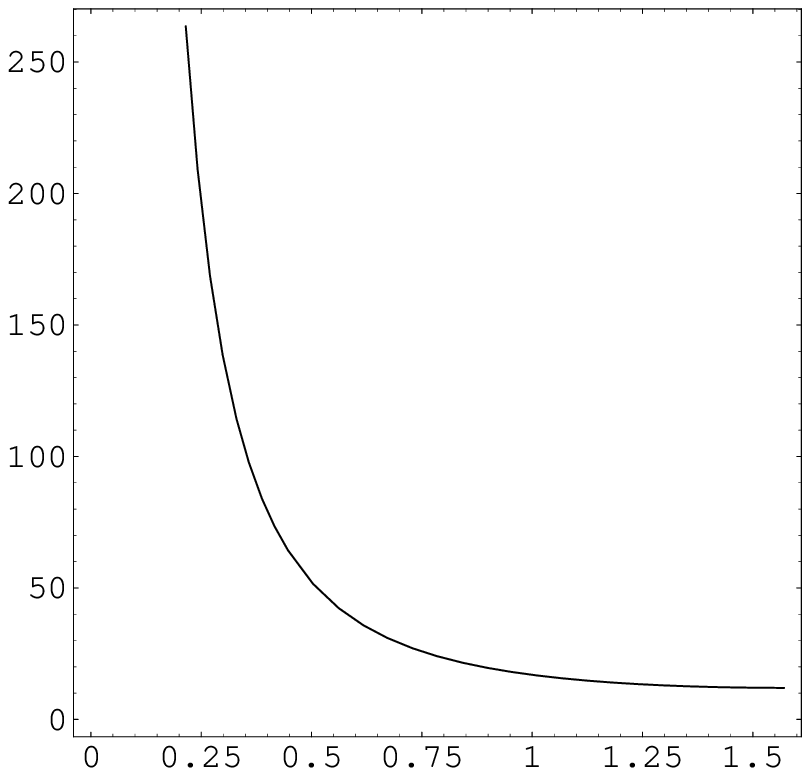, width=1.5in, height=1.5in}
    $\quad$
    \epsfig{file=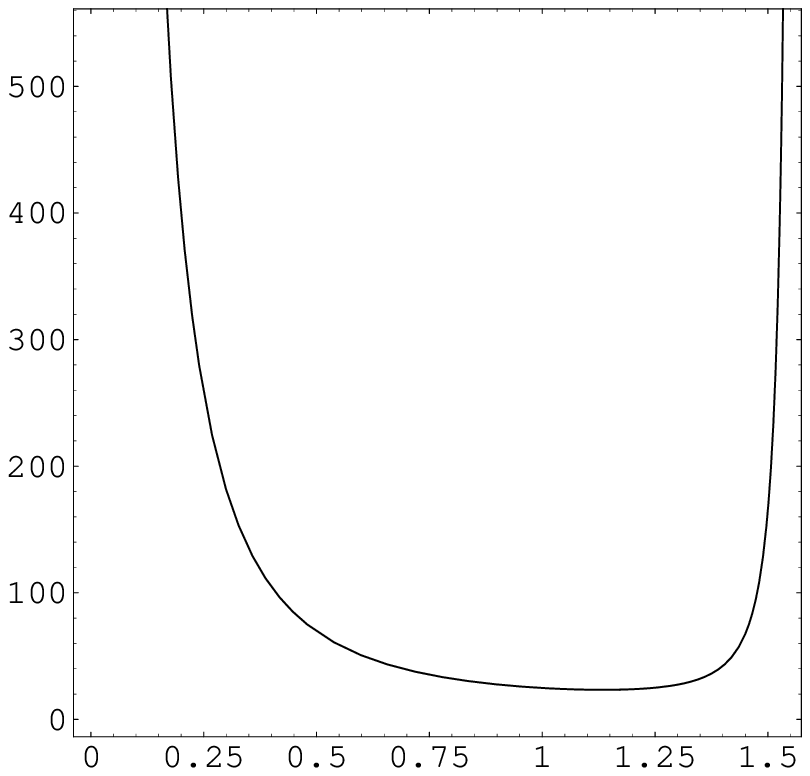, width=1.5in, height=1.5in}
\caption{Potentials for AdS scalar type perturbations in dimensions $d=4$, $d=5$, $d=6$ and $d=7$, respectively.}
}

\noindent
In this case, the potential looks like the infinite well potential with rounded edges \textit{only} at $d \geq 7$. In $d=4, 5, 6$ the potential is different \textit{and} changing. For the illustrative plots of Figure \ref{4567dadss} we shall take $|\LL| = 1$ and $\ell = 2$ (as one should recall that for scalar type perturbations $\ell \geq 2$), and sequentially $d=4$, $d=5$, $d=6$ and $d=7$. Observe that as $x \to \frac{\pi}{2\sqrt{|\LL|}}$ both in four dimensions (the first figure on the left) and in six dimensions (the third figure) the potential does \textit{not} go to zero. Rather, we have

\begin{eqnarray*}
\lim_{x \to \frac{\pi}{2\sqrt{|\LL|}}} \left. V_{\mathsf{S}} (x) 
\right|_{d=4} &=& |\LL|\ \ell \left( \ell + 1 \right), \\
\lim_{x \to \frac{\pi}{2\sqrt{|\LL|}}} \left. V_{\mathsf{S}} (x) 
\right|_{d=6} &=& |\LL|\ \left( \ell \left( \ell + 3 \right) + 2 
\right).
\end{eqnarray*}

\noindent
The potentials in $d=4$ and $d=6$, will not produce substantial changes in the results for the normal modes. But the $d=5$ potential will, and in fact will lead to a continuous normal mode spectrum, which turns out to be an unexpected result. Let us next proceed with the actual computation.

It turns out that the AdS wave equation can be reduced to a hypergeometric equation, in which case it pays off not to start with the potentials as written above, but in the original $r$ coordinate. In this case, for the pure AdS spacetime the potential governing generic $d$--dimensional gravitational perturbations may be written as

$$
V(r) = \left( 1 + |\LL| r^{2} \right) \left( a |\LL| + \frac{b \left( b + 1 \right)}{r^{2}} \right),
$$

\noindent
where, in the expression above, 

$$
b = \frac{d+2\ell-4}{2},
$$

\noindent
and $a$ depends on the type of perturbation under study. Explicitly:
 
$$
 a = \left\{ \begin{array}{lllll}
\frac{1}{4} d \left( d-2 \right), & {\mathsf{\, \ Tensor \, Type \, Perturbations}}, \\
& \\
\frac{1}{4} \left( d-2 \right) \left( d-4 \right), &{\mathsf{\, \ Vector \, Type \, Perturbations}}, \\
& \\
\frac{1}{4} \left( d-4 \right) \left( d-6 \right), &{\mathsf{\, \ Scalar \, Type \, Perturbations}}.
\end{array}\right.
$$ 

\noindent
Changing to the new variable $z = \sin \left( {\sqrt{|\LL|}x} \right)^{2}$, with $x$ the tortoise coordinate, the wave equation gets transformed into a  hypergeometric type of differential equation,

\begin{equation}\label{hipergeral}
4 z \left( 1-z \right) \frac{d^{2}\Phi}{dz^{2}} + 2 \left( 1-2z \right) \frac{d\Phi}{dz} + \frac{4 \hat{\omega}^{2} z \left( 1-z \right) - 4az - 4b \left( b+1 \right) \left( 1-z \right)}{4z \left( 1-z \right)} \Phi = 0,
\end{equation}

\noindent
where we have introduced the notation $\hat {\omega} \equiv \frac{\omega}{\sqrt{|\LL|}}$, and where the new coordinate, $z$, ranges from $0$ to $1$. Because one knows the solution to the hypergeometric differential equation, this yields the solution for an arbitrary perturbation in the $d$--dimensional AdS spacetime. We shall now make a separate analysis of this equation, for each of the different type of perturbations.

We begin with the analysis of normal modes for tensor type perturbations. For these perturbations, it is not too hard to find that the solution to (\ref{hipergeral}) is 

\begin{eqnarray}
\Phi (z) &=& C_{1} \times \left. _{2}F_{1} \left[ \frac{\ell-\hat{\omega}}{2}, \frac{\ell+\hat{\omega}}{2}, \frac{-1+d+2\ell}{2} \right| z \right]\ \left( z-1 \right)^{\frac{2-d}{4}}\ z^{\frac{-2+d+2\ell}{4}} + \nonumber \\
&&
+ C_{2} \times \left. _{2}F_{1} \left[ \frac{3-d-\ell-\hat{\omega}}{2}, \frac{3-d-\ell+\hat{\omega}}{2}, \frac{5-d-2\ell}{2} \right| z \right]\ \left( z-1 \right)^{\frac{2-d}{4}}\ z^{\frac{4-d-2\ell}{4}},
\label{hipertensor}
\end{eqnarray}

\noindent
where $C_{1}, C_{2} \in {\mathbb{C}}$ are constants and $\left. _{2}F_{1} \left[\alpha, \beta, \gamma \right| z \right]$ is the standard hypergeometric function. Two basic properties of $\left. _{2}F_1 \left[\alpha, \beta, \gamma \right| z \right]$, which shall be needed in the following, are

\begin{eqnarray}\label{prop2}
\lim_{z \to 0}\ \left. _{2}F_1 \left[\alpha, \beta, \gamma \right| z \right] &=& 1, \nonumber \\
\lim_{z \to 1}\ \left. _{2}F_1 \left[\alpha, \beta, \gamma \right| z \right] &=& \frac{\Gamma[\gamma]\ \Gamma[\gamma-\alpha-\beta]}{\Gamma[\gamma-\alpha]\ \Gamma[\gamma-\beta]}.
\end{eqnarray}

\noindent
Next, one imposes boundary conditions for normal modes. As we are restricting ourselves to dimension $d \geq 4$, the solution to (\ref{hipertensor}) which is regular at the origin, $z=0$, is the first term on the right hand side of (\ref{hipertensor}). One thus sets $C_{2} = 0$. In order to satisfy boundary conditions at the other boundary, $z=1$, one still needs to demand for a vanishing wave function at this point. This can only be achieved if

$$
\left. _{2}F_1 \left[ \frac{\ell-\hat{\omega}}{2}, \frac{\ell+\hat{\omega}}{2}, \frac{-1+d+2\ell}{2} \right| 1\right] = 0,
$$

\noindent
as the poles of the Gamma function will dominate over the power law singularity in $(z-1)^{\frac{2-d}{4}}$. It is simple to observe, using (\ref{prop2}), that the condition above is satisfied when the frequency satisfies the quantization condition

$$
\hat{\omega} = 2n + d + \ell -1, \qquad n = 0, 1, 2, \ldots .
$$

\noindent
Let us note that we have chosen only positive values for the frequency. It thus follows that the normal frequencies for tensor type perturbations of $d$--dimensional AdS spacetime are given by
 
$$
\omega = \sqrt{|\LL|}\ \left( 2n + d + \ell - 1 \right), \qquad n = 0, 1, 2, \ldots .
$$

\noindent 
This result agrees with the one in \cite{burgess-lutken}, which is a result for scalar fields. This is expected as the Klein--Gordon evolution equation is actually identical to the one for tensorial gravitational perturbations.

Let us proceed with the analysis of normal modes for vector type perturbations. This follows along very similar lines to the previous case. In $d=4$ the parameter $a$ in (\ref{hipergeral}) actually vanishes, and this equation has for solution (this is the general solution when $a=0$, we have not yet set $d=4$)

\begin{eqnarray}\label{azero}
\Phi (z) &=& C_{1} \times \left. _{2}F_{1} \left[ \frac{-2+d+2\ell-2\hat{\omega}}{4}, \frac{-2+d+2\ell+2\hat{\omega}}{4}, \frac{-1+d+2\ell}{2} \right| z \right]\ z^{\frac{-2+d+2\ell}{4}} + \nonumber \\
&&
+ C_{2} \times \left. _{2}F_{1} \left[ \frac{4-d-2\ell-2\hat{\omega}}{4}, \frac{4-d-2\ell+2\hat{\omega}}{4}, \frac{5-d-2\ell}{2} \right| z \right]\ z^{\frac{4-d-2\ell}{4}}.
\end{eqnarray}

\noindent
This four dimensional case has actually been already solved in \cite{ckl}. A similar line of thought to the one above leads to choosing $C_{2} = 0$ and to the quantization of frequency as

\begin{equation}\label{azerofreq}
\hat{\omega} = 2n + \frac{d}{2} + \ell, \qquad n = 0, 1, 2, \dots .
\end{equation}

\noindent
Thus, the normal frequencies for vector type perturbations of four dimensional AdS spacetime follow as

$$
\omega = \sqrt{|\LL|}\ \left( 2n + \ell + 2 \right), \qquad n = 0, 1, 2, \ldots .
$$

\noindent
For any other dimension $d>4$, the solution for vector type perturbations is given by

\begin{eqnarray*}
\Phi (z) &=& C_{1} \times \left. _{2}F_1 \left[ \frac{1+\ell-\hat{\omega}}{2}, \frac{1+\ell+\hat{\omega}}{2}, \frac{-1+d+2\ell}{2} \right| z \right]\ (z-1)^{\frac{4-d}{4}}\ z^{\frac{-2+d+2\ell}{4}} + \\
&&
+ C_{2} \times \left. _{2}F_1 \left[\frac{4-d-\ell-\hat{\omega}}{2}, \frac{4-d-\ell+\hat{\omega}}{2}, \frac{5-d-2\ell}{2} \right| z \right]\ (z-1)^{\frac{4-d}{4}}\ z^{\frac{4-d-2\ell}{4}},
\end{eqnarray*}

\noindent
and an analogous line of reasoning leads to setting $C_{2} = 0$ and to the normal frequencies for vector type perturbations of $d$--dimensional AdS spacetime as

$$
\omega = \sqrt{|\LL|}\ \left( 2n + d + \ell - 2 \right), \qquad n = 0, 1, 2, \ldots .
$$

\noindent
If we set $d=4$ in the equation above we obtain the result we had previously found concerning the four dimensional case. Thus, even though the normal mode eigenfunctions differ from $d=4$ to $d>4$, the normal frequencies are all given by the above formula.

To end the analysis of AdS spacetime, we now discuss the normal modes for scalar type perturbations, again via the same line of reasoning. As we have seen in our description of AdS effective potentials, the parameter $a$ in (\ref{hipergeral}) will vanish when either $d=4$ or $d=6$. The solution to the wave equation in these cases will then be given by (\ref{azero}), with the prescribed frequencies (\ref{azerofreq}) (the four dimensional case was dealt with in \cite{ckl}). So, in dimension $d=4$ the normal frequencies of scalar type perturbation in AdS spacetime are

$$
\omega = \sqrt{|\LL|}\ \left( 2n + \ell + 2 \right), \qquad n = 0, 1, 2, \ldots, 
$$

\noindent
while in $d=6$ they are given by

$$
\omega = \sqrt{|\LL|}\ \left( 2n + \ell + 3 \right), \qquad n = 0, 1, 2, \ldots . 
$$

\noindent
Let us proceed with the study of the wave equation when $d>6$ (we shall address the special $d=5$ case at the end). In this case the solution to the wave equation is

\begin{eqnarray*}
\Phi(z) &=& C_{1} \times \left. _{2}F_1 \left[ \frac{2+\ell-\hat{\omega}}{2}, \frac{2+\ell+\hat{\omega}}{2}, \frac{-1+d+2\ell}{2} \right| z \right]\ (z-1)^{\frac{6-d}{4}}\ z^{\frac{-2+d+2\ell}{4}} + \nonumber \\
&&
+ C_{2} \times \left. _{2}F_1 \left[ \frac{5-d-\ell-\hat{\omega}}{2}, \frac{5-d-\ell+\hat{\omega}}{2}, \frac{5-d-2\ell}{2} \right| z \right]\ (z-1)^{\frac{6-d}{4}}\ z^{\frac{4-d-2\ell}{4}},
\end{eqnarray*}

\noindent
and the standard analysis leads to $C_{2} = 0$ and to the normal frequencies for scalar type perturbations of $d$--dimensional AdS spacetime as ($d>6$)

$$
\omega = \sqrt{|\LL|}\ \left( 2n + d + \ell - 3 \right), \qquad n = 0, 1, 2, \ldots,
$$ 

\noindent
where as usual we have only selected the positive values for the frequency. We also observe that the $d=6$ case is included in the previous expression, but the $d=4$ case is not (and thus one needs to apply the expression we have previously obtained). In order to conclude our analysis, one must still address the case where $d=5$. As we have seen in the description of AdS effective potentials, this dimension seems rather special as the effective potential is negative in certain regions of spacetime. This will lead to an unexpected result for scalar type perturbations in five dimensions. If we use the solution above (where, as we have already seen, $C_{2} = 0$) in the particular case of $d=5$, we find

$$
\Phi(z) = C_{1} \times \left. _{2}F_1 \left[ \frac{2+\ell-\hat{\omega}}{2}, \frac{2+\ell+\hat{\omega}}{2}, \frac{4+2\ell}{2} \right| z \right]\ (z-1)^{\frac{1}{4}}\ z^{\frac{3+2\ell}{4}}.
$$

\noindent
One immediately realizes that given the well behaved solution at the origin, it automatically satisfies the other boundary condition of $\Phi (z=1) =0$. Thus, in five dimensions, the spectrum of AdS scalar type perturbations is \textit{continuous}, \textit{i.e.}, $\omega \in {\mathbb{R}}$. This is a rather unexpected result. We believe that the consequences of this result should deserve further studies. Let us also point out that it was shown in \cite{ckl}, in the particular $d=4$ case, that as one progressively decreases the size of a Schwarzschild AdS black hole, its quasinormal frequencies approach the pure AdS frequencies. We have made use of this property, as well as of our results above, in section 3, when discussing Schwarzschild AdS quasinormal frequencies.


\subsection{The de Sitter Solution}


The last case of exact solutions we wish to address is that of the pure dS spacetime solution. The analysis of wave propagation in $d$--dimensional dS spacetime closely parallels that for AdS, but with one crucial difference: there is now a cosmological horizon and this will have defining consequences in the boundary conditions one should pick. Indeed, one now chooses quasinormal mode boundary conditions at the cosmological horizon, \textit{i.e.}, an out--going plane wave solution. Given this, while in pure AdS we only found normal modes, in pure dS we shall only find quasinormal modes. The question of pure dS quasinormal modes has actually been recently subject to some discussion, and also to some confusion, in the literature. For the case of dimension $d=4$, the analysis we shall follow was presented in detail in \cite{bclp, choudhury-padmanabhan}. We believe this is a clear and rigorous derivation, and the authors found that there are no quasinormal modes in four dimensional dS spacetime. The open issue is what happens as we extend these results to $d$--dimensions. Some authors \cite{acl} claim there will always be dS quasinormal modes. This is in contradiction with the results in \cite{bclp, choudhury-padmanabhan} and a closer inspection at \cite{acl} reveals that the authors do not actually impose quasinormal mode boundary conditions at the cosmological horizon (they demand for vanishing of the wave function), thus falsifying their conclusions. Some other authors \cite{myung-kim} claim there will never be dS quasinormal modes. While this result seems not to contradict any other earlier authors, truth is that upon closer inspection the authors of \cite{myung-kim} are actually looking for \textit{real} frequencies, which is clearly not the case for quasinormal frequencies, thus falsifying their conclusions. Further authors have recently addressed quasinormal frequencies of massive scalar fields in pure de Sitter spacetime, in $d$ dimensions \cite{dws}. We hope our results will help to settle the dust, as we shall show that, in general, even dimensional pure dS spacetime never has quasinormal modes, but odd dimensional pure dS always has quasinormal modes.

As in the previous section, we begin with a quick ``tour'' of the different Schr\"odinger--type effective potentials one finds in dS spacetime. The dS solution has parameter $\LL > 0$ (with $\MM = 0 = \QQ$), so that $f(r) = 1 - \LL r^{2}$. In any dimension $d$ the tortoise is

\begin{eqnarray*}
x &=& \int \frac{d\rho}{f(\rho)} = \frac{1}{\sqrt{\LL}}\; {\mathrm{arctanh}} \left( \sqrt{\LL}\ r \right), \\
r &=& \frac{1}{\sqrt{\LL}} \tanh \left( \sqrt{\LL}\ x \right).
\end{eqnarray*}

\noindent
This time around $r < R_{C} = \frac{1}{\sqrt{\LL}}$ (see the appendix) so 
that the tortoise is in ${\mathbb{R}}^{+}$ as $\lim_{r \to 0} x[r] = 0$ and $\lim_{r \to R_{C}} x[r] = + \infty$. The potential for tensor type perturbations is thus written as (see appendix B)

$$
V_{\mathsf{T}} (x) = \frac{\LL \left( 2 \ell \left( \ell+d-3 
\right) + \left( d-2 \right)^{2} + 2 \Big( \ell \left( \ell+d-3 
\right) - \left( d-2 \right) \Big) \cosh \left( 2 \sqrt{\LL}\ x 
\right) \right)}{4 \sinh^{2} \left( \sqrt{\LL}\ x \right) \cosh^{2} 
\left( \sqrt{\LL}\ x \right)}.
$$

\FIGURE[ht]{\label{5ddst}
	\epsfig{file=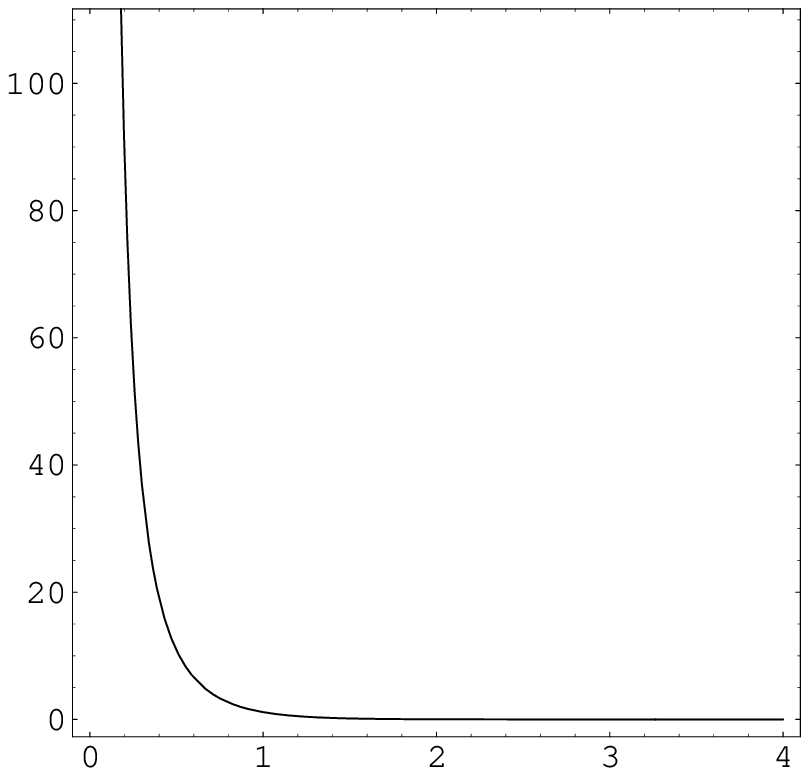, width=1.5in, height=1.5in}
\caption{Potential for dS tensor type perturbations in $d=5$.}
}

\noindent
In its range $x \in {\mathbb{R}}^{+}$, this potential looks like the one plotted in Figure \ref{5ddst} (where we set $\LL = 1$, $d=5$ and $\ell = 1$, as one should recall that for tensor type perturbations $\ell \geq 1$). The shape of the potential does not change much as we change its parameters, and at infinity the potential decays to zero. For vector type perturbations the potential is

$$
V_{\mathsf{V}} (x) = \frac{\LL \left( 2 \ell \left( \ell+d-3 
\right) + \left( d-2 \right) \left( d-4 \right) + 2 \ell \left( 
\ell+d-3 \right) \cosh \left( 2 \sqrt{\LL}\ x \right) \right)}{4 
\sinh^{2} \left( \sqrt{\LL}\ x \right) \cosh^{2} \left( \sqrt{\LL}\ x 
\right)}.
$$

\noindent
In its range $x \in {\mathbb{R}}^{+}$, this potential looks very much like the previous one. Again, the shape of the potential is basically unaffected as we change its parameters and at infinity the potential decays to zero. Finally, the potential for scalar type perturbations is (see appendix B)

$$
V_{\mathsf{S}} (x) = \frac{\LL \left( 2 \ell \left( \ell+d-3 
\right) + \left( d-4 \right)^{2} + 2 \Big( \ell \left( \ell+d-3 
\right) + \left( d-4 \right) \Big) \cosh \left( 2 \sqrt{\LL}\ x 
\right) \right)}{4 \sinh^{2} \left( \sqrt{\LL}\ x \right) \cosh^{2} 
\left( \sqrt{\LL}\ x \right)}.
$$

\noindent
In the range $x \in {\mathbb{R}}^{+}$ this potential again looks like the previous ones, at infinity decaying to zero. We learn that all potentials are very much alike and they do not change substantially as we change dimension (unlike the AdS counterpart). Let us then proceed with the quasinormal mode computation. 

As in the AdS case, for pure dS spacetime the wave equation can be reduced to a hypergeometric equation, and again it pays off to start with the potentials as written in the original $r$ coordinate. In pure dS spacetime, the potential governing $d$--dimensional gravitational perturbations may be written as

$$
V(r) = \left( 1 - {\LL} r^{2} \right) \left( - a {\LL} + \frac{b \left( b+1 \right)}{r^{2}} \right),
$$

\noindent
where $a$ and $b$ were defined in the previous section. The dS problem is slightly simpler than the AdS one, as the factorization $\Phi (r) = r^{b+1} \left( 1- \LL r^{2} \right)^{i\frac{{\hat \omega}}{2}} \Psi (r)$ transforms the dS wave equation for $\Phi (r)$ in a hypergeometric equation for $\Psi (r)$, namely

$$
r \left( 1 - \LL r^{2} \right) \frac{d^{2} \Psi}{dr^{2}} + 2 \left( 1 + \left( 1 - \LL r^{2} \right) b - \left( 2 + i \hat{\omega} \right) \LL r^{2} \right) \frac{d\Psi}{dr} - \LL r \left( 2 - a + b^{2} + b \left( 3 + 2 i \hat{\omega} \right) + 3 i \hat{\omega} - \hat{\omega}^{2} \right) \Psi = 0.
$$

\noindent
Given the hypergeometric solution for $\Psi (r)$, it immediately follows the general solution for $\Phi (r)$ as

\begin{eqnarray*}
\Phi (r) &=& C_{1} \times \left. _{2}F_1\left[ \frac{3-\sqrt{1+4a}+2b+2i\hat{\omega}}{4}, \frac{3+\sqrt{1+4a}+2b+2i\hat{\omega}}{4}, \frac{3+2b}{2} \right| \LL r^{2} \right]\ r^{b+1} \left( 1 - \LL r^{2} \right)^{i \frac{\hat{\omega}}{2}} + \\
&&
+ C_{2} \times \left. _{2}F_1\left[ \frac{1-\sqrt{1+4a}-2b+2i\hat{\omega}}{4}, \frac{1+\sqrt{1+4a}-2b+2i\hat{\omega}}{4}, \frac{1-2b}{2} \right| \LL r^{2} \right]\ r^{-b} \left( 1 - \LL r^{2} \right)^{i \frac{\hat{\omega}}{2}}.
\end{eqnarray*}

\noindent
Interestingly enough, with the allowed values for $a$ (which depend on the type of gravitational perturbation), one will always get rid of the square root. In particular, one obtains the general expression

\begin{eqnarray*}
\Phi (r) &=& C_{1} \times \left. _{2}F_1\left[ \frac{3-d+j+2b+2i\hat{\omega}}{4}, \frac{3+d-j+2b+2i\hat{\omega}}{4}, \frac{3+2b}{2} \right| \LL r^{2} \right]\ r^{b+1} \left( 1 - \LL r^{2} \right)^{i \frac{\hat{\omega}}{2}} + \\
&&
+ C_{2} \times \left. _{2}F_1\left[ \frac{1-d+j-2b+2i\hat{\omega}}{4}, \frac{1+d-j-2b+2i\hat{\omega}}{4}, \frac{1-2b}{2} \right| \LL r^{2} \right]\ r^{-b} \left( 1 - \LL r^{2} \right)^{i \frac{\hat{\omega}}{2}}.
\end{eqnarray*}

\noindent
where $j=1, 3, 5$, for tensorial, vectorial and scalar type gravitational perturbations, respectively. The fact that $j$ is always an odd number will have significant consequences in the final result, as we shall see shortly.

Next, one needs to impose quasinormal mode boundary conditions. At the origin, there is no horizon whatsoever, and one demands for a regular solution. This immediately sets $C_{2} = 0$. We are thus left with the first term, and the quasinormal mode boundary condition of out--going plane waves near the cosmological horizon. At the cosmological horizon, $r \sim R_{C}$, one has $\left( 1 - \LL r^{2} \right) \to 0$ and $\LL r^{2} \to 1$, while the hypergeometric function is particularly simple when its argument goes to zero, \textit{i.e.}, $\lim_{z \to 0} \left. _{2}F_1 \left[ \alpha, \beta, \gamma \right| z \right] = 1$. Therefore, in order to conveniently analyze our solution near $r \sim R_{C}$, it is helpful to make use of the following transformation law for hypergeometric functions

\begin{eqnarray*}
&&
\left( 1 - \LL r^{2} \right)^{i\frac{{\hat{\omega}}}{2}} \left. _{2}F_1 \left[ \frac{3-d+j+2b+2i\hat{\omega}}{4}, \frac{3+d-j+2b+2i\hat{\omega}}{4}, \frac{3+2b}{2} \right| \LL r^{2} \right] = \\
&&
= \frac{\left( 1 - \LL r^{2} \right)^{-i\frac{\hat{\omega}}{2}} 
\Gamma \left[ i\hat{\omega} \right]\ \Gamma \left[ \frac{3+2b}{2} \right]}
{\Gamma \left[ \frac{3-d+j+2b+2i\hat{\omega}}{4} \right]\ \Gamma \left[ \frac{3+d-j+2b+2i\hat{\omega}}{4} \right]} \left. _{2}F_1 \left[ \frac{3+d-j+2b-2i\hat{\omega}}{4}, \frac{3-d+j+2b-2i\hat{\omega}}{4}, 1-i\hat{\omega} \right| 1 - \LL r^{2} \right] + \\
&&
+ \frac{\left( 1 - \LL r^{2} \right)^{+i\frac{{\hat{\omega}}}{2}} 
\Gamma \left[ -i\hat{\omega} \right]\ \Gamma \left[ \frac{3+2b}{2} \right]}{\Gamma \left[ \frac{3+d-j+2b-2i\hat{\omega}}{4} \right]\ \Gamma \left[ \frac{3-d+j+2b-2i\hat{\omega}}{4} \right]} \left. _{2}F_1 \left[ \frac{3-d+j+2b+2i\hat{\omega}}{4}, \frac{3+d-j+2b+2i\hat{\omega}}{4}, 1+i\hat{\omega} \right| 1 - \LL r^{2} \right].
\end{eqnarray*}

\noindent
Near the cosmological horizon $\left( 1 - \LL r^{2} \right) \to 0$ and $\lim_{z \to 0} \left. _{2}F_1 \left[ \alpha, \beta, \gamma \right| z \right] = 1$, so that making use of this identity the hypergeometric piece is no longer relevant and one can concentrate only on the remaining factors. At the cosmological horizon the tortoise is $x \to + \infty$ and one can write

\begin{eqnarray*}
\left( 1 - \LL r^{2} \right)^{-i \frac{\hat{\omega}}{2}} &=& \left( \frac{1}{\cosh \left( \sqrt{\LL} x \right)} \right)^{-i \hat{\omega}} \to e^{i \omega x}, \\
\left( 1 - \LL r^{2} \right)^{+i \frac{\hat{\omega}}{2}} &=& \left( \frac{1}{\cosh \left( \sqrt{\LL} x \right)} \right)^{+i \hat{\omega}} \to e^{-i \omega x}.
\end{eqnarray*}

\noindent
Out--going waves at infinity (waves that behave as $e^{-i\omega x}$), satisfying dS quasinormal mode boundary conditions, must thus have the form $\Phi (r) \sim \left( 1 - \LL r^{2} \right)^{i \frac{\hat{\omega}}{2}}$. In other words, one can ensure quasinormal mode boundary conditions by requiring that the in--going wave term is zero and that the out--going term is non--vanishing. One is thus lead to the conditions

\begin{equation}
\frac{\Gamma \left[ i\hat{\omega} \right]\ \Gamma \left[ \frac{3+2b}{2} \right]}{\Gamma \left[ \frac{3-d+j+2b+2i\hat{\omega}}{4} \right]\ \Gamma \left[ \frac{3+d-j+2b+2i\hat{\omega}}{4} \right]} = 0 \qquad {\mathrm{and}} \qquad 
\frac{\Gamma \left[ -i\hat{\omega} \right]\ \Gamma \left[ \frac{3+2b}{2} \right]}{\Gamma \left[ \frac{3+d-j+2b-2i\hat{\omega}}{4} \right]\ \Gamma \left[ \frac{3-d+j+2b-2i\hat{\omega}}{4} \right]} \ne 0.
\label{van}
\end{equation}

\noindent
The first of the conditions in (\ref{van}) implies that one of the Gamma functions in the denominator must have a pole. This is actually not enough: if one of the Gamma functions in the denominator has pole, then so will $\Gamma \left[ i \hat{\omega} \right]$ have a pole (of the same degree), and so they will cancel each other out. One must thus require \textit{both} Gamma functions in the denominator to have a pole. This will guarantee the vanishing equality. 
Recalling that $2b = d + 2\ell - 4$, it is simple to see that one will have the required poles whenever

\begin{eqnarray*}
2i\hat{\omega} + 2\ell + j - 1 &=& - 4n, \\
2i\hat{\omega} + 2d + 2\ell - j - 1 &=& - 4m,
\end{eqnarray*}

\noindent
with both $n,m \in {\mathbb{N}}$. On the one hand both these equations imply that $i \hat{\omega}$ will be a negative integer (recall that $j$ is an odd number) so that the Gamma function in the numerator will have a pole, as we have stated previously. On the other hand, both these equations imply that none of the Gamma functions on the second condition in (\ref{van}) can have poles in such a way that that expression could vanish. The requirement of out--going boundary conditions is thus satisfied. We still have to deal with the fact that there are \textit{two} conditions for the \textit{one} unknown, the frequency. These equations are compatible, and yield a solution for the quasinormal frequency, if and only if 

$$
d - j = 2 \left( n - m \right) \in 2 {\mathbb{Z}}.
$$

\noindent
Because $j=1,3,5$, $d-j$ can only be even if $d$ is odd. One is led to the conclusion that the pure dS spacetime will have quasinormal modes \textit{only} in odd dimensions. In this case, the quasinormal frequencies for tensor ($j=1$), vector ($j=3$) and scalar ($j=5$) type perturbations are given by

$$
\omega = i \sqrt{\LL} \left( 2n + \ell + \frac{j-1}{2} \right), \qquad n = 0, 1, 2, \ldots .
$$

\noindent
One might worry about this result for odd spacetime dimension, as in this case the solution to the wave equation is not given by the general expression above, involving two hypergeometric functions. In fact, the two corresponding solutions became dependent\footnote{We thank Bin Wang for pointing this out to us.}, and the second hypergeometric function must be replaced by a Meijer $G$--function in order to generate the space of solutions. However, the Meijer $G$--function has the same singular behavior at the origin as the hypergeometric function it is replacing, and hence it must be discarded. The remaining hypergeometric function is exactly the same as we had above and thus our reasoning remains unchanged, leading to the final result above. Observe that this result is certainly in agreement with the recent work in \cite{choudhury-padmanabhan}, where it was shown that four dimensional dS spacetime has no quasinormal modes. It would be very interesting to obtain numerical evidence in support of our results, both in this dS case and our previous study of AdS.


\section{Applications to Quantum Gravity}


The quest for asymptotic quasinormal frequencies was subject to increased momentum after the works of Hod \cite{hod}, showing how these frequencies could be used to infer on the quantum of area at the black hole event horizon, and of Dreyer \cite{dreyer}, showing how the Barbero--Immirzi parameter \cite{barbero, immirzi, immirzi-2} in loop quantum gravity could be fixed starting from these frequencies. Having computed quasinormal frequencies in a wide variety of situations, we would now like to understand whether the ideas of \cite{hod, dreyer} are only valid for Schwarzschild four dimensional black holes, or if they have a broader scope. Let us point out that it is the scalar type perturbations of the gravitational mode which seem to be the most relevant for applications in quantum gravity, as it is this type of perturbations which corresponds to changes in the geometry of the black hole event horizon \cite{kodama-ishibashi-3}. As we have also shown that asymptotic quasinormal frequencies for scalar type perturbations always coincide with the ones for tensor type, this will actually not be an issue. We will next swiftly review these different approaches to the study of quantum gravity from asymptotic quasinormal frequencies and see what we can  further learn by making use of our own results.


\subsection{The Quantum of Area from Quasinormal Frequencies}


It has been long advocated that black holes should have a discrete area spectrum \cite{bekenstein-mukhanov, bekenstein} and that one could expect, on general grounds, horizon area to be quantized as $A_{n} = f(n)$, with $n \in {\mathbb{N}}$. The authors of \cite{bekenstein-mukhanov, bekenstein} actually make the case for a linear function, with the proposed non--extremal black hole discrete spectrum for area being of the harmonic oscillator type (in $G_{d} = c = \hbar = 1$ units)

$$
A_{n} = n \Omega, \quad n \in {\mathbb{N}},
$$

\noindent
where $\Omega$ was an undetermined number. Observe that \cite{bekenstein-mukhanov} if the energy at level $n$ is degenerate, with multiplicity $g(n)$, one can further identify the black hole entropy at this level with $\log g(n)$. So, if one makes use of the Bekenstein--Hawking (BH) entropy, $S = \frac{A}{4}$, one can immediately obtain

$$
g(n) = g(1) \exp \frac{\Omega}{4} \left( n-1 \right),
$$

\noindent
where we have appropriately normalized $g(n)$. Because $g(n)$ is an integer, this further constraints $\Omega = 4 \log k$ with $k = 2, 3, 4, \ldots$. The choice of $k$ is, however, subject to some debate.

Along these lines, \cite{hod} made a pertinent observation which allowed for a first principles determination of the above value for $k$ (and, in some ways, for an independent check of the above arguments concerning horizon area quantization), at least in the Schwarzschild case. What \cite{hod} did was to use asymptotic quasinormal frequencies in order to determine the unknown $\Omega$. We have seen that these frequencies are located in the complex plan and that the real part of the quasinormal frequencies approaches a constant value as the mode $n$ is increased. Making use of Bohr's correspondence principle which states that ``\textit{at large quantum numbers, the [quantum] transition frequencies should be equal to the classical oscillation frequencies}'', one thus expects that the asymptotic behavior of the quasinormal frequencies may contain relevant information about the quantum black hole (the problem with this approach to black hole quantum mechanics is, of course, to correctly identify the physically relevant oscillation frequencies). For the Schwarzschild black hole, the asymptotic behavior for the highly damped frequencies is \cite{motl, motl-neitzke, birmingham}

$$
\frac{\omega_{n}}{T_{H}} = \log 3 + 2 \pi i \left( n + \frac{1}{2} \right) + {\mathcal{O}} \left( \frac{1}{\sqrt{n+1}} \right),
$$

\noindent
for any type of perturbation and in any dimension $d$, with $T_{H}$ the Hawking temperature at the Schwarzschild horizon. Clearly, $\re (\omega)$ depends only on the black hole parameters. One thus interprets the highly damped quasinormal frequencies (specifically, $\lim_{n \to +\infty} \re (\omega_{n})$) as a characteristic of the black hole itself.

In order to remain fully general, we now proceed with a slight deviation from the arguments in \cite{hod}. Let us consider the first law of black hole mechanics \cite{bch}, which states that a static black hole, with mass $M$, charge $Q$, surface gravity $k$, and electric surface potential $\Phi$, when perturbed so that $M \to M + d M$ and $Q \to Q + d Q$, reacts as

$$
d M = \frac{k}{8\pi} dA + \left. \Phi \right|_{H} d Q,
$$

\noindent
basically stating ``energy = heat + work''. If one makes use of Bohr's principle as suggested in \cite{hod}, $dM = \re (\omega_{\infty})$, and further understanding that $dQ=0$, it immediately follows 

$$
dA = \frac{4}{T_{H}} \re (\omega_{\infty}) = 4 \log 3.
$$

\noindent
All together, via a quasinormal mode calculation the unknown number $\Omega$ in the horizon quantum mechanical area spectrum formula for a given black hole has now been pinpointed to be $\Omega = 4 \log 3$. Attempts to extend these ideas to other black holes were first performed in \cite{setare-1, setare-2} but, as we have computed in the present work, the asymptotic frequencies used in those papers turned out not to be the correct ones to use.

When moving to the general case (which is the main focus of this paper), some questions concerning the previous procedure come to mind. Let us first address the question of the value of $dQ$, when in the Einstein--Maxwell situation. Recall that we are computing perturbations of both the gravitational and electromagnetic modes at the \textit{electrovacuum}, and that these perturbations are themselves \textit{uncharged}. So, any application of Bohr's principle we may make will always have to satisfy $dQ = 0$. In this case, for an asymptotic quasinormal frequency to be of any relevance in the determination of the area spectrum as described above, it better be of the form $\re (\omega_{\infty}) = T_{H} \log x$, with $x \in {\mathbb{N}}$. While the proportionality to the temperature is somewhat generic from dimensional analysis, obtaining the term in $\log x$ is a rather special event. Another source of confusion are situations with more than one horizon. Again we recall that one is always working with massless, uncharged, perturbations of the electrovacuum, in which case the application of the first law as above still holds. Indeed, generically, if one has two horizons---name them $1$ and $2$---then $k_{1} dA_{1} = k_{2} dA_{2}$ and it is irrelevant which horizon we choose to concentrate on. With these points cleared and out of the way, let us see what happens in the different situations.

Actually not much happens, besides the simple Schwarzschild situation which we have just reviewed. Indeed, for RN, Schwarzschild dS, or RN dS, there is no algebraic solution for the asymptotic quasinormal frequencies and thus no ``logarithm'' to use in the arguments above. For Schwarzschild AdS and RN AdS, there is an algebraic solution for the asymptotic frequencies (depending on the parameter $x_{0}$), but this solution does \textit{not} have the required logarithmic structure to be used in the arguments above. Finally, in the extremal RN situation, there is both a solution and the required logarithmic structure. Unfortunately, the argument of the logarithm is only an integer in dimensions $d=4$ and $d=5$ where it takes the values $1$ and $0$, respectively, thus again invalidating the arguments above. Thus, it does not seem likely that the arguments in \cite{hod} will have application away from the realm of the Schwarzschild spacetime.


\subsection{The Barbero--Immirzi Parameter in Loop Quantum Gravity}


The entropy of a large class of four dimensional\footnote{Let us stress that this subsection is strictly four dimensional.} non--extremal black holes has been computed within the formalism of loop quantum gravity \cite{abck, abk}. While this calculation reproduces the expected scaling of entropy with area, $S \propto A$, it fails to reproduce the numerical factor of $\frac{1}{4}$ up to an arbitrary factor, the Barbero--Immirzi parameter, $\gamma$ \cite{barbero, immirzi, immirzi-2}. This parameter appears in the definition of the [real] phase space variables and is thus present in the quantization, fixing the spectrum of the area operator. Because it is arbitrary (positive real), it can be adjusted so that the entropy calculation is correct, but then this result is no longer a prediction of loop quantum gravity. It has been suggested that one should regard this parameter as an analogue of the $\theta$ parameter in QCD \cite{abk}. It has also been suggested that one could use quasinormal mode considerations in order to fix this parameter semi--classically \cite{dreyer}. Let us see how this parameter comes about and how it can be fixed for the four dimensional Schwarzschild black hole using quasinormal frequencies. In order to remain fully general in what respects the spectrum of the area operator \cite{abck, krasnov, ling-smolin-1} (for the upcoming analysis of our results), we proceed with a slight deviation from the arguments in \cite{dreyer} (including at the same time the supersymmetric results of \cite{ling-zhang}).

In loop quantum gravity the degrees of freedom associated to a four dimensional quantum black hole are described by a Chern--Simons gauge theory living at the event horizon \cite{abck, abk}. One constructs surface states with the aid of spin networks such that when an edge of the spin network intersects and punctures a surface it associates to it a quantum of area. In order to compute the black hole entropy one starts by computing the dimension of the boundary Hilbert space which describes the surface states, $\dim {\mathcal{H}}_{\Sigma}$, and then simply writes $S = \log \dim {\mathcal{H}}_{\Sigma}$. For a large number of punctures, Chern--Simons theory yields \cite{abck}

$$
\dim {\mathcal{H}}_{\Sigma} = \prod_{p=1}^{N} \left( 2 j_{p} + 1 \right),
$$

\noindent
where there are $N$ punctures at the event horizon and where $j_{p}$ labels the $SU(2)$ representation associated to the spin network edge piercing at puncture $p$ (see \cite{abk} for further details). A similar result holds for ${\mathcal{N}} = 1$ supersymmetric spin networks, where this time around \cite{ling-smolin-2}

$$
\dim {\widehat{\mathcal{H}}}_{\Sigma} = \prod_{p=1}^{N} \left( 4 j_{p} + 1 \right),
$$

\noindent
and $j_{p}$ now labels representations of $Osp(1|2)$. For large horizon area, one further expects the counting of surface states to be dominated by states where all the spin network edges at the horizon punctures carry the same [fundamental] representation $j$ \cite{abck, abk}. In this case, for bosonic spin networks

$$
S = N \log \left( 2j+1 \right),
$$

\noindent
while for ${\mathcal{N}}=1$ supersymmetric spin networks

$$
\widehat{S} = N \log \left( 4j+1 \right).
$$

\noindent
Clearly, one still needs to estimate the number of horizon punctures, $N$, and to determine the fundamental representation, $j$. This can be accomplished via use of the area operator.

In loop quantum gravity the area operator is quantized. However, it happens that in the classical geometrical theory there is more than one operator describing area \cite{krasnov}. Upon quantization, this fact leads to two different spectra of eigenvalues: there is a standard area operator \cite{abck, abk} which leads to a quantum of area---for a surface intersected by \textit{one} edge of a spin network with label $j$---with eigenvalues (again, we use units where all physical constants are set to unity, so that $\ell_{P}=1$)

\begin{equation} \label{area-1}
A (j) = 8 \pi \gamma \sqrt{j \left( j+1 \right)},
\end{equation}

\noindent
but there is also a different area operator, introduced in \cite{krasnov}, which leads to the alternative eigenvalue spectrum

\begin{equation} \label{area-2}
\widetilde{A} (j) = 8 \pi \widetilde{\gamma} j.
\end{equation}

\noindent
In the expressions above $\gamma$, $\widetilde{\gamma}$ is the Barbero--Immirzi parameter, finally making its appearance, and the total area of the black hole event horizon is given by a sum over all the punctures, $A = \sum_{p=1}^{N} A(j_{p})$. While it is operator (\ref{area-1}) which is used in most applications, there is no \textit{a priori} reason to rule out operator (\ref{area-2}). Yet another area operator appears when dealing with supersymmetric spin networks, which has eigenvalue spectrum given by \cite{ling-smolin-1}

\begin{equation} \label{area-susy}
\widehat{A} (j) = 8 \pi \widehat{\gamma} \sqrt{j \left( j+\frac{1}{2} \right)}.
\end{equation}

\noindent
Here $\widehat{\gamma}$ is the supersymmetric Barbero--Immirzi parameter and $j$ is now understood to label $Osp(1|2)$ representations instead of $SU(2)$ ones. To keep all possibilities together, we shall proceed with the following expression for the area spectrum

$$
A_{n} (j) = 8 \pi \gamma_{n} \sqrt{j \left( j+n \right)},
$$

\noindent
where notation should be clear. When $n=0, \frac{1}{2}$ and $1$, one obtains expressions (\ref{area-2}), (\ref{area-susy}) and (\ref{area-1}), respectively. Going back to our calculation of the entropy, recall that one expects the counting to be dominated by states where all spin network edges carry the same fundamental representation $j$. Thus, for a black hole with total area $A$ and quantum of area $A(j)$, the number of punctures $N$ can be computed simply as $N = \frac{A}{A(j)}$, where we still do not know $j$. The entropy for bosonic and supersymmetric spin networks then follows as 

$$
S = \frac{\log \left( 2j + 1 \right)}{2 \pi \gamma_{n} \sqrt{j \left( j + n \right)}}\ \frac{A}{4} \qquad {\mathrm{and}} \qquad 
\widehat{S} = \frac{\log \left( 4j + 1 \right)}{2 \pi \widehat{\gamma} \sqrt{j \left( j + \frac{1}{2} \right)}}\ \frac{A}{4}.
$$

\noindent
The standard lore \cite{abck, abk} now states that the physically fundamental representation $j$ should be the mathematically fundamental one, and so $j=\frac{1}{2}$. In this case

$$
S = \frac{\log 2}{\gamma_{n} \pi \sqrt{2n+1}}\ \frac{A}{4}, \qquad \widehat{S} = \frac{\log 3}{\widehat{\gamma} \pi \sqrt{2}}\ \frac{A}{4},
$$

\noindent
and one can obtain $S = \frac{A}{4}$ with an adequate choice of the (super) Barbero--Immirzi parameter $\gamma$. On the other hand, one would much rather go without the final arbitrary choice of $\gamma$ (even though one should point out that the very \textit{same} choice of Barbero--Immirzi parameter holds for the large class of black holes studied in \cite{abck, abk}, including charged black holes and cosmological situations).

An alternative path, which we wish to study, was presented in \cite{dreyer} (and extended to supersymmetric spin networks in \cite{ling-zhang}). The point of departure from the previous reasoning lies in the computation of the number of punctures, $N$. Dreyer \cite{dreyer} uses instead Hod's \cite{hod} result for the quantum of area,

$$
dA = \frac{4}{T_{H}} \re (\omega_{\infty}),
$$

\noindent
and from $dA = A(j)$ (with an unknown fundamental representation $j$) it follows a direct expression for the Barbero--Immirzi parameter

$$
\gamma_{n} = \frac{\re (\omega_{\infty})}{2 \pi T_{H} \sqrt{j \left( j + n \right)}}.
$$

\noindent
At the same time, from $N=\frac{A}{dA}$, it also follows a direct and independent expression for the entropy,

$$
S = \frac{T_{H} \log \left( 2j+1 \right)}{\re (\omega_{\infty})}\ \frac{A}{4} \qquad {\mathrm{and}} \qquad 
\widehat{S} = \frac{T_{H} \log \left( 4j+1 \right)}{\re (\omega_{\infty})}\ \frac{A}{4}.
$$

\noindent
The interest of these expressions is best appreciated in the Schwarzschild case, where they become

$$
S = \frac{\log \left( 2j+1 \right)}{\log 3}\ \frac{A}{4} \qquad {\mathrm{and}} \qquad \widehat{S} = \frac{\log \left( 4j+1 \right)}{\log 3}\ \frac{A}{4}.
$$

\noindent
The fundamental representation for the bosonic spin network must then be chosen as $j=1$ \cite{dreyer} while the one for the supersymmetric spin network must be $j=\frac{1}{2}$ \cite{ling-zhang}. This also leads to an \textit{independent} result for the Barbero--Immirzi parameter. These calculations have initiated a debate on the correct choice of gauge group at the spin network edges, but we do not wish to proceed into that. Instead, we just note that if it is \textit{not} the case that $\re (\omega_{\infty}) = T_{H} \log x$, with $x \in {\mathbb{N}}$, then the previous argument will fail and one is back at the initial arbitrary choice for $\gamma$.

It so happens that the previous calculations apply to a broad range of horizons, including cosmological horizons \cite{abck, abk}. This is an immediate problem, as we have shown that, with the exception of the Schwarzschild case, one cannot find solutions with the required logarithmic structure and with $x$ in the integers. So, it does not seem likely that the arguments of \cite{dreyer} will have application away from the realm of the Schwarzschild spacetime and, generically, one will thus not be able to fix the arbitrary Barbero--Immirzi parameter using asymptotic quasinormal frequencies. Yet another problem has recently emerged directly within loop quantum gravity \cite{domagala-lewandowski, meissner}. These authors have shown that the initial calculation of black hole entropy in loop quantum gravity \cite{abck, abk} was actually incorrect: there was a miscounting of the microscopic states contributing to the entropy. In fact, the lore (mentioned above) that the physically fundamental representation $j$ should be the mathematically fundamental one, of $j=\frac12$, is actually not true, as states labeled by higher spins also contribute for the entropy calculation \cite{domagala-lewandowski, meissner}. This has clear implications in the derivation of the BH black hole entropy within loop quantum gravity; of particular interest to our discussion above is the fact that it affects the derivation of the Barbero--Immirzi parameter for Schwarzschild black holes. Indeed, instead of having

$$
\gamma_1 = \frac{\log 2}{\pi\sqrt{3}},
$$

\noindent
for the Barbero--Immirzi parameter, one actually has that $\gamma$ obeys a complicated functional relationship and is a number which lies in the range \cite{meissner}

$$
\frac{\log 2}{\pi} \leq \gamma_1 \leq \frac{\log 3}{\pi}.
$$

\noindent
However, it does not seem that this number can be written as a logarithm \cite{domagala-lewandowski, meissner} which would even invalidate an attempt to match it against the Schwarzschild asymptotic quasinormal frequencies. There has been some attempts to try to keep the Schwarzschild quasinormal relation alive in \cite{smolin, alexandrov, dms}, but at the moment the dust is far from settled. It thus seems that asymptotic quasinormal frequencies will not be of much help in loop quantum gravity. Their future role in quantum gravity is at this moment unknown.


\section{Conclusions and Future Directions}


In this paper we have learned how to analytically compute asymptotic quasinormal frequencies in a wide variety of black hole spacetimes. This made use of a monodromy technique which was first introduced in \cite{motl-neitzke} for asymptotically flat spacetimes, and later extended to non--asymptotically flat spacetimes in \cite{cns}. Such a complete classification as the one we have performed in this paper certainly proves the analytical power of the monodromy technique and opens way to many new results in the calculation of quasinormal modes. The directions in which one can proceed are multiple. Perhaps the first question to address should be the Kerr black hole spacetime, as one would like to know whether the monodromy technique we have developed at length in this work may be extended to non--static spacetimes. Another question to address, and closer to the discussion in this paper, would be to compute asymptotic quasinormal frequencies of  the many extremal situations which arise in our setting and that we have not considered. To do this, one should bear in mind our example of the extremal RN black hole, \textit{i.e.}, the extremal solution calculation may need to be considered from scratch, independently of the non--extremal one. Also, the full significance of the continuous spectrum for scalar type gravitational perturbations in dimension $d=5$, for asymptotically AdS spacetimes, which we have found in this paper, should be further explored.

On what concerns quantum gravity our results are not very encouraging, at least on a first reading. Indeed, they seem to indicate that the results in \cite{hod, dreyer} are nothing but numerical coincidences. In spite of this, we believe that the calculation we have performed in this paper is opening the door for the calculation of asymptotic greybody factors, and these will certainly have a key role to play on the road to quantum gravity. Thus, one can view this paper as the first step towards a complete classification of greybody factors for black hole spacetimes. These greybody factors will describe scattering at high imaginary frequencies and may yet have a role to play on the identification of the dual CFT which microscopically describe the several black holes we have considered. The greybody factors may also provide for additional information on the statistical mechanical behavior of these black hole spacetimes.

In spite of the fact that there is an intrinsic general relativistic interest in the computation and classification of asymptotic quasinormal frequencies, the goal of late on this line of problems has been to find new applications on what concerns quantum gravity. It thus seems like a natural step to take this calculation and apply it to string theoretic black holes. Here, one is immediately faced with a wide range of open problems, starting from higher derivative stringy corrections to the Einstein equations, proceeding towards many $D$--brane constructions of higher dimensional black holes, and passing through the understanding of the asymptotic quasinormal frequencies of the Schwarzschild AdS black hole in terms of the dual finite temperature gauge theory.

If one is to make the step towards string theory, we should finish by mentioning that there is another line of research which deals with the so--called non--quasinormal frequencies \cite{birmingham-carlip} and the study of horizon CFT  \cite{strominger, carlip-1, carlip-2, carlip-3}. Here, it seems that it is frequencies closely related to the quasinormal frequencies that are actually able to determine the correct quantization of the Virasoro operators associated to the horizon CFT, at least in a dilute gas regime, thus yielding the correct BH entropy via use of the Cardy formula \cite{cardy, birmingham-carlip, carlip-3}. Let us thus finish by reviewing one last application of quasinormal modes to quantum gravity, in this context of Cardy's formula for the asymptotic growth of states in CFT \cite{cardy}. To put what we wish to do in perspective, let us start by reviewing a remark by Strominger \cite{strominger}, itself based on an earlier observation of Brown and Henneaux \cite{brown-henneaux}. For the moment, we concentrate on three dimensional quantum gravity. What \cite{brown-henneaux} showed is that any theory of quantum gravity in AdS$_{3}$ must be described by a CFT with central charge (as before, we use units where all physical constants are set to unity)

$$
c = \frac{3}{2 \sqrt{|\LL|}}.
$$

\noindent
\cite{strominger} takes this remarkable observation one step further by making use of Cardy's formula for the asymptotic growth of states in this particular CFT, thus being able to [microscopically] compute black hole entropy. Because the results in \cite{brown-henneaux} guarantee that black hole states belong to the Hilbert space of the given CFT, one may concentrate in the BTZ case \cite{btz} and show that mass $M$ and angular momentum $J$ can be related to the Virasoro operators, so as to yield the eigenvalues of the Virasoro operators themselves \cite{strominger},

$$
L_{0} = \frac{M + \sqrt{|\LL|} J}{2 \sqrt{|\LL|}} \qquad {\mathrm{and}} \qquad \overline{L}_{0} = \frac{M - \sqrt{|\LL|} J}{2 \sqrt{|\LL|}},
$$

\noindent
where one should take into account that these semi--classical considerations hold only when the central charge is $c \gg 1$. Now Cardy's formula for the asymptotic growth of the number of states, $\rho ( L_{0}, \overline{L}_{0} )$, of a given CFT  states that (when $L_{0} | 0 \rangle = 0$, $| 0 \rangle$ being the ground state)

$$
\log \rho ( L_{0}, \overline{L}_{0} ) \sim 2\pi \sqrt{\frac{c L_{0}}{6}} + 2\pi \sqrt{\frac{c \overline{L}_{0}}{6}},
$$

\noindent
and the entropy can be computed from $S = \log \rho ( L_{0}, \overline{L}_{0} )$. Direct application to the BTZ black hole yields

$$
S = \pi \sqrt{\frac{M + \sqrt{|\LL|} J}{2 |\LL|}} + \pi \sqrt{\frac{M - \sqrt{|\LL|} J}{2 |\LL|}}
$$

\noindent
in precise agreement with the BH result for this black hole. This line of reasoning is actually quite generic (see \cite{strominger} for further details) but applies only to black holes whose near--horizon geometry is three dimensional. What one would wish for would be similar arguments valid in any dimension. 

Such an approach has been suggested by Carlip \cite{carlip-1, carlip-2, carlip-3}, in an attempt to generalize the previous arguments to $d$--dimensional black holes. In \cite{carlip-1, carlip-2} one concentrates on the symmetries at the horizon of an arbitrary $d$--dimensional black hole. Using the ADM formalism one studies the algebra of constraints of general relativity in the presence of a boundary and finds that the presence of such boundary actually leads to a central extension of the algebra of spacetime diffeomorphisms. By then concentrating in a subalgebra isomorphic to ${\mathrm{Diff}} \left({\mathbb{S}}^{1} \right)$, one obtains a Virasoro algebra where one can then apply Cardy's formula as before. The algebra in question, that of black hole horizon surface deformations, is the algebra of deformations of the $\left( t, r \right)$ plane which leave the horizon fixed. In this ``horizon CFT'' holomorphic and anti--holomorphic functions correspond to functions of $t \pm x$, where $x$ is the tortoise coordinate \cite{carlip-3}. Working within classical general relativity alone, \cite{carlip-1, carlip-2} showed that the Virasoro algebra in question should have (with $A$ the horizon area)

$$
c = \frac{3A}{2\pi} \qquad {\mathrm{and}} \qquad L_{0} = \frac{A}{16\pi},
$$

\noindent
with the right moving modes having vanishing central charge and Virasoro operator. Applying Cardy's formula one then immediately obtains $S = \frac{A}{4}$. It is important to realize one of the main messages in \cite{carlip-1, carlip-2, carlip-3}: that by making use of general relativity's classical symmetries alone, one may be able to determine the black hole entropy, \textit{independently} of any microscopic analysis. \cite{birmingham-carlip} now proposes a semi--classical way to fix the eigenvalues of the Virasoro operator $L_{0}$ with the help of quasinormal frequencies.

Let us review \cite{birmingham-carlip}, where the authors centered upon the concept of non--quasinormal frequencies. These non--quasinormal frequencies are defined using  monodromies at inner and outer horizons alone, rather than involving any condition at infinity. Recall that quasinormal mode boundary conditions are defined in terms of the behavior of perturbations at the horizon and at infinity. It so happens that for the three dimensional BTZ black hole these conditions can be recast in terms of monodromy conditions at the inner and outer event horizons alone \cite{birmingham-carlip} (further insight into the origin of these non--quasinormal frequencies has appeared recently in \cite{gupta-sen}). One is thus naturally lead to a definition of non--quasinormal frequencies, for $d$--dimensional black holes, only in terms of these monodromies, via ${\mathfrak{M}} \left( R_{H}^{+} \right) {\mathfrak{M}} \left( R_{H}^{-} \right) = 1$. This is accomplished by choosing ingoing boundary conditions at the outer horizon and a general solution at the inner horizon. This monodromy requirement for solutions leads to two non--quasinormal frequencies, $\omega_{\pm}$. The idea of \cite{birmingham-carlip} is then to relate these frequencies, $\omega_{\pm}$, to the quantization of the Virasoro operators, $L_{0}$ and $\overline{L}_{0}$, via a natural application of the correspondence principle to this situation. Surprisingly, in a dilute gas regime, these non--quasinormal frequencies lead to the correct quantization of the Virasoro operators as well as the correct quantizations of energy and angular momentum \cite{birmingham-carlip}. It should be very interesting to find an extension of those results to a broader situation, such as the one in this paper. It may be that quasinormal modes hold more information about quantum gravity than expected at first!

To finish, let us stress that the study of asymptotic quasinormal modes is an active area of present research and has also led to many other applications besides the ones we have studied in this paper. Some recent references one should keep in mind include \cite{tamaki-nomura, kiefer, fernando, kkm, chen-jing, cbkz, konoplya-zhidenko-2, extra-ref-1, extra-ref-2, extra-ref-3}.

\section*{Acknowledgements}
We would like to thank Carlos Herdeiro, Christopher Herzog, Hideo Kodama, Bin Wang and Shijun Yoshida for comments or discussions, at different stages of this research project, and specially Vitor Cardoso for initial collaboration in this work. JN is partially supported by FCT/POCTI/FEDER. RS is supported in part by funds provided by the Funda\c c\~ao para a Ci\^encia e a Tecnologia, under the grants SFRH/BPD/7190/2001 and FEDER--POCTI/FNU/38004/2001.

\vfill

\eject

\appendix


\section{List of Conventions for Black Hole Spacetimes}


In this paper we consider spherically symmetric, static, black holes 
in $d$--dimensional spacetime, with mass $M$, charge $Q$ and 
background cosmological constant $\Lambda$. This set $(M, Q, \Lambda)$ 
of parameters thus describes the black hole. In this appendix we 
review and set notation on basic facts for $d$--dimensional $(M, Q, 
\Lambda)$ black holes. The Einstein--Maxwell equations describing our 
backgrounds are the well known

\begin{eqnarray*}
R_{\mu\nu} &=& \frac{2}{d-2} \Lambda g_{\mu\nu} + 8\pi G_{d} \left( 
T_{\mu\nu} - \frac{1}{d-2} T g_{\mu\nu} \right), \\
\nabla^{\lambda} F_{\lambda\mu} &=& 0,
\end{eqnarray*}

\noindent
where $G_{d}$ is the $d$--dimensional Newton constant, and the 
energy--momentum tensor for the electromagnetic field strength $F = dA$ 
is given by

$$
T_{\mu\nu} = {F_{\mu}}^{\lambda} F_{\nu\lambda} - \frac{1}{4} 
g_{\mu\nu} F_{\alpha\beta} F^{\alpha\beta}.
$$

\noindent
The background spacetime metric $g$, solution to the previous 
Einstein--Maxwell system, is given by

\begin{equation} \label{metric}
g = - f(r)\ dt \otimes dt + {f(r)}^{-1}\ dr \otimes dr + r^{2} d 
\Omega_{d-2}^{2},
\end{equation}

\noindent
where the function $f(r)$ is the well known (we only consider $d>3$)

\begin{equation} \label{f(r)}
f(r) = 1 - \frac{2\MM}{r^{d-3}} + \frac{\QQ^{2}}{r^{2d-6}} - \LL r^{2}. 
\end{equation}

\noindent
In terms of the background parameters $\MM$, $\QQ$ and $\LL$ in the 
metric, one can further compute the ADM mass, $M$, of the black hole as

\begin{equation} \label{mass}
M = \frac{\left( d-2 \right) {\mathcal{A}}_{d-2}}{8\pi G_{d}} \MM,
\end{equation}

\noindent
where ${\mathcal{A}}_{n}$ is the area of an unit $n$--sphere,

\begin{equation} \label{area-sphere}
{\mathcal{A}}_{n} = \frac{2\pi^{\frac{n+1}{2}}}{\Gamma \left( 
\frac{n+1}{2} \right)}.
\end{equation}

\noindent
Likewise one can obtain the charge, $Q$, of the black hole as

\begin{equation} \label{charge}
Q^{2} = \frac{\left( d-2 \right)\left( d-3 \right)}{8\pi G_{d}} \QQ^{2}, 
\end{equation}

\noindent
and the value of the cosmological constant, $\Lambda$, as

\begin{equation} \label{lambda}
\Lambda = \frac{1}{2} \left( d-1 \right)\left( d-2 \right) \LL.
\end{equation}

\noindent
These parameters $\MM$, $\QQ$ and $\LL$, are thus a convenient way of 
keeping track of the black hole defining parameters, $M$, $Q$ and 
$\Lambda$, in the $d$--dimensional case \cite{kodama-ishibashi-1, 
kodama-ishibashi-3, myers-perry}.

Let us denote---for the moment---the black hole background spacetime 
metric by $\hat{g}_{\mu\nu}$, and let us perturb it with some disturbance 
$h_{\mu\nu}$ (where $\| h_{\mu\nu} \| \ll 1$ in some appropriate sense). 
Quantities with a hat on top will here arise from the background metric, 
while un--hatted quantities will arise from the linear perturbation. The 
metric to consider is thus $g_{\mu\nu} = \hat{g}_{\mu\nu} + h_{\mu\nu}$. 
While the background $\hat{g}_{\mu\nu}$ satisfies Einstein's equations 
(or the Einstein--Maxwell equations, depending on the conditions one is 
considering), one still needs to compute the linearized equations of motion 
describing $h_{\mu\nu}$. The perturbed components of the Levi--Civita 
connection are

$$
\Gamma^{\sigma}_{\mu\nu} = {\widehat{\Gamma}}^{\sigma}_{\mu\nu} + 
\delta \Gamma^{\sigma}_{\mu\nu} + {\mathcal{O}} (h^{2}),
$$

\noindent
where the variation of the connection is actually a tensor,

$$
\delta \Gamma^{\sigma}_{\mu\nu} = \frac{1}{2} \hat{g}^{\sigma\rho} 
\left( \widehat{\nabla}_{\mu} h_{\nu\rho} + \widehat{\nabla}_{\nu} 
h_{\mu\rho} - \widehat{\nabla}_{\rho} h_{\mu\nu} \right).
$$

\noindent
From this result one immediately has the variation of the Ricci tensor

$$
\delta R_{\mu\nu} = \widehat{\nabla}_{\nu} \delta 
\Gamma^{\sigma}_{\mu\sigma} - \widehat{\nabla}_{\sigma} \delta 
\Gamma^{\sigma}_{\mu\nu}
$$

\noindent
and, for vacuum solutions, Einstein's equations reduce to $\delta 
R_{\mu\nu} = 0$ (in the Einstein--Maxwell case there will be other terms). 
These are linear differential equations for the 
perturbation, $h_{\mu\nu}$, which can still be further simplified 
using the spherical symmetry of the background $\hat{g}_{\mu\nu}$. For 
$d$--dimensional spacetimes these equations were recently studied by 
Ishibashi and Kodama in \cite{kodama-ishibashi-1, kodama-ishibashi-2, 
kodama-ishibashi-3} and we shall return to their results in appendix B. 

The black hole backgrounds defined by the function $f(r)$ can, 
generically, have both event horizons and cosmological horizons, which 
shall be denoted by $R_{H}$ and $R_{C}$, respectively. They will 
be simply given by the zeros of $f(r)$, \textit{i.e.}, the locations 
in which $f(r)$ changes sign. Also, the parameters $\MM$, $\QQ$ and 
$\LL$ in the background metric cannot generically be taken as 
arbitrary, but are constrained if one is not to consider naked 
singularities. The general analysis can be quite intricate 
\cite{kodama-ishibashi-3}, and we shall in here mainly list the results 
of these authors, referring the reader to \cite{kodama-ishibashi-3} for 
the details. The several possible cases are:

\medskip

\noindent
\underline{\textsf{The Schwarzschild Solution:}} In here $f(r) = 1 - 
\frac{2\MM}{r^{d-3}}$, and the event horizon, determined by $f(r) = 0$, is located at

$$
R_{H} = \left( 2\MM \right)^{\frac{1}{d-3}},
$$

\noindent
so that $f(r) \geq 0$ for $r \geq R_{H}$. The parameter $\MM$ can take 
any positive value and there is no cosmological horizon.

\medskip

\noindent
\underline{\textsf{The RN Solution:}} In 
here $f(r) = 1 - \frac{2\MM}{r^{d-3}} + \frac{\QQ^{2}}{r^{2d-6}}$, and 
the event horizons, determined by $f(r) = 0$, correspond to 

$$
R_{H}^{\pm} = \left( \MM \pm \sqrt{\MM^{2} - \QQ^{2}} 
\right)^{\frac{1}{d-3}},
$$

\noindent
where one is usually interested in the outer horizon, $R_{H}^{+}$. In 
this case $f(r) \geq 0$ for $r \geq R_{H}^{+}$. This result also leads 
to the constraint $\QQ^{2} \le \MM^{2}$; the extremal solution 
corresponding to $\QQ^{2} = \MM^{2}$ in which case there is only one 
solution for the horizon, 

$$
R_{H}^{+} \equiv R_{H}^{-} = \MM^{\frac{1}{d-3}}.
$$

\noindent
Again, there is no cosmological horizon.

\medskip

\noindent
\underline{\textsf{Solutions with a Cosmological Constant:}} For 
solutions with cosmological constant alone, $f(r) = 1 - \LL r^{2}$, 
there is only a cosmological horizon, corresponding to 

$$
R_{C} = \frac{1}{\sqrt{\LL}}.
$$

\noindent
It follows that $f(r) \geq 0$ for $r \leq R_{C}$. This clearly constraints 
$\LL > 0$, so that the AdS solution has no cosmological 
horizon---it will only exist for the dS solution. Because there 
is no black hole, there is no event horizon.

\medskip

\noindent
\underline{\textsf{The Schwarzschild AdS Solution:}} In here $f(r) = 1 
- \frac{2\MM}{r^{d-3}} - \LL r^{2}$ and $\LL < 0$. There is an event 
horizon implicitly defined by $f(r) = 0$, \textit{i.e.}, by the 
unique real and positive root of

$$
| \LL | r^{d-1} + r^{d-3} - 2 \MM = 0.
$$

\noindent
There is no cosmological horizon. While $\MM$ can take any positive value, 
$\LL$ can take any negative value.

\medskip

\noindent
\underline{\textsf{The RN AdS Solution:}} In here $f(r) = 1 - 
\frac{2\MM}{r^{d-3}} + \frac{\QQ^{2}}{r^{2d-6}} - \LL r^{2}$ and $\LL 
< 0$. There are event horizons, implicitly defined by $f(r) = 0$,
\textit{i.e.}, by 

$$
| \LL | r^{2d-4} + r^{2d-6} - 2\MM r^{d-3} + \QQ^{2} = 0,
$$

\noindent
and no cosmological horizon. This equation generically has two real 
positive zeros (corresponding to an inner and an outer horizon, as 
is familiar in the RN class of solutions) and $f(r) > 0$ for $r \ge 
R_{H}^{+}$ (as usual, one is interested in the outer horizon). The 
background parameters are constrained to obey $\QQ^{2} < \MM^{2}$ and 
the cosmological constant is negative but bounded from below 
\cite{kodama-ishibashi-3},

$$
0 > \LL \ge - 2^{\frac{2}{d-3}}\ \frac{d-3}{d-2} \frac{\sqrt{\left( d-1 
\right)^{2} \MM^{2} - 4 \left( d-2 \right) \QQ^{2}} - \left( d-3 \right) 
\MM}{\left( \left( d-1 \right) \MM - \sqrt{\left( d-1 \right)^{2} 
\MM^{2} - 4 \left( d-2 \right) \QQ^{2}} \right)^{\frac{d-1}{d-3}}}.
$$

\noindent
The condition $\QQ^{2} = \MM^{2}$ cannot be achieved in here as that 
would imply $\LL = 0$ and the solution would no longer be AdS. Rather, 
the extremal solution for this case is when the cosmological constant 
reaches its minimal value

$$
\LL = - 2^{\frac{2}{d-3}}\ \frac{d-3}{d-2} \frac{\sqrt{\left( d-1 
\right)^{2} \MM^{2} - 4 \left( d-2 \right) \QQ^{2}} - \left( d-3 \right) 
\MM}{\left( \left( d-1 \right) \MM - \sqrt{\left( d-1 \right)^{2} 
\MM^{2} - 4 \left( d-2 \right) \QQ^{2}} \right)^{\frac{d-1}{d-3}}}.
$$

\noindent
From \cite{kodama-ishibashi-3}, one finds that at extremality of the 
AdS cosmological constant, inner and outer horizons will coincide. 
This point of coincidence also corresponds to a minimal value for the 
[outer] horizon radius, and is located at

$$
R_{H}^{+} \equiv R_{H}^{-} = \left( \frac{2 \left( d-2 \right) 
\QQ^{2}}{\left( d-1 \right) \MM + \sqrt{\left( d-1 \right)^{2} \MM^{2} 
- 4 \left( d-2 \right) \QQ^{2}}} \right)^{\frac{1}{d-3}}.
$$

\noindent
As the AdS cosmological constant grows from extremality to zero, the 
[outer] radius of the RN AdS black hole grows from this minimum to its 
maximum, which corresponds to the pure RN solution.

\medskip

\noindent
\underline{\textsf{The Schwarzschild dS Solution:}} In here $f(r) = 1 
- \frac{2\MM}{r^{d-3}} - \LL r^{2}$ and $\LL > 0$. There is both an 
event horizon and a cosmological horizon, implicitly defined by 
$f(r) = 0$, \textit{i.e.}, by

$$
\LL r^{d-1} - r^{d-3} + 2 \MM = 0.
$$

\noindent
This equation generically has two real and positive zeros (the event 
horizon and the cosmological horizon) and $f(r) \ge 0$ for $R_{H} \le r 
\le R_{C}$. While $\MM$ can take any positive value, $\LL$ is constrained 
to be positive but smaller than

$$
0 < \LL \le \frac{d-3}{d-1} \frac{1}{\left[ \left( d-1 \right) \MM 
\right]^{\frac{2}{d-3}}}.
$$

\noindent
Because of this constraint it follows that there is an extremal 
solution in this case, precisely corresponding to the case where 
$\LL$ attains its maximum value,

$$
\LL = \frac{d-3}{d-1} \frac{1}{\left[ \left( d-1 \right) \MM 
\right]^{\frac{2}{d-3}}}.
$$

\noindent
At this extremal point, the event horizon and the cosmological 
horizon will actually coincide. The solution at this point can be 
found to be

$$
R_{H} \equiv R_{C} = \left[ \left( d-1 \right) \MM 
\right]^{\frac{1}{d-3}}.
$$

\noindent
Observe that this is a quite particular situation, as there is now no 
range of $r$ where $f(r) > 0$.

\medskip

\noindent
\underline{\textsf{The RN dS Solution:}} In here $f(r) = 1 - 
\frac{2\MM}{r^{d-3}} + \frac{\QQ^{2}}{r^{2d-6}} - \LL r^{2}$ and $\LL 
> 0$. There are both event horizons and a cosmological horizon, 
implicitly defined by $f(r) = 0$, \textit{i.e.}, by

$$
\LL r^{2d-4} - r^{2d-6} + 2\MM r^{d-3} - \QQ^{2} = 0.
$$

\noindent
This equation generically has three positive real roots (inner and outer 
event horizons plus the cosmological horizon) and $f(r) \ge 0$ for 
$R_{H}^{+} \le r \le R_{C}$. The constraints on the background parameters 
are a bit more intricate now \cite{kodama-ishibashi-3}: one can choose 
$\QQ^{2} \le \MM^{2}$ in which case $\LL$ is constrained to be positive 
but smaller than

$$
0 < \LL \le 2^{\frac{2}{d-3}}\ \frac{d-3}{d-2} \frac{\left( d-3 
\right) \MM + \sqrt{\left( d-1 \right)^{2} \MM^{2} - 4 \left( d-2 
\right) \QQ^{2}}}{\left( \left( d-1 \right) \MM + \sqrt{\left( d-1 
\right)^{2} \MM^{2} - 4 \left( d-2 \right) 
\QQ^{2}}\right)^{\frac{d-1}{d-3}}}. 
$$

\noindent
On this range of parameters the choice of extremality corresponds to 
making $\LL$ extremal (the standard cosmological constant extremality 
condition---observe that the standard RN extremality condition, of 
$\QQ^{2} = \MM^{2}$, is in general \textit{not} extremal for $\LL \not = 
0$). One thus requires

$$
\LL = 2^{\frac{2}{d-3}}\ \frac{d-3}{d-2} \frac{\left( d-3 
\right) \MM + \sqrt{\left( d-1 \right)^{2} \MM^{2} - 4 \left( d-2 
\right) \QQ^{2}}}{\left( \left( d-1 \right) \MM + \sqrt{\left( d-1 
\right)^{2} \MM^{2} - 4 \left( d-2 \right) 
\QQ^{2}}\right)^{\frac{d-1}{d-3}}}. 
$$

\noindent
Instead of three positive real roots one now only finds two positive 
roots, corresponding to the dS extremal condition in which the [outer] 
event horizon and the cosmological horizon coincide. The solution at 
this point is \cite{kodama-ishibashi-3}

$$
R_{H}^{+} \equiv R_{C} = \left( \frac{\left( d-1 \right) \MM + 
\sqrt{\left( d-1 \right)^{2} \MM^{2} - 4 \left( d-2 \right) \QQ^{2}}}{2} 
\right)^{\frac{1}{d-3}}.
$$

\noindent
As we have said previously, in this RN dS case the constraints on the 
background parameters are more intricate. Thus, besides the choice we 
have just studied, one can also study the case where \cite{kodama-ishibashi-3}

$$
\MM^{2} < \QQ^{2} \le \frac{1}{4}\ \frac{\left( d-1 \right)^{2}}{d-2} 
\MM^{2},
$$

\noindent
in which case $\LL$ is constrained to be a positive value bounded both 
from above and below as

$$
\LL_{-} \le \LL \le \LL_{+},
$$

\noindent
where

$$
\LL_{\pm} = 2^{\frac{2}{d-3}}\ \frac{d-3}{d-2} \frac{\left( d-3 
\right) \MM \pm \sqrt{\left( d-1 \right)^{2} \MM^{2} - 4 \left( d-2 
\right) \QQ^{2}}}{\left( \left( d-1 \right) \MM \pm \sqrt{\left( d-1 
\right)^{2} \MM^{2} - 4 \left( d-2 \right) 
\QQ^{2}}\right)^{\frac{d-1}{d-3}}}.
$$

\noindent
In this case, the equation for the horizon radius will generically 
have three real zeros (again corresponding to both an inner and an 
outer event horizons plus the cosmological horizon) where $f(r)>0$ in 
$R_{H}^{+} \le r \le R_{C}$. There now seem to be several choices for 
extremality. The first choice we will make, exploring extremality 
conditions on $\QQ$, is to make the extremal choice of

$$
\QQ^{2} = \frac{1}{4}\ \frac{\left( d-1 \right)^{2}}{d-2} \MM^{2}.
$$

\noindent
This situation is actually interesting as in this case

$$
\LL_{+} \equiv \LL_{-} = 2^{\frac{2}{d-3}}\ \frac{\left( d-3 
\right)^{2}}{\left( d-2 \right) \left( d-1 \right)^{\frac{d-1}{d-3}}}\ 
\frac{1}{\MM^{\frac{2}{d-3}}},
$$

\noindent
so that this automatically also makes the cosmological constant 
extremal (at the above value). In this case, one finds a single real 
and positive root, corresponding to a situation where inner event 
horizon, outer event horizon and cosmological horizon, all coincide 
at the very same point. This is the dS type of extremal condition, 
and the point of coincidence of the three possible horizons is at

$$
R_{H}^{+} \equiv R_{H}^{-} \equiv R_{C} = \left( \frac{\left( d-1 
\right) \MM}{2} \right)^{\frac{1}{d-3}}.
$$

\noindent
The other possible extremal choices one can make leave $\QQ$ in its 
parameter range, and saturate $\LL$ at either $\LL_{-}$ or $\LL_{+}$. These 
extremal conditions on $\LL$ will make two of the three real positive 
solutions for the horizon radius coincide. If one saturates $\LL = 
\LL_{-}$, one will only find two real positive roots, and this 
corresponds to a RN type of extremality, where inner and outer event 
horizons coincide. On the other hand, if one saturates $\LL = \LL_{+}$, 
one will again only find two real positive roots, but this will now 
correspond to a dS type of extremality, where the outer event horizon 
and the cosmological horizon coincide. Observe that this is coherent 
with the previous picture of full saturation at $\LL_{+} = \LL_{-}$ where 
all three horizons coincide at a point. At the $\LL = \LL_{-}$ extremal 
point, inner and outer horizons coincide at the point

$$
R_{H}^{+} \equiv R_{H}^{-} = \left( \frac{\left( d-1 \right) \MM - 
\sqrt{\left( d-1 \right)^{2} \MM^{2} - 4 \left( d-2 \right) 
\QQ^{2}}}{2} \right)^{\frac{1}{d-3}}.
$$

\noindent
Likewise, at the $\LL = \LL_{+}$ extremal point, outer event horizon 
and cosmological horizon coincide at

$$
R_{H}^{+} \equiv R_{C} = \left( \frac{\left( d-1 \right) \MM + 
\sqrt{\left( d-1 \right)^{2} \MM^{2} - 4 \left( d-2 \right) 
\QQ^{2}}}{2} \right)^{\frac{1}{d-3}}.
$$

\noindent
Recall that situations where the [outer] event horizon and the 
cosmological horizon coincide are very particular, as there will be no 
range of $r$ where $f(r) > 0$.


\section{Formulae on Ishibashi--Kodama Master Equations} \label{appendixB}


A set of equations describing the perturbations to background black 
hole $d$--dimensional spacetimes were derived by Ishibashi and Kodama in 
\cite{kodama-ishibashi-1, kodama-ishibashi-2, kodama-ishibashi-3}. The 
most general case including $(M,Q,\Lambda)$ non--vanishing parameters 
is studied in \cite{kodama-ishibashi-3} and in this appendix we shall 
schematically review the results in that paper for the Schr\"odinger--like 
equations describing the quasinormal modes. In $d$ dimensions, the IK master 
equations present a gauge invariant formalism for perturbations 
to the background spacetime, and are of the Schr\"odinger--like form 
(\ref{schrodinger}). These variables are grouped in three types, 
according to their tensorial behavior on the ${\mathbb{S}}^{d-2}$ sphere: 
tensor type perturbations, vector type perturbations and scalar type 
perturbations \cite{kodama-ishibashi-1, kodama-ishibashi-3}. Perturbations 
of the gravitational metric field will include all these three types 
of perturbations, while perturbations of the electromagnetic vector 
field will only include vector type and scalar type perturbations 
\cite{kodama-ishibashi-3}. As the Einstein--Maxwell system couples 
gravitational and electromagnetic fields, one expects the master 
equations to be coupled as well. What \cite{kodama-ishibashi-3} shows 
is that one can always decouple the master equations and these are 
the relevant gauge invariant equations to be used when studying quasinormal modes.
Because quasinormal modes are studied as perturbations to the electrovacuum, the 
IK master equations will all be homogeneous. Let us list the IK 
equations and respective Schr\"odinger--like potentials:

\medskip

\noindent
\underline{\textsf{Tensor Type Perturbations:}} Only the gravitational 
metric field displays this type of perturbations. The IK master equation 
is of the Schr\"odinger--like form (as for the previous scalar wave 
equation case), where the potential to consider is now the following 
\cite{kodama-ishibashi-1, kodama-ishibashi-3}

\begin{equation} \label{tensor}
V_{\mathsf{T}} (r) = f(r) \left( \frac{\ell \left( \ell+d-3 
\right)}{r^{2}} + \frac{\left( d-2 \right) \left( d-4 \right) f(r)}{4 
r^{2}} + \frac{\left( d-2 \right) f'(r)}{2r} \right).
\end{equation}

\noindent
One immediately realizes that this is precisely the same potential 
as in the case of the massless, uncharged, scalar wave equation. It 
thus follows at once that quasinormal modes of tensor type perturbations of the 
gravitational metric field will coincide with the quasinormal modes one obtains 
from the scalar wave equation. In this situation there is also a 
special mode with $\ell = 0$, corresponding to a constant [tensorial] 
eigenfunction of the laplacian \cite{kodama-ishibashi-1, 
kodama-ishibashi-3}. One thus disregards this mode and takes 
$\ell = 1, 2, 3, \ldots$.

\medskip

\noindent
\underline{\textsf{Vector Type Perturbations:}} Both gravitational metric 
field and electromagnetic vector field display this type of perturbations, 
and one thus expects the master equations to be coupled. It was shown 
in \cite{kodama-ishibashi-3} that these equations can actually be 
decoupled so that the IK master equations for these fields are again of 
the Schr\"odinger--like form. In the case where $Q=0$ there is of 
course only a gravitational metric field to worry about. Let us see 
both cases in turn. In the \textsf{uncharged} $Q=0$ case there is 
only one IK master equation \cite{kodama-ishibashi-1}, with potential

\begin{eqnarray} \label{vector-uncharged}
V_{\mathsf{V}} (r) &=& f(r) \left( \frac{\ell \left( \ell+d-3 
\right)}{r^{2}} + \frac{\left( d-2 \right) \left( d-4 \right) f(r)}{4 
r^{2}} - \frac{\left( d-1 \right) \left( d-2 \right) \MM}{r^{d-1}} 
\right) = \nonumber \\
&&
= f(r) \left( \frac{\ell \left( \ell+d-3 \right)}{r^{2}} + \frac{\left( 
d-2 \right) \left( d-4 \right) f(r)}{4 r^{2}} - \frac{r f'''(r)}{2 
\left( d-3 \right)} \right).
\end{eqnarray}

\noindent
In the \textsf{charged} $Q \not = 0$ case, there are two decoupled IK 
master equations \cite{kodama-ishibashi-3}, with potentials

\begin{equation} \label{vector-charged}
V_{\mathsf{V}^{\pm}} (r) =  f(r) \left( \frac{\ell \left( \ell+d-3 
\right)}{r^{2}} + \frac{\left( d-2 \right) \left( d-4 \right) f(r)}{4 
r^{2}}  - \frac{\left( d-1 \right) \MM}{r^{d-1}} + \frac{\left( d-2 
\right)^{2} \QQ^{2}}{r^{2d-4}} \pm \frac{\Delta}{r^{d-1}} \right),
\end{equation}

\noindent
where the $\Phi^{+}$ equation represents the electromagnetic mode and 
the $\Phi^{-}$ equation represents the gravitational mode. Thus, 
when $\QQ = 0$, one has that $V_{\mathsf{V}^{-}} (r) = 
V_{\mathsf{V}} (r)$. In the expression above

$$
\Delta \equiv \sqrt{\left( d-1 \right)^{2} \left( d-3 \right)^{2} 
\MM^{2} + 2 \left( d-2 \right) \left( d-3 \right) \Big( \ell \left( 
\ell+d-3 \right) - \left( d-2 \right) \Big) \QQ^{2}}.
$$

\noindent
These potentials are clearly different from the previous ones 
characterizing tensor type perturbations of the gravitational metric 
field and characterizing the scalar wave equation. However, in some 
cases, they will have the same leading singularities at the origin as 
those other potentials. For vector type perturbations there are two 
special modes \cite{kodama-ishibashi-1, kodama-ishibashi-3}. One is $\ell 
= 0$, and corresponds to a constant [vectorial] eigenfunction of the 
laplacian. The other special mode has $\ell = 1$, whose solution 
corresponds to addition of a small rotation to the black hole background, 
and which leads to no gravitational dynamical degrees of freedom. One 
therefore disregards both these modes and takes $\ell = 2, 3, 4, \ldots$.

\medskip

\noindent
\underline{\textsf{Scalar Type Perturbations:}} Again, both 
gravitational metric field and electromagnetic vector field will 
display this type of perturbations, so that the master equations will 
be coupled. Also in this case it was shown that these equations can 
be decoupled \cite{kodama-ishibashi-3} and that the IK master 
equations for these fields are of Schr\"odinger--like form. In the 
$Q=0$ case one clearly only needs to be concerned about the 
gravitational metric field. Let us see both cases in turn. In the 
\textsf{uncharged} $Q=0$ case there is only one IK master equation 
\cite{kodama-ishibashi-1}, with potential

\begin{equation} \label{scalar-uncharged}
V_{\mathsf{S}} (r) = \frac{f(r) U(r)}{16 r^{2} H^{2}(r)},
\end{equation}

\noindent
where

$$
H(r) = \ell \left( \ell+d-3 \right) - \left( d-2 \right) + 
\frac{\left( d-1 \right) \left( d-2 \right) \MM}{r^{d-3}},
$$

\noindent
and 

\begin{eqnarray*}
U(r) &=& - \Bigg[ 4 d  \left( d-1 \right)^{2} \left( d-2 \right)^{3} 
\frac{\MM^{2}}{r^{2d-6}} - 24 \left( d-1 \right) \left( d-2 
\right)^{2} \left( d-4 \right) \Big\{ \ell \left( \ell+d-3 \right) - 
\left( d-2 \right) \Big\} \frac{\MM}{r^{d-3}} + \\
&&
+ 4 \left( d-4 \right) \left( d-6 \right) \Big\{ \ell \left( \ell+d-3 
\right) - \left( d-2 \right) \Big\}^{2} \Bigg] \LL r^{2} + 8 \left( d-1 
\right)^{2} \left( d-2 \right)^{4} \frac{\MM^{3}}{r^{3d-9}} + 4 \left( d-1 
\right) \left( d-2 \right) \cdot \\
&&
\cdot \Bigg[ 4 \left( 2d^{2}-11d+18 \right) \Big\{ \ell \left( \ell+d-3 
\right) - \left( d-2 \right) \Big\} + \left( d-1 \right) \left( d-2 \right) 
\left( d-4 \right) \left( d-6 \right) \Bigg] \frac{\MM^{2}}{r^{2d-6}} - 
24 \left( d-2 \right) \cdot \\
&&
\cdot \Bigg[ \left( d-6 \right) \Big\{ \ell \left( \ell+d-3 \right) - 
\left( d-2 \right) \Big\} + \left( d-1 \right) \left( d-2 \right) \left( 
d-4 \right) \Bigg] \Big\{ \ell \left( \ell+d-3 \right) - \left( d-2 \right) 
\Big\} \frac{\MM}{r^{d-3}} + \\
&&
+ 16 \Big\{ \ell \left( \ell+d-3 \right) - \left( d-2 \right) 
\Big\}^{3} + 4 d \left( d-2 \right) \Big\{ \ell \left( \ell+d-3 \right) - 
\left( d-2 \right) \Big\}^{2}.
\end{eqnarray*}

\noindent
In the \textsf{charged} $Q \not = 0$ case, there are two decoupled IK 
master equations \cite{kodama-ishibashi-3}, with potentials

\begin{equation} \label{scalar-charged}
V_{\mathsf{S}^{\pm}} (r) = \frac{f(r) U_{\pm} (r)}{64 r^{2} 
H_{\pm}^{2} (r)},
\end{equation}

\noindent
where

\begin{eqnarray*}
H_{+} (r) &=& 1 + \frac{\left( d-1 \right) \left( d-2 \right) \left( 
1 - \Omega \right)}{2 \Big\{ \ell \left( \ell+d-3 \right) - \left( d-2 
\right) \Big\}}\ \frac{\MM}{r^{d-3}}, \\
H_{-} (r) &=& \ell \left( \ell+d-3 \right) - \left( d-2 \right) + 
\frac{1}{2} \left( d-1 \right) \left( d-2 \right) \left( 1 + \Omega 
\right) \frac{\MM}{r^{d-3}},
\end{eqnarray*}

\noindent
with the definition, here and below, that

$$
\Omega = \sqrt{1 + \frac{4 \Big\{ \ell \left( \ell+d-3 \right) - 
\left( d-2 \right) \Big\}}{\left( d-1 \right)^{2}}\ 
\frac{\QQ^{2}}{\MM^{2}}},
$$

\noindent
and where

\begin{eqnarray*}
U_{+} (r) &=& \Bigg[ - 4 \frac{d \left( d-1 \right)^{2} \left( d-2 
\right)^{3} \left( 1 - \Omega \right)^{2}}{\Big\{ \ell \left( \ell+d-3 
\right) - \left( d-2 \right) \Big\}^{2}}\ \frac{\MM^{2}}{r^{2d-6}} 
+ 48 \frac{\left( d-1 \right) \left( d-2 \right)^{2} \left( d-4 
\right) \left( 1 - \Omega \right)}{\ell \left( \ell+d-3 \right) - \left( 
d-2 \right)}\ \frac{\MM}{r^{d-3}} + \\
&&
- 16 \left( d-4 \right) \left( d-6 \right) \Bigg] \LL r^{2} + \frac{\left( 
d-1 \right)^{4} \left( d-2 \right)^{3} \left( 3d-8 \right) \left( 1 - 
\Omega \right)^{3} \left( 1 + \Omega \right)}{\Big\{ \ell \left( \ell+d-3 
\right) - \left( d-2 \right) \Big\}^{3}}\ \frac{\MM^{4}}{r^{4d-12}} + \\
&&
+ \frac{8 \left( d-1 \right)^{2} \left( d-2 \right)^{2} \left( 1 - \Omega 
\right)^{2}}{\Big\{ \ell \left( \ell+d-3 \right) - \left( d-2 \right) 
\Big\}^{2}} \Bigg[ - \frac{1}{2} \left( d-1 \right) \left( 3d-8 \right) 
\left( 1 - \Omega \right) + 4 d^{2} - 15 d + 12 \Bigg] 
\frac{\MM^{3}}{r^{3d-9}} + \\
&&
+ \frac{8 \left( d-1 \right) \left( 1 - \Omega \right)}{\ell \left( 
\ell+d-3 \right) - \left( d-2 \right)} \Bigg[ \frac{\left( d-1 \right) 
\left( d-4 \right) \left( d-6 \right) \left( \ell \left( \ell+d-3 \right) 
- \left( d-2 \right) + \left( d-2 \right)^{2} \right) \left( 1 - \Omega 
\right)}{2 \Big\{ \ell \left( \ell+d-3 \right) - \left( d-2 \right) 
\Big\}} + \\
&&
+ 7 d^{3} - 49 d^{2} + 126 d - 120 \Bigg] \frac{\MM^{2}}{r^{2d-6}} + \\
&&
- \Bigg[ \frac{16 \left( d-1 \right) \left( 3 \left( d-2 \right)^{2} 
\left( d-4 \right) - 4 \Big\{ \ell \left( \ell+d-3 \right) - \left( d-2 
\right) \Big\} \right) \left( 1 - \Omega \right)}{\ell \left( \ell+d-3 
\right) - \left( d-2 \right)} + \\
&&
+ 32 \left( d-4 \right) \left( 3d-8 \right) \Bigg] \frac{\MM}{r^{d-3}} 
+ 64 \Big\{ \ell \left( \ell+d-3 \right) - \left( d-2 \right) \Big\} + 
16 d \left( d-2 \right), \\
U_{-} (r) &=& \Bigg[ - 4 d \left( d-1 \right)^{2} \left( d-2 \right)^{3} 
\left( 1 + \Omega 
\right)^{2} \frac{\MM^{2}}{r^{2d-6}} + 48 \left( d-1 \right) \left( d-2 
\right)^{2} \left( d-4 \right) \Big\{ \ell \left( \ell+d-3 \right) - 
\left( d-2 \right) \Big\} \cdot \\
&&
\cdot \left( 1 + \Omega \right) \frac{\MM}{r^{d-3}} - 16 \left( d-4 
\right)
\left( d-6 \right) \Big\{ \ell \left( \ell+d-3 \right) - \left( 
d-2 \right) \Big\}^{2} \Bigg] \LL r^{2} + \frac{\left( d-1 \right)^{4} 
\left( d-2 \right)^{3} \left( 3d-8 \right)}{\ell \left( \ell+d-3 \right) 
- \left( d-2 \right)} \cdot \\
&&
\left( 1 - \Omega \right) \left( 1 + \Omega \right)^{3} 
\frac{\MM^{4}}{r^{4d-12}} + 8 \left( d-1 \right)^{2} \left( d-2 \right)^{2} 
\left( 1 + \Omega \right)^{2} \Bigg[ \frac{1}{2} \left( d-1 \right) \left( 
3d-8 \right) \left( 1 - \Omega \right) + \\
&&
+ \left( d-2 \right)^{2} \Bigg] \frac{\MM^{3}}{r^{3d-9}} + 8 
\left( d-1 \right) \left( 1 + \Omega \right) \Bigg[ - \frac{1}{2} 
\left( d-1 \right) \left( d-4 \right) \left( d-6 \right) \Bigg( \ell 
\left( \ell+d-3 \right) - \left( d-2 \right) + \\
&&
+ \left( d-2 \right)^{2} \Bigg) \left( 1 - \Omega 
\right) + 4 \left( d-2 \right) \left( 2 d^{2} - 11 d + 18 \right) \Big\{ 
\ell \left( \ell+d-3 \right) - \left( d-2 \right) \Big\} + \\
&&
+ \left( d-1 \right) \left( d-2 
\right)^{2} \left( d-4 \right) \left( d-6 \right) \Bigg] 
\frac{\MM^{2}}{r^{2d-6}} - 32 \Big\{ \ell \left( \ell+d-3 \right) - 
\left( d-2 \right) \Big\} \cdot \\
&&
\cdot \Bigg[ - \frac{1}{2} \left( d-1 \right) \left( 3 \left( d-2 \right)^{2} 
\left( d-4 \right) - 4 \Big\{ \ell \left( \ell+d-3 \right) - \left( d-2 
\right) \Big\} \right) \left( 1 - \Omega \right) + \\
&&
+ 3 \left( d-2 \right) \left( d-6 \right) \Big\{ \ell \left( \ell+d-3 
\right) - \left( d-2 \right) \Big\} + 3 \left( d-1 \right) \left( d-2 
\right)^{2} \left( d-4 \right) \Bigg] \frac{\MM}{r^{d-3}} + \\
&&
+ 64 \Big\{ \ell \left( \ell+d-3 \right) - \left( d-2 \right) 
\Big\}^{3} + 16 d \left( d-2 \right) \Big\{ \ell \left( \ell+d-3 \right) - 
\left( d-2 \right) \Big\}^{2}.
\end{eqnarray*}

\noindent
In these expressions, the $\Phi^{+}$ equation represents the 
electromagnetic mode and the $\Phi^{-}$ equation represents the 
gravitational mode. Indeed, when $\QQ = 0$, one observes that

\begin{eqnarray*}
H_{-} (r) \Big|_{\QQ=0} &=& H (r), \\
U_{-} (r) \Big|_{\QQ=0} &=& 4 U (r),
\end{eqnarray*}

\noindent
so that $V_{\mathsf{S}^{-}} (r) = V_{\mathsf{S}} (r)$ at $\QQ = 0$ as 
expected. This is simple to check as when $\QQ = 0$ one has $\Omega = 1$. 
These scalar potentials are different from the previous ones characterizing 
tensor and vector type perturbations of the gravitational and 
electromagnetic fields. However, in some cases, they will have the same 
leading singularities at the origin as those other potentials. Also in 
here there are two special modes \cite{kodama-ishibashi-1, 
kodama-ishibashi-3}. The mode with $\ell = 0$ corresponds 
to a constant [scalar] eigenfunction of the laplacian and represents a 
small change in the background parameters, $\MM$, $\QQ$ and $\LL$. The 
other mode, with $\ell = 1$, leads to an ill--defined master equation 
involving gauge dependent quantities, and to no gravitational dynamical 
degrees of freedom. One again naturally disregards both these modes 
and takes $\ell = 2, 3, 4, \ldots$.


\section{The Tortoise and the Master Equation Potentials} \label{appendixC}


As explained elsewhere in this paper, the master equation for perturbations takes a Schr\"odinger--like form (\ref{schrodinger}) when written in terms of the tortoise coordinate. This coordinate explicitly depends upon the background spacetime in consideration and at first sight it would seem useful to obtain closed form expressions---relating the tortoise coordinate $x$ with the standard $r$ coordinate appearing in the spacetime metric---for each of the black holes we consider in this paper. This we shall do in the following. However, such analytic expressions are not needed for the calculation we carry out. As explained in the main text, the monodromy method we use requires only information around the region at infinity, the region at the origin, and the region at the event horizons (nevertheless we obtain such analytic expressions for completeness).

Another point which needs to be taken into account concerns the master equation potentials described in the previous appendix. As we have just said, the monodromy method only makes use of their values in three specific regions: infinity, the origin and the horizons. Their full expressions, as presented earlier, are not required except for an exact analysis of the quasinormal modes (rather than an asymptotic analysis as the one we perform). Thus, one needs to list the values of the master potentials in the different regions of interest, both in terms of the metric coordinate $r$ and the tortoise coordinate $x$. In this appendix we make an exhaustive study of the tortoise coordinate for the spacetimes considered in this paper, alongside with a list of the master potential values in the regions of interest for the monodromy calculation.

Starting with the master equation potentials, it is simple to obtain their value at the black hole horizon: they all vanish. Indeed, as can be seen from the previous appendix, for all types of perturbations, tensor, vector and scalar, and both for charged and uncharged black holes, all master potentials are given by the product of $f(r)$ by a function which is regular at the horizon $R_H$. Thus, since $f(R_H)$ vanishes, so do all master potentials, $V(R_{H})=0$. Let us next analyze all the cases we address in the main text:

\medskip

\noindent
\underline{\textsf{The Schwarzschild Solution:}} The horizons are defined via $r^{d-3}-2\MM=0$ with the roots

$$
R_{n} = \left| \left( 2\MM \right)^{\frac{1}{d-3}} \right| \exp \left( \frac{2\pi i}{d-3}\ n \right), \qquad n=0,1,\ldots,d-4,
$$

\noindent
so that one finds $(d-3)$ complex horizons (only one real positive root). The arguments of the horizons are those of the $(d-3)$ roots of unity. Given these roots one can proceed and factorize $f(r)$. Then, all one has to do in order to find the tortoise coordinate is an integration of a rational function. Choosing $x[r=0]=0$ it follows

$$
x[r] = \int \frac{dr}{f(r)} = r + \sum_{n=0}^{d-4} \frac{1}{2k_{n}}\, \log \left( 1 - \frac{r}{R_{n}} \right),
$$

\noindent
where $k_{n} = \frac{1}{2} f'(R_{n})$ is the surface gravity at the given horizon and the sum goes through all possible horizons. Two special regions are of interest for our calculation: near $r = 0$ one finds

$$
x[r] \sim - \frac{1}{2\MM} \int dr\ r^{d-3} = - \frac{r^{d-2}}{2 \left( d-2 \right) \MM},
$$

\noindent
while near $r = \infty$ one finds

$$
x[r] \sim \int dr\ 1 = r.
$$

\noindent
Recalling that quasinormal modes live in the region $r > R_{H}$, one can study the asymptotics of the tortoise and find

\begin{eqnarray*}
\lim_{r \searrow {R_{H}}} x[r] &=& - \infty, \\
\lim_{r \to + \infty} x[r] &=& + \infty.
\end{eqnarray*}

Next we turn to the master equation potentials, as listed in the previous appendix. Using the general formulae presented earlier it is simple to obtain the following asymptotics. Near $r = 0$ one finds

\begin{eqnarray*}
V_{\mathsf{T}} (r) &\sim& - \left( d-2 \right)^{2} 
\frac{\MM^{2}}{r^{2d-4}} = - \frac{1}{4 x^{2}} = \frac{0^2-1}{4 x^{2}}, \\
V_{\mathsf{V}} (r) &\sim& 3 \left( d-2 \right)^{2} 
\frac{\MM^{2}}{r^{2d-4}} = \frac{3}{4 x^{2}} = \frac{2^2-1}{4 x^{2}}, \\
V_{\mathsf{S}} (r) &\sim& - \left( d-2 \right)^{2} 
\frac{\MM^{2}}{r^{2d-4}} = - \frac{1}{4 x^{2}} = \frac{0^2-1}{4 x^{2}},
\end{eqnarray*}

\noindent
while near $r = \infty$ one finds

\begin{eqnarray*}
V_{\mathsf{T}} (r) &\sim& 0, \\
V_{\mathsf{V}} (r) &\sim& 0, \\
V_{\mathsf{S}} (r) &\sim& 0.
\end{eqnarray*}

\medskip

\noindent
\underline{\textsf{The RN Solution:}} The horizons are defined via $r^{2d-6} - 2\MM r^{d-3} + \QQ^2 = 0$ with the roots

$$
R^{\pm}_{n} = \left| \left( \MM \pm \sqrt{\MM^2 - \QQ^2} \right)^{\frac{1}{d-3}} \right| \exp \left( \frac{2\pi i}{d-3}\ n \right), \qquad n=0,1,\ldots,d-4,
$$

\noindent
so that one finds $(2d-6)$ complex horizons (only two real positive roots). On what concerns the angular direction in the complex $r$--plane, the horizons are located at the $(d-3)$ roots of unity. Given these roots one can proceed and factorize $f(r)$, and the tortoise again follows from integration of a rational function. Choosing $x[r=0]=0$ one obtains

$$
x[r] = \int \frac{dr}{f(r)} = r + \sum_{n=0}^{d-4} \frac{1}{2k^{+}_{n}}\, \log \left( 1 - \frac{r}{R^{+}_{n}} \right) + \sum_{n=0}^{d-4} \frac{1}{2k^{-}_{n}}\, \log \left( 1 - \frac{r}{R^{-}_{n}} \right),
$$

\noindent
where $k^\pm_{n} = \frac{1}{2} f'(R^\pm_{n})$ is the surface gravity at the given horizon and the sum goes through all possible horizons. Two special regions are of interest for our calculation: near $r = 0$ one finds

$$
x[r] \sim \frac{1}{\QQ^2} \int dr\ r^{2d-6} = \frac{r^{2d-5}}{\left( 2d-5 \right) \QQ^{2}},
$$

\noindent
while near $r = \infty$ one finds

$$
x[r] \sim \int dr\ 1 = r.
$$

\noindent
Recalling that quasinormal modes live in the region $r > R_{H}^{+}$, one can study the asymptotics of the tortoise and find

\begin{eqnarray*}
\lim_{r \searrow {R_{H}^{+}}} x[r] &=& - \infty, \\
\lim_{r \to + \infty} x[r] &=& + \infty.
\end{eqnarray*}

\noindent
Note that when addressing the extremal RN solution, the above formulae for the tortoise will change. First of all, the horizons will now be defined via $r^{d-3} - \MM = 0$ (at extremality $\MM=\QQ$), with the roots

$$
R_{n} = \left| \MM^{\frac{1}{d-3}} \right| \exp \left( \frac{2\pi i}{d-3}\ n \right), \qquad n=0,1,\ldots,d-4,
$$

\noindent
so that one finds $(d-3)$ complex horizons (only one real positive root). The arguments of the horizons are those of the $(d-3)$ roots of unity. Given these roots one factorizes $f(r)$ and the tortoise again follows from integration of a rational function. Choosing $x[r=0]=0$ one obtains

$$
x[r] = \int \frac{dr}{f(r)} = r - \sum_{n=0}^{d-4} \frac{2}{f''(R_{n})}\, \frac{1}{r-R_{n}} + \frac{d-2}{(d-3)^{2}} \sum_{n=0}^{d-4} R_{n}\, \log \left( 1 - \frac{r}{R_{n}} \right),
$$

\noindent
except in $d=4$ where one still has to subtract $\MM$ to the above expression in order to satisfy $x[r=0]=0$.

Next we turn to the master equation potentials, as listed in the previous appendix. Using the general formulae presented earlier it is simple to obtain the following asymptotics. Near $r = 0$ one finds

\begin{eqnarray*}
V_{\mathsf{T}} (r) &\sim& - \frac{1}{4} \left( d-2 
\right) \left( 3d-8 \right) \frac{\QQ^{4}}{r^{4d-10}} = - \frac{\left( d-2 
\right) \left( 3d-8 \right)}{4 \left( 2d-5 \right)^{2} x^{2}} = \frac{j_{\mathsf{T}}^{2}-1}{4 x^{2}}, \\
V_{\mathsf{V}^{\pm}} (r) &\sim& \frac{1}{4} \left( d-2 
\right) \left( 5d-12 \right) \frac{\QQ^{4}}{r^{4d-10}} = \frac{\left( 
d-2 \right) \left( 5d-12 \right)}{4 \left( 2d-5 \right)^{2} x^{2}} = \frac{j_{\mathsf{V}^{\pm}}^{2}-1}{4 x^{2}}, \\
V_{\mathsf{S}^{\pm}} (r) &\sim& - \frac{1}{4} \left( d-2 
\right) \left( 3d-8 \right) \frac{\QQ^{4}}{r^{4d-10}} = - \frac{\left( d-2 \right) \left( 3d-8 \right)}{4 \left( 2d-5 \right)^{2} x^{2}} = \frac{j_{\mathsf{S}^{\pm}}^{2}-1}{4 x^{2}}.
\end{eqnarray*}

\noindent
where $j_{\mathsf{T}} = j_{\mathsf{S}^{\pm}} = \frac{d-3}{2d-5}$ and $j_{\mathsf{V}^{\pm}} = \frac{3d-7}{2d-5} = 2 - j_{\mathsf{T}}$. Near $r = \infty$ one finds

\begin{eqnarray*}
V_{\mathsf{T}} (r) &\sim& 0, \\
V_{\mathsf{V^{\pm}}} (r) &\sim& 0, \\
V_{\mathsf{S^{\pm}}} (r) &\sim& 0.
\end{eqnarray*}

\medskip

\noindent
\underline{\textsf{The Schwarzschild dS Solution:}} The horizons are defined via $- \LL r^{d-1} + r^{d-3} - 2\MM = 0$, and there is no general analytic solutions for the roots. In even dimension there is an odd number of roots yielding $(d-1)$ complex horizons (only two real positive roots). If we denote by $R_H$ and $R_C$ the black hole and cosmological horizons, respectively, the roots are

$$
R_{n} =  R_H, R_C, \gamma_{1}, \overline{\gamma}_{1}, \ldots, \gamma_{\frac{d-4}{2}}, \overline{\gamma}_{\frac{d-4}{2}}, \widetilde{R},
$$

\noindent
where $\widetilde{R} = - \left( R_H + R_C + \sum_{i=1}^{\frac{d-4}{2}} \left( \gamma_{i} + \overline{\gamma}_{i} \right) \right)$ (as the roots must add up to zero). Determination of the $\gamma_{i}$ and $\overline{\gamma}_{i}$ must be numerical for each given case. In odd dimension there is an even number of roots again yielding $(d-1)$ complex horizons (only two real positive roots). With the same conventions as before, the roots are

$$
R_{n} =  R_H, R_C, -R_H, -R_C, \gamma_{1}, \overline{\gamma}_{1}, \ldots, \gamma_{\frac{d-5}{2}}, \overline{\gamma}_{\frac{d-5}{2}},
$$

\noindent
where again the roots add up to zero (in fact cancel in pairs). Determination of the $\gamma_{i}$ and $\overline{\gamma}_{i}$ must be numerical for each given case. Even without explicit analytic solutions for the roots one can proceed and factorize $f(r)$ so that again the tortoise follows from integration of a rational function. Choosing $x[r=0]=0$ one obtains

$$
x[r] = \int \frac{dr}{f(r)} = \sum_{n=1}^{d-1} \frac{1}{2k_{n}}\, \log \left( 1 - \frac{r}{R_{n}} \right),
$$

\noindent
where $k_{n} = \frac{1}{2} f'(R_{n})$ is the surface gravity at the given horizon and the sum goes through all possible horizons. Two special regions are of interest for our calculation: near $r = 0$ one finds

$$
x[r] \sim - \frac{1}{2\MM} \int dr\ r^{d-3} = - \frac{r^{d-2}}{2 \left( d-2 \right) \MM},
$$

\noindent
just like in the pure Schwarzschild case. Near $r = \infty$ one finds

$$
x[r] \sim - \frac{1}{\LL} \int \frac{dr}{r^2} = x_{0} + \frac{1}{\LL r}.
$$

\noindent
Observe that the constant $x_{0} \in {\mathbb{C}}$ is required so that we keep the choice of $x[r=0]=0$ fixed. As quasinormal modes live in the region $R_{H} < r < R_{C}$, one can study the asymptotics of the tortoise to find

\begin{eqnarray*}
\lim_{r \searrow {R_{H}}} x[r] &=& - \infty, \\
\lim_{r \nearrow R_{C}} x[r] &=& + \infty.
\end{eqnarray*}

Next we turn to the master equation potentials, as listed in the previous appendix. Using the general formulae presented earlier it is simple to obtain the following asymptotics. Near $r = 0$ one finds

\begin{eqnarray*}
V_{\mathsf{T}} (r) &\sim& - \left( d-2 \right)^{2} 
\frac{\MM^{2}}{r^{2d-4}} = - \frac{1}{4 x^{2}} = \frac{0^2-1}{4 x^{2}}, \\
V_{\mathsf{V}} (r) &\sim& 3 \left( d-2 \right)^{2} 
\frac{\MM^{2}}{r^{2d-4}} = \frac{3}{4 x^{2}} = \frac{2^2-1}{4 x^{2}}, \\
V_{\mathsf{S}} (r) &\sim& - \left( d-2 \right)^{2} 
\frac{\MM^{2}}{r^{2d-4}} = - \frac{1}{4 x^{2}} = \frac{0^2-1}{4 x^{2}},
\end{eqnarray*}

\noindent
just like in the pure Schwarzschild case. Near $r = \infty$ one finds

\begin{eqnarray*}
V_{\mathsf{T}} (r) &\sim& \frac{1}{4} d(d-2) \LL^2 r^2 = \frac{d(d-2)}{4 (x-x_0)^2} = \frac{(d-1)^2-1}{4 (x-x_0)^2}, \\
V_{\mathsf{V}} (r) &\sim& \frac{1}{4} (d-2)(d-4) \LL^2 r^2 = \frac{(d-2)(d-4)}{4 (x-x_0)^2} = \frac{(d-3)^2-1}{4 (x-x_0)^2}, \\
V_{\mathsf{S}} (r) &\sim& \frac{1}{4} (d-4)(d-6) \LL^2 r^2 = \frac{(d-4)(d-6)}{4 (x-x_0)^2} = \frac{(d-5)^2-1}{4 (x-x_0)^2}.
\end{eqnarray*}

\medskip

\noindent
\underline{\textsf{The RN dS Solution:}} The horizons are defined via $- \LL r^{2d-4} + r^{2d-6} - 2 \MM r^{d-3} + \QQ^{2} = 0$, and there is no general analytic solutions for the roots. In even dimension there is an even number of roots yielding $(2d-4)$ complex horizons (only three real positive roots). If we denote the black hole [outer] radius by $R_H^+$, the inner horizon by $R_H^{-}$ and the cosmological horizon by $R_{C}$, the roots are

$$
R_{n} =  R_H^+, R_H^{-}, R_{C}, \gamma_{1}, \overline{\gamma}_{1}, \ldots, \gamma_{d-4}, \overline{\gamma}_{d-4}, \widetilde{R},
$$

\noindent
where $\widetilde{R} = - \left( R_H^+ + R_H^{-} + R_{C} + \sum_{i=1}^{d-4} \left( \gamma_{i} + \overline{\gamma}_{i} \right) \right)$ (as the roots must add up to zero). Determination of the $\gamma_{i}$ and $\overline{\gamma}_{i}$ must be numerical for each given case. In odd dimension there is an even number of roots again yielding $(2d-4)$ complex horizons (only three real positive roots). With the same conventions as before, the roots are

$$
R_{n} =  R_H^+, -R_H^+, R_H^{-}, -R_H^{-}, R_{C}, -R_{C}, \gamma_{1}, \overline{\gamma}_{1}, \ldots, \gamma_{d-5}, \overline{\gamma}_{d-5},
$$

\noindent
where again the roots add up to zero (in fact cancel in pairs). Determination of the $\gamma_{i}$ and $\overline{\gamma}_{i}$ must be numerical for each given case. Even without explicit analytic solutions for the roots one can proceed and factorize $f(r)$ so that again the tortoise follows from integration of a rational function. Choosing $x[r=0]=0$ one obtains

$$
x[r] = \int \frac{dr}{f(r)} = \sum_{n=1}^{2d-4} \frac{1}{2k_{n}}\, \log \left( 1 - \frac{r}{R_{n}} \right),
$$

\noindent
where $k_{n} = \frac{1}{2} f'(R_{n})$ is the surface gravity at the given horizon and the sum goes through all possible horizons. Two special regions are of interest for our calculation: near $r = 0$ one finds

$$
x[r] \sim \frac{1}{\QQ^2} \int dr\ r^{2d-6} = \frac{r^{2d-5}}{\left( 2d-5 \right) \QQ^{2}},
$$

\noindent
just like in the pure RN case. Near $r = \infty$ one finds

$$
x[r] \sim - \frac{1}{\LL} \int \frac{dr}{r^2} = x_{0} + \frac{1}{\LL r}.
$$

\noindent
The constant $x_{0} \in {\mathbb{C}}$ is required so that we keep the choice of $x[r=0]=0$ fixed. As quasinormal modes live in the region $R_{H}^{+} < r < R_{C}$, one can study the asymptotics of the tortoise to find

\begin{eqnarray*}
\lim_{r \searrow {R_{H}^{+}}} x[r] &=& - \infty, \\
\lim_{r \nearrow R_{C}} x[r] &=& + \infty.
\end{eqnarray*}

Next we turn to the master equation potentials, as listed in the previous appendix. Using the general formulae presented earlier it is simple to obtain the following asymptotics. Near $r = 0$ one finds

\begin{eqnarray*}
V_{\mathsf{T}} (r) &\sim& - \frac{1}{4} \left( d-2 
\right) \left( 3d-8 \right) \frac{\QQ^{4}}{r^{4d-10}} = - \frac{\left( d-2 
\right) \left( 3d-8 \right)}{4 \left( 2d-5 \right)^{2} x^{2}} = \frac{j_{\mathsf{T}}^{2}-1}{4 x^{2}}, \\
V_{\mathsf{V}^{\pm}} (r) &\sim& \frac{1}{4} \left( d-2 
\right) \left( 5d-12 \right) \frac{\QQ^{4}}{r^{4d-10}} = \frac{\left( 
d-2 \right) \left( 5d-12 \right)}{4 \left( 2d-5 \right)^{2} x^{2}} = \frac{j_{\mathsf{V}^{\pm}}^{2}-1}{4 x^{2}}, \\
\lim_{r \to 0} V_{\mathsf{S}^{\pm}} (r) &\sim& - \frac{1}{4} \left( d-2 
\right) \left( 3d-8 \right) \frac{\QQ^{4}}{r^{4d-10}} = - \frac{\left( d-2 \right) \left( 3d-8 \right)}{4 \left( 2d-5 \right)^{2} x^{2}} = \frac{j_{\mathsf{S}^{\pm}}^{2}-1}{4 x^{2}}.
\end{eqnarray*}

\noindent
where $j_{\mathsf{T}} = j_{\mathsf{S}^{\pm}} = \frac{d-3}{2d-5}$ and $j_{\mathsf{V}^{\pm}} = \frac{3d-7}{2d-5} = 2 - j_{\mathsf{T}}$. This is just like in the pure RN case. Near $r = \infty$ one finds as before

\begin{eqnarray*}
V_{\mathsf{T}} (r) &\sim& \frac{1}{4} d(d-2) \LL^2 r^2 = \frac{d(d-2)}{4 (x-x_0)^2} = \frac{(d-1)^2-1}{4 (x-x_0)^2}, \\
V_{\mathsf{V}^{\pm}} (r) &\sim& \frac{1}{4} (d-2)(d-4) \LL^2 r^2 = \frac{(d-2)(d-4)}{4 (x-x_0)^2} = \frac{(d-3)^2-1}{4 (x-x_0)^2}, \\
V_{\mathsf{S}^{\pm}} (r) &\sim& \frac{1}{4} (d-4)(d-6) \LL^2 r^2 = \frac{(d-4)(d-6)}{4 (x-x_0)^2} = \frac{(d-5)^2-1}{4 (x-x_0)^2}.
\end{eqnarray*}

\medskip

\noindent
\underline{\textsf{The Schwarzschild AdS Solution:}} The horizons are defined via $|\LL| r^{d-1} + r^{d-3} - 2\MM = 0$, and there is no general analytic solutions for the roots. In even dimension there is an odd number of roots yielding $(d-1)$ complex horizons (only one real positive root). If we denote the black hole radius by $R_H$, the roots are

$$
R_{n} =  R_H, \gamma_{1}, \overline{\gamma}_{1}, \ldots, \gamma_{\frac{d-2}{2}}, \overline{\gamma}_{\frac{d-2}{2}},
$$

\noindent
and again must add up to zero. Determination of the $\gamma_{i}$ and $\overline{\gamma}_{i}$ must be numerically for each given case. In odd dimension there is an even number of roots again yielding $(d-1)$ complex horizons (only one real positive root). With the same conventions as before, the roots are

$$
R_{n} =  R_H, -R_H, \gamma_{1}, \overline{\gamma}_{1}, \ldots, \gamma_{\frac{d-3}{2}}, \overline{\gamma}_{\frac{d-3}{2}},
$$

\noindent
where again the roots add up to zero (in fact cancel in pairs). Determination of the $\gamma_{i}$ and $\overline{\gamma}_{i}$ must be numerical for each given case. One case of interest is that of large Schwarzschild black holes in AdS, where one basically drops the $1$ in the expression for $f(r)$. For these large black holes the horizons are defined via $|\LL| r^{d-1} - 2\MM = 0$, with the roots

$$
R_{n} = \left| \left( \frac{2\MM}{|\LL|} \right)^{\frac{1}{d-1}} \right| \exp \left( \frac{2\pi i}{d-1}\ n \right), \qquad n=0,1,\ldots,d-2,
$$

\noindent
so that one finds $(d-1)$ complex horizons (only one real positive root). The arguments of the horizons are those of the $(d-1)$ roots of unity. Even without an analytic solution for the roots, in the general case, one can proceed and factorize $f(r)$ so that the tortoise will follow from integration of a rational function, as usual. Choosing $x[r=0]=0$ we get

$$
x[r] = \int \frac{dr}{f(r)} = \sum_{n=1}^{d-1} \frac{1}{2k_{n}}\, \log \left( 1 - \frac{r}{R_{n}} \right),
$$

\noindent
where $k_{n} = \frac{1}{2} f'(R_{n})$ is the surface gravity at the given horizon and the sum goes through all possible horizons. Two special regions are of interest for our calculation: near $r = 0$ one finds

$$
x[r] \sim - \frac{1}{2\MM} \int dr\ r^{d-3} = - \frac{r^{d-2}}{2 \left( d-2 \right) \MM},
$$

\noindent
just like in the pure Schwarzschild case. Near $r = \infty$ one finds

$$
x[r] \sim \frac{1}{|\LL|} \int \frac{dr}{r^2} = x_{0} - \frac{1}{|\LL| r}.
$$

\noindent
The constant $x_{0} \in {\mathbb{C}}$ is required so that we keep the choice of $x[r=0]=0$ fixed. As quasinormal modes live in the region $r > R_{H}$, one can study the asymptotics of the tortoise to find\footnote{Here one has to use a different choice of branch cuts resulting in a \textit{real} value for $x_0$.}

\begin{eqnarray*}
\lim_{r \searrow {R_{H}}} x[r] &=& - \infty, \\
\lim_{r \to + \infty} x[r] &=& x_0.
\end{eqnarray*}

Next we turn to the master equation potentials, as listed in the previous appendix. Using the general formulae presented earlier it is simple to obtain the following asymptotics. Near $r = 0$ one finds

\begin{eqnarray*}
V_{\mathsf{T}} (r) &\sim& - \left( d-2 \right)^{2} 
\frac{\MM^{2}}{r^{2d-4}} = - \frac{1}{4 x^{2}} = \frac{0^2-1}{4 x^{2}}, \\
V_{\mathsf{V}} (r) &\sim& 3 \left( d-2 \right)^{2} 
\frac{\MM^{2}}{r^{2d-4}} = \frac{3}{4 x^{2}} = \frac{2^2-1}{4 x^{2}}, \\
V_{\mathsf{S}} (r) &\sim& - \left( d-2 \right)^{2} 
\frac{\MM^{2}}{r^{2d-4}} = - \frac{1}{4 x^{2}} = \frac{0^2-1}{4 x^{2}},
\end{eqnarray*}

\noindent
just like in the pure Schwarzschild case. Near $r = \infty$ one finds

\begin{eqnarray*}
V_{\mathsf{T}} (r) &\sim& \frac{1}{4} d(d-2) \LL^2 r^2 = \frac{d(d-2)}{4 (x-x_0)^2} = \frac{(d-1)^2-1}{4 (x-x_0)^2}, \\
V_{\mathsf{V}} (r) &\sim& \frac{1}{4} (d-2)(d-4) \LL^2 r^2 = \frac{(d-2)(d-4)}{4 (x-x_0)^2} = \frac{(d-3)^2-1}{4 (x-x_0)^2}, \\
V_{\mathsf{S}} (r) &\sim& \frac{1}{4} (d-4)(d-6) \LL^2 r^2 = \frac{(d-4)(d-6)}{4 (x-x_0)^2} = \frac{(d-5)^2-1}{4 (x-x_0)^2}.
\end{eqnarray*}

\medskip

\noindent
\underline{\textsf{The RN AdS Solution:}} The horizons are defined via $|\LL| r^{2d-4} + r^{2d-6} - 2 \MM r^{d-3} + \QQ^{2} = 0$, and there is no general analytic solutions for the roots. In even dimension there is an even number of roots yielding $(2d-4)$ complex horizons (only two real positive roots). If we denote the black hole (outer) radius by $R_H^+$ and denote by $R_H^{-}$ the inner horizon, the roots are

$$
R_{n} =  R_H^+, R_H^{-}, \gamma_{1}, \overline{\gamma}_{1}, \ldots, \gamma_{d-3}, \overline{\gamma}_{d-3},
$$

\noindent
where again the roots must add up to zero. Determination of the $\gamma_{i}$ and $\overline{\gamma}_{i}$ must be numerical for each given case. In odd dimension there is an even number of roots again yielding $(2d-4)$ complex horizons (only two real positive roots). With the same conventions as before, the roots are

$$
R_{n} =  R_H^+, -R_H^+, R_H^{-}, -R_H^{-}, \gamma_{1}, \overline{\gamma}_{1}, \ldots, \gamma_{d-4}, \overline{\gamma}_{d-4},
$$

\noindent
where again the roots add up to zero (and in fact cancel in pairs). Determination of the $\gamma_{i}$ and $\overline{\gamma}_{i}$ must be numerical for each given case. Even without explicit analytic solutions for the roots one can proceed and factorize $f(r)$ so that the tortoise follows from integration of a rational function, as usual. Choosing $x[r=0]=0$ one obtains

$$
x[r] = \int \frac{dr}{f(r)} = \sum_{n=1}^{2d-4} \frac{1}{2k_{n}}\, \log \left( 1 - \frac{r}{R_{n}} \right),
$$

\noindent
where $k_{n} = \frac{1}{2} f'(R_{n})$ is the surface gravity at the given horizon and the sum goes through all possible horizons. Two special regions are of interest for our calculation: near $r = 0$ one finds

$$
x[r] \sim \frac{1}{\QQ^2} \int dr\ r^{2d-6} = \frac{r^{2d-5}}{\left( 2d-5 \right) \QQ^{2}},
$$

\noindent
just like in the pure RN case. Near $r = \infty$ one finds

$$
x[r] \sim \frac{1}{|\LL|} \int \frac{dr}{r^2} = x_{0} - \frac{1}{|\LL| r}.
$$

\noindent
The constant $x_{0} \in {\mathbb{C}}$ keeps the choice of $x[r=0]=0$ fixed. As quasinormal modes live in the region $r > R_{H}^{+}$, one can study the asymptotics of the tortoise to find\footnote{Here one has to use a different choice of branch cuts resulting in a \textit{real} value for $x_0$.}

\begin{eqnarray*}
\lim_{r \searrow {R_{H}^{+}}} x[r] &=& - \infty, \\
\lim_{r \to +\infty} x[r] &=& x_0.
\end{eqnarray*}

Next we turn to the master equation potentials, as listed in the previous appendix. Using the general formulae presented earlier it is simple to obtain the following asymptotics. Near $r = 0$ one finds

\begin{eqnarray*}
V_{\mathsf{T}} (r) &\sim& - \frac{1}{4} \left( d-2 
\right) \left( 3d-8 \right) \frac{\QQ^{4}}{r^{4d-10}} = - \frac{\left( d-2 
\right) \left( 3d-8 \right)}{4 \left( 2d-5 \right)^{2} x^{2}} = \frac{j_{\mathsf{T}}^{2}-1}{4 x^{2}}, \\
V_{\mathsf{V}^{\pm}} (r) &\sim& \frac{1}{4} \left( d-2 
\right) \left( 5d-12 \right) \frac{\QQ^{4}}{r^{4d-10}} = \frac{\left( 
d-2 \right) \left( 5d-12 \right)}{4 \left( 2d-5 \right)^{2} x^{2}} = \frac{j_{\mathsf{V}^{\pm}}^{2}-1}{4 x^{2}}, \\
\lim_{r \to 0} V_{\mathsf{S}^{\pm}} (r) &\sim& - \frac{1}{4} \left( d-2 
\right) \left( 3d-8 \right) \frac{\QQ^{4}}{r^{4d-10}} = - \frac{\left( d-2 \right) \left( 3d-8 \right)}{4 \left( 2d-5 \right)^{2} x^{2}} = \frac{j_{\mathsf{S}^{\pm}}^{2}-1}{4 x^{2}}.
\end{eqnarray*}

\noindent
where $j_{\mathsf{T}} = j_{\mathsf{S}^{\pm}} = \frac{d-3}{2d-5}$ and $j_{\mathsf{V}^{\pm}} = \frac{3d-7}{2d-5} = 2 - j_{\mathsf{T}}$, just like in pure RN. Near $r = \infty$

\begin{eqnarray*}
V_{\mathsf{T}} (r) &\sim& \frac{1}{4} d(d-2) \LL^2 r^2 = \frac{d(d-2)}{4 (x-x_0)^2} = \frac{(d-1)^2-1}{4 (x-x_0)^2}, \\
V_{\mathsf{V}^{\pm}} (r) &\sim& \frac{1}{4} (d-2)(d-4) \LL^2 r^2 = \frac{(d-2)(d-4)}{4 (x-x_0)^2} = \frac{(d-3)^2-1}{4 (x-x_0)^2}, \\
V_{\mathsf{S}^{\pm}} (r) &\sim& \frac{1}{4} (d-4)(d-6) \LL^2 r^2 = \frac{(d-4)(d-6)}{4 (x-x_0)^2} = \frac{(d-5)^2-1}{4 (x-x_0)^2}.
\end{eqnarray*}


\vfill

\eject

\bibliographystyle{plain}

\begin{thebibliography}{10}

\bibitem{bch}
J.~M.~Bardeen, B.~Carter and S.~W.~Hawking,
\textit{The Four Laws of Black Hole Mechanics},
Commun.\ Math.\ Phys.\ \textbf{31} (1973) 161.

\bibitem{hawking}
S.~W.~Hawking,
\textit{Particle Creation by Black Holes},
Commun.\ Math.\ Phys.\ \textbf{43} (1975) 199.

\bibitem{regge-wheeler}
T.~Regge and J.~A.~Wheeler,
\textit{Stability of a Schwarzschild Singularity},
Phys.\ Rev.\ \textbf{108} (1957) 1063.
    
\bibitem{zerilli-1}
F.~J.~Zerilli,
\textit{Gravitational Field of a Particle Falling in a Schwarzschild Geometry Analyzed in Tensor Harmonics},
Phys.\ Rev.\ \textbf{D2} (1970) 2141.

\bibitem{zerilli-2}
F.~J.~Zerilli,
\textit{Perturbation Analysis for Gravitational and Electromagnetic Radiation in a Reissner--Nordstr\"om Geometry},
Phys.\ Rev.\ \textbf{D9} (1974) 860.

\bibitem{nollert}
H-P.~Nollert,
\textit{Quasinormal Modes: The Characteristic ``Sound'' of Black Holes and Neutron Stars},
Class.\ Quant.\ Grav.\ \textbf{16} (1999) R159.

\bibitem{kokkotas-schmidt}
K.~D.~Kokkotas and B.~G.~Schmidt,
\textit{Quasinormal Modes of Stars and Black Holes},
Living\ Rev.\ Rel.\ \textbf{2} (1999) 2,
\texttt{[gr-qc/9909058]}.

\bibitem{kodama-ishibashi-1}
A.~Ishibashi and H.~Kodama,
\textit{A Master Equation for Gravitational Perturbations of Maximally Symmetric Black Holes in Higher Dimensions},
Prog.\ Theor.\ Phys.\ \textbf{110} (2003) 701, 
\texttt{[hep-th/0305147]}.

\bibitem{kodama-ishibashi-2}
A.~Ishibashi and H.~Kodama,
\textit{Stability of Higher Dimensional Schwarzschild Black Holes},
Prog.\ Theor.\ Phys.\  \textbf{110} (2003) 901,
\texttt{[hep-th/0305185]}.

\bibitem{kodama-ishibashi-3}
A.~Ishibashi and H.~Kodama,
\textit{Master Equations for Perturbations of Generalized Static Black Holes with Charge in Higher Dimensions},
Prog.\ Theor.\ Phys.\  \textbf{111} (2004) 29,
\texttt{[hep-th/0308128]}.

\bibitem{hod}
S.~Hod,
\textit{Bohr's Correspondence Principle and the Area Spectrum of Quantum Black Holes},
Phys.\ Rev.\ Lett.\ \textbf{81} (1998) 4293,
\texttt{[gr-qc/9812002]}.

\bibitem{dreyer}
O.~Dreyer,
\textit{Quasinormal Modes, the Area Spectrum, and Black Hole Entropy},
Phys.\ Rev.\ Lett.\ \textbf{90} (2003) 081301,
\texttt{[gr-qc/0211076]}.

\bibitem{motl}
L.~Motl,
\textit{An Analytical Computation of Asymptotic Schwarzschild Quasinormal Frequencies},
Adv.\ Theor.\ Math.\ Phys.\ \textbf{6} (2003) 1135, 
\texttt{[gr-qc/0212096]}.

\bibitem{motl-neitzke}
L.~Motl and A.~Neitzke,
\textit{Asymptotic Black Hole Quasinormal Frequencies},
Adv.\ Theor.\ Math.\ Phys.\ \textbf{7} (2003) 307, 
\texttt{[hep-th/0301173]}.

\bibitem{konoplya-2}
R.~A.~Konoplya,
\textit{Gravitational Quasinormal Radiation of Higher--Dimensional Black Holes},
Phys.\ Rev.\ \textbf{D68} (2003) 124017,
\texttt{hep-th/0309030}.

\bibitem{cns}
V.~Cardoso, J.~Nat\'ario and R.~Schiappa
\textit{Asymptotic Quasinormal Frequencies for Black Holes in Non--Asymptotically Flat Spacetimes},
J.\ Math.\ Phys.\ \textbf{45} (2004) 4698,
\texttt{[hep-th/0403132]}.

\bibitem{neitzke}
A.~Neitzke,
\textit{Greybody Factors at Large Imaginary Frequencies},
\texttt{[hep-th/0304080]}.

\bibitem{krasnov-solodukhin}
K.~Krasnov and S.~N.~Solodukhin,
\textit{Effective Stringy Description of Schwarzschild Black Holes},
\texttt{[hep-th/0403046]}.

\bibitem{cardoso-lemos-extra}
V.~Cardoso and J.~P.~S.~Lemos,
\textit{Scalar, Electromagnetic and Weyl Perturbations of BTZ Black Holes: Quasinormal Modes},
Phys.\ Rev.\ \textbf{D63} (2001) 124015, 
\texttt{[gr-qc/0101052]}.

\bibitem{birmingham-extra}
D.~Birmingham,
\textit{Choptuik Scaling and Quasinormal Modes in the AdS/CFT Correspondence},
Phys.\ Rev.\ \textbf{D64} (2001) 064024,
\texttt{[hep-th/0101194]}.

\bibitem{chandra}
S.~Chandrasekhar,
\textit{The Mathematical Theory of Black Holes},
Oxford University Press (1998).

\bibitem{starinets}
A.~O.~Starinets,
\textit{Quasinormal Modes of Near Extremal Black Branes},
Phys.\ Rev.\ \textbf{D66} (2002) 124013,
\texttt{[hep-th/0207133]}.

\bibitem{nunez-starinets}
A.~N\'u\~nez and A.~O.~Starinets,
\textit{AdS/CFT Correspondence, Quasinormal Modes, and Thermal Correlators in ${\mathcal{N}}=4$ SYM},
Phys.\ Rev.\ \textbf{D67} (2003) 124013,
\texttt{[hep-th/0302026]}.

\bibitem{fhks}
L.~Fidkowski, V.~Hubeny, M.~Kleban and S.~Shenker,
\textit{The Black Hole Singularity in AdS/CFT},
JHEP\ \textbf{0402} (2004) 014,
\texttt{[hep-th/0306170]}.

\bibitem{musiri-siopsis}
S.~Musiri and G.~Siopsis,
\textit{Asymptotic Form of Quasinormal Modes of Large AdS Black Holes},
Phys.\ Lett.\ \textbf{B576} (2003) 309,
\texttt{[hep-th/0308196]}.

\bibitem{burgess-lutken}
C.~P.~Burgess and C.~A.~L\"utken,
\textit{Propagators and Effective Potentials in Anti--de Sitter Space},
Phys.\ Lett.\ \textbf{B153} (1985) 137.

\bibitem{ckl}
V.~Cardoso, R.~Konoplya and J.~P.~S.~Lemos,
\textit{Quasinormal Frequencies of Schwarzschild Black Holes in Anti--de Sitter Spacetimes: A Complete Study on the Overtone Asymptotic Behavior},
Phys.\ Rev.\ \textbf{D68} (2003) 044024, 
\texttt{[gr-qc/0305037]}.

\bibitem{cdl}
V.~Cardoso, O.~J.~C.~Dias and J.~P.~S.~Lemos,
\textit{Gravitational Radiation in $d$--Dimensional Spacetimes},
Phys.\ Rev.\ \textbf{D67} (2003) 064026, 
\texttt{[hep-th/0212168]}.

\bibitem{mmv-1}
A.~J.~M.~Medved, D.~Martin and M.~Visser,
\textit{Dirty Black Holes: Quasinormal Modes},
Class.\ Quant.\ Grav.\ \textbf{21} (2004) 1393,
\texttt{[gr-qc/0310009]}.

\bibitem{padmanabhan}
T.~Padmanabhan,
\textit{Quasinormal Modes: A Simple Derivation of the Level Spacing of the Frequencies},
Class.\ Quant.\ Grav.\ \textbf{21} (2004) L1,
\texttt{[gr-qc/0310027]}.

\bibitem{mmv-2}
A.~J.~M.~Medved, D.~Martin and M.~Visser,
\textit{Dirty Black Holes: Quasinormal Modes for ``Squeezed'' Horizons},
Class.\ Quant.\ Grav.\ \textbf{21} (2004) 2393,
\texttt{[gr-qc/0310097]}.

\bibitem{choudhury-padmanabhan}
T.~R.~Choudhury and T.~Padmanabhan,
\textit{Quasinormal Modes in Schwarzschild--de Sitter Spacetime: A Simple Derivation of the Level Spacing of the Frequencies},
Phys.\ Rev.\ \textbf{D69} (2004) 064033,
\texttt{[gr-qc/0311064]}.

\bibitem{choudhury-padmanabhan-2}
T.~R.~Choudhury and T.~Padmanabhan,
\textit{Concept of Temperature in Multi--Horizon Spacetimes: Analysis of Schwarzschild--de Sitter Metric},
\texttt{[gr-qc/0404091]}.

\bibitem{kunstatter}
G.~Kunstatter,
\textit{$d$--Dimensional Black Hole Entropy Spectrum from Quasinormal Modes},
Phys.\ Rev.\ Lett.\ \textbf{90} (2003) 161301,
\texttt{[gr-qc/0212014]}.

\bibitem{birmingham}
D.~Birmingham,
\textit{Asymptotic Quasinormal Frequencies of $d$--Dimensional Schwarzschild Black Holes},
Phys.\ Lett.\ \textbf{B569} (2003) 199, 
\texttt{[hep-th/0306004]}.

\bibitem{cly}
V.~Cardoso, J.~P.~S.~Lemos and S.~Yoshida,
\textit{Quasinormal Modes of Schwarzschild Black Holes in Four and Higher Dimensions},
Phys.\ Rev.\ \textbf{D69} (2004) 044004,
\texttt{[gr-qc/0309112]}.

\bibitem{konoplya-1}
R.~A.~Konoplya,
\textit{Quasinormal Behavior of the $d$--Dimensional Schwarzshild Black Hole and Higher Order WKB Approach},
Phys.\ Rev.\ \textbf{D68} (2003) 024018,
\texttt{gr-qc/0303052}.

\bibitem{andersson-onozawa}
N.~Andersson and H.~Onozawa,
\textit{Quasinormal Modes of Nearly Extreme Reissner--Nordstr\"om Black Holes},
Phys.\ Rev.\ \textbf{D54} (1996) 7470,
\texttt{[gr-qc/9607054]}.

\bibitem{berti-kokkotas-2}
E.~Berti and K.~D.~Kokkotas,
\textit{Asymptotic Quasinormal Modes of Reissner--Nordstr\"om and Kerr Black Holes},
Phys.\ Rev.\ \textbf{D68} (2003) 044027, 
\texttt{[hep-th/0303029]}.

\bibitem{andersson-howls}
N.~Andersson and C.~J.~Howls,
\textit{The Asymptotic Quasinormal Mode Spectrum of Non--Rotating Black Holes},
Class.\ Quant.\ Grav.\ \textbf{21} (2004) 1623,
\texttt{[gr-qc/0307020]}.

\bibitem{omoi}
H.~Onozawa, T.~Mishima, T.~Okamura and H.~Ishihara,
\textit{Quasinormal Modes of Maximally Charged Black Holes},
Phys.\ Rev.\ \textbf{D53} (1996) 7033,
\texttt{[gr-qc/9603021]}.

\bibitem{oomi}
H.~Onozawa, T.~Okamura, T.~Mishima and H.~Ishihara,
\textit{Perturbing Supersymmetric Black Holes},
Phys.\ Rev.\ \textbf{D55} (1997) 4529,
\texttt{[gr-qc/9606086]}.

\bibitem{das-shankaranarayanan}
S.~Das and S.~Shankaranarayanan
\textit{High Frequency Quasinormal Modes for Black Holes with Generic Singularities},
\texttt{[hep-th/0410209]}.

\bibitem{berti}
E.~Berti,
\textit{Black Hole Quasinormal Modes: Hints of Quantum Gravity?},
\texttt{[gr-qc/0411025]}.

\bibitem{gibbons-hawking}
G.~W.~Gibbons and S.~W.~Hawking,
\textit{Cosmological Event Horizons, Thermodynamics and Particle 
Creation},
Phys.\ Rev.\ \textbf{D15} (1977) 2738. 

\bibitem{ssv}
M.~Spradlin, A.~Strominger and A.~Volovich,
\textit{Les Houches Lectures in de Sitter Space},
Les Houches 2001 ``Gravity, Gauge Theories and Strings'' (2001) 423, 
\texttt{[hep-th/0110007]}.

\bibitem{bclp}
P.~Brady, C.~Chambers, W.~Laarakkers and E.~Poisson,
\textit{Radiative Falloff in Schwarzschild--de Sitter Spacetime},
Phys.\ Rev.\ \textbf{D60} (1999) 064003,
\texttt{[gr-qc/9902010]}.

\bibitem{cardoso-lemos}
V.~Cardoso and J.~P.~S.~Lemos,
\textit{Quasinormal Modes of the Near Extreme Schwarzschild--de Sitter 
Black Hole},
Phys.\ Rev.\ \textbf{D67} (2003) 084020,
\texttt{[gr-qc/0301078]}.

\bibitem{molina}
C.~Molina,
\textit{Quasinormal Modes of $d$--Dimensional Spherical Black Holes with Near Extreme Cosmological Constant},
Phys.\ Rev.\ \textbf{D68} (2003) 064007, 
\texttt{[gr-qc/0304053]}.

\bibitem{suneeta}
V.~Suneeta,
\textit{Quasinormal Modes for the SdS Black Hole: An Analytic 
Approximation Scheme},
Phys.\ Rev.\ \textbf{D68} (2003) 024020, 
\texttt{[gr-qc/0303114]}.

\bibitem{castellobranco-abdalla}
K.~H.~C.~Castello--Branco and E.~Abdalla,
\textit{Analytic Determination of the Asymptotic Quasinormal Mode Spectrum of Schwarzschild--de Sitter Black Holes},
\texttt{[gr-qc/0309090]}.

\bibitem{medved-martin}
A.~J.~M.~Medved and D.~Martin,
\textit{A Note on Quasinormal Modes: A Tale of Two Treatments},
\texttt{[gr-qc/0311086]}.

\bibitem{yoshida-futamase}
S.~Yoshida and T.~Futamase,
\textit{Numerical Analysis of Quasinormal Modes in Nearly Extremal Schwarzschild de Sitter Spacetimes},
Phys.\ Rev.\ \textbf{D69} (2004) 064025,
\texttt{[gr-qc/0308077]}.

\bibitem{konoplya-zhidenko}
R.~A.~Konoplya and A.~Zhidenko,
\textit{High Overtones of Schwarzschild de Sitter Quasinormal Spectrum},
JHEP\ \textbf{0406} (2004) 037,
\texttt{[hep-th/0402080]}.

\bibitem{jing}
J.~Jing,
\textit{Dirac Quasinormal Modes of the Reissner--Nordstr\"om de Sitter Black Hole},
Phys.\ Rev.\ \textbf{D69} (2004) 084009, 
\texttt{[gr-qc/0312079]}.

\bibitem{ais}
S.~J.~Avis, C.~J.~Isham and D.~Storey,
\textit{Quantum Field Theory in Anti--de Sitter Space},
Phys.\ Rev.\ \textbf{D18} (1978) 3565.

\bibitem{chan-mann}
S.~F.~J.~Chan and R.~B.~Mann,
\textit{Scalar Wave Falloff in Asymptotically Anti--de Sitter Backgrounds},
Phys.\ Rev.\ \textbf{D55} (1997) 7546,
\texttt{[gr-qc/9612026]}.

\bibitem{horowitz-hubeny}
G.~T.~Horowitz and V.~E.~Hubeny,
\textit{Quasinormal Modes of $AdS$ Black Holes and the Approach to 
Thermal Equilibrium},
Phys.\ Rev.\ \textbf{D62} (2000) 024027,
\texttt{[hep-th/9909056]}.

\bibitem{cardoso-lemos-2}
V.~Cardoso and J.~P.~S.~Lemos,
\textit{Quasinormal Modes of Schwarzschild Anti--de Sitter Black Holes: Electromagnetic and Gravitational Perturbations},
Phys.\ Rev.\ \textbf{D64} (2001) 084017,
\texttt{[gr-qc/0105103]}.

\bibitem{giammatteo-jing}
M.~Giammatteo and J.~Jing,
\textit{Dirac Quasinormal Frequencies in Schwarzschild AdS Spacetime},
\texttt{[gr-qc/0403030]}.

\bibitem{siopsis}
G.~Siopsis,
\textit{Large Mass Expansion of Quasinormal Modes in AdS$_{5}$},
Phys.\ Lett.\ \textbf{B590} (2004) 105,
\texttt{[hep-th/0402083]}.

\bibitem{siopsis-2}
G.~Siopsis,
\textit{On Quasinormal Modes and the AdS$_{5}$/CFT$_{4}$ Correspondence},
\texttt{[hep-th/0407157]}.

\bibitem{berti-kokkotas-1}
E.~Berti and K.~D.~Kokkotas,
\textit{Quasinormal Modes of Reissner--Nordstr\"om Anti--de Sitter 
Black Holes: Scalar, Electromagnetic and Gravitational Perturbations},
Phys.\ Rev.\ \textbf{D67} (2003) 064020, 
\texttt{[gr-qc/0301052]}.

\bibitem{konoplya-3}
R.~A.~Konoplya,
\textit{On Quasinormal Modes of Small Schwarzschild Anti--de Sitter Black Hole},
Phys.\ Rev.\ \textbf{D66} (2002) 044009,
\texttt{hep-th/0205142}.

\bibitem{gubser-mitra-1}
S.~S.~Gubser and I.~Mitra,
\textit{Instability of Charged Black Holes in Anti--de Sitter Space},
\texttt{[hep-th/0009126]}.

\bibitem{gubser-mitra-2}
S.~S.~Gubser and I.~Mitra,
\textit{The Evolution of Unstable Black Holes in Anti--de Sitter Space},
JHEP\ \textbf{0108} (2001) 018, 
\texttt{[hep-th/0011127]}.

\bibitem{hubeny-rangamani}
V.~E.~Hubeny and M.~Rangamani,
\textit{Unstable Horizons},
JHEP\ \textbf{0205} (2002) 027, 
\texttt{[hep-th/0202189]}.

\bibitem{wlm}
B.~Wang, C.-Y.~Lin and C.~Molina,
\textit{Quasinormal Behavior of Massless Scalar Field Perturbation in Reissner--Nordstr\"om Anti--de Sitter Spacetimes},
Phys.\ Rev.\ \textbf{D70} (2004) 064025,
\texttt{[hep-th/0407024]}.

\bibitem{mezincescu-townsend}
L.~Mezincescu and P.~K.~Townsend,
\textit{Stability at a Local Maximum in Higher Dimensional Anti--de Sitter Space and Applications to Supergravity},
Annals\ Phys.\ \textbf{160} (1985) 406.

\bibitem{acl}
E.~Abdalla, K.~H.~C.~Castello--Branco and A.~Lima--Santos,
\textit{Support of dS/CFT Correspondence from Spacetime Perturbations},
Phys.\ Rev.\ \textbf{D66} (2002) 104018, 
\texttt{[hep-th/0208065]}.

\bibitem{myung-kim}
Y.~S.~Myung and N.~J.~Kim,
\textit{Difference between AdS and dS Spaces: Wave Equation Approach},
Class.\ Quant.\ Grav.\ \textbf{21} (2004) 63,
\texttt{[hep-th/0304231]}.

\bibitem{dws}
D.-P.~Du, B.~Wang and R.-K.~Su,
\textit{Quasinormal Modes in Pure de Sitter Spacetimes},
\texttt{[hep-th/0404047]}.

\bibitem{barbero}
J.~F.~Barbero,
\textit{Real Ashtekar Variables for Lorentzian Signature Spacetimes},
Phys.\ Rev.\ \textbf{D51} (1995) 5507,
\texttt{[gr-qc/9410014]}.

\bibitem{immirzi}
G.~Immirzi,
\textit{Real and Complex Connections for Canonical Gravity},
Class.\ Quant.\ Grav.\ \textbf{14} (1997) L177,
\texttt{[gr-qc/9612030]}.

\bibitem{immirzi-2}
G.~Immirzi,
\textit{Quantum Gravity and Regge Calculus},
Nucl.\ Phys.\ Proc.\ Suppl.\ \textbf{57} (1997) 65,
\texttt{[gr-qc/9701052]}.

\bibitem{bekenstein-mukhanov}
J.~D.~Bekenstein and V.~F.~Mukhanov,
\textit{Spectroscopy of the Quantum Black Hole},
Phys.\ Lett.\ \textbf{B360} (1995) 7,
\texttt{[gr-qc/9505012]}.

\bibitem{bekenstein}
J.~D.~Bekenstein,
\textit{Quantum Black Holes as Atoms},
Proceedings of the VIII Marcel Grossmann Meeting on General Relativity, 
World Scientific (1999) 9,
\texttt{[gr-qc/9710076]}.

\bibitem{setare-1}
M.~R.~Setare,
\textit{Area Spectrum of Extremal Reissner--Nordstr\"om Black Holes from Quasinormal Modes},
Phys.\ Rev.\ \textbf{D69} (2004) 044016,
\texttt{hep-th/0312061}.

\bibitem{setare-2}
M.~R.~Setare,
\textit{Near Extremal Schwarzschild--de Sitter Black Hole Area Spectrum from Quasinormal Modes},
\texttt{hep-th/0401063}.

\bibitem{abck}
A.~Ashtekar, J.~Baez, A.~Corichi and K.~Krasnov,
\textit{Quantum Geometry and Black Hole Entropy},
Phys.\ Rev.\ Lett.\ \textbf{80} (1998) 904,
\texttt{[gr-qc/9710007]}.

\bibitem{abk}
A.~Ashtekar, J.~Baez, and K.~Krasnov,
\textit{Quantum Geometry of Isolated Horizons and Black Hole Entropy},
Adv.\ Theor.\ Math.\ Phys.\ \textbf{4} (2000) 1,
\texttt{[gr-qc/0005126]}.

\bibitem{krasnov}
K.~Krasnov,
\textit{The Area Spectrum in Quantum Gravity},
Class.\ Quant.\ Grav.\ \textbf{15} (1998) L47,
\texttt{[gr-qc/9803074]}.

\bibitem{ling-smolin-1}
Y.~Ling and L.~Smolin,
\textit{Supersymmetric Spin Networks and Quantum Supergravity},
Phys.\ Rev.\ \textbf{D61} (2000) 044008,
\texttt{[hep-th/9904016]}.

\bibitem{ling-zhang}
Y.~Ling and H.~Zhang,
\textit{Quasinormal Modes Prefer Supersymmetry?},
Phys.\ Rev.\ \textbf{D68} (2003) 101501,
\texttt{[gr-qc/0309018]}.

\bibitem{ling-smolin-2}
Y.~Ling and L.~Smolin,
\textit{Holographic Formulation of Quantum Supergravity},
Phys.\ Rev.\ \textbf{D63} (2001) 064010,
\texttt{[hep-th/0009018]}.

\bibitem{domagala-lewandowski}
M.~Domagala and J.~Lewandowski
\textit{Black Hole Entropy from Quantum Geometry},
Class.\ Quant.\ Grav.\ \textbf{21} (2004) 5233,
\texttt{[gr-qc/0407051]}.

\bibitem{meissner}
K.~A.~Meissner,
\textit{Black Hole Entropy in Loop Quantum Gravity},
Class.\ Quant.\ Grav.\ \textbf{21} (2004) 5245,
\texttt{[gr-qc/0407052]}.

\bibitem{smolin}
L.~Smolin,
\textit{An Invitation to Loop Quantum Gravity},
\texttt{[hep-th/0408048]}.

\bibitem{alexandrov}
S.~Alexandrov,
\textit{On the Counting of Black Hole States in Loop Quantum Gravity},
\texttt{[gr-qc/0408033]}.

\bibitem{dms}
O.~Dreyer, F.~Markopoulou and L.~Smolin,
\textit{Symmetry and Entropy of Black Hole Horizons},
\texttt{[hep-th/0409056]}.

\bibitem{birmingham-carlip}
D.~Birmingham and S.~Carlip,
\textit{Non--Quasinormal Modes and Black Hole Physics},
Phys.\ Rev.\ Lett.\ \textbf{92} (2004) 111302,
\texttt{[hep-th/0311090]}.

\bibitem{strominger}
A.~Strominger,
\textit{Black Hole Entropy from Near Horizon Microstates},
JHEP\ \textbf{9802} (1998) 009,
\texttt{[hep-th/9712251]}.

\bibitem{carlip-1}
S.~Carlip,
\textit{Black Hole Entropy from Conformal Field Theory in Any Dimension},
Phys.\ Rev.\ Lett.\ \textbf{82} (1999) 2828,
\texttt{[hep-th/9812013]}.

\bibitem{carlip-2}
S.~Carlip,
\textit{Entropy from Conformal Field Theory at Killing Horizons},
Class.\ Quant.\ Grav.\ \textbf{16} (1999) 3327,
\texttt{[gr-qc/9906126]}.

\bibitem{carlip-3}
S.~Carlip,
\textit{Black Hole Entropy from Horizon Conformal Field Theory},
Nucl.\ Phys.\ Proc.\ Suppl.\ \textbf{88} (2000) 10,
\texttt{[gr-qc/9912118]}.

\bibitem{cardy}
J.~Cardy,
\textit{Operator Content of Two Dimensional Conformallly Invariant Theories},
Nucl.\ Phys.\ \textbf{B270} (1986) 186.

\bibitem{brown-henneaux}
J.~Brown and M.~Henneaux,
\textit{Central Charges in the Canonical Realization of Asymptotic Symmetries: An Example from Three Dimensional Gravity},
Commun.\ Math.\ Phys.\ \textbf{104} (1986) 207.

\bibitem{btz}
M.~Banados, C.~Teitelboim and J.~Zanelli,
\textit{The Black Hole in Three Dimensional Spacetime},
Phys.\ Rev.\ Lett.\ \textbf{69} (1992) 1849,
\texttt{[hep-th/9204099]}.

\bibitem{gupta-sen}
K.~S.~Gupta and S.~Sen,
\textit{Geometric Finiteness and Non--Quasinormal Modes of the BTZ Black Hole},
\texttt{[hep-th/0504175]}.

\bibitem{tamaki-nomura}
T.~Tamaki and H.~Nomura,
\textit{The Universal Area Spectrum in Single--Horizon Black Holes},
Phys.\ Rev.\ \textbf{D70} (2004) 044041,
\texttt{[hep-th/0405191]}.

\bibitem{kiefer}
C.~Kiefer,
\textit{Hawking Temperature from Quasinormal Modes},
Class.\ Quant.\ Grav.\ \textbf{21} (2004) L123,
\texttt{[gr-qc/0406097]}.

\bibitem{fernando}
S.~Fernando,
\textit{Gravitational Perturbation and Quasinormal Modes of Charged Black Holes in Einstein--Born--Infeld Gravity},
\texttt{[hep-th/0407062]}.

\bibitem{kkm}
J.~Kettner, G.~Kunstatter and A.~J.~M.~Medved,
\textit{Quasinormal Modes for Single Horizon Black Holes in Generic 2d Dilaton Gravity},
\texttt{[gr-qc/0408042]}.

\bibitem{chen-jing}
S.~Cheng and J.~Jing,
\textit{Asymptotic Quasinormal Modes of a Coupled Scalar Field in the Garfinkle--Horowitz--Strominger Dilaton Spacetime},
\texttt{[gr-qc/0409013]}.

\bibitem{cbkz}
K.~H.~C.~Castello--Branco, R.~A.~Konoplya and A.~Zhidenko,
\textit{High Overtones of Dirac Perturbations of a Schwarzschild Black Hole},
\texttt{hep-th/0411055}.

\bibitem{konoplya-zhidenko-2}
R.~A.~Konoplya and A.~V.~Zhidenko,
\textit{Decay of Massive Scalar Field in a Schwarzschild Background},
\texttt{gr-qc/0411059}.

\bibitem{extra-ref-1}
J.~Jing and Q.~Pan,
\textit{Dirac Quasinormal Frequencies of Reissner--Nordstr\"om Black Hole in Anti--de Sitter Spacetime},
\texttt{gr-qc/0502011}.

\bibitem{extra-ref-2}
J.~Jing,
\textit{Dirac Quasinormal Modes of Schwarzschild Black Hole},
\texttt{gr-qc/0502023}.

\bibitem{extra-ref-3}
H.~Nomura and T.~Tamaki,
\textit{Continuous Area Spectrum in Regular Black Hole},
\texttt{hep-th/0504059}.

\bibitem{myers-perry}
R.~C.~Myers and M.~J.~Perry,
\textit{Black Holes in Higher Dimensional Spacetimes},
Annals\ Phys.\ \textbf{172} (1986) 304. 

\end{thebibliography}

\end{document}